\documentclass[12pt]{article}
\ifdefined\pdfminorversion\pdfminorversion=7\fi

\usepackage[margin=1in]{geometry}
\usepackage{amsmath,amssymb,amsfonts,mathtools}
\usepackage[scr=dutchcal]{mathalpha}

\usepackage{hyperref}
\hypersetup{hidelinks}
\usepackage{graphicx}
\usepackage{placeins}
\usepackage{booktabs}
\usepackage{caption}
\usepackage{cite}
\usepackage{xcolor}

\numberwithin{equation}{section} 

\newcommand{\dd}{\mathrm d}
\newcommand{\ls}{\ell_s}
\newcommand{\muIFS}{\mu}
\newcommand{\GammaSplit}{\Gamma_{\rm split}}

\newcommand{\be}{\begin{equation}}
\newcommand{\ee}{\end{equation}}

\newcommand{\maybeincludegraphics}[2][]{%
\IfFileExists{#2}{%
\includegraphics[#1]{#2}%
}{%
\fbox{\begin{minipage}[c][0.23\textheight][c]{0.84\textwidth}
\centering
\textbf{Missing figure file:} \texttt{\detokenize{#2}}\\[0.3cm]
Figure file not included in this source tree.
\end{minipage}}%
}%
}
\begin{document}

\begin{center}
{\Large\bfseries Instant Folded Strings: Fractal Decay \\ \vspace{2mm} and ANEC Saturation}

\vspace{6mm}

\textit{Nissan Itzhaki ${}^{(1)}$ and Uri Peleg ${}^{(2)}$ }
\break \break\break
(1)\  School of Physics and Astronomy, Tel Aviv University, Ramat Aviv, 69978, Israel\\ \vspace{2mm}
(2)  Department of Physics, Princeton University, Princeton, NJ 08544, USA

\end{center}

\vspace{6mm}

\begin{abstract}

Instant Folded Strings (IFSs) present a puzzle. On the one hand, they appear to be an intrinsic part of string theory: they are fundamental closed strings nucleated classically in backgrounds in which the string coupling grows toward the future, and they have been argued to play an important role in black-hole interiors and cosmology. On the other hand, their energy-momentum tensor violates the Averaged Null Energy Condition (ANEC), placing IFSs in tension with standard causality and unitarity arguments. We show that finite string coupling resolves this tension by triggering a decay cascade that saturates the ANEC in Minkowski spacetime. The remnant of the original IFS persists indefinitely, continuously
seeding a self-similar fragmentation cascade that produces infinitely
many strings over its full history. In the semiclassical regime, the cascade is governed by a single Malthusian exponent,
$
\kappa_{\rm IFS}=2\sqrt{2}-2,
$
which controls the branching growth, pressure decay, cutoff-ANEC convergence, and anomalous non-integer dependence on the string coupling. The complete state has vanishing total energy and momentum while remaining highly excited and carrying negative integrated pressure on every finite time slice. This provides a concrete example of genuinely stringy low-energy dynamics in which microscopic null-energy consistency coexists with macroscopic negative pressure.

\end{abstract}

\baselineskip18pt

\newpage
\section{Introduction and summary of results}
\label{sec:introduction-summary}

The reason it is difficult to find observational signatures that are
truly stringy, rather than effects that can be mimicked by particle
effective field theory, is simple: light strings look like particles.
 The situation can be much more interesting, however, in time-dependent
backgrounds.
In particular, when the string coupling grows with time, string theory also admits light strings that do not look at all like particles: Instant Folded Strings (IFSs) \cite{Itzhaki:2018xik}. As their name suggests, these are closed folded strings nucleated at an instant. After creation, they expand, with their folds asymptotically approaching null trajectories. IFSs are light while remaining intrinsically non-particle-like. Their macroscopic spacetime extent, instantaneous creation, and folded structure have no natural counterpart in a particle effective field theory. They therefore raise the exciting possibility that time-dependent settings, and cosmology in particular, may exhibit genuinely stringy low-energy signatures.

The evidence for IFSs now goes well beyond the classical solution. The exact conformal field theory (CFT) description of the $SL(2,\mathbb{R})/U(1)$ black hole was exploited in \cite{Giveon:2019gfk,Giveon:2019twx} to identify the worldsheet operator associated with an IFS and investigate its role in the Lorentzian version of FZZ duality~\cite{Giveon:2019gfk}. A complementary perspective on this duality and the role of folded strings was developed in Ref.~\cite{StringyEREPR}, highlighting a possible connection to the ER=EPR proposal~\cite{EREPR}. In a timelike linear-dilaton CFT, an exact worldsheet description was
constructed in \cite{Hashimoto:2022iqp}, making the IFS production rate calculable.
As shown in Ref.~\cite{Itzhaki:2024cxh},
the resulting rate agrees with a calculation based on the timelike Liouville
partition function of Ref.~\cite{Giribet:2011tn}. A second check comes from the
$SL(2,\mathbb{R})/U(1)$ black hole
\cite{Witten:1991yr,Elitzur:1990ubs,Mandal:1991tz,Dijkgraaf:1991ba}: in its
future-increasing-coupling interior, the number of IFSs encountered by an
infalling observer scales parametrically with the Bekenstein--Hawking entropy
\cite{Hashimoto:2022iqp}. In parallel, the collective backreaction of IFSs
in homogeneous time-dependent dilaton-gravity backgrounds has been developed
in Refs.~\cite{Itzhaki:2024vaw,Itzhaki:2025cyclic}.

Despite this progress, IFSs present a basic consistency
question. The same stress that underlies the proposed applications to
black-hole interiors and cosmology also gives a macroscopic violation of the
ANEC~\cite{Attali:2018goq}. To illustrate this, we begin in $1+1$ dimensions, where the
energy-momentum tensor of an IFS
nucleated at $(t,x)=(0,0)$ is
\footnote{We neglect finite dilaton gradient effects which can be found in \cite{Attali:2018goq}.}
\begin{equation}
\label{eq:ifs_tmunu}
\begin{aligned}
    T_{uu} &= -\mu v\,\Theta(v)\delta(u),\\
    T_{vv} &= -\mu u\,\Theta(u)\delta(v),\\
    T_{uv} &= \mu\,\Theta(u)\Theta(v),
\end{aligned}
\end{equation}
where \(\mu=1/(2\pi\alpha')\), \(u=t-x\), and \(v=t+x\).


The first two terms are localized on the folds and carry \emph{negative} null
energy, whose magnitude grows with time. The third term fills the interior of the folded string and carries
its traceful bulk stress.  Although these contributions are individually
nonzero, they cancel in the total Minkowski energy and momentum:
\begin{equation}
    E^{\rm tot}(t)
    \equiv
    \int dx\,T_{00}(t,x)
    =0,
    \qquad
    P_x^{\rm tot}(t)
    \equiv
    \int dx\,T^{\,01}(t,x)
    =0.
    \label{eq:intro-zero-charges}
\end{equation}
They do not, however, cancel in the integrated pressure:
\begin{equation}
    p^{\rm tot}(t)
    \equiv
    \int dx\,T_{11}(t,x)
    =-8\mu t \Theta(t),
    \label{eq:intro-negative-pressure}
\end{equation}
which is negative for $t>0$ and grows linearly in magnitude. This unusual combination of vanishing total energy and negative pressure is the origin of the possible importance
of IFSs in cosmology and black hole interiors. At the same time,
Eq.~\eqref{eq:ifs_tmunu} implies a violation of the ANEC,
\begin{equation}
 \mathcal{A}(v) \equiv  \int_{-\infty}^{\infty}du\,T_{uu}(u,v)
    =
    -\mu \Theta(v) v,
    \label{eq:intro-bare-anec}
\end{equation}
with an analogous result under $u\leftrightarrow v$. 

This behavior is markedly different from that of an ordinary QFT in \(1+1\) dimensions, where any state with vanishing total energy and momentum, normally the vacuum, must saturate the ANEC \cite{Verch:1999nt}. To see this, note that
energy-momentum conservation implies
\be
\frac{d\mathcal{A}(v)}{dv}
=
-\left[T_{uv}(u,v)\right]_{u=-\infty}^{u=+\infty}.
\ee
For conventional QFT states, the traceful component \(T_{uv}\) vanishes at the asymptotic endpoints of the null ray, and so \(\mathcal{A}(v)\) is independent of \(v\). If the total energy and momentum vanish, this constant is necessarily zero, and the ANEC is therefore saturated.

An IFS evades this conclusion since its stress tensor retains traceful support at null infinity, so \(T_{uv}\) does not vanish at the endpoint of the null ray. As a result, \(\mathcal{A}(v)\) need not be constant, and the IFS can violate the ANEC despite carrying vanishing total energy and momentum.

In flat-space quantum field theory, the ANEC is closely tied to causality \cite{Hartman:2016lgu}. It can also be derived using monotonicity of relative entropy \cite{Faulkner:2016mzt}, within the relative-entropy framework emphasized in Ref.~\cite{Casini:2008cr}. The tension is particularly sharp because IFS production requires only a locally increasing string coupling: it can occur at arbitrarily weak coupling and arbitrarily small curvature, precisely in the regime where one would expect flat-space positivity arguments to remain valid. Moreover, the violation in \eqref{eq:intro-bare-anec} is macroscopic. Its magnitude is controlled by the extent of the folded string along the null ray, rather than by the local curvature scale.

\begin{figure}[t!]
\centering
\includegraphics[width=16cm]{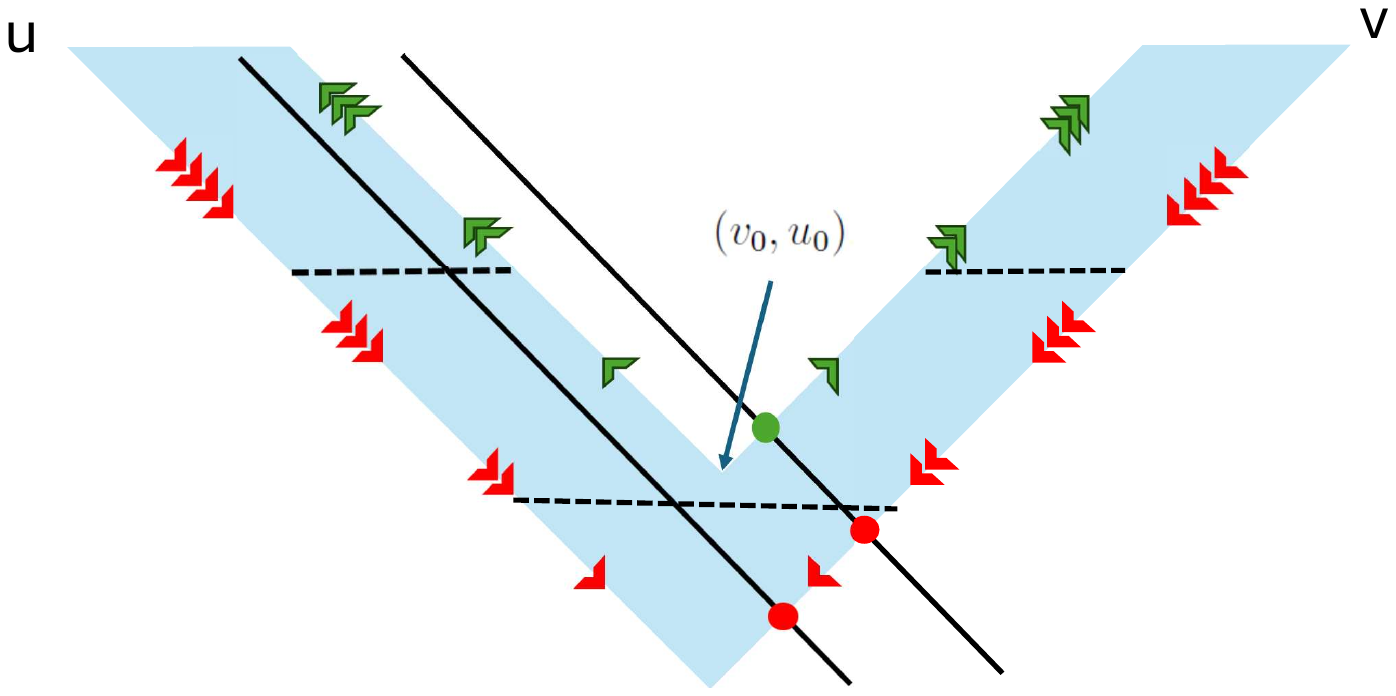}
\vspace{-2cm}
\caption{A single split configuration. The solid lines illustrate why the violation of the ANEC stops growing after the split. The growth of the negative contribution from the red dot is balanced by that of the green dot. The dashed segments indicate the portions of the strings contributing
to the negative integrated pressure; their total length remains
constant after the split.
   }
\label{fig:single-split}
\end{figure}

For some time, we took a conservative view of this puzzle. Although we did not understand how it was resolved, the accumulating evidence indicated that IFSs are genuine objects in string theory. If so, the resolution should arise intrinsically within string theory, rather than through an additional constraint imposed from outside — for example, by simply declaring that IFSs are not part of the theory. Surprisingly, the resolution of this tension is straightforward: at finite string coupling, an IFS, like any other string, can split.
At first sight, it is hard to imagine how IFS splitting can be relevant for the ANEC puzzle.  The local splitting probability,
\be
dP_{\rm split}
=
\Gamma_{\rm split}dudv,
\qquad
\Gamma_{\rm split}\sim\frac{g_s^2}{\ell_s^2},
\label{eq:Gamma-convention}
\ee
vanishes in the weak-coupling limit.  One might therefore expect splitting to give only a small perturbative correction to the classical IFS, far too small to repair the macroscopic ANEC violation in \eqref{eq:intro-bare-anec}.
Examining a single split seems to support this. Suppose that an IFS breaks at $(v_0, u_0)$ with $v_0,u_0 >0$. The split produces two segments moving in opposite directions at the
speed of light (see Fig.~\ref{fig:single-split}). Each has a new inner fold carrying positive
null energy that grows with time. This affects the negative pressure and the ANEC violation in a way that seems correlated:\\
${\it (i)}$ The IFS pressure stops growing for $t>t_{0}=(v_0+u_0)/2$,
\be\label{l}
p^{\rm total}= \begin{cases}
        0,
        &t<0,
        \\[1mm]
        -8\mu t ,
        &0< t <t_0,
        \\[1mm]
        -8\mu t_0 ,&
        t_0<t.
    \end{cases}
\ee
 The reason is that the size of the two segments is fixed after the splitting, and the growth of the negative null energy at the external fold is canceled by the growth of the positive null energy in the internal fold (the green arrows in Fig.~\ref{fig:single-split}).\\
${\it (ii)}$  ANEC violation stops growing as well for $v>v_{0}$,
\be\label{i}
\mathcal{A}(v)=\begin{cases}
        0,
        &v<0,
        \\[1mm]
        -\mu v ,
        &0< v <v_0,
        \\[1mm]
        -\mu v_0 ,&
        v_0<v,
    \end{cases} 
\ee
 as follows from the cancellation between the contribution of the external fold (the red dot) and the internal fold (the green dot).

This suggests a simple interpretation: the negative pressure ${\it (i)}$
and the ANEC violation  ${\it (ii)}$ are two manifestations of the same physics viewed from two different perspectives: in Minkowski time as negative pressure, and in light-cone time as ANEC violation.  It is therefore natural to expect that subsequent splittings modify the late-time evolution of the negative pressure and the ANEC violation, while preserving the qualitative link between them.

Despite its plausibility, the expectation that negative pressure
and ANEC violation would continue to go hand in hand is incorrect.
The distinction becomes clear when we consider the full decay.
At any finite Minkowski time, only finitely many splittings have
occurred, whereas a complete null ray samples contributions from
arbitrarily many splits extending to arbitrarily late times.

\begin{figure}[t!]
\centering
\includegraphics[width=0.96\textwidth,trim=43bp 94bp 75.5bp 148.5bp,clip]{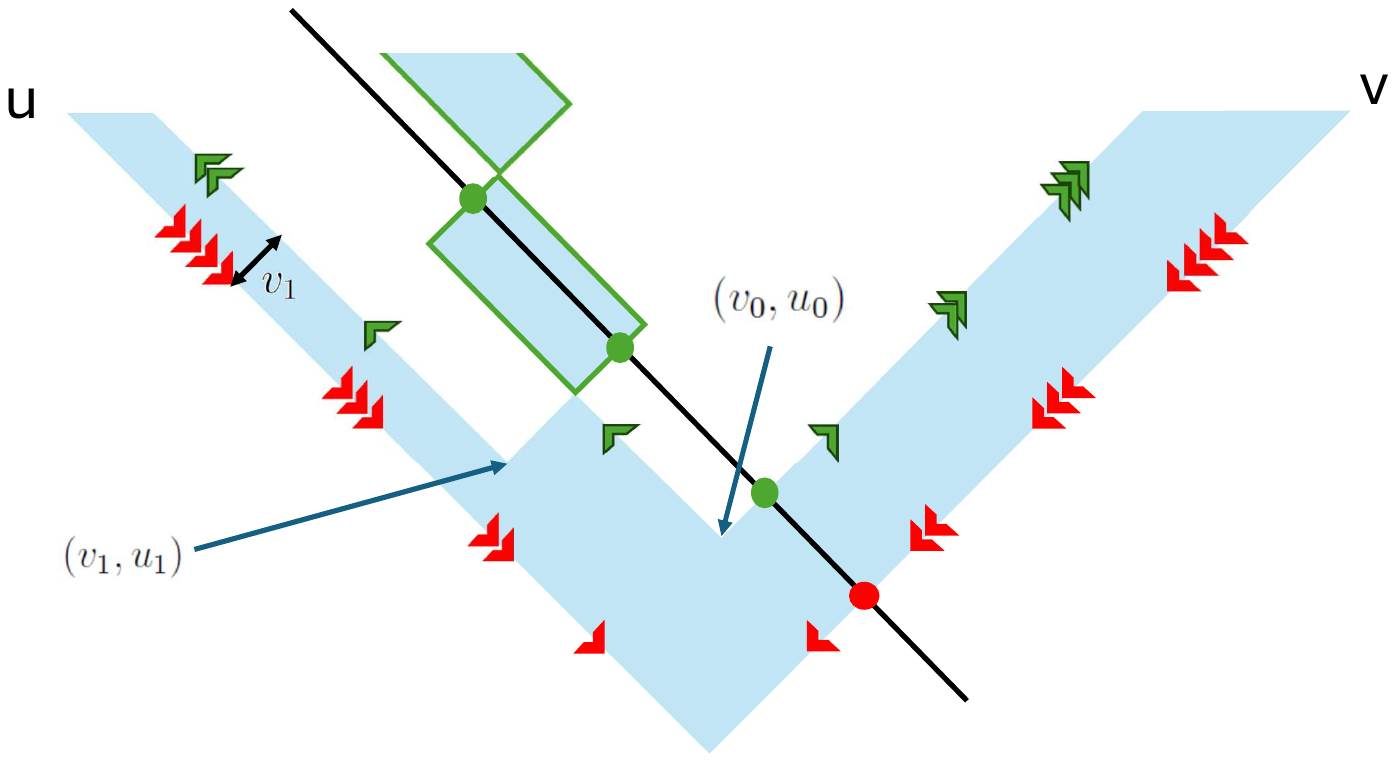}
\caption{Successive emissions build the completed null
cancellation. Positive folds (green) compensate the original
negative fold (red). Later yo-yos, not shown, account for the
remaining deficit marked by $v_1$.}
\label{fig:intro-anec-cascade}
\end{figure}

The perpetual nature of the decay is reflected in the
statistical boost symmetry,
\be u\to\lambda u,\quad v\to\lambda^{-1}v,\ee
of both the stress tensor \eqref{eq:ifs_tmunu} and the splitting rate \eqref{eq:Gamma-convention}.
At fixed $uv$, a boost can move a splitting event to arbitrarily late Minkowski time, $t=(u+v)/2$. 
Thus, a single split, or any finite sequence of splits, can describe part of the decay, but cannot capture the complete boost-invariant decay process.
Put differently, boost invariance implies that a complete decay of an IFS must take an infinite amount of time, and must involve an infinite number of decay products.
The decay products are yo-yos  \cite{Bardeen:1975gx,Bardeen:1976yt} that are detached 
from the remnant of the original IFS (see Fig.~\ref{fig:intro-anec-cascade}). These yo-yos can split again, and their
daughters can do the same. Their fragmentation ultimately reaches the string scale, where the classical yo-yo description gives way to the light-string modes of the terminal sector.

Our main claim is that, once all descendants
and the terminal radiation sector are included, the complete decay
cascade of an IFS saturates the ANEC:
\begin{equation}
  \mathcal{A}(v)=  \int_{-\infty}^{\infty}du\,T_{uu}(u,v)=0,
    \label{eq:intro-anec-saturation}
\end{equation}
for every $v$ (and similarly with $u\leftrightarrow v$).

There are two complementary ways to prove this claim.  The first is a direct fold-by-fold construction, illustrated schematically in Fig.~\ref{fig:intro-anec-cascade}. Each split creates positive-energy null folds. As the remnant continues to split, the contributions from the emitted yo-yos form a sum that compensates the negative null energy carried by the external fold of the original IFS.

\begin{figure}[t!]
 \centering
 \includegraphics[width=0.96\textwidth]{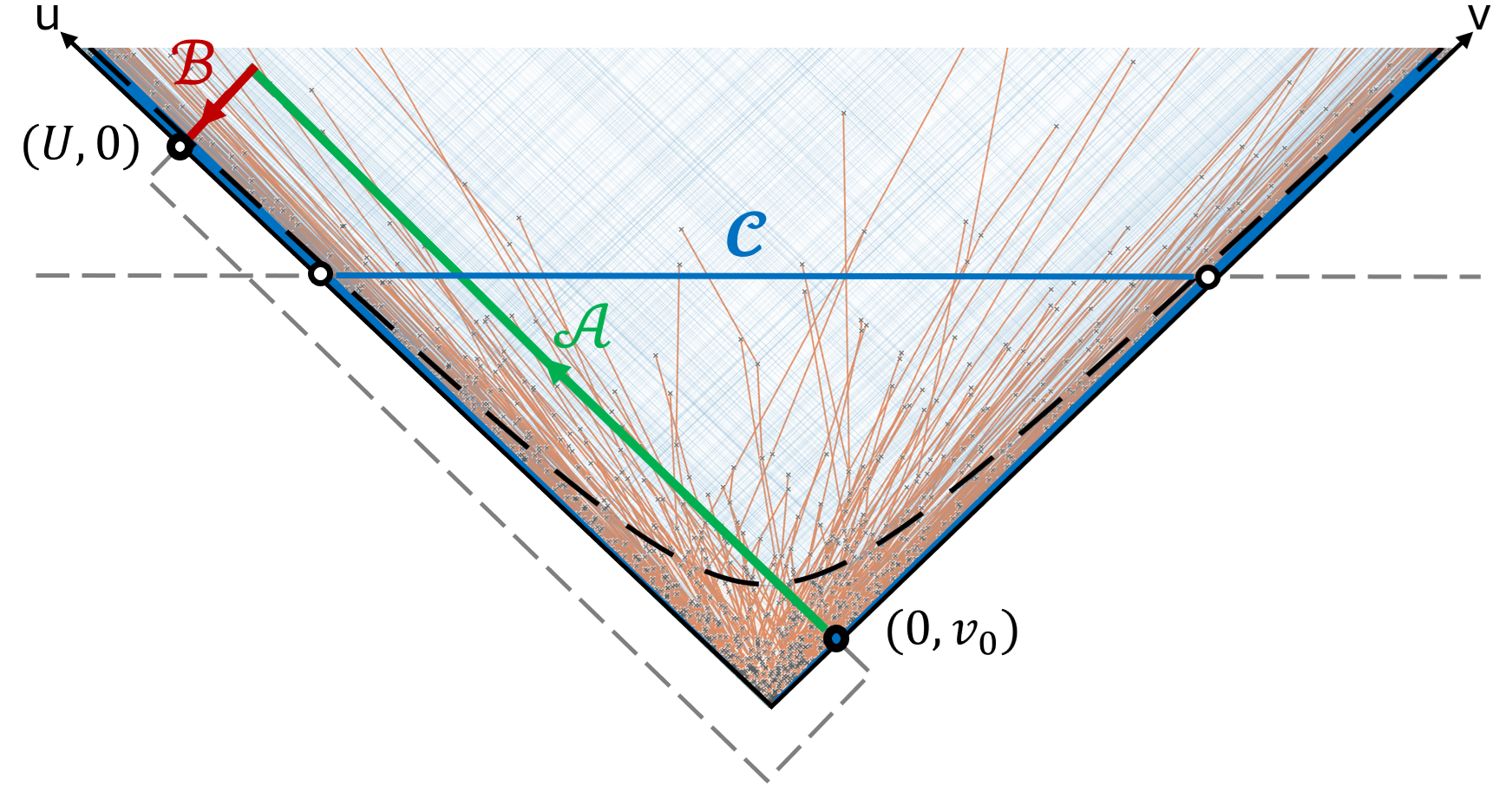}
 \caption{The three integration cuts $\mathcal {A,B}$ and $\mathcal{C}$, used in the conservation and pressure arguments, are shown superimposed over a numerical simulation of the complete decay cascade.
 The dashed hyperbola illustrates the stochastic boost invariance of the cascade. 
 Green $\mathcal A$ is the ray $v=v_0$, continued into the vacuum at
 $u<0$ and truncated at $u=U$.
 Red $\mathcal B$ lies at $u=U$; its integral runs from $v=0$ to
 $v=v_0$, opposite to the red arrow.
 Blue $\mathcal C$ is a constant-Minkowski-time slice.}
 \label{fig:intro-anec-trace-cuts}
\end{figure}

The second is a conservation argument, illustrated in 
Fig.~\ref{fig:intro-anec-trace-cuts}.
Since the stress vanishes before nucleation,
$\partial_vT_{uu}+\partial_uT_{uv}=0$ relates the green null cut
$\mathcal A$ to the red segment $\mathcal B$:
\begin{equation}
 \mathcal A(v_0;U)
 \equiv\int_{\mathcal A}du\,T_{uu}
 =-\int_{\mathcal B}dv\,T_{uv}.
 \label{eq:intro-cutoff-anec}
\end{equation}
Statistical boost invariance gives
$\langle T_{uv}(u,v)\rangle=f(uv)\geq0$.
Writing $I(z)\equiv\int_0^z d\zeta\,f(\zeta)$, we obtain
\begin{equation}
 \langle\mathcal A(v_0;U)\rangle
 =-\int_0^{v_0}dv\,f(Uv)
 =-\frac{I(Uv_0)}{U}.
 \label{eq:intro-integrated-endpoint}
\end{equation}
Saturation along each fixed null ray therefore requires only
$I(z)/z\to0$.

The physical reason this condition holds is that splitting prevents
the integrated trace from growing without bound.
Let $W_{\rm tot}(t)\equiv-\int_{\mathcal C}dx\,T^\mu{}_\mu$ denote
the trace weight on the blue slice $\mathcal C$.
A single split already halts its linear growth.
More strongly, if the first split occurs at $t_s$, all subsequent
splittings and yo-yo oscillations obey
$W_{\rm tot}(t)\leq8\muIFS t_s$.\footnote{For a first split at
$(u_s,v_s)$, the two branches inherit conserved null charges
$(E-P)_R=2\muIFS u_s$ and $(E+P)_L=2\muIFS v_s$.
Each charge excludes its negative outer fold and bounds the
positive bulk energy, including all descendants.
Since $W=2E_{\rm bulk}$, their sum gives
$W_{\rm tot}\leq4\muIFS(u_s+v_s)=8\muIFS t_s$.
Here $P=\int dx\,T^{01}$.}
Since $\langle t_s\rangle=O(\GammaSplit^{-1/2})$, this gives
$\langle W_{\rm tot}(t)\rangle\lesssim
\muIFS/\sqrt{\GammaSplit}$ without any assumption about the
subsequent decay law.

Positivity and boost invariance translate this bounded trace into
geometric dilution:
\begin{equation}
 \frac{I(z)}{z}
 \leq\frac{\langle W_{\rm tot}(\sqrt z)\rangle}{4\sqrt z},
 \label{eq:intro-Tuv-boundary}
\end{equation}
where changing variables from $x$ to $\zeta=t^2-x^2$ along $C$ gives
$\langle W_{\rm tot}(t)\rangle\geq 4I(t^2)/t$.
Because the numerator is bounded, $I(z)/z$ falls at least as
$1/\sqrt z=1/\sqrt{uv}$, even if the trace weight itself does not decay.
This geometric dilution suffices for ANEC saturation without a
particular decay law or terminal conversion.
The stronger terminal result $I(\infty)<\infty$ improves the
convergence to $O(U^{-1})$.

The same figure explains why the integrated pressure can nevertheless
remain negative. Since $W_{\rm tot}=E_{\rm tot}-p_{\rm tot}$ and the
complete state has $E_{\rm tot}=0$,
\begin{equation}
 \langle p_{\rm tot}(t)\rangle
 =-\langle W_{\rm tot}(t)\rangle
 =-4\int_{\mathcal C}dx\,f(t^2-x^2)<0,\qquad t>0.
 \label{eq:intro-pressure-profile}
\end{equation}
The null folds carry no trace. A completed null ray collects their
compensating flux, while a finite spacelike slice still intersects
the traceful remnant and live descendants.

In summary, the complete IFS decay cascade exhibits a remarkable combination of properties:

\nopagebreak[4]
\begin{center}
\itshape
It saturates the ANEC and has vanishing total energy and momentum,\\
yet retains negative integrated pressure at every finite time.
\end{center}


This raises many questions about the IFS decay cascade and its possible applications. For example:\\
\noindent{\it (i)} \textit{How fast is the ANEC saturated?}
Saturation of the complete null integral does not tell us how rapidly the negative flux disappears as the cutoff is increased.  A benchmark is provided by the geometric bound \eqref{eq:intro-Tuv-boundary}, which gives
\begin{equation}
    \bigl\langle\mathcal A(v_0;U)\bigr\rangle
    \gtrsim
    -\frac{\muIFS}{\sqrt{\GammaSplit}}
    \sqrt{\frac{v_0}{U}}.
    \label{eq:intro-A-no-yoyo}
\end{equation}
For every fixed $v_0$, the residual approaches zero as $U\to\infty$, but the convergence is not uniform in $v_0$.  This is particularly important for homogeneous cosmological settings, where coarse-graining involves an unbounded family of translated null rays.  Consequently, completing each null ray and taking the homogeneous limit need not commute.  This motivates investigating how subsequent yo-yo splittings modify the rate of ANEC saturation and its dependence on $v_0$.

Another motivation comes from the gravitational time advance generated by the IFS decay cascade, which is proportional to $\bigl\langle\mathcal A(v_0;U)\bigr\rangle$. Since the effective cutoff depends on the wavelength of the probing matter, $U\sim \lambda$, the time advance may be strongly wavelength-dependent. This could play an important role in determining how different infalling observers experience the interior of the $SL(2)/U(1)$ black hole, where IFSs are expected to be abundant.

\par\smallskip
\noindent{\it (ii)} \textit{What is the total integrated pressure as a function of time?}
The identity \(p_{\rm tot}=-W_{\rm tot}\) leaves open how rapidly
fragmentation removes the surviving tension. We determine the time and
coupling dependence of \(p_{\rm tot}\) and its crossover to the
string-scale terminal regime. This one-event response is also the input to
the homogeneous IFS cosmologies of
Refs.~\cite{Itzhaki:2024vaw,Itzhaki:2025cyclic}.

\par\smallskip
\noindent{\it (iii) What are the spatial moments of the stress tensor of an
IFS cascade?}  For the homogeneous evolution, the pressure is enough, but perturbations require
more information: how localized is the traceful stress, and how do its
spatial moments grow?  These moments determine the leading nonlocal
response of an IFS cascade and are therefore relevant for cosmological
perturbations and possible observational imprints.
\par\smallskip
\noindent{\it (iv)}  \textit{What happens to the ANEC saturation in higher dimensions?} In $D>2$ the situation appears to be more complicated. For example, the light ray associated with the ANEC
and the IFS need not be aligned, and similarly the compensating radiation can pass through
neighboring rays. 
\par\smallskip
\noindent{\it (v)} \textit{Does ANEC saturation persist in cosmological settings?}
In particular, does the fact that the IFS decay cascade saturates the ANEC in flat spacetime imply that, at finite coupling, IFSs cannot mediate a cosmological bounce?

To address these questions, it is useful to introduce logarithmic Milne
coordinates in the future wedge, in which boosts act as translations:
\be
    \eta
    =
    \frac12\ln (uv),
    \qquad
    \phi
    =
    \frac12\ln (u/v),
    \qquad
    ds^2=e^{2\eta}(-d\eta^2+d\phi^2),
\ee
with
\be
    e^\eta=\sqrt{uv}
\ee
being the invariant Milne scale of the cascade, while a boost acts as a
translation of \(\phi\).

\begin{figure}[t!]
\centering
\maybeincludegraphics[width=0.96\textwidth]{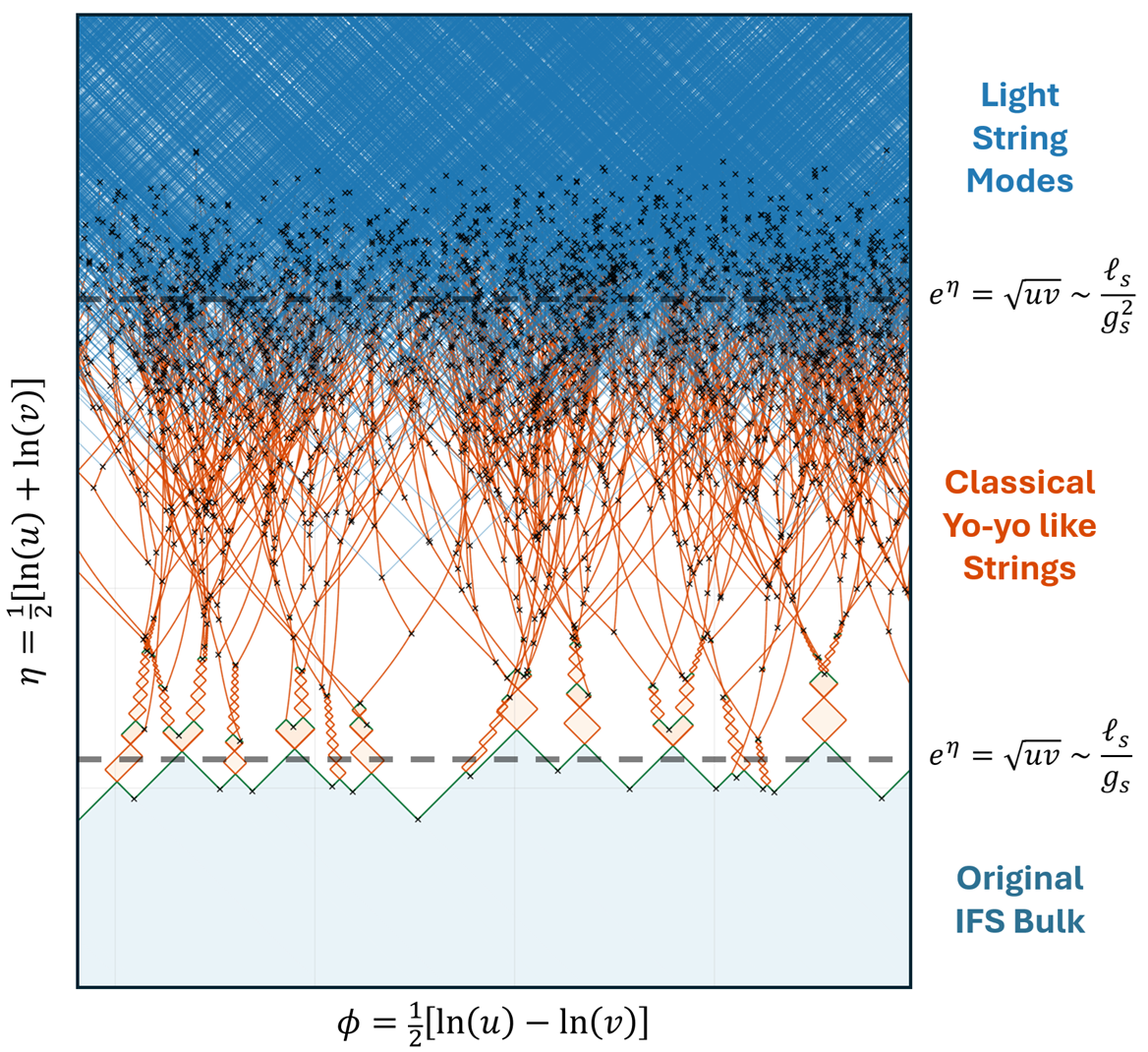}
\caption{Finite-coupling IFS decay cascade in logarithmic future-wedge
coordinates
\(\eta\!=\!\tfrac12[\ln u+\ln v]\),
\(\phi\!=\!\tfrac12[\ln u-\ln v]\).
Translations in \(\phi\) are boosts, so boost invariance of the local
decay process appears as statistical translation invariance in
rapidity.  The lower light-blue shaded region is the traceful original
IFS bulk/remnant, the orange structures are finite classical yo-yo-like
descendants, the black crosses mark splitting and terminal events, and
the blue upper region consists of light string modes.  The upper band depicts terminal conversion, not
a sharp surface beyond which every history has zero trace.  The
ensemble-averaged trace decreases toward the future null endpoint;
the integrated condition for ANEC saturation is given by Eq.~\eqref{eq:intro-integrated-endpoint}.}
\label{fig:rindler-cascade-anec}
\end{figure}

Fig.~\ref{fig:rindler-cascade-anec} summarizes our findings. The
horizontal direction is rapidity and the vertical direction is
logarithmic Milne time.  A particular decay history contains random
split points, but the ensemble is statistically translation invariant in
\(\phi\).  The physically meaningful crossovers are therefore horizontal
bands, set by \(e^\eta=\sqrt{uv}\).
The figure displays three regimes.  At early Milne time,
\be
    e^\eta
    \lesssim
    \Gamma_{\rm split}^{-1/2}
    \sim
    \frac{\ell_s}{g_s},
\ee
the expected number of splitting events in a causal patch is still of
order one.  The configuration is close to the unbroken classical IFS.
The first active splits form a causal staircase: moving from left to right, successive split points have increasing \(u\) and decreasing \(v\).  This
staircase cuts the original traceful bulk into remnant pieces and finite
yo-yos.  It supplies the initial data for the later cascade, but it does
not eliminate the traceful stress.

The next regime is the semiclassical yo-yo cascade,
\be
    \frac{\ell_s}{g_s}
    \ll
    e^\eta
    \ll
    \frac{\ell_s}{g_s^2},
    \label{eq:intro-yoyo-window}
\ee
where the finite strings are large compared with the string scale and are
well described as classical yo-yos.  Each yo-yo contains traceful bulk
stress, oscillates periodically, and can split into smaller yo-yos.  
Since there is no additional scale in this window, the decay
becomes self-similar: trace weight is redistributed among
progressively smaller descendants and gradually depleted.
The orange network in Fig.~\ref{fig:rindler-cascade-anec}
is the visual form of this self-similar trace transport.\footnote{We neglect gravitational emission during this semiclassical
fragmentation stage.  For a yo-yo of rest-frame size \(L\gg\ell_s\),
the fractional energy lost to long-wavelength four-dimensional
gravitational radiation before the next split is parametrically
\(O((\ell_s/L)^2[1+\ln(L/\ell_s)])\).
Quantum suppression of the high-frequency tail is expected to preserve
this hierarchy.  Appendix~\ref{app:gravitational-emission} derives the
estimate and discusses its ultraviolet scope.}

Finally, for
\be
    e^\eta
    \gtrsim
    \frac{\ell_s}{g_s^2},
\ee
finite descendants have typically reached the string scale and decay into
light string modes.  A string-scale yo-yo has period \(O(\ell_s)\) and
splits or decays with probability \(O(g_s^2)\) per oscillation, so its
terminal lifetime is \(O(\ell_s/g_s^2)\).  Above this scale the finite
traceful descendants have mostly become traceless light string modes,
\(T_{uv}=0\), shown as the blue region in the figure. This is an asymptotic conversion regime, rather than
a simultaneous end to all traceful evolution.  The persistent remnant
continues to seed new trees near the outgoing folds.  These regimes determine the approach to ANEC saturation and the decay of the surviving negative pressure, quantified below.

The Milne picture also suggests a useful analogy between a single IFS
and a microscopic $1+1$-dimensional big bang. Defining a Milne time $
    \tau_{\rm M}\equiv e^\eta=\sqrt{uv},$
we see that a single IFS is nucleated at the tip \(\tau_{\rm M}=0\) of an open
Milne universe, while boosts act as translations of the comoving coordinate
\(\phi\). Statistical boost invariance is therefore spatial homogeneity in
this cosmological description, and the three horizontal bands in
Fig.~\ref{fig:rindler-cascade-anec} may
be viewed as successive epochs.
The growing multiplicity of decay products also suggests
coarse-grained entropy production analogous to reheating. The first finite yo-yos are emitted
when \(\Gamma_{\rm split}uv=O(1)\), so their mass is of order
$m_s/g_s$.
Each
subsequent split produces daughters of smaller invariant mass, so along each
branch the cascade flows from \(m\sim m_s/g_s\), through progressively
lighter extended strings, to \(m=O(m_s)\), where the semiclassical yo-yo
description crosses over to massless, traceless light-string excitations.
In this sense, a single IFS realizes a
microscopic \(1+1\)-dimensional big-bang history: a Milne origin, a first
generation of extended strings with invariant mass $m_s/g_s$, a self-similar
fragmentation era, and a final radiative phase.

In the semiclassical yo-yo regime, the answers to questions {\it (i)--(iii)} are
controlled by a single exponent,
\be
    q_{\rm IFS}
    =
    3-2\sqrt2
    \simeq
    0.17157 .
    \label{eq:intro-qP}
\ee
Equivalently, the self-similar branching has Malthusian exponent
\be
    \kappa_{\rm IFS}
    =
    1-q_{\rm IFS}
    \simeq
0.82843 .
    \label{eq:intro-kappa}
\ee
This exponent characterizes the fractal-like part of the cascade: it is
the power governing the growth of the number of terminal leaves produced
by a large finite yo-yo.

The answer to question {\it (i)} is 
\be
    \langle\mathcal A(v;U)\rangle
    \propto
    -U^{\,\alpha}v^{\,\beta},
    \qquad \mbox{with}~~~
    \alpha=\sqrt2-2,
    \qquad
    \beta=\sqrt2-1 .
    \label{eq:intro-A-yoyo}
\ee
Conservation and boost invariance fix $\beta-\alpha=1$, while
\be
\beta+\alpha=-q_{\rm IFS},
\ee
relates the dynamical decay of the surviving tension to the
approach to null cancellation, as derived in
Sec.~\ref{sec:anec_moments}. 

The answer to question {\it(ii)} is that the pressure, again  in the yo-yo  regime, behaves as
\be
    p_{\rm tot}(t)
    \sim
    -\frac{\mu}{\sqrt{\Gamma_{\rm split}}}
    \bigl(\sqrt{\Gamma_{\rm split}}\,t\bigr)^{-q_{\rm IFS}},
    \qquad
    1\ll \sqrt{\Gamma_{\rm split}}\,t \ll g_s^{-1}.
    \label{eq:intro-pressure-yoyo-Gamma}
\ee
Using \(\Gamma_{\rm split}\sim g_s^2/\ell_s^2\), this may be written as
\be
    p_{\rm tot}(t)
    \sim
    -\frac{1}{\ell_s}\,
    g_s^{-1-q_{\rm IFS}}
    \bigl(t/\ell_s\bigr)^{-q_{\rm IFS}},
    \qquad
    \frac{\ell_s}{g_s}
    \ll
    t
    \ll
    \frac{\ell_s}{g_s^2},
    \label{eq:intro-pressure-yoyo}
\ee
which, interestingly, exhibits a non-integer power also of the string coupling,
\(g_s^{-2+\kappa_{\rm IFS}}\).  This anomalous coupling dependence is
another manifestation of the fractal cascade: it arises from resumming arbitrarily many splitting histories and therefore cannot be associated
with any fixed order in the usual genus expansion.  In a sense, the ordinary \(g_s^{-2}\) scaling is dressed by the Malthusian exponent of the self-similar decay cascade.

This is the intermediate semiclassical power law generated by
self-similar trace-weight transport.
Terminal conversion makes the invariant integrated trace profile finite,
\be
    I_\infty\equiv\int_0^\infty dz\,f(z)<\infty.
    \label{eq:intro-terminal-integrated-trace}
\ee
As derived in Sec.~\ref{sec:terminal-decay}, this same quantity fixes
both the terminal null residual and the surviving negative pressure:
\begin{align}
    \langle\mathcal A(v;U)\rangle
    &\sim-\frac{I_\infty}{U},
    \qquad Uv\gg\frac{\ell_s^2}{g_s^4},
    \label{eq:intro-A-terminal}\\
    \langle p_{\rm tot}(t)\rangle
    &\sim-\frac{4I_\infty}{t},
    \qquad t\gg\frac{\ell_s}{g_s^2}.
    \label{eq:intro-pressure-terminal}
\end{align}
Thus the complete null integral vanishes while the pressure
remains negative at every finite time. The terminal bound
$0\leq-\langle\mathcal A(v;U)\rangle\leq I_\infty/U$ is uniform in
$v>0$ at a fixed IFS origin. It does not, however, provide an
integrable bound over an infinite distribution of source positions,
which translate both null coordinates. Null completion and homogeneous
coarse graining therefore need not commute. This distinction, and the
retarded stress relevant to cosmology, are discussed in
Sec.~\ref{sec:cosmological-implications}.

For question {\it (iii)}, boost invariance fixes the relative
coefficients of the spatial trace moments as well as their powers.
Define
\be
    M_n^{(W)}(t)
    =\int dx\,x^n\bigl(T_{00}-T_{11}\bigr)(x,t).
\ee
In the semiclassical regime, the trace has a self-similar spatial
profile spread across the light-cone scale $|x|\sim t$. Hence, for the
left-right symmetric ensemble,
\be
    \bigl\langle M_{2n}^{(W)}(t)\bigr\rangle
    \propto t^{2n}\langle W_{\rm tot}(t)\rangle
    \propto t^{2n-q_{\rm IFS}},
    \qquad
    \bigl\langle M_{2n+1}^{(W)}(t)\bigr\rangle=0.
    \label{eq:intro-trace-moments-yoyo}
\ee
The relative coefficients are generally different from unity in
this regime. After terminal conversion, the surviving trace instead
concentrates near the outgoing folds, so each even moment approaches
$t^{2n}$ times the total trace weight.
This terminal localization underlies the narrow-band description used
later for cosmology.

Energy moments obey different constraints.  Define
\be
    E_n(t)
    =
    \int dx\,x^nT_{00}(x,t),
\ee
then conservation gives
\be
    E_0=0,
    \qquad
    E_1=0
\ee
since the IFS is nucleated with zero total energy, momentum,
and center-of-energy moment.  The first nontrivial energy moment obeys
\be
    \frac{d^2E_2}{dt^2}
    =
    2p_{\rm tot}(t),
    \label{eq:intro-E2-pressure}
\ee
and therefore, in the semiclassical regime,
\be
    E_2(t)
    \propto
    -t^{2-q_{\rm IFS}}.
    \label{eq:intro-E2-yoyo}
\ee
The derivation and the higher moment generalizations are given in
Appendix~\ref{app:stress-moments-terminal-anec}.


Question {\it (iv)} is addressed in Section 7, where we show that transverse momentum conservation implies that an IFS is nucleated in a state that saturates the ANEC for all light rays. We also discuss the amount of ANEC violation associated with other, more localized states. Question {\it (v)} is the subject of Section 8. There we derive the leading homogeneous transport equations for the highly boosted strings and the remnant source, and show that the IFS decay cascade can violate both the NEC and ANEC in a contracting universe. This violation, in turn, can mediate a cosmological bounce. 

\section{Geometry of the finite-coupling cascade}
\label{sec:cascade_algorithm}

We construct the decay directly on the worldsheet, showing how local breaks
divide the expanding IFS into a persistent remnant and detached strings.
Throughout this section we work in the flat-space, locally constant-coupling
approximation. The sole local input is a uniform splitting probability per
unit active worldsheet area, where ``active'' denotes traceful parent bulk
that has not been removed by an earlier split. A split removes the causal
future of its point from the parent bulk and emits positive null folds. Only
the causal-minimal splits affect the surviving parent stress; their records
form a monotone staircase.

The construction separates two classes of objects: the unbounded remnant
retains the original null arms, while crossings of neighboring positive
folds emit finite yo-yos. Their invariant
mass is set by the product of their null extents and their boost by
the ratio. The sampler below determines these quantities without
generating irrelevant candidate splits.

\paragraph{Fixed-background decay model.}
We condition on nucleation in a growing-dilaton background and evolve
the leading small-gradient stress at fixed $g_s$, neglecting gravitational
backreaction. Complete-null limits refer to this fixed-background decay
model. In a varying background it describes a finite history while the
coupling and curvature change little. Writing
$Q=\partial_t\ln g_s$, this requires $Qt\ll1$, and in particular
$Qt_\times\ll1$ to reach the terminal crossover within the approximation.

We work in null coordinates \(u=t-x\), \(v=t+x\).
The physical starting point is the IFS splitting argument~\cite{Itzhaki:2024cxh}:
splitting is local and, away from the folds, does not distinguish
one position along the string from another.

At locally constant coupling, we describe these equivalent bulk
patches by a uniform splitting probability per unit worldsheet area.
The closed-string splitting amplitude is of order \(g_s\), giving
a probability of order \(g_s^2\) per string-scale area.  Restoring
dimensions gives
\be
    dP_{\rm split}=\GammaSplit\,du\,dv .
    \label{eq:local_split_rate}
\ee
Here \(\GammaSplit\sim g_s^2/\ell_s^2\); its normalization
absorbs the numerical coefficient and the area convention.
We implement the local law by independent splitting events, so the
splitting points form a homogeneous Poisson process on any active
bulk region.  Thus, for a region \(\mathcal A\), the
probability that no split occurs in \(\mathcal A\) is
\be
    P_0(\mathcal A)
    =
    \exp[-\GammaSplit |\mathcal A|].
    \label{eq:poisson_void_probability}
\ee

The growing IFS exposes an area of order \(t^2\), so the
first split becomes likely when \(\GammaSplit t^2=O(1)\).
This recovers the scale \(t\sim\ell_s/g_s\) estimated in
\cite{Itzhaki:2024cxh}.  The construction below follows the
subsequent splittings and their causal removal of active bulk.

For a parent worldsheet domain \(\mathcal D\), a split at \(p=(u_p,v_p)\) creates two
positive-energy null folds and removes the causal future
\be
    J^+(p)
    =
    \{(u,v)\,|\,u>u_p,\;v>v_p\}
\ee
from the bulk of the parent object. For several splits in \(\mathcal D\), the surviving parent
bulk is therefore
\be
    \mathcal D_{\rm parent}
    =
    \mathcal D\setminus\bigcup_iJ^+(p_i).
\ee
Equivalently, the sequential active splitting process may be coupled to a homogeneous
Poisson cloud on the original parent domain \(\mathcal D\). Points lying in the causal future of
an earlier causal-minimal split are candidates in regions that have already been removed,
and therefore do not affect the parent stress tensor. By the independent-increments
property of the Poisson process, retaining only the causal-minimal points gives the same
distribution for the physical parent history as sampling sequentially on the active region.
Regions lying in the future of more than one split are removed only once. In the numerical
construction, this is implemented as a mask on the parent contribution to \(T_{uv}\).

\begin{figure}[t!]
\centering
\maybeincludegraphics[width=0.9\textwidth]{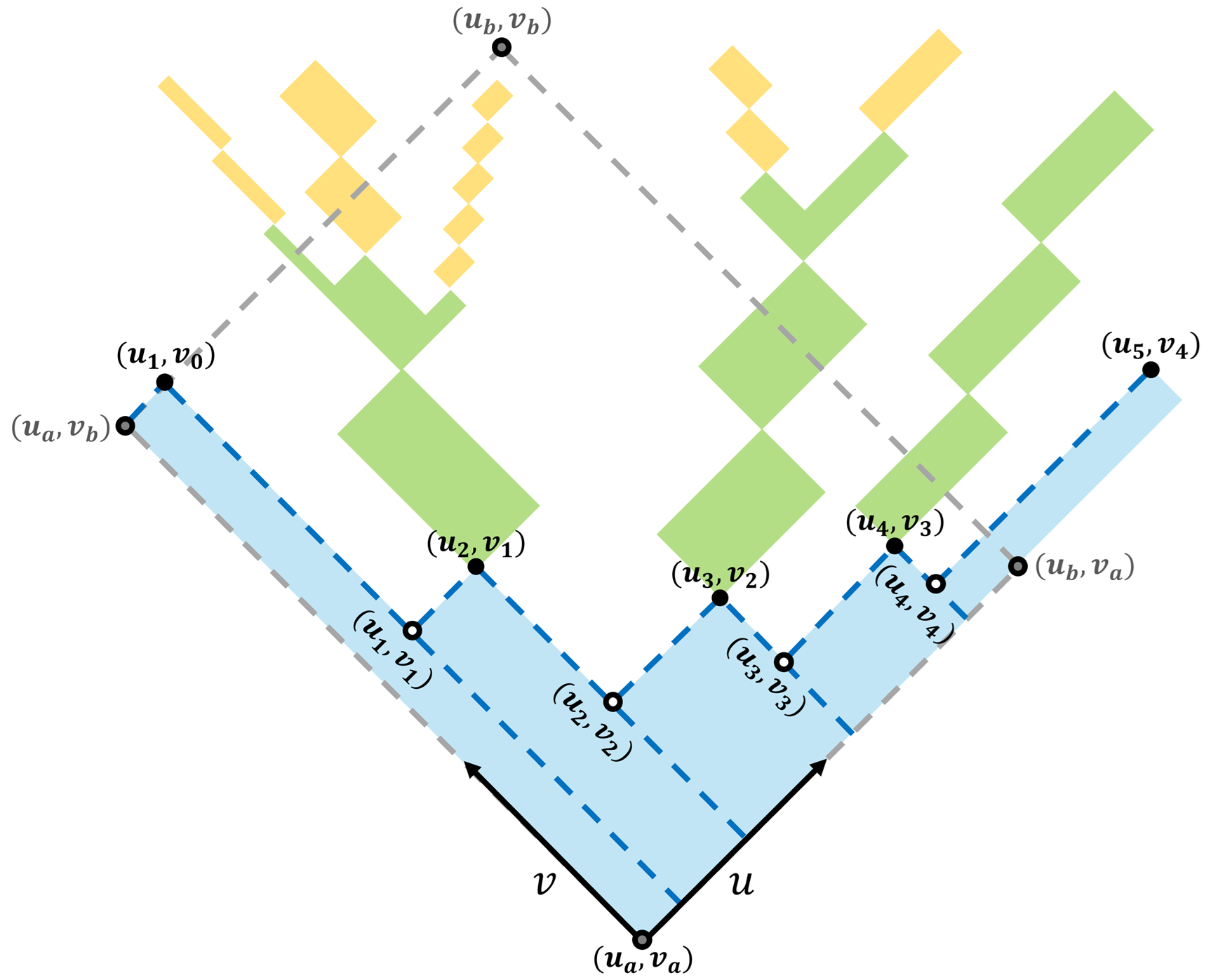}
\caption{
The persistent IFS remnant (blue) and its decay cascade.
Causal-minimal splits \(\{(u_i,v_i)\}\) leave a staircase attached
to the original null arms. Neighboring positive folds cross to emit
first-generation yo-yos (green), whose own splits produce later
generations (yellow). A finite sampling window does not terminate
the remnant.
}
\label{fig:cascade_staircase}
\end{figure}

The boundary left by these causal removals forms a monotone staircase.
The positive folds emitted from neighboring staircase vertices
intersect, producing finite yo-yos.  Each yo-yo may then split by the
same mechanism, generating the recursive cascade shown in
Fig.~\ref{fig:cascade_staircase}.

\subsection{Causal records and the common staircase construction}
\label{subsec:record_sampler}

Consider the null rectangle
\(\mathcal D=[u_a,u_b]\times[v_a,v_b]\), a causal diamond in
\((t,x)\). With \(p\prec q\) when \(u_p<u_q\) and \(v_p<v_q\),
the physical records are the Poisson points with no other point in
their causal past. Ordering them as \(p_i=(u_i,v_i)\),
\(i=1,\ldots,N\), by increasing \(u\) gives
\be
    u_1<u_2<\cdots<u_N,\qquad
    v_1>v_2>\cdots>v_N .
    \label{eq:record_ordering}
\ee
Set \((u_0,v_0)=(u_a,v_b)\) and \(u_{N+1}=u_b\) to complete
the sampling boundary; these are not additional splitting events.
The surviving parent bulk is the union of non-overlapping rectangles
\be
    \mathcal S
    =\bigcup_{i=0}^{N}R_i,
    \qquad
    R_i=[u_i,u_{i+1}]\times[v_a,v_i].
    \label{eq:staircase_rectangles}
\ee
Thus \(T_{uv}^{\rm stair}=\muIFS\chi_{\mathcal S}\), where
\(\chi_{\mathcal S}\) is the indicator of the surviving region.
Overlapping removed futures are counted only once.

The same staircase also determines the detached yo-yos. Whenever
consecutive vertices carry physical folds, the fold from \(p_i\)
at fixed \(v_i\) meets the fold from \(p_{i+1}\) at fixed \(u_{i+1}\)
and emits a yo-yo with null extents
\be
    \Delta u_i=u_{i+1}-u_i,\qquad
    \Delta v_i=v_i-v_{i+1}.
    \label{eq:daughter_extents}
\ee
Its birth event and opposite cross-corner are
\be
    b_i=(u_{i+1},v_i),\qquad
    \bar b_i=(u_i,v_{i+1}),
    \label{eq:daughter_cross_corners}
\ee
which span an invariant rectangle of area
\(A_i^Y=\Delta u_i\Delta v_i\). For the remnant, this rule
applies to adjacent physical records; artificial sampling edges do not
emit daughters. For a finite yo-yo, the physical turning points also
carry folds and close the staircase, as described below. The bulk
construction is common to both; their boundary folds differ.

\paragraph{Direct sampling of the records.}
Suppose that the current vertex is \((u_i,v_i)\). The event
\(u_{i+1}>U\) is equivalent to finding no split in
\([u_i,U]\times[v_a,v_i]\), so
Eq.~\eqref{eq:poisson_void_probability} gives
\be
    P(u_{i+1}\geq U\,|\,u_i,v_i)
    =\exp\!\left[-\GammaSplit (U-u_i)(v_i-v_a)\right].
    \label{eq:next_u_survival}
\ee
The horizontal step is therefore exponential, and the next height
is uniform over the surviving vertical interval:
\begin{align}
 \Delta u_i&\sim{\rm Exp}\!\left(\GammaSplit(v_i-v_a)\right),
 \label{eq:u_semp}\\
 v_{i+1}&=v_a+Y_i(v_i-v_a),\qquad Y_i\sim{\cal U}(0,1).
 \label{eq:v_semp}
\end{align}
If the trial \(u_i+\Delta u_i\) exceeds \(u_b\), there is no
further record and the sampler stops at the final bulk rectangle.
Otherwise the point is accepted and the rule is repeated.
This one-dimensional Markov chain samples the causal-minimal records
without generating the full Poisson cloud. Their number grows only
logarithmically with the expected number of candidates in a large
domain. Ordering by \(u\) is a sampling convention; the underlying
process is symmetric under \(u\leftrightarrow v\).

\subsection{The persistent remnant and its emission scale}
\label{subsec:original_remnant_first_generation}
\label{subsec:record_area_statistics}

The original IFS retains its two negative-energy null folds while
neighboring positive folds detach as first-generation yo-yos
(Fig.~\ref{fig:cascade_staircase}). Its full staircase is unbounded;
the rectangle \(\mathcal D\) is only a sampling window, extended
by continuing the same record construction. Within that window the
bulk is given by Eq.~\eqref{eq:staircase_rectangles}.
The remnant is an emitting subsystem: at daughter births it transfers
null momentum to the detached sector,
\be
    \partial_vT_{uu}^{(0)}+\partial_uT_{uv}^{(0)}=-J_u,
    \qquad
    \partial_uT_{vv}^{(0)}+\partial_vT_{uv}^{(0)}=-J_v.
    \label{eq:remnant_conservation}
\ee
The descendants have the opposite divergences, so the complete stress
is conserved, including at emission. The fold stresses and vertex
currents \(J_u,J_v\) are given in
App.~\ref{app:open-staircase-stress}.

The source's invariant scale follows from the record statistics.
With the IFS origin at \((u_a,v_a)=(0,0)\), the continuing
process defined by Eqs.~\eqref{eq:u_semp}--\eqref{eq:v_semp}
has independent marks
\be
    X_i\equiv\GammaSplit v_i(u_{i+1}-u_i)\sim{\rm Exp}(1),
    \qquad
    Y_i\equiv\frac{v_{i+1}}{v_i}\sim{\cal U}(0,1),
    \label{eq:XY_definition}
\ee
also independent of the earlier staircase history. Here \(X_i\)
is the newly swept area in units of \(\GammaSplit^{-1}\), and
\(Y_i\) is the fraction of the previous height retained.
The dimensionless record area obeys the affine recursion
\be
    A_i\equiv\GammaSplit u_iv_i,
    \qquad
    A_{i+1}=Y_i(A_i+X_i).
    \label{eq:A_staircase_recursion}
\ee
Fresh area replenishes the process, while the contraction erases its
initial condition: two trajectories driven by the same marks have a
separation proportional to \(\prod_{i<N}Y_i\), which decays
exponentially because \(\langle\ln Y\rangle=-1\).
Here \(\langle\cdot\rangle\) denotes stochastic expectation.

For the stationary Laplace transform
\(\Phi_A(s)=\langle e^{-sA}\rangle\), independence gives
\begin{align}
 \Phi_A(s)&=\int_0^1\frac{\Phi_A(sy)}{1+sy}\,dy
       =\frac1s\int_0^s\frac{\Phi_A(t)}{1+t}\,dt,
 \label{eq:area_laplace_integral}\\
 \Phi_A'(s)&=-\frac{\Phi_A(s)}{1+s},\qquad
 \Phi_A(0)=1\quad\Longrightarrow\quad\Phi_A(s)=\frac1{1+s}.
 \nonumber
\end{align}
Thus the unique stationary law and its area moments are
\be
 A=\GammaSplit uv\sim{\rm Exp}(1),\qquad
 \langle(uv)^m\rangle=\frac{m!}{\GammaSplit^m},\qquad
 \langle uv\rangle=\sigma(uv)=\GammaSplit^{-1}.
 \label{eq:uv_exponential_distribution}
\ee
This law describes the continuing remnant staircase away from the
edges of a finite sampling window. It has order-one relative area
fluctuations and restores the \(u\leftrightarrow v\) symmetry
hidden by the coordinate-ordered sampler.

For adjacent records in this regime, the daughter rule
\eqref{eq:daughter_extents} yields
\be
    \GammaSplit A_i^Y=X_i(1-Y_i),
    \qquad
    \left\langle(\GammaSplit A_i^Y)^m\right\rangle
    =\frac{m!}{m+1}.
    \label{eq:direct_yoyo_area}
\ee
In particular,
\be
    \left\langle A_i^Y\right\rangle
    =\frac{1}{2\GammaSplit}\sim\frac{\ell_s^2}{g_s^2}.
    \label{eq:mean_direct_yoyo_area}
\ee
The mass--area relation~\eqref{eq:yoyo_mass} therefore gives a typical
first-generation mass of order \(m_s/g_s\). Splitting selects this invariant scale,
rather than a last emission time: later records lie farther along the
outgoing null arms and emit increasingly boosted yo-yos. This is the
geometric origin of the continuing source in
Fig.~\ref{fig:rindler-cascade-anec}.

\subsection{Finite yo-yos: physical endpoints and conservation}
\label{subsec:yoyo_splitting}

\begin{figure}[t!]
\centering
\maybeincludegraphics[width=0.6\textwidth]{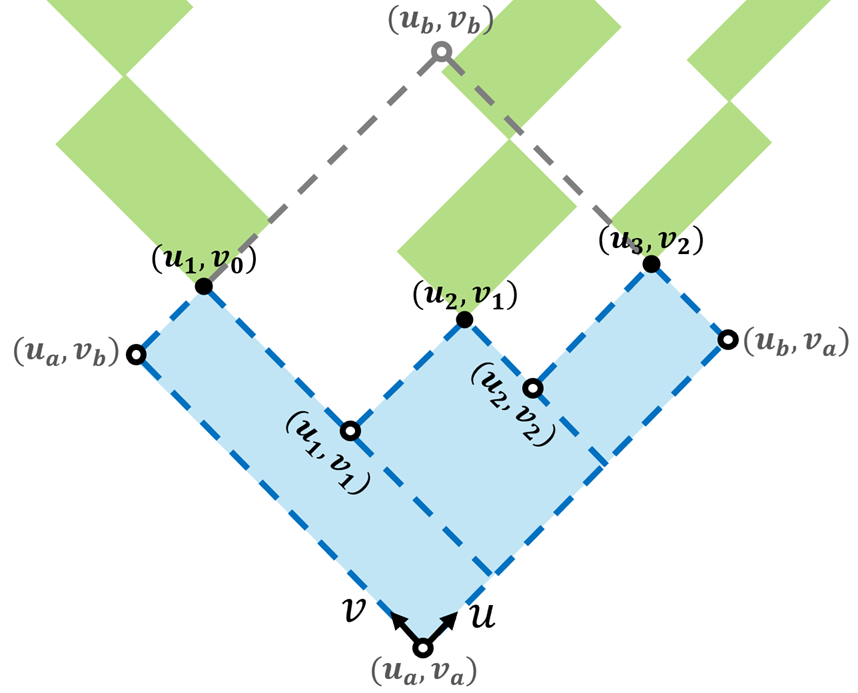}
\caption{
Splitting during one yo-yo oscillation. The diagonal corners are
fold-coincidence events; the other two corners are physical turning
points. Together with the interior split vertices, the turning points
close the staircase around the surviving blue bulk. The turning-point
folds supply the end daughters; adjacent interior records supply the others.
Thus \(N\) splits give \(N+1\) outgoing yo-yos (green). These physical
endpoints distinguish this construction from a truncated remnant window.
}
\label{fig:yoyo_staircase}
\end{figure}

A finite yo-yo sweeps the bounded rectangle shown in
Fig.~\ref{fig:yoyo_staircase},
\(\mathcal D_Y=[u_a,u_b]\times[v_a,v_b]\) during one
oscillation, with extents \(\Delta u=u_b-u_a\) and
\(\Delta v=v_b-v_a\). Its diagonal vertices
\(c_-=(u_a,v_a)\) and \(c_+=(u_b,v_b)\) are consecutive
fold-coincidence events, at which its spatial extent vanishes.
The other two corners are its physical turning points. They close
the staircase by supplying the endpoint vertices
\be
    p_0=(u_a,v_b),\qquad p_{N+1}=(u_b,v_a).
    \label{eq:yoyo_boundary_vertices}
\ee
Unlike the edges of a remnant sampling window, these endpoints carry
physical folds and participate in the daughter construction. They are
not stochastic splits; only the \(N\) interior records are.
The bounded and unbounded staircases refer to worldsheet domains;
both the yo-yo and the IFS are closed strings.

The surviving bulk is again Eq.~\eqref{eq:staircase_rectangles},
and the daughter extents and birth events follow from
Eqs.~\eqref{eq:daughter_extents}--\eqref{eq:daughter_cross_corners},
now for \(i=0,\ldots,N\). Thus \(N\geq1\) physical splits
produce \(N+1\) daughters. For \(N=0\), the single outgoing
piece is the same unsplit parent continuing its oscillation, not a
new generation. The next oscillation sweeps a translated rectangle
with the same extents.

The physical endpoints make conservation transparent: the daughter
extents telescope to those of the parent,
\be
    \sum_{i=0}^{N}\Delta u_i=\Delta u,
    \qquad
    \sum_{i=0}^{N}\Delta v_i=\Delta v.
    \label{eq:yoyo_telescoping}
\ee
For any yo-yo, the stress tensor in
App.~\ref{app:yoyo-staircase-stress} gives
\begin{align}
    E_Y&=\muIFS(\Delta u+\Delta v),\qquad
    P_Y=\muIFS(\Delta v-\Delta u),
    \label{eq:yoyo_energy_momentum}\\
    m_Y&=\sqrt{E_Y^2-P_Y^2}
        =2\muIFS\sqrt{\Delta u\Delta v}.
    \label{eq:yoyo_mass}
\end{align}
We use \(P=\int dx\,T^{01}=\int dx\,(T_{uu}-T_{vv})\);
with \(u=t-x\), \(v=t+x\), a right-moving null packet
contributes to \(T_{uu}\). Applying the same relations to each
daughter immediately gives
\be
    \sum_{i=0}^{N}E_i=E_Y,\qquad
    \sum_{i=0}^{N}P_i=P_Y.
    \label{eq:daughter_conservation}
\ee
Energy and momentum conservation are therefore encoded in the
staircase geometry, without a separate branching rule.

\subsection{Recursive sampling and the dilute single-split limit}
\label{subsec:recursive_sampling}

Applying the same construction to each daughter closes the
recursive cascade. Every yo-yo samples
Eqs.~\eqref{eq:u_semp}--\eqref{eq:v_semp} in its own oscillation
rectangle, of invariant area \(A_Y=\Delta u\Delta v\).
Deep in the cascade,
\be
    \Lambda_Y
    \equiv
    \GammaSplit A_Y
    \ll1 .
\ee
Most oscillations then contain no split. A more efficient implementation can therefore
skip empty oscillations analytically and invoke the staircase sampler only in the first
occupied oscillation. This is the cycle-skipping rule used in the numerical implementation.

For the analytic discussion below, a key simplification comes from the dilute single-split
limit. The probability that one oscillation contains no split is
\be
    P_0(Y)
    =
    e^{-\GammaSplit A_Y}.
    \label{eq:yoyo_no_split_probability}
\ee
If \(K\) is the number of complete unsplit oscillations before the first occupied one, then
\be
    P(K=k)
    =
    e^{-k\Lambda_Y}
    \left(1-e^{-\Lambda_Y}\right),
    \qquad
    k=0,1,2,\ldots .
    \label{eq:empty_yoyo_periods}
\ee
Let \(N_Y\sim{\rm Poisson}(\Lambda_Y)\) be the number of Poisson splitting points
in one oscillation. For \(\Lambda_Y\ll 1\), an occupied oscillation contains exactly
one Poisson point, and hence one causal-minimal record, with probability
\(1-O(\Lambda_Y)\), since
\[
P(N_Y\geq 2\mid N_Y\geq 1)
=
\frac{\Lambda_Y}{2}+O(\Lambda_Y^2).
\]

Moreover, the position of the first split in an occupied oscillation is
approximately uniform over the yo-yo worldsheet,
\be
    \rho(u_s,v_s\,|\,N_Y\geq1)
    =
    \frac{1}{A_Y}
    \left[1+\mathcal O(\Lambda_Y)\right].
    \label{eq:uniform_dilute_split}
\ee
Conditioned on there being exactly one split, the position is uniformly
distributed over \(\mathcal D_Y\) for any value of \(\Lambda_Y\).

Thus the late semiclassical cascade is dominated by long sequences of
empty oscillations, separated by isolated splitting periods containing a
single, nearly uniform split.  This is the single-split regime used in
the renewal analysis below, while the numerical construction retains the
full staircase whenever more than one causal-minimal split occurs.

\section{Kinematics and trace-weight flow in the decay cascade}
\label{sec:yoyo_dynamics}

Section~\ref{sec:cascade_algorithm} gives the stochastic geometry of the
finite-coupling cascade. The pressure is governed by a smaller set of
kinematic variables, which we introduce here.  Each finite yo-yo conserves energy and momentum
when it splits.  The pressure is controlled by the trace weight
\be
    W=-\int dx\,T^\mu{}_{\mu},
\ee
which, for the complete zero-energy IFS decay state, is equal to minus
the integrated pressure.  The remnant injects trace weight into highly
boosted first-generation yo-yos; later splittings transport this weight
through the hierarchy and deplete it when the descendants reach finite
boost.

Let \(T_{\mu\nu}^{(0)}\) denote the remnant bulk and folds of
Sec.~\ref{subsec:original_remnant_first_generation}, with each positive
boundary fold ending at its first daughter-birth vertex.
Its directly emitted yo-yos form generation $\mathcal Y_1$; each
subsequent split advances the generation by one. The semiclassical stress is
\be
    T_{\mu\nu}^{\rm tot}
    =T_{\mu\nu}^{(0)}+
    \sum_{g=1}^{\infty}\sum_{Y\in\mathcal Y_g^{\rm live}}T_{\mu\nu}^{Y}.
    \label{eq:generational_stress_decomposition}
\ee
The sum includes only yo-yos whose live intervals intersect the time
slice under consideration.  Contributions belonging to distinct objects
are added even when their spacetime supports overlap.  Terminal
string-scale decay products will be included separately when the
semiclassical cascade is cut off.

The equations below distinguish exact instantaneous quantities from the
slowly varying, coarse-grained ones used in the renewal analysis.  The
identity \(p_{\rm tot}=-W_{\rm tot}\) is exact for the complete
zero-energy cascade.  The renewal calculation is first formulated for
ensemble- and period-averaged yo-yo weights; Appendix~\ref{app:geometric-matching}
then shows that the exact instantaneous ensemble pressure has the same
leading semiclassical asymptotic.  The boost-spectrum calculation retains
the period-averaged weight unless stated otherwise.

\subsection{Finite-yo-yo kinematics}
\label{subsec:yoyo_kinematics}

The null extents introduced above determine both the area exposed to
splitting and the elapsed Minkowski time:
\be
    A_Y=\Delta u\Delta v,
    \qquad
    \mathcal T_Y=\frac{\Delta u+\Delta v}{2}.
    \label{eq:yoyo_area_period}
\ee
Here \(\mathcal T_Y\) is the time between consecutive
fold-coincidence events. One such interval sweeps one rectangle with
splitting intensity \(\GammaSplit\,du\,dv\). Following labeled
folds for two rectangles doubles both the time and the area, leaving
the splitting rate unchanged.

Using the energy, momentum and mass in
Eqs.~\eqref{eq:yoyo_energy_momentum}--\eqref{eq:yoyo_mass}, the
boost and velocity are
\be
    \gamma_Y
    =
    \frac{E_Y}{m_Y}
    =
    \frac{\Delta u+\Delta v}
         {2\sqrt{\Delta u\Delta v}},
    \qquad
    \beta_Y
    =
    \frac{P_Y}{E_Y}
    =
    \frac{\Delta v-\Delta u}{\Delta u+\Delta v}.
    \label{eq:yoyo_gamma_beta}
\ee
The sign agrees with the center-of-mass displacement
between consecutive fold coincidences:
\(\Delta x=(\Delta v-\Delta u)/2\) and
\(\Delta t=(\Delta u+\Delta v)/2\).
The invariant mass is fixed by the area \(\Delta u\Delta v\), whereas
the boost is fixed by the aspect ratio \(\Delta u/\Delta v\).  This is
the reason the first-generation source can emit yo-yos with an
approximately fixed invariant scale while their boosts grow with the
emission time.

\subsection{Trace weight and pressure}
\label{subsec:yoyo_pressure}

Energy moves between a yo-yo's bulk and folds during each
oscillation, so its instantaneous pressure oscillates. The cascade
calculation follows the period average. Integrating the stress over
one oscillation gives the particle-like relation
\be
    \bar p_Y
    =
    \gamma_Ym_Y-\frac{m_Y}{\gamma_Y}
    =
    \gamma_Ym_Y\beta_Y^2
    =
    \frac{P_Y^2}{E_Y}.
\ee
\paragraph{Origin of the negative integrated pressure.}
Thus an isolated positive-energy yo-yo has non-negative period-averaged pressure,
\(\bar p_Y=P_Y^2/E_Y\geq 0\). The negative integrated pressure belongs to the combined zero-energy
state of the remnant and its descendants.

For any contribution \(\mathcal O\), define its trace weight
\be
    W_{\mathcal O}
    \equiv
    -\int_{\mathcal O}dx\,T^\mu{}_{\mu}
    =
    \int_{\mathcal O}dx\,
    \bigl(T_{00}-T_{11}\bigr)
    =
    4\int_{\mathcal O}dx\,T_{uv}.
    \label{eq:W_general_definition}
\ee
The null folds do not contribute to this combination.  It measures the
traceful bulk content carried by the remnant and the live finite yo-yos.
For a yo-yo at oscillation phase \(\tau\), the instantaneous weight is
\be
    W_Y^{\rm inst}(\tau)=E_Y-p_Y(\tau),
    \label{eq:instantaneous_yoyo_W}
\ee
and its period average is
\be
    W_Y
    \equiv
    \frac{1}{\mathcal T_Y}
    \int_0^{\mathcal T_Y}d\tau\,
    W_Y^{\rm inst}(\tau)
    =
    E_Y-\bar p_Y
    =
    \frac{m_Y}{\gamma_Y}
    =
    4\muIFS
    \frac{\Delta u\Delta v}{\Delta u+\Delta v}.
    \label{eq:W-def}
\ee
An unsplit oscillation has maximum bulk trace
\(4\muIFS\min(\Delta u,\Delta v)\), so
\be
 0\leq W_Y^{\rm inst}(t)\leq2W_Y.
 \label{eq:instantaneous-yoyo-trace-bound}
\ee
Removing bulk by fragmentation cannot increase this parent contribution.
The branching process transports $W_Y=m_Y/\gamma_Y$.
The exact instantaneous trace still oscillates within an individual
yo-yo period; Appendix~\ref{app:geometric-matching} controls its
ensemble asymptotic directly from physical split events, so no separate
assumption about phase randomization is needed.

The corresponding weight of the zeroth-generation remnant is
\be
    W_0(t)
    \equiv
    \int dx\,
    \bigl(T_{00}^{(0)}-T_{11}^{(0)}\bigr),
    \label{eq:remnant_W_definition}
\ee
and the exact total weight is
\be
    W_{\rm tot}(t)
    =
    W_0(t)
    +
    \sum_{g=1}^{\infty}
    \sum_{Y\in\mathcal Y_g^{\rm live}}
    W_Y^{\rm inst}(t).
    \label{eq:total_W_decomposition}
\ee
The complete IFS decay state has vanishing total energy.  Therefore
\be
    p_{\rm tot}(t)
    \equiv
    \int dx\,T_{11}(t,x)
    =
    E_{\rm tot}(t)-W_{\rm tot}(t)
    =
    -W_{\rm tot}(t).
    \label{eq:pressure_trace_relation}
\ee
This identity is the origin of the negative integrated pressure: the
positive energy carried by finite yo-yos is cancelled by the remnant and
fold sector, while the trace weight remains positive.

The trace weight \(W\) therefore measures the tension that remains after
cancellation of the large positive and negative energy contributions. The
decay problem is reduced to determining how rapidly fragmentation removes
this traceful content.

The renewal calculation uses the period-averaged weights \(W_Y\);
the exact time-slice identity above uses \(W_Y^{\rm inst}\).

\subsection{The remnant and its trace-weight source}
\label{subsec:zeroth_generation_remnant}
\label{subsec:first_generation_source}
The zeroth-generation remnant contributes directly to the pressure and continuously supplies
the first-generation yo-yos which seed the branching cascade. A point \((u,v)\) remains in the
original IFS bulk if and only if its causal past rectangle contains no splitting point. For
\(u,v>0\), the Poisson void probability derived in
Sec.~\ref{sec:cascade_algorithm} therefore gives
\be
    \left\langle
        T_{uv}^{(0)}(u,v)
    \right\rangle
    =
    \muIFS e^{-\GammaSplit uv}.
    \label{eq:mean_remnant_Tuv}
\ee
On a constant-time slice, \(u=t-x\), \(v=t+x\), and hence \(uv=t^2-x^2\). At late times the
corresponding remnant weight is
\be
    \left\langle W_0(t)\right\rangle
    =
    \frac{4\muIFS}{\GammaSplit t}
    \left[
        1+
        \mathcal O\!\left(
            \frac{1}{\GammaSplit t^2}
        \right)
    \right].
    \label{eq:remnant_weight_asymptotic}
\ee
The direct remnant contribution therefore decays as \(t^{-1}\).  Since
the yo-yo cascade will decay more slowly in the semiclassical scaling
regime, the remnant is subleading as a direct pressure contribution.
Its more important role is as a persistent source.

At late emission time \(\GammaSplit t_0^2\gg1\), the two null
arms supply asymptotically two first-generation yo-yos per unit
\(\ln t_0\):
\be
    \dot N_1(t_0)\sim\frac{2}{t_0}.
    \label{eq:first_generation_rate}
\ee
Their typical invariant mass remains of order
\(\muIFS/\sqrt{\GammaSplit}\), as established in
Sec.~\ref{subsec:original_remnant_first_generation}, while their
boost grows as \(\gamma_1=O(\sqrt{\GammaSplit}\,t_0)\).
Later emissions have increasingly elongated null rectangles at the
same invariant scale, suppressing the trace \(W_1=m_1/\gamma_1\)
carried by each yo-yo. The fixed-emission-time source measure derived
in App.~\ref{app:direct-source-measure} gives
\be
    \left\langle W_1\mid t_0\right\rangle
    \sim\frac{2\muIFS}{\GammaSplit t_0}.
    \label{eq:mean_first_generation_W}
\ee
This boost suppression supplies a second inverse power of time to
the trace-weight injection rate:
\be
    \dot W_1^{\rm src}(t_0)
    =\dot N_1(t_0)\left\langle W_1\mid t_0\right\rangle
    \sim\frac{4\muIFS}{\GammaSplit t_0^2}.
    \label{eq:first_generation_W_source}
\ee
Appendix~\ref{app:direct-source-measure} gives the full birth
distribution and the subleading corrections to these source laws.

These are direct remnant emissions. Their first splittings can be
non-dilute, so the population entering the later binary cascade has
different mass, boost and birth-time statistics. The entrance
construction in App.~\ref{app:threshold-entrance-checks} controls
this intermediate fragmentation; the renewal calculation keeps that
entrance population distinct from the first-generation source.

\subsection{Transfer of trace weight between later generations}
\label{subsec:finite_boost_map}

The later generations redistribute \(W=m/\gamma\) among descendants. In the semiclassical dilute window
\be
    \ell_s^2\ll\Delta u\Delta v\ll\GammaSplit^{-1},
    \label{eq:dilute-descendant-window}
\ee
Sec.~\ref{subsec:recursive_sampling} gives a single, uniformly
distributed split \((u_s,v_s)\) in an occupied oscillation to
leading order in \(\Lambda_Y=\GammaSplit\Delta u\Delta v\).
We write
\be
    \alpha_U
    \equiv
    \frac{u_s-u_a}{\Delta u},
    \qquad
    \alpha_V
    \equiv
    \frac{v_b-v_s}{\Delta v},
    \qquad
    \alpha_U,\alpha_V\sim{\cal U}(0,1),
    \label{eq:alpha-split-fractions}
\ee
with \(\alpha_U\) and \(\alpha_V\) independent.  The daughter extents
are
\be
    (\Delta u_1,\Delta v_1)
    =
    (\alpha_U\Delta u,\alpha_V\Delta v),
    \qquad
    (\Delta u_2,\Delta v_2)
    =
    \bigl((1-\alpha_U)\Delta u,
          (1-\alpha_V)\Delta v\bigr).
    \label{eq:binary_daughters}
\ee
This is the binary limit of the exact staircase construction derived in
Sec.~\ref{subsec:yoyo_splitting}.

The daughter mass fractions are
\be
    \frac{m_1}{m}
    =
    \sqrt{\alpha_U\alpha_V},
    \qquad
    \frac{m_2}{m}
    =
    \sqrt{(1-\alpha_U)(1-\alpha_V)},
    \label{eq:mass_fractions}
\ee

Define $q_A=W_A/W$. The component expressions for the boosts and trace
fractions are collected in Appendix~\ref{app:binary-component-map}; their
physical content follows directly from conservation.

The trace lost in a split measures the relative motion of the
daughters.  Write \(E=\sum_AE_A\), \(P=\sum_AP_A\),
\(v_A=P_A/E_A\), and \(\bar v=P/E\).  Since
\(W=m^2/E=E-P^2/E\), energy and momentum conservation give
\be
 W-\sum_AW_A
 =\sum_A\frac{P_A^2}{E_A}-\frac{P^2}{E}
 =\sum_AE_A(v_A-\bar v)^2\geq0.
 \label{eq:trace-velocity-variance}
\ee
For two daughters the loss is $E_1E_2(v_1-v_2)^2/E$.
Equal fractions leave both daughters at the parent's velocity
and preserve the trace.  Unequal fractions convert part of the trace
weight into kinematic pressure while conserving energy and momentum.

By reflection take \(\Delta u\geq\Delta v\). For a highly boosted parent, \(\gamma\gg1\),
\be 
q_1\longrightarrow\alpha_V, \qquad q_2\longrightarrow1-\alpha_V, \qquad q_1+q_2\longrightarrow1,\label{eq:large-boost-q-limit} 
\ee 
up to corrections that vanish as \(\gamma\to\infty\).
Highly boosted splittings therefore transport the total trace weight almost conservatively. 
The motion of this nearly conserved weight is most transparent in logarithmic
boost. Define
\be
    \Delta\ln\gamma_A
    \equiv
    \ln\frac{\gamma_A}{\gamma} .
\ee
At large boost,
\be
    \Delta\ln\gamma_1
    =
    \frac12\ln\frac{\alpha_U}{\alpha_V}
    +\mathcal O(\gamma^{-2}),
    \qquad
    \Delta\ln\gamma_2
    =
    \frac12\ln\frac{1-\alpha_U}{1-\alpha_V}
    +\mathcal O(\gamma^{-2}) .
    \label{eq:large-boost-boost-jumps}
\ee
Mass weighting has no preferred sign for these jumps, by
\(\alpha_U\leftrightarrow\alpha_V\) symmetry. Trace weighting favors the
less boosted daughter because \(W=m/\gamma\).

Using Eq.~\eqref{eq:large-boost-q-limit}, the trace-weighted drift is
\begin{align}
    \left\langle \Delta\ln\gamma \right\rangle_W
    &\equiv
    \left\langle
        \sum_A q_A\Delta\ln\gamma_A
    \right\rangle
    =
    \frac12
    \left\langle
        \alpha_V\ln\frac{\alpha_U}{\alpha_V}
        +(1-\alpha_V)
        \ln\frac{1-\alpha_U}{1-\alpha_V}
    \right\rangle
    =
    -\frac14 .
    \label{eq:W-weighted-boost-drift}
\end{align}
Thus highly boosted splits transport trace almost conservatively,
with mean drift toward smaller boost.  At finite boost,
Eq.~\eqref{eq:trace-velocity-variance} gives the depletion of this transported weight.

The binary branching rule is
\be
    (\ln\gamma,W)
    \longrightarrow
    (\ln\gamma_1,q_1W)
    +
    (\ln\gamma_2,q_2W).
    \label{eq:W_branching_kernel}
\ee
\section{Trace transport and the semiclassical scaling regime}
\label{sec:pressure}

The slow trace decay follows from the two effects identified in
Sec.~\ref{sec:yoyo_dynamics}. The remnant feeds highly boosted strings into the cascade.
Their splittings transport trace almost conservatively toward lower
boost, where relative daughter motion depletes it. But a daughter with
less trace also splits more slowly. The late pressure reflects the
balance between this slowing clock and the multiplication of branches.

We use the period-averaged yo-yo weight
$W=m/\gamma$ to expose the branching clock in a covariant dilute
binary reference process. For the entrance population constructed in
Appendix~\ref{appendix:first_generation_source}, its large-time limit describes the intermediate window
$1\ll\sqrt{\GammaSplit}\,t\ll g_s^{-1}$.  Appendix~\ref{app:geometric-matching}
then restores the full geometric multi-split dynamics and proves the
same leading result for the instantaneous ensemble pressure through the
physical split-event measure. This section studies the unabsorbed
semiclassical cascade; Appendix~\ref{app:gravitational-emission}
estimates the gravitational losses neglected in the model.

\subsection{Branch multiplication and the slowing clock}
\label{subsec:large-boost-selection}

A yo-yo sweeps worldsheet area $A_Y$ in period $\mathcal T_Y$.
Since $W_Y=2\muIFS A_Y/\mathcal T_Y$, its splitting rate and mean
laboratory lifetime are
\be
 \lambda_Y=\frac{\GammaSplit A_Y}{\mathcal T_Y}
 =\frac{\GammaSplit}{2\muIFS}W_Y,
 \qquad
 \tau(W_Y)=\frac{2\muIFS}{\GammaSplit W_Y}.
 \label{eq:W-lifetime}
\ee
Thus the weight that measures the surviving trace also sets the clock.
Each daughter inherits no more than the parent's trace and has a
correspondingly longer mean lifetime.

First set aside these waits and count the available mass hierarchy.
The independent uniform null fractions give daughter mass fractions
\be
 z_1=\sqrt{\alpha_U\alpha_V},\qquad
 z_2=\sqrt{(1-\alpha_U)(1-\alpha_V)}.
\ee
Their branching moment is
\be
 M(p)=\langle z_1^p+z_2^p\rangle
     =\frac{2}{(1+p/2)^2}.
 \label{eq:branching-mass-moment}
\ee
A split doubles the branch count, $M(0)=2$, while the expected sum
of rest masses decreases, $M(1)=8/9$. Energy is conserved; relative
motion accounts for the difference. The intermediate power which
balances multiplication and shrinking masses satisfies
\be
 M(\kappa_{\rm IFS})=1,\qquad
 \kappa_{\rm IFS}=2(\sqrt2-1),\qquad
 \left\langle m_1^{\kappa_{\rm IFS}}+
 m_2^{\kappa_{\rm IFS}}\mid m\right\rangle=m^{\kappa_{\rm IFS}}.
 \label{eq:malthusian-condition}
\ee
The expected $\kappa_{\rm IFS}$-weighted mass is therefore conserved
across generations.
Since $M(p)$ decreases strictly, its positive balancing exponent is unique. For a hierarchy stopped at a lower mass cutoff, with initial
mass-to-cutoff ratio $\mathcal C$, Appendix~\ref{app:mass-hierarchy-count}
then gives
\be
 \mathcal N(\mathcal C)\sim C_N\mathcal C^{\kappa_{\rm IFS}}.
 \label{eq:malthusian-node-growth}
\ee

The first source-scale strings can undergo several splits in one
oscillation. As fragmentation reduces their mass, the worldsheet area
per period decreases and such events become rare. Deep generations
therefore approach the dilute binary law. The earlier evolution supplies
the strings entering this regime and fixes their overall contribution;
the repeated binary law determines the exponent and normalized limiting
boost profile. Appendix~\ref{appendix:first_generation_source} derives
this entrance population, and Appendix~\ref{app:geometric-matching}
establishes its matching to the full geometric cascade.

\subsection{From the branching hierarchy to pressure}
\label{subsec:chronological-pressure}

The physical argument has four steps. The mass hierarchy supplies
the power $\kappa_{\rm IFS}$. The splitting clock turns that hierarchy
into an accumulated number of events growing as
$t^{\kappa_{\rm IFS}}$. Averaging over source boosts preserves the
same power. Finally, the instantaneous splitting rate measures the
surviving trace, giving
$W_{\rm live}\sim t^{\kappa_{\rm IFS}-1}$.
A late daughter's birth time contains the correlated waits of all its
ancestors. Retaining those waits determines the normalization as well
as the scaling law.

\paragraph{Split count and surviving trace.}
Each live yo-yo splits at rate
$\GammaSplit W_Y/(2\muIFS)$. Summing these rates gives,
within the period-averaged binary model,
\be
 \frac{d}{dt}\langle N_{\rm split}(t)\rangle
 =\frac{\GammaSplit}{2\muIFS}W_{\rm live}(t),
 \label{eq:chronological-split-rate}
\ee
where $N_{\rm split}(t)$ counts descendant splitting events before
$t$. More surviving tension gives a larger mean splitting rate.
Moreover, every dilute node contributes the same expected
time-integrated trace,
\be
 W_Y\tau_Y=\frac{2\muIFS}{\GammaSplit},
 \label{eq:node-integrated-trace-area}
\ee
independently of its mass and boost. Equation~\eqref{eq:node-integrated-trace-area} links branch counting directly to the surviving trace.

\paragraph{The inherited splitting clock.}
The inherited waiting times simplify when each node is measured in its own
lifetime units. For a node of mass $m$ and signed velocity
rapidity $y$, define
\be
 U=\frac{\GammaSplit m}{2\muIFS}e^y u,
 \qquad
 V=\frac{\GammaSplit m}{2\muIFS}e^{-y}v .
 \label{eq:main-chronological-birth-coordinates}
\ee
If the parent waits a dimensionless proper time
$\epsilon\sim{\rm Exp}(1)$ and a selected daughter receives null
momentum fractions $a,b$, so that its mass fraction is
$z=\sqrt{ab}$, then its inherited birth coordinates obey the exact
affine recursion
\be
 (U,V)\longmapsto
 \bigl(a(U+\epsilon),\,b(V+\epsilon)\bigr).
 \label{eq:main-chronological-affine-recursion}
\ee
Thus each fresh waiting time enters at order unity in the daughter's
own lifetime units, while the accumulated timing information inherited
from earlier generations is multiplied by shrinking daughter fractions.
Appendix~\ref{app:chronological-contraction} makes this loss of memory
precise and proves that the rescaled birth-coordinate process approaches
a stationary law with the inverse-radius moments needed below.  The
ancestral waits therefore renormalize the coefficient, but introduce no
new leading power or logarithm.

\paragraph{Boost averaging and the scaling exponent.}
An event at invariant radius $\rho=\sqrt{uv}$ lies on the boost
orbit $u=\rho e^\varphi$, $v=\rho e^{-\varphi}$, with laboratory time
$t=\rho\cosh\varphi$. The contribution of the complete boost orbit
to a laboratory-time event rate is exactly
\be
 K_\rho(t)
 \equiv\int_{-\infty}^{\infty}d\varphi\,
       \delta(t-\rho\cosh\varphi)
 =\frac{2\Theta(t-\rho)}{\sqrt{t^2-\rho^2}},
 \label{eq:main-chronological-boost-kernel}
\ee
and its Mellin transform is
\be
 \int_\rho^\infty dt\,t^{-\zeta}K_\rho(t)
 =\mathrm B\!\left(\frac{\zeta}{2},\frac12\right)\rho^{-\zeta}.
 \label{eq:main-chronological-boost-mellin}
\ee
The beta function is finite and analytic at
$\zeta=\kappa_{\rm IFS}$.  Hence boost averaging changes the
normalization but not the selected power.  The same is true of the
stationary chronological clock: the only nonanalytic factor at the
leading exponent comes from repeated branching.

\paragraph{The branching pole and the resulting pressure.}
Set $t_*=\GammaSplit^{-1/2}$ and $T=t/t_*$, and define the
dimensionless live-trace response
$\mathcal W(T)=\sqrt{\GammaSplit}\,W_{\rm live}(t_*T)/(2\muIFS)$.
Appendix~\ref{appendix:chronological-renewal} shows that its
Mellin transform,
$\widetilde{\mathcal W}(\zeta)
\equiv\int_0^\infty dT\,T^{-\zeta}\mathcal W(T)$,
has the form
\be
 \widetilde{\mathcal W}(\zeta)
 =\mathrm B\!\left(\frac{\zeta}{2},\frac12\right)
 \left[\frac{C(\zeta)}{1-M(\zeta)}+H(\zeta)\right],
 \qquad C(\kappa_{\rm IFS})>0,
 \label{eq:chronological-pressure-pole}
\ee
where $C$ and $H$ are analytic at the Malthusian line.  Thus the only
leading nonanalyticity is the simple branching pole selected by
$M(\kappa_{\rm IFS})=1$; the waiting-time and boost integrals change its
residue but cannot generate a leading logarithm.  Equivalently,
\be
 \langle N_{\rm split}(t_*T)\rangle
 \sim\frac{D}{\kappa_{\rm IFS}}T^{\kappa_{\rm IFS}},
 \qquad
 \mathcal W(T)\sim D T^{\kappa_{\rm IFS}-1}.
 \label{eq:chronological-fixed-time-scaling}
\ee
Hence
\be
 q_{\rm IFS}=1-\kappa_{\rm IFS}=3-2\sqrt2,
 \qquad
 W_{\rm live}(t)=C_P\muIFS\GammaSplit^{-1/2}
       (\sqrt{\GammaSplit}\,t)^{-q_{\rm IFS}}[1+o(1)].
 \label{eq:pressure-law}
\ee
The amplitude depends on the entrance population, while the power does
not.  For a fixed mass-threshold entrance, the direct remnant and the
pre-entrance trace are both $O(t^{-1})$ and are subleading.

Appendix~\ref{app:geometric-matching} proves that the same simple pole
survives the exact geometric cascade.  In particular, the local
physical split intensity is proportional to the instantaneous
$\langle T_{uv}\rangle$; an exact Abel relation between the radial
split-event measure and a constant-time slice then gives the same
leading coefficient for the instantaneous ensemble pressure.

\subsection{Boost distribution of the surviving trace}
\label{subsec:finite-boost-profile}

The total trace fixes the surviving weight, while its boost distribution
identifies the strings that carry it. Write
$\rho_W(t,\gamma)$ for the trace density per unit $\ln\gamma$, so that
$\rho_W(t,\gamma)\,d\ln\gamma$ is the trace in a logarithmic boost bin. Each string contributes $m/\gamma$; its birth history and waiting time
determine whether it is still present on the slice.

Let $y$ denote a string's signed velocity rapidity, $\gamma=\cosh y$.
It need not equal the spacetime rapidity of its birth event.
Consequently a boost-invariant entrance source does not give a
flat velocity histogram on a Minkowski-time slice.
Appendix~\ref{app:chronological-snapshot-profile} derives a
parameter-free occupation factor
$\mathcal F_\kappa^{\rm occ}(w)$ from the stationary birth history
and final wait, with $\kappa=\kappa_{\rm IFS}$; its explicit
definition is Eq.~\eqref{eq:snapshot-occupation-correction}.
The normalized trace profile is
\be
 \mathcal R_\kappa(\gamma)=
 \frac{2}{\mathcal Z_\kappa}
 \frac{\gamma^{-\kappa}}{\sqrt{1-\gamma^{-2}}}
 \mathcal F_\kappa^{\rm occ}(w_\gamma),
 \qquad
 w_\gamma=\frac{1+\sqrt{1-\gamma^{-2}}}{2},
 \label{eq:main-normalized-snapshot-profile}
\ee
Here $w_\gamma=(1+|\beta|)/2$ depends only on speed.
The constant $\mathcal Z_\kappa$ fixes
$\int_1^\infty\mathcal R_\kappa(\gamma)\,d\ln\gamma=1$. The result is
\be
 \frac{\rho_W(t,\gamma)\,d\ln\gamma}{W_{\rm live}(t)}
 \Longrightarrow\mathcal R_\kappa(\gamma)\,d\ln\gamma.
 \label{eq:main-snapshot-weak-limit}
\ee
Because $m=\gamma W$, the corresponding sum-of-rest-masses spectrum is
\be
 \frac{dM}{d\ln\gamma}(t,\gamma)
 \sim W_{\rm live}(t)\gamma\mathcal R_\kappa(\gamma).
 \label{eq:numerical-mass-profile}
\ee

The three factors have distinct meanings. The branching weight gives
$\gamma^{-\kappa}$, and conversion from trace to mass changes this to
$\gamma^{1-\kappa}=\gamma^{q_{\rm IFS}}$. The denominator is the
Jacobian from signed rapidity to logarithmic boost. The occupation
factor accounts for the correlated birth histories and lifetimes;
it approaches a finite positive constant at large boost.
Normalizing the profile by the total trace removes the entrance strength.

At small boost, the rapidity Jacobian gives the integrable enhancement
$\mathcal R_\kappa(\gamma)\propto(\gamma-1)^{-1/2}$.
Equation~\eqref{eq:main-snapshot-weak-limit} fixes the distribution on
each fixed boost interval. The total mass also samples the growing
source region at $\gamma_{\rm src}\sim\sqrt{\GammaSplit}t$;
Appendix~\ref{appendix:moment-normalization} treats its contribution.

\subsection{Physical consequences and numerical tests}
\label{subsec:pressure-mass-spectrum}

Energy, summed rest mass, and trace probe different parts of the
population: one string contributes $\gamma m$, $m$, and $m/\gamma$,
respectively. A large boost enhances its energy but suppresses its
trace relative to its rest mass. Here $M_{\rm inv}=\sum_Ym_Y$
denotes the sum of live descendant rest masses, not the invariant
mass of their total four-momentum.

The live boost distribution fixes the mass spectrum through
\(m=\gamma W\).  With \(1-\kappa_{\rm IFS}=q_{\rm IFS}\), the limiting
profile has the large-boost scaling

\begin{align}
 \rho_W(t,\gamma)&\propto t^{-q_{\rm IFS}}\gamma^{-\kappa_{\rm IFS}},
 \qquad
 \frac{\dd M}{\dd\ln\gamma}=\gamma\rho_W(t,\gamma)
 \propto t^{-q_{\rm IFS}}\gamma^{q_{\rm IFS}}.
 \label{eq:dMdlngamma-scaling}
\end{align}
The temporal pressure exponent and the boost-space mass-spectrum
slope are therefore the same.  The occupation factor changes the
finite-boost shape while preserving this exponent relation.

\begin{figure}[t!]
\centering
\includegraphics[width=0.98\textwidth]{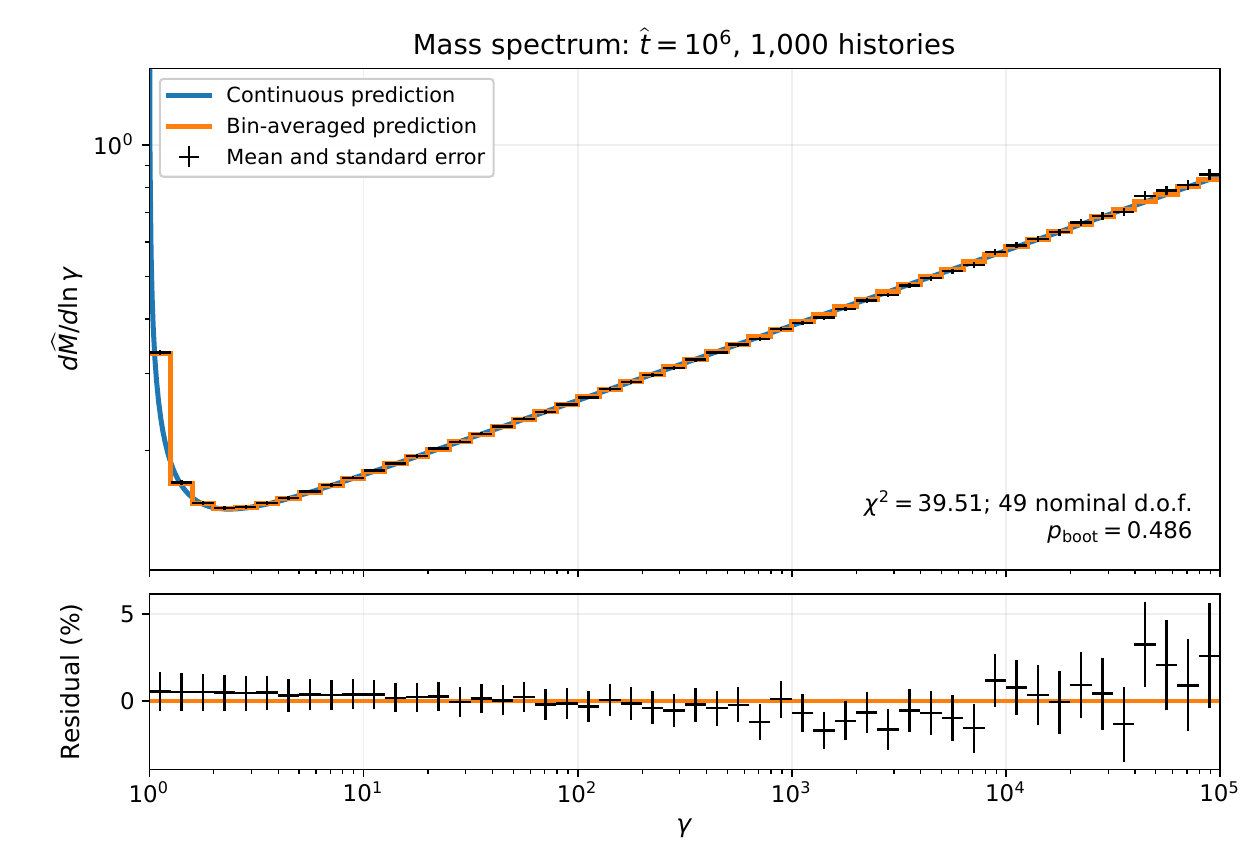}
\caption{Sum-of-rest-masses spectrum at
$\sqrt{\GammaSplit}t=10^6$, averaged over 1,000 histories without
terminal absorption, in 50 logarithmic bins over $1\leq\gamma\leq10^5$.
Here $\widehat M=\sqrt{\GammaSplit}M/(2\muIFS)$.
The blue curve is the predicted density, the orange staircase its
bin averages, and the black crosses the data. The lower panel gives
percentage residuals from the bin prediction. Error bars are correlated
standard errors of the mean. The fit gives $p_{\rm boot}=0.486$;
its template and statistical procedure are given in
App.~\ref{app:mass-histogram-methods}.}
\label{fig:renewal-mass-profile}
\end{figure}

Figure~\ref{fig:renewal-mass-profile} compares the predicted mass spectrum
with 1,000 unabsorbed cascade histories at dimensionless time $\widehat t=\sqrt{\GammaSplit}\,t=10^6$.
With only the overall amplitude fitted, the calculated shape describes both the
low-boost enhancement and the rising spectrum across five decades in
boost. The upper edge, $\gamma=10^5$, lies a factor of ten below the
characteristic source scale. Appendix~\ref{app:mass-histogram-methods}
gives the correlated fit and its sensitivity checks. An independent
binary-cascade simulation also tests the chronological profile and trace
amplitude without a fitted normalization
(App.~\ref{app:binary-chronological-benchmark}).

The complete population also extends to the moving source scale
\be
    \gamma_{\rm src}(t)
    \sim
    \sqrt{\GammaSplit}\,t
    \label{eq:gamma-source-scale}
\ee
Energy and total invariant mass sample this growing boost range.
Their integrated behavior is constrained separately by conservation
and the full source history.

\paragraph{Energy, summed rest mass, and pressure.}
The different weights allow descendant energy to grow while trace
decays. The binary calculation also predicts a plateau in summed rest
mass under the source conditions of App.~\ref{appendix:moment-normalization}:
\begin{align}
 E_{\rm yoyo}(t)
 &=2\muIFS t-\frac{2\muIFS}{\GammaSplit t}+\mathcal O(t^{-3}),
 \label{eq:three-moment-energy-summary-main}\\
 M_{\rm inv}(t)
 &=M_\infty-B\widehat t^{-q_{\rm IFS}}
                  +o(\widehat t^{-q_{\rm IFS}}),
 \label{eq:main-invariant-mass-plateau}\\
 |p_{\rm tot}(t)|
 &=C_W(\sqrt{\GammaSplit}t)^{-q_{\rm IFS}}
   [1+o(1)].
 \label{eq:three-moment-pressure-summary-main}
\end{align}
The first relation is an energy sum rule: later splittings redistribute energy among
descendants but do not change the energy injected by the first-generation
source. The remnant and fold sector balances this positive descendant energy,
preserving $E_{\rm tot}=0$ for the complete state.

In the chronological binary model,
Appendix~\ref{appendix:moment-normalization} proves the plateau and its
leading approach from below under its additional mass-moment and source-tail
conditions. In that model the surviving tension and remaining mass deficit
share an exponent, and Eq.~\eqref{eq:mass-trace-amplitude-ratio} predicts
\(B/A\simeq3.08\) when \(W_{\rm live}\sim A\widehat t^{-q_{\rm IFS}}\).
The geometric simulations below test these integrated-mass predictions.
An analytic extension of the mass coefficient to the geometric process
requires matching its mass-loss weight and controlling the pre-entrance
mass remainder, as specified in that appendix.
Since \(q_{\rm IFS}<1\), the live cascade dominates the pressure tail
over the direct remnant contribution of order \(t^{-1}\).

\begin{figure}[t!]
\centering
\includegraphics[width=0.98\textwidth]{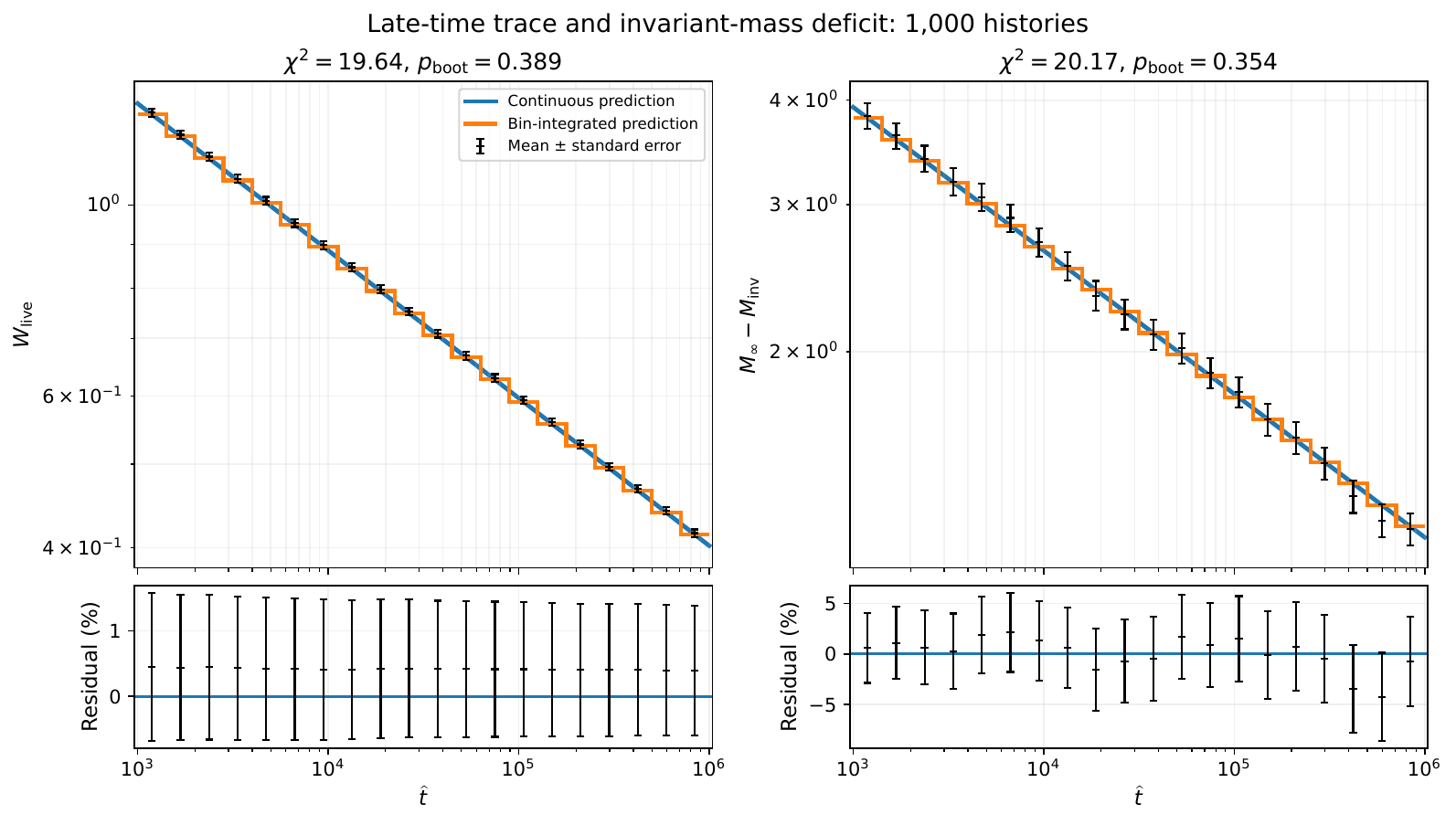}
\caption{Live-trace decay and invariant-mass deficit in
1,000 histories, with \(q=3-2\sqrt2\) fixed and
\(\muIFS=\GammaSplit=1\). The horizontal coordinate is
\(\widehat t=\sqrt{\GammaSplit}t\); the vertical coordinates show
\(W\) and \(M\) in these physical units.
The left panel fits \(W_{\rm live}=A\widehat t^{-q}+C/\widehat t\);
the right shows the mass deficit from the fitted plateau.
Blue curves are fitted predictions, orange staircases their time-bin
averages, and black crosses the data with correlated standard errors.
The lower panels show percentage residuals. Fit statistics appear in
Table~\ref{tab:direct-time-predictions}, with the averaging and error
analysis in App.~\ref{app:time-domain-fits}.
The live trace excludes the direct remnant contribution.}
\label{fig:renewal-moment-diagnostics}
\end{figure}

\paragraph{Direct tests of the time dependence.}
We test these predictions with \(q=q_{\rm IFS}\) fixed, fitting
\be
 W_{\rm live}=A\widehat t^{-q}+C/\widehat t,
 \qquad M_{\rm inv}=M_\infty-B\widehat t^{-q}.
 \label{eq:direct-time-fit-models}
\ee
The comparison uses 1,000 histories over
$10^3\leq\widehat t\leq10^6$, with results in
Table~\ref{tab:direct-time-predictions} and
Fig.~\ref{fig:renewal-moment-diagnostics}.
Appendix~\ref{app:time-domain-fits} gives the interval averaging,
correlated fits, and amplitudes.

\begin{table}[t!]
\centering
\renewcommand{\arraystretch}{1.14}
\begin{tabular}{lrrr}
\toprule
Fixed-exponent test & $\chi^2$ & Nominal d.o.f. & $p_{\rm boot}$\\
\midrule
Live trace, leading power only & 36.10 & 19 & 0.013\\
\addlinespace[0.22em]
Live trace, including $C/\widehat t$ & 19.64 & 18 & 0.389\\
\addlinespace[0.22em]
Invariant mass, fitted plateau & 20.17 & 18 & 0.354\\
\addlinespace[0.22em]
Joint, corrected trace and predicted $B/A$ & 41.45 & 37 & 0.355\\
\bottomrule
\end{tabular}
\caption{Direct tests with $q_{\rm IFS}=3-2\sqrt2$ fixed, using
1,000 histories and 20 logarithmic time intervals. The joint row
imposes $B/A=3.0774$ and fits $A,C,M_\infty$. The probabilities
are calibrated by whole-history resampling.}
\label{tab:direct-time-predictions}
\end{table}

The increased precision resolves a small negative finite-time correction
to the trace. Including it gives agreement with the predicted leading
exponent and mass-to-trace amplitude ratio. The fitted $C/\widehat t$
term describes the observed window; the asymptotic subleading power
remains undetermined. Spatial moments also resolve small departures
from their leading ratios (App.~\ref{app:spatial-moment-fits}).

Fluctuations in the early remnant leave a persistent imprint on the
late cascade: histories with weaker early emission retain a smaller
trace at later times. Most of the trace variance is a common amplitude
fluctuation, so changes of shape are measured more precisely than
individual levels. Appendix~\ref{app:source-fluctuations} quantifies
this memory of the source.

These simulations omit terminal absorption and test the semiclassical
predictions in the intermediate string regime
\(1\ll\sqrt{\GammaSplit}t\ll g_s^{-1}\).

The cascade has now determined how quickly the surviving tension decreases
on a time slice. Conservation relates this same tension to the positive flux
still missing from a truncated null ray, so the pressure law also fixes the
approach to null cancellation.

\section{Cutoff ANEC convergence}\label{sec:anec_moments}

A truncated null ray has crossed the original negative fold but has
not yet collected all the compensating positive flux.  Its residual
measures how much cancellation is still to come.  We determine that
residual from the surviving trace, using the pressure law of
Sec.~\ref{sec:pressure}.  The numerical diagnostics are ensemble
averages; the related equal-time moment tests are collected in
App.~\ref{app:stress-moments-terminal-anec}.

\subsection{From a time slice to a null segment}
\label{subsec:cutoff_anec}

The pressure and the null deficit sample the same positive trace
profile on different cuts. The cutoff observable is
\be
 \mathcal A(v;U)=\int_{-\infty}^{U}du\,T_{uu}(u,v).
\ee
Here $U$ advances along a null ray and $v$ labels which ray is
measured. Conservation gives
\be
 \partial_v\mathcal A(v;U)=-T_{uv}(U,v).
 \label{eq:Iuu-conservation}
\ee
Since the stress vanishes before nucleation and $T_{uv}\geq0$,
$\mathcal A(v;U)\leq0$ history by history. Thus the numerical
absolute-value diagnostic satisfies
$\langle|\mathcal A|\rangle=-\langle\mathcal A\rangle$.

Write $\langle T_{uv}(u,v)\rangle=f(uv)$ and
$I(S)=\int_0^Sf(z)\,dz$. Conservation fixes the null residual,
$\langle\mathcal A(v;U)\rangle=-I(Uv)/U$.
The pressure measures this same profile on $z=t^2-x^2$.
Changing variables on that time slice gives
\be
 W(t)\equiv-\langle p_{\rm tot}(t)\rangle
 =4\int_0^{t^2}\frac{f(z)}{\sqrt{t^2-z}}\,dz.
 \label{eq:pressure-abel-trace}
\ee
The square-root factor is the spatial length occupied by each
interval of invariant age. Applying the same kernel a second time
removes that geometric weight:
\be
 I(S)=\frac{1}{4\pi}\int_0^S
             \frac{W(\sqrt r)}{\sqrt{S-r}}\,dr,\qquad S>0.
 \label{eq:pressure-abel-inversion}
\ee
Indeed, exchanging the two nonnegative integrals leaves
$\int_z^S[(r-z)(S-r)]^{-1/2}dr=\pi$.
Pressure therefore determines the accumulated trace and the null
deficit without requiring reconstruction of the local density $f$.

Put $q=q_{\rm IFS}$. The pressure law is
$W(t)\sim A_Wt^{-q}$, with
$A_W=C_P\muIFS\GammaSplit^{-(1+q)/2}>0$.
Substituting $r=Sa$ gives
\be
 I(S)\sim\frac{A_W}{4\pi}
       {\rm B}(1-q/2,1/2)S^{(1-q)/2}.
 \label{eq:integrated-trace-from-pressure}
\ee
The limit follows directly by separating a fixed early interval,
whose contribution is $O(S^{-1/2})$, from the late tail. On the tail
the pressure asymptotic is uniform, and
$a^{-q/2}(1-a)^{-1/2}$ is integrable at both endpoints.
Consequently
\be
 \langle\mathcal A(v;U)\rangle
 \sim-C_A U^{-(1+q)/2}v^{(1-q)/2},\qquad
 C_A=\frac{A_W}{4\pi}{\rm B}(1-q/2,1/2).
 \label{eq:anec-pressure-amplitude}
\ee
The limit is $Uv\to\infty$ in the unabsorbed boost-invariant model.
For finite-coupling strings, it describes the intermediate
invariant-age window and inherits the pressure theorem's entrance
and averaging assumptions.

For comparison with the two-variable numerical fit, define
\be
 \langle\mathcal A(v;U)\rangle\sim-C_AU^\alpha v^\beta,\qquad
 \alpha=\sqrt2-2,\quad\beta=\sqrt2-1.
 \label{eq:ANEC-predicted-exponents}
\ee
Their difference is fixed by the boost weight and their sum by the
decay of trace:
\be
 \beta-\alpha=1,\qquad \alpha+\beta=-q_{\rm IFS}.
 \label{eq:beta-alpha-relation}
\ee
The same branching exponent also explains
$\beta=\kappa_{\rm IFS}/2$: the invariant radius is $\sqrt{Uv}$.
Growth with the ray label $v$ and decay with the advancing cutoff
$U$ are therefore compatible. Equations~\eqref{eq:anec-pressure-amplitude}
and \eqref{eq:ANEC-predicted-exponents} fix both the scaling and its
normalization for the numerical test.

\begin{figure}[!b]
\centering
\includegraphics[width=0.98\textwidth]{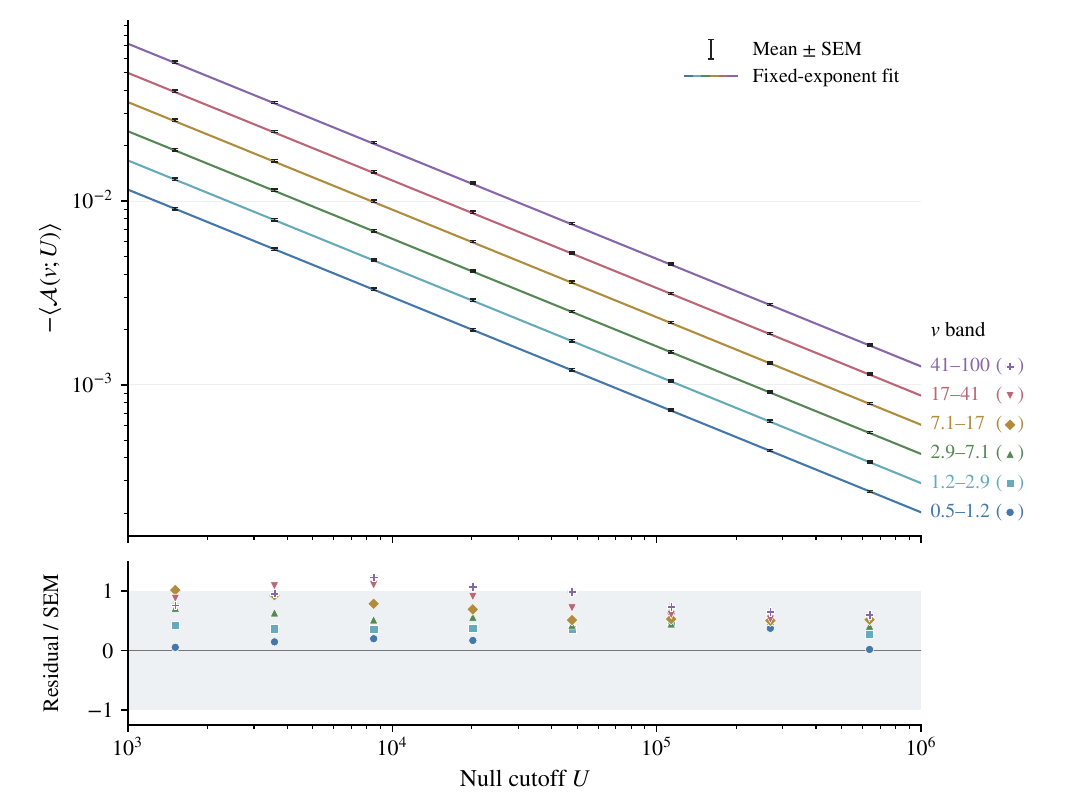}
\caption{
Cutoff ANEC convergence in the unabsorbed semiclassical cascade,
with $\muIFS=\GammaSplit=1$. The 1,000-history ensemble is averaged over
six logarithmic $v$ bands and eight logarithmic $U$ intervals.
Black crosses mark bin means and their marginal standard errors;
the short horizontal strokes locate the means and do not denote
horizontal uncertainty. Blue curves show
$C_AU^{-(1+q)/2}v^{(1-q)/2}$ averaged over each $v$ band, with
$q=3-2\sqrt2$ fixed and one common fitted amplitude $C_A=0.72978$.
Each mean is plotted at the effective cutoff where the continuous
curve equals its bin-averaged prediction; data values and errors are
not rescaled. Both this finer 48-cell fit and the fiducial 30-cell comparison use the full covariance; their statistics appear in
App.~\ref{app:anec-fit-calibration}.}
\label{fig:broad-anec-fit}
\end{figure}

The cancellation also holds history by history. After the original
negative fold is crossed, each subsequent positive contribution reduces
the deficit. Its vanishing ensemble limit therefore implies saturation
with probability one. Appendix~\ref{app:longitudinal-completion}
gives the argument simultaneously for all longitudinal rays.

\subsection{\texorpdfstring{Numerical test of the ANEC prediction}{Numerical test of the ANEC prediction}}
\label{subsec:anec-numerical-test}

We test Eq.~\eqref{eq:anec-pressure-amplitude} on 1,000 independent
cascade histories evolved without terminal absorption, using
$\muIFS=\GammaSplit=1$. The fiducial fit averages the data into five
logarithmic $v$ bands over $0.5\leq v\leq100$ and six logarithmic
$U$ intervals over $10^3\leq U\leq10^6$. We apply the same averaging
to the prediction, fix $q=3-2\sqrt2$, and fit only $C_A$ using the
full covariance of the 30 averages. App.~\ref{app:anec-fit-calibration}
reports the amplitude and fit statistics; the fiducial whole-history fit gives \(p_{\rm boot}=0.797\).
Figure~\ref{fig:broad-anec-fit} shows a finer binning of the same
histories. All tested windows and binnings remain compatible with the
prediction; the procedure and checks are given in
App.~\ref{app:anec-fit-calibration}.

Allowing $q$ to vary while retaining the boost relation
$\beta-\alpha=1$ gives $q=0.17135$, with a 95\% bootstrap interval
$[0.17084,0.17183]$, containing the predicted $0.17157$.
The improvement is only $\Delta\chi^2=0.802$, with nested-bootstrap
$p_\Delta=0.392$. Thus these data support the predicted exponent
without requiring an additional fitted slope.

The pressure fit supplies a further test of the normalization.
Using $A_W=4.30203$ from the corrected trace fit in
App.~\ref{app:time-domain-fits},
Eq.~\eqref{eq:anec-pressure-amplitude} predicts
$C_A^{(W)}=0.72372$. The fiducial ANEC fit gives
\be
 \frac{C_A}{C_A^{(W)}}=1.010,
 \qquad 95\%\text{ paired-bootstrap interval: }[0.991,1.026].
 \label{eq:anec-pressure-normalization-test}
\ee
The paired amplitudes agree within their uncertainty. They probe
different age ranges: the ANEC endpoints have
invariant ages $\sqrt{Uv}\leq10^4$, whereas the pressure reference is
fitted over $10^3\leq\widehat t\leq10^6$.

\section{Terminal decay in
\texorpdfstring{\(1+1\)}{1+1} dimensions}
\label{sec:terminal-decay}

The semiclassical cascade eventually reaches masses
\(m=O(m_s)\), with \(m_s\sim\alpha'^{-1/2}\), where microscopic
light-string dynamics takes over. We allow a finite-width crossover
localized near this scale. The \(1/t\) pressure result applies to terminal
completions for which each finite tree---one emitted yo-yo together with all
its descendants---has a finite integrated trace response and the final
products are traceless.

The remnant continues to emit new trees at arbitrarily late times, even
after earlier trees have decayed, supplying an order-one number per
logarithmic interval of time. Under the scale-covariance and source-tail
conditions below, convolving this continuing source with the finite response
of its trees gives a \(1/t\) pressure tail. The terminal dynamics determines
its coefficient and the properties of the emitted radiation.

Stable free massive products instead belong to the broader completed-flux
class treated in Sec.~\ref{sec:higher-dimensional-extension}: they retain
trace $m_j^2/E_j$ indefinitely and therefore do not satisfy the finite-response
assumption used for the pressure law.

The analysis proceeds from the response of one tree to its convolution with the continuing source and finally to the crossover in a sharp absorbing reference model.
Unless stated otherwise, tree responses and late-time pressure laws
are ensemble-averaged over cascade histories and period-averaged over
yo-yo oscillations; we suppress ensemble brackets on asymptotic
pressure laws.

\subsection{Node counting and finite-tree response}
\label{subsec:terminal-finite-tree-response}

The splitting clock~\eqref{eq:W-lifetime} gives each dilute
yo-yo the same expected integrated trace:
\be
    W\tau(W)
    =
    \frac{2\muIFS}{\GammaSplit}.
    \label{eq:universal-node-trace-lifetime}
\ee
A node with less trace weight waits proportionally longer
before splitting.  The two effects cancel in its time integral:
every node in the dilute self-similar part of the cascade contributes
the same expected integrated trace weight, independently of its mass
and boost.  Summing a complete tree's response therefore amounts
to counting its splitting nodes.
If the early stages of each tree supply a finite, nonzero
\(\kappa_{\rm IFS}\)-weighted entrance mass, independent of the
terminal hierarchy, they modify the coefficient of the complete-tree
response.

Section~\ref{subsec:chronological-pressure} used waiting times
to determine which descendants are reached by a given observation
time.  Here we integrate each tree over its full history.  The cutoff
is instead the terminal mass: we count the semiclassical nodes
produced before the string-scale crossover. Write
\(\xi=m/m_{\rm term}\), with \(m_{\rm term}=O(m_s)\).
The complete-tree response, with \(T\) measured from the root's birth, is
\be
    \mathcal R_{\rm tree}(\xi)
    \equiv
    \int_0^\infty dT\,
    \bigl\langle W_{\rm tree}(T;\xi)\bigr\rangle .
    \label{eq:terminal-tree-response-definition}
\ee
Here \(W_{\rm tree}\) is the spatially integrated trace of that one
tree at age $T$, including the trace carried during terminal evolution
before conversion. Thus $\mathcal R_{\rm tree}$ measures its response
accumulated over its entire lifetime. In the sharp absorbing model only live
semiclassical nodes contribute, and the expected number
\(\mathcal N_{\rm int}(\xi)\) of above-threshold splitting nodes is
the stopped-tree count of App.~\ref{app:mass-hierarchy-count}.
Its recurrence and Laplace transform are derived there; the result is
\be
 \mathcal N_{\rm int}(\xi)\sim C_N^{\rm term}\xi^{\kappa_{\rm IFS}},
 \qquad \kappa_{\rm IFS}=2(\sqrt2-1)=1-q_{\rm IFS}.
 \label{eq:terminal-multiplicity-exponent}
\ee
Multiplying this stopped-tree count by the response per node gives
\be
 \mathcal R_{\rm tree}^{\rm sharp}(\xi)
 =\frac{2\muIFS}{\GammaSplit}\mathcal N_{\rm int}(\xi),\qquad
 \mathcal R_{\rm tree}(\xi)
 \sim\frac{2\muIFS}{\GammaSplit}C_{\rm term}\xi^{\kappa_{\rm IFS}}.
 \label{eq:terminal-tree-response-scaling}
\ee
The smooth-crossover form assumes the finite entrance weight above
and a terminal law localized near $\xi=O(1)$, with no additional
parametric lifetime hierarchy. Terminal dwell times then change
$C_{\rm term}$ while preserving the hierarchy exponent.
Appendix~\ref{app:terminal-kernel} calculates their contribution in a
finite-rate example and gives the renewal equation for competing rates.

\subsection{Persistent source and the universal
\texorpdfstring{\(1/t\)}{1/t} tail}
\label{subsec:terminal-persistent-source}

Summing these tree responses over successive emissions of one IFS remnant gives the complete descendant response. Time $t$ is measured from the IFS nucleation, while $t_0$
is the birth time of an emitted yo-yo. At late $t_0$, the remnant produces
first-generation trees at the asymptotic rate
\be
    \dot N_1(t_0)
    =
    \frac{2}{t_0}
    \left[1+o(1)\right].
    \label{eq:terminal-logarithmic-source}
\ee
Their invariant-mass distribution approaches a time-independent form,
while their trace weight scales as \(t_0^{-1}\).  Their complete
response is therefore scale covariant: increasing the emission time by
a factor \(\lambda\) stretches the complete tree history by the same
factor and reduces its instantaneous trace weight by \(\lambda^{-1}\).
After averaging over the masses and boosts supplied by this
first-generation source, the response of one emitted tree takes the form
\be
    \left\langle
        W_{\rm tree}(t;t_0)
    \right\rangle_{\rm src}
    =
    \frac{1}{t_0}
    \mathcal F\!\left(\frac{t-t_0}{t_0}\right),
    \qquad
    \int_0^\infty dy\,\mathcal F(y)
    =
    \left\langle\mathcal R_{\rm tree}\right\rangle_{\rm src}.
    \label{eq:terminal-scale-covariant-response}
\ee
The variable \(y=(t-t_0)/t_0\) measures a tree's age relative
to its emission time. Summing over all emission times gives
$W_{\rm desc}(t)$, the total descendant trace of this single IFS.
Changing the integration variable to $y$ gives $dt_0/t_0^2=-dy/t$,
so the scale-covariant late-source contribution depends only on the
integrated tree response:
\be
\begin{aligned}
    W_{\rm desc}(t)
    &\simeq
    \int^t dt_0\,
    \dot N_1(t_0)
    \left\langle W_{\rm tree}(t;t_0)\right\rangle_{\rm src}
    \\
    &=\frac{2}{t}\int_0^\infty\mathcal F(y)\,dy+o(t^{-1})
    \\
    &=
    \frac{2}{t}
    \left\langle\mathcal R_{\rm tree}\right\rangle_{\rm src}
    +o(t^{-1}).
\end{aligned}
    \label{eq:terminal-descendant-tail}
\ee
Under the stated finite-response and source-tail assumptions,
the continuing source fixes the $1/t$ power; the integrated response
of its trees fixes the coefficient.

The same $1/t$ behavior follows directly from the late-time spatial geometry.
A finite invariant-profile integral $I_\infty$ places most of the surviving trace at bounded
invariant age $z=t^2-x^2$. On a late time slice these regions lie near
the outgoing folds, at distance $t-|x|=z/(t+|x|)$; their widths shrink
as $1/t$. The pressure integral then gives
\be
 W(t)\sim\frac{4I_\infty}{t}.
 \label{eq:terminal-abel-limit}
\ee
Appendix~\ref{app:terminal-pressure-endpoint} bounds the contribution
from large invariant ages and establishes this limit. The tree convolution
and the invariant-profile argument are two representations of the same
asymptotic coefficient: the former separates remnant emission from descendant
response, while the latter combines them in $I_\infty$.

The characteristic first-generation mass hierarchy is
\be
    \xi_*
    \equiv
    \frac{2\muIFS}{m_{\rm term}\sqrt{\GammaSplit}}
    \sim
    g_s^{-1}.
    \label{eq:terminal-source-hierarchy}
\ee
Averaging Eq.~\eqref{eq:terminal-tree-response-scaling} over the broad
but time-independent source distribution gives
\be
    \left\langle\mathcal R_{\rm tree}(\xi_*)\right\rangle_{\rm src}
    \sim
    \frac{2\muIFS}{\GammaSplit}
    C_{\rm src}\,\xi_*^{\kappa_{\rm IFS}},
    \qquad \xi_*\to\infty.
\ee
The order-one coefficient $C_{\rm src}$ includes the full
first-generation mass distribution and the terminal response.

Including the direct remnant, the complete zero-energy state has,
at fixed hierarchy \(\xi_*\),
\be
    p_{\rm tot}(t)
    =-\frac{1}{t}\left[
       \frac{4\muIFS}{\GammaSplit}
       +2\left\langle\mathcal R_{\rm tree}(\xi_*)\right\rangle_{\rm src}
      \right]+o_t(t^{-1}).
    \label{eq:terminal-pressure-main}
\ee
At weak coupling the descendant term dominates.  Using
\(\GammaSplit\alpha'\propto g_s^2\) and
\(\kappa_{\rm IFS}=2\sqrt2-2\),
\be
    |p_{\rm tot}(t)|
    \propto
    \frac{g_s^{-2\sqrt2}}{t}.
    \label{eq:terminal-pressure-gs-main}
\ee

The noninteger coupling power counts the extra generations exposed
at weak coupling.  The source starts at mass \(O(m_s/g_s)\), and a
tree produces \(O(g_s^{-\kappa_{\rm IFS}})\) semiclassical nodes
before reaching the string scale.  Each contributes an integrated
trace of order \(g_s^{-2}\).  Their product gives
\(g_s^{-2-\kappa_{\rm IFS}}=g_s^{-2\sqrt2}\).
The coupling exponent assumes
that the large-hierarchy binary fragmentation law remains valid down to
a scale-local terminal crossover and that the terminal dynamics
introduces no new power of \(g_s\).

The transition from the intermediate \(t^{-q_{\rm IFS}}\) law is
resolved in Sec.~\ref{subsec:terminal-crossover}. The marginal terminal
power \(q=1\) has two immediate
consequences. Stress-tensor conservation gives
\(M_2^{(E)},M_2^{(p)}\propto-t\log(t/t_\times)\) at leading order.
The even trace-weight moments have a common terminal
coefficient: \(M_{2k}^{(W)}/(t^{2k}M_0^{(W)})\to1\) for each fixed
\(k\), as derived in App.~\ref{app:stress-moments-terminal-anec}.
This expresses the concentration of the remaining trace near the
original folds on the scale \(|x|/t\).

In the collinear massless completion, the terminal products
carry the conserved right- and left-moving null charges of the
absorbed branch and have $T_{uv}^{\rm rad}=0$. With the momentum convention of
Sec.~\ref{subsec:yoyo_splitting}, their total energies are
\(E_R=(E+P)/2=\muIFS\Delta v\) and
\(E_L=(E-P)/2=\muIFS\Delta u\).
The endpoint is controlled by the integrated trace of the entire
continuing cascade, rather than by termination of each finite tree
separately.  Boost invariance and conservation give
\be
    \langle\mathcal A(v;U)\rangle
    =-\frac{I(Uv)}{U}
    =-\frac{I_\infty}{U}+o(U^{-1})
    \longrightarrow0,
    \label{eq:terminal-collinear-anec}
\ee
at fixed \(v>0\). Here \(I\) includes the remnant and all live
descendants, and \(I_\infty=\int_0^\infty f(z)\,dz<\infty\)
under the finite-response assumptions above, as quantified in
App.~\ref{subsec:app-malthusian-moments}.
In particular, \(0\leq-\langle\mathcal A(v;U)\rangle\leq I_\infty/U\).
This controls the persistent source as well as its finite trees and
establishes the terminal \(1+1\)-dimensional null balance.

\subsection{The terminal crossover}
\label{subsec:terminal-crossover}

The source argument fixes the late-time power. Its connection to the
semiclassical regime can also be seen directly in the boost spectrum.
At fixed time, the string-scale cutoff removes the lower-mass, lower-boost
part of the population. In the sharp absorbing binary reference, the
stationary clock of App.~\ref{app:chronological-snapshot-profile} gives,
at large boost,
\be
 \frac{(dW/d\ln\gamma)_{\rm sharp}}
      {(dW/d\ln\gamma)_{\rm sc}}
 \simeq\mathcal S_\alpha(T/\gamma),\qquad
 \mathcal S_\alpha(x)=\frac{\Gamma(\alpha,x)}{\Gamma(\alpha)},
 \quad \alpha=3-\sqrt2,
 \label{eq:terminal-moving-cutoff}
\ee
where \(\Gamma(\alpha,x)\) is the upper incomplete gamma function and
\(T=t/t_\times\). The crossover scale is
\be
 t_\times=\frac{2\muIFS}{\GammaSplit m_{\rm term}},\qquad
 \sqrt{\GammaSplit}\,t_\times=\xi_*,\qquad
 t_\times\sim\frac{\ell_s}{g_s^2}.
 \label{eq:terminal-crossover-time}
\ee
For \(T\gg1\), the surviving trace lies at boosts of order \(T\) and
above. Integrating the semiclassical spectrum
\(dW/d\ln\gamma\propto t^{\kappa_{\rm IFS}-1}\gamma^{-\kappa_{\rm IFS}}\)
against this suppression factor supplies \(T^{-\kappa_{\rm IFS}}\),
giving \(W\propto t^{-1}\). Terminal decay therefore removes the slow
descendants while increasingly boosted strings continue to carry the trace.
The cutoff moves through the spectrum with a fixed width in
\(\ln\gamma\); weak coupling separates it from the source edge without
making this width grow.

To follow the same IFS through the full crossover, we integrate its
total descendant trace up to the observation time. The joint stationary
clock and boost projection include moderate boosts as well as the tail.
Appendix~\ref{app:terminal-kernel}
constructs a positive reference function \(\Psi_{\rm sharp}\) such that,
in the large-hierarchy limit at fixed \(T\),
\be
 \begin{aligned}
 \mathscr R_{\rm desc}^{\rm sharp}(t)
 &\equiv\int_0^t W_{\rm desc}^{\rm sharp}(s)\,ds
 \sim\mathcal A_{\rm term}\Psi_{\rm sharp}(T),\\
 \Psi_{\rm sharp}(T)&=
 \begin{cases}
 C_{\rm early}T^{\kappa_{\rm IFS}}[1+o(1)],&T\ll1,\\
 2\ln T+C_{\rm sharp}+o(1),&T\gg1.
 \end{cases}
 \end{aligned}
 \label{eq:terminal-crossover-main}
\ee
The logarithmic growth counts the continuing arrival of new trees;
each tree still has a finite lifetime-integrated response.

\begin{figure}[ht!]
\centering
\includegraphics[width=\textwidth]{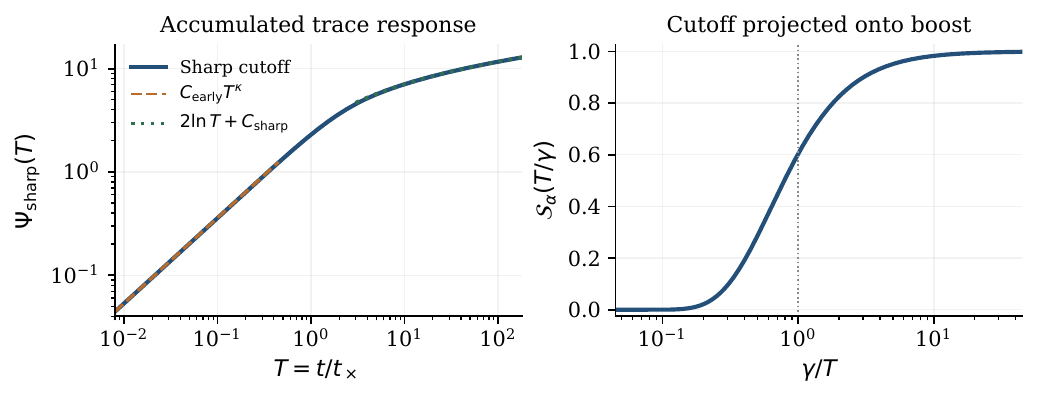}
\caption{Sharp-cutoff binary reference in the large-hierarchy limit.
Left: the accumulated response function \(\Psi_{\rm sharp}(T)\), with its
early power and late logarithmic asymptotes. Right: suppression of the
high-boost trace spectrum by the mass cutoff, Eq.~\eqref{eq:terminal-moving-cutoff}.
The surviving trace moves to \(\gamma\sim T\) with a fixed logarithmic
transition width. The curve on the left is evaluated from the stationary
joint clock in App.~\ref{app:terminal-kernel}; its normalization is separate
from the entrance-dependent amplitude \(\mathcal A_{\rm term}\).}
\label{fig:terminal-crossover}
\end{figure}
Here \(\mathcal A_{\rm term}\) is fixed by the entrance mass moment,
as in Eq.~\eqref{eq:terminal-kernel-normalization}; the early limit is
taken after entrance relaxation. Figure~\ref{fig:terminal-crossover}
shows this calculated interpolation. The persistent remnant contributes
its additional response from Sec.~\ref{subsec:terminal-persistent-source}.

A finite-width terminal layer changes the crossover shape and its
normalization. With finite response and the delay moments specified in
Appendix~\ref{app:terminal-kernel}, the semiclassical power and late
logarithmic accumulation persist. The same appendix gives the convergence
estimates for the sharp reference and more general terminal laws.

\FloatBarrier
\section{Higher-dimensional extension}
\label{sec:higher-dimensional-extension}

In higher dimensions, the situation is more complicated. The null ray
entering the ANEC need not lie in the plane of the IFS. Moreover, even
when it does, terminal radiation can be emitted in arbitrary directions
and need not remain in that plane. In either case, the positive flux
that compensates the negative-energy fold can miss the null ray,
leading to ANEC violation. Thus a transversely localized IFS need not
saturate the ANEC on every null ray.

However, IFSs are not nucleated in generic states. For an unconditioned
state produced from the vacuum by spatially translation-invariant
dynamics, momentum conservation places the complete IFS state,
including all descendants, in the exact zero-spatial-momentum sector.
The state is therefore spatially translation invariant. In the next
subsection we combine this symmetry with the complete cascade's flux
balance to establish ANEC saturation on every null ray. We then turn
to a transversely localized IFS and quantify the residual after
smearing over a finite transverse scale.

\subsection{Transverse zero modes and ANEC}
\label{subsec:terminal-radiation-anec}

We first establish the flux balance integrated over the transverse
screen, then use translation invariance to determine the light-ray
expectation on each generator.

Let $\hat k=(1,\hat{\mathbf n})$, with $\hat{\mathbf n}^{\,2}=1$, be
a future-directed null vector. A displacement $b=(0,\mathbf b)$ with
$\mathbf b\cdot\hat{\mathbf n}=0$ labels neighboring generators.
For a reference point $x_0$, their complete light-ray expectation is
\be
    \mathcal A_{\hat k}(\mathbf b)
    \equiv\int_{-\infty}^{\infty}d\lambda\,
    \bigl\langle T_{\mu\nu}(x_0+\lambda\hat k+b)\bigr\rangle
    \hat k^\mu\hat k^\nu.
    \label{eq:higher-dimensional-light-ray}
\ee
For a localized IFS this can depend on $\mathbf b$ and be negative.
The conserved current $J^\mu_{\hat k}=-T^\mu{}_{\nu}\hat k^\nu$
has Cauchy-surface flux
\be
    \mathscr Q_{\hat k}
    =\int d^{D-1}x\,\langle J^0_{\hat k}\rangle
    =\langle P^0-\mathbf P\cdot\hat{\mathbf n}\rangle,
    \qquad P^\mu=\int d^{D-1}x\,T^{0\mu}.
    \label{eq:null-momentum-flux}
\ee
Evaluating this current on $t-\hat{\mathbf n}\cdot\mathbf x=s>0$
relates it to the flux integrated over the complete screen, provided
all the compensating flux is collected.

It remains to establish that no finite amount of compensating flux
is permanently delayed beyond the cutoff, since the cascade continues
to emit at arbitrarily late times. For directions not parallel to the
original folds, both negative folds cross the null plane in finite
time. The region still ahead of the plane then contains only positive
stress. Conservation and causal support bound its remaining null
charge by a quantity that tends to zero as the cutoff is removed.
For longitudinal directions, the first-split bound applied to the
transversely integrated longitudinal stress gives the same conclusion.
Appendix~\ref{subsec:app-complete-null-flux} establishes both limits
directly for the full cascade. Consequently,
\be
    \int d^{D-2}b\,\mathcal A_{\hat k}(\mathbf b)
    =\langle P^0-\mathbf P\cdot\hat{\mathbf n}\rangle=0,
    \label{eq:light-ray-zero-mode-sum-rule}
\ee
using the vanishing total energy and momentum of the complete state.
This balance does not require finite integrated trace or a specified
terminal lifetime.

To implement a normalized translation average, retain this completed
flux balance in a periodic transverse regulator of volume $V_\perp$
and average uniformly over translations of the localized response.
Translation invariance makes the averaged response independent of
$\mathbf b$, so
\be
    V_\perp\,\mathcal A_{\hat k}
    =\int_{V_\perp}d^{D-2}b\,\mathcal A_{\hat k}(\mathbf b)=0.
    \label{eq:regulated-generator-ANEC}
\ee
Thus the ANEC is saturated on every generator in this
completed, translation-invariant construction.

The same reasoning applies directly to a translation-invariant
quantum state. Let $\mathbf P_\perp$ denote the momentum conjugate to
translations within the screen. An exact zero-momentum state obeys
\be
 \mathbf P_\perp|\Psi_{\rm IFS}\rangle=0,
 \qquad
 e^{-i\mathbf a_\perp\cdot\mathbf P_\perp}
 |\Psi_{\rm IFS}\rangle=|\Psi_{\rm IFS}\rangle.
 \label{eq:ifs-exact-zero-momentum}
\ee
Its stress expectation, and hence its complete light-ray expectation,
is independent of the transverse position. Combined with the
completed flux balance, this gives a vanishing integral on each
generator. A normalized positional mixture provides the corresponding
semiclassical realization of the translation average.

This average keeps the total IFS number fixed. A homogeneous
population at fixed number density contains more events as the volume
grows and can retain negative averaged null energy. The stronger
finite-number estimate uniform in the longitudinal cutoff is given in
Appendix~\ref{app:positional-average}; the cosmological population is
discussed in Sec.~\ref{sec:cosmological-implications}.

\subsection{Finite transverse smearing}
\label{subsec:finite-transverse-smearing}

For a localized IFS, with collinear preterminal evolution, smear the complete light-ray integral over the
screen of a fixed null plane $s=-k\cdot x>0$:
\be
    \mathcal A_{k,L}(s,\mathbf b_0)
    =\int d^{D-2}b\,K_L(\mathbf b_0-\mathbf b)\,
      \mathcal A_k(s,\mathbf b),
    \qquad K_L(\mathbf b)=L^{-(D-2)}K(\mathbf b/L).
    \label{eq:main-smeared-light-ray}
\ee
The normalized kernel $K$ and its first two derivatives are bounded.
Positive and negative flux at nearby screen positions then receive
nearly equal weights. The residual is controlled by their separation,
weighted by the compensating flux.

The first-split trace bound controls the mean distance of the
collinear positive energy from the folds. Terminal radiation adds
angular spreading, but its mass budget is bounded by that of the
first-generation yo-yos. Appendix~\ref{app:finite-transverse-smearing}
combines these observations to give the sufficient smearing scale
\be
    L_{\rm smear}\equiv\GammaSplit^{-1/2}
    \sim\frac{\ell_s}{g_s}.
    \label{eq:main-sufficient-smearing-scale}
\ee
For the boost-invariant cascade with local terminal conversion into
free products of positive energy, the bound is
\be
    \sup_{\mathbf b_0}|\mathcal A_{k,L}(s,\mathbf b_0)|
    \leq C_K\frac{\mathscr F_k(s)}{L^{D-2}}
             \frac{L_{\rm smear}}{L},
    \qquad s>0,\quad\text{every normalized }k.
    \label{eq:main-finite-smearing-result}
\ee
Here $\mathscr F_k(s)$ is the magnitude of the negative fold flux,
equal to the total compensating positive flux. The constant is
independent of the null direction, including grazing directions:
nearly parallel products can travel far but carry little test flux.
A fractional residual at most $\varepsilon$ therefore requires only
$L\gtrsim C_KL_{\rm smear}/\varepsilon$. Its sign at finite resolution
can be either positive or negative. No terminal node count or finite
$I_\infty$ is needed for this estimate.

\section{Cosmological implications}
\label{sec:cosmological-implications}

The cosmological interest in instant folded strings (IFSs) rests on
an unusual property of their energy-momentum tensor. An isolated IFS has vanishing total
energy in Minkowski space but a large negative integrated pressure. A
homogeneous population can consequently violate the null energy condition
(NEC), $\rho+p\geq0$, and make $\dot H$ positive in a contracting universe.
On the expanding branch, IFS production can oppose the rolling dilaton
and sustain a dark-energy-like stress
\cite{Itzhaki:2024vaw,Itzhaki:2025cyclic}. In the cyclic construction,
the dilaton crosses an enhanced symmetry point (ESP), where nonadiabatically
produced particles grow heavier as the field moves away. Their restoring
force turns the dilaton back through the moduli-trapping mechanism of
\cite{Kofman:2004}. IFS production then begins while the universe is
still contracting, and its negative pressure can turn the geometry around.

The finite-coupling decay raises an apparent obstacle to this picture.
The averaged null energy condition (ANEC) constrains the stress integrated
along a \emph{complete} null ray and has a firmer basis in Minkowski
spacetime than the pointwise NEC
\cite{Verch:1999nt,Hartman:2016lgu,Faulkner:2016mzt,KellyWall:2014}.
The preceding sections show that the IFS cascade saturates ANEC along
complete longitudinal rays in $1+1$ dimensions: positive folds and
terminal radiation eventually compensate the negative null flux.
Section~\ref{sec:higher-dimensional-extension} identifies the corresponding
completed transverse zero mode in higher dimensions. Yet a regular,
nonstatic, two-sided null-complete flat-FRW cosmology has negative complete
affine ANEC under the assumptions of
Sec.~\ref{subsec:cosmology-energy-conditions}. In a cyclic universe,
the affine integral is already negative over each full cycle bounded by
equivalent zeros of $H$, even when $a$ grows from cycle to cycle.
How can the same decay process restore the Minkowski null-energy balance
and still support a bounce?

The finite-lifetime approximation used in
Refs.~\cite{Itzhaki:2024vaw,Itzhaki:2025cyclic} already illustrates this
compatibility. In Minkowski space, let the traceful string bulk convert
locally into collinear null products within a finite time\footnote{At fixed total four-momentum, isotropically averaged null
products have $p_{\rm rad}=\rho_{\rm rad}/3$ independently of their
emission directions. Collinearity therefore leaves the homogeneous
cosmological equations of Refs.~\cite{Itzhaki:2024vaw,Itzhaki:2025cyclic}
unchanged.}. The complete longitudinal ANEC integral then vanishes already in this approximation: by Eq.~\eqref{eq:intro-cutoff-anec}, it equals the flux of \(-T_{uv}\) through the closing segment, which vanishes once that segment lies entirely after conversion.
For the translation-invariant single-event state considered in
Sec.~\ref{subsec:terminal-radiation-anec}, the completed null-flux
balance extends this saturation to every null ray.

The effect of contraction or expansion follows directly from energy
conservation. To isolate this effect, hold the dilaton
fixed and follow a homogeneous, isotropically averaged cohort of IFSs
and their decay products. In a
comoving volume of physical size $V\propto a^3$, its energy $E=\rho V$
obeys $dE=-p\,dV$. While the pressure is negative, contraction lowers
the energy of a cohort born with $E=0$, whereas expansion raises it.
After conversion to null products, the cohort therefore retains a
negative or positive energy balance, respectively.\footnote{This balance
refers to the energy left in the cohort. The complete ANEC also weights
the history by the affine measure, so an energy surplus after decay
alone does not determine its sign; see
Sec.~\ref{subsec:cosmology-energy-conditions}.}

A rolling dilaton also exchanges energy with the traceful strings.
The cosmological equations of
Refs.~\cite{Itzhaki:2024vaw,Itzhaki:2025cyclic} include this work and
explicitly admit bouncing solutions. The finite-lifetime description
thus already illustrates how completed Minkowski compensation can
coexist with a cosmological bounce.

In the full cascade, the persistent remnant continually produces
traceful descendants, while completed branches deposit compensating
positive null energy over a broad range of delays. The trace controls
the force on the dilaton. When the response is strong enough to balance
the driving force with little field motion, that balance fixes the trace
required by the background. The IFS energy density and pressure then follow from conservation
and the initial stress. We first derive the bounce and expanding solutions from force balance
and conservation, then use the retarded cascade response to test the
bounce and estimate the rolling rate on the expanding branch.

\subsection{Force balance and the cosmological stress}
\label{subsec:cosmology-two-fluid}

The quantity followed through the decay cascade is also the one that
opposes the dilaton's motion. For a homogeneous, isotropically averaged
population, let $W=\rho_{\rm IFS}-3p_{\rm IFS}$, and write
$g_s=g_0e^{\phi/(\sqrt2M_{\rm Pl})}$ and $\Pi=\dot\phi$.
Orientation averaging preserves the event trace, since
$3p=p_\parallel$; the accumulated event responses therefore determine
$W$. Traceless terminal products contribute to the energy and pressure,
but the dilaton force is set by the remaining trace:
\be
 \dot\Pi+3H\Pi+V_{,\phi}=-\frac{W}{\sqrt2M_{\rm Pl}}.
 \label{eq:cosmology-general-dilaton}
\ee
For $V_{,\phi}<0$, the potential drives the field toward stronger
coupling, where positive rolling produces IFSs. Increasing production
increases the opposing force. If this response is sufficiently strong
that scalar acceleration and Hubble drag are small compared with the
two forces, the required trace is
\be
 W\simeq W_{V,\rm bal}\equiv-\sqrt2M_{\rm Pl}V_{,\phi}.
 \label{eq:cosmology-potential-balance-trace}
\ee
The potential slope thus fixes the required trace; the cascade
determines the rolling rate needed to supply it. When ESP particles are
present, their force replaces $V_{,\phi}$ by
$V_{,\phi}+\partial_\phi\rho_\chi$. Whether the delayed response
can maintain this balance is a dynamical question, tested below for
the homogeneous bounce.

The trace fixes the dilaton force, while the geometry also depends on
the IFS energy density. In spatially flat Einstein-frame FLRW,
$ds^2=-dt^2+a^2(t)d\mathbf x^2$,
\be
 3M_{\rm Pl}^2H^2=\rho,\qquad
 \dot H=-\frac{\rho+p}{2M_{\rm Pl}^2},
 \label{eq:cosmology-friedmann-raychaudhuri}
\ee
where $\rho$ and $p$ include all sectors. Applied to this total stress,
the Minkowski values $\rho=0$, $p<0$ describe the instant of a bounce:
$H=0$ and $\dot H>0$.\footnote{For discussions of stability and
strong-coupling constraints on local scalar-field realizations of NEC
violation, see
Refs.~\cite{Rubakov:2014NEC,BattefeldPeter:2015,BrandenbergerPeter:2017,
Creminelli:2006NEC,IjjasSteinhardt:2016,IjjasSteinhardt:2017,
Kobayashi:2016Horndeski}.}
The singular ratio $p/\rho$ conceals a regular
cancellation between positive and negative contributions whose subsequent
evolution differs.

The trace and null contributions can be kept separately as
\be
 \rho_{\rm IFS}=\frac W4+\rho_N,\qquad
 p_{\rm IFS}=-\frac W4+\frac{\rho_N}{3},\qquad
 \rho_{\rm IFS}+p_{\rm IFS}=\frac43\rho_N.
 \label{eq:cosmology-trace-traceless-decomposition}
\ee
The first component, $-Wg_{\mu\nu}/4$, carries the full trace and has
zero null contraction. The signed, traceless remainder $\rho_N$
carries all the homogeneous IFS null energy. In the Minkowski state,
$\rho_N=-W/4$: the energy densities cancel while the pressures add
to $-W/3$. This split stays regular through zero total density.
Both components describe the combined stress of strings, folds, and
decay products; $W/4$ is not separately the live-string energy.

Energy conservation, including the work exchanged with the dilaton, gives
\be
\begin{aligned}
 \dot\rho_{\rm IFS}+4H\rho_{\rm IFS}
 &=\left(H+\frac{\Pi}{\sqrt2M_{\rm Pl}}\right)W,\\
 \dot\rho_N+4H\rho_N
 &=\frac{\Pi W}{\sqrt2M_{\rm Pl}}-\frac{\dot W}{4}.
\end{aligned}
 \label{eq:ifs-ward-cosmology}
\ee
For nearly constant $W$ and negligible dilaton work, the trace component
is vacuum-like and $\rho_N$ scales freely as $a^{-4}$. Expansion dilutes
an initially negative null component, leaving positive vacuum-like
stress; contraction amplifies it. A nearly constant potential slope
can therefore sustain an approximately constant IFS density through
Eq.~\eqref{eq:cosmology-potential-balance-trace}, even though each
individual event's trace decays.

A changing potential slope can be described along the trajectory by
\be
 \epsilon_{V'}\equiv-\frac12\frac{d\ln|V_{,\phi}|}{d\ln a},
 \qquad V_{,\phi}[\phi(a)]\simeq V_{,\phi,*}
       \left(\frac a{a_*}\right)^{-2\epsilon_{V'}}.
 \label{eq:cosmology-slope-variation}
\ee
The power law applies when $\epsilon_{V'}$ is approximately constant;
this is a running of the slope along the trajectory, distinct from
$\epsilon_V=M_{\rm Pl}^2(V_{,\phi}/V)^2/2$. In force balance,
$W$ follows the same power law. Neglecting dilaton work, integration
of Eq.~\eqref{eq:ifs-ward-cosmology} gives, for $\epsilon_{V'}\ne2$,
the \emph{dual-fluid prescription},
\be
\begin{aligned}
 \rho_{\rm IFS}&=\rho_{\rm IDE}+\rho_R,
 &p_{\rm IFS}&=p_{\rm IDE}+\frac{\rho_R}{3},\\
 \rho_{\rm IDE}&=\frac{W}{4-2\epsilon_{V'}},
 &p_{\rm IDE}&=\left(-1+\frac{2\epsilon_{V'}}3\right)\rho_{\rm IDE}.
\end{aligned}
 \label{eq:cosmology-power-law-trace}
\ee
Here $\rho_R=C_R(a_*/a)^4$, with $C_R$ fixed by the initial stress.
The label IDE refers to \emph{instant dark energy}; this component is
vacuum-like only when $\epsilon_{V'}=0$. For a changing slope, it
carries the stress that follows the driving force. The freely scaling radiation-like
term is $\rho_R$; the pure null remainder is
$\rho_N=\rho_R+\epsilon_{V'}\rho_{\rm IDE}/2$.
For example, a force proportional to $a^{-3}$ gives
$\rho_{\rm IDE}=W$ and $p_{\rm IDE}=0$. This is the matter-like
IFS tracking contribution relevant to the ESP bounce.

Appendix~\ref{app:cosmology-expansion-estimates} includes dilaton
work through a string-frame scaling prescription and derives the
transfer between the effective components when its exponent varies.
At the onset of production, where a scaling exponent need not be
defined, the trace/null equations above remain regular.

\subsection{ESP turnaround and the IFS-mediated bounce}
\label{subsec:terminal-ESP-bounce}
\label{subsec:cosmology-bounce-matching}

The cyclic scenario requires two turnarounds: first the dilaton, then
the geometry. Initially the universe contracts with $\Pi<0$, so no
new IFSs are produced. Passage through the ESP creates $\chi$ particles
with $m_\chi=\lambda_\chi|\phi-\phi_{\rm ESP}|$ and conserved
comoving number density. Away from the production interval they become
nonrelativistic, and their increasing mass supplies a restoring force
toward the ESP \cite{Kofman:2004,Itzhaki:2025cyclic}.
We use this nonrelativistic description, take the particles to be
stable over the bounce, and neglect the ordinary dilaton potential.
On the weak-coupling side, define the constant comoving force
$F_\chi\equiv\lambda_\chi n_{\chi,*}>0$ and $P=a^3\Pi$. Then
\be
 \rho_\chi=F_\chi a^{-3}(\phi_{\rm ESP}-\phi),\qquad
 \dot P=F_\chi-\frac{a^3W}{\sqrt2M_{\rm Pl}}.
 \label{eq:cosmology-ESP-canonical-equation}
\ee
With no inherited IFSs, $W=0$ while $\Pi<0$, and $\dot P=F_\chi>0$
brings the initially negative momentum to zero during contraction.\footnote{The
incoming dilaton velocity must be small enough for this turnaround to
precede a cosmological crunch; see Ref.~\cite{Itzhaki:2025cyclic}.}

Once $\Pi>0$, IFS production begins and its trace opposes the ESP
force. Instantaneous balance would require
\be
 W_{\chi,\rm bal}\equiv\sqrt2M_{\rm Pl}F_\chi a^{-3}.
 \label{eq:cosmology-ESP-balance-trace}
\ee
The enhanced cascade response allows slow positive rolling to generate
a large opposing force: faster motion produces more IFSs, which push
back on the dilaton. The response is delayed, however, and contraction
amplifies residual kinetic motion. The dynamical question is whether
the accumulated response keeps the field close to its first turn long
enough for the geometry to bounce.

The $a^{-3}$ dependence of the ESP force makes its balancing IFS
stress matter-like. In the nearly fixed-dilaton limit, the dual-fluid
prescription gives
\[
 \rho_{\rm IDE}\simeq W_{\chi,\rm bal}
       =\frac{\sqrt2M_{\rm Pl}F_\chi}{a^3},\qquad
 p_{\rm IDE}\simeq0,\qquad \rho_R=-\frac{c_r}{a^4}.
\]
The trace and its associated energy thus grow together, while the
negative radiation term grows faster during contraction. Let $t_m$
be the first dilaton turn, with $d_m=\phi_{\rm ESP}-\phi_m$.
Holding the field near $\phi_m$ also gives
$\rho_\chi\simeq F_\chi d_m/a^3$. Neglecting the small dilaton
kinetic energy, the total density therefore gives the reference trajectory
\be
 \begin{aligned}
 3M_{\rm Pl}^2H^2 &\simeq\frac{c_m}{a^3}-\frac{c_r}{a^4},
 &c_m=F_\chi(d_m+\sqrt2M_{\rm Pl}),\qquad
 c_r=c_ma_m-3M_{\rm Pl}^2a_m^4H_m^2.
 \end{aligned}
 \label{eq:cosmology-ESP-two-fluid}
\ee
At the first turn $\Pi_m=0$ and no IFSs have yet been produced.
With no additional radiation, matching to the zero IFS entrance
density fixes $c_r=\sqrt2M_{\rm Pl}F_\chi a_m>0$.
The reference's two IFS contributions therefore cancel initially.
As contraction amplifies the negative radiation term relative to the
positive matter-like contributions, the total density reaches zero
at $a_b=c_r/c_m$. In the fiducial example below,
$a_b/a_m\simeq0.956$: only about $4\%$ further contraction is
needed after the dilaton turns.

The trace starts from zero at $t_m$, so establishing the balancing
force takes time. The reference remains predictive if this adjustment
produces little field motion. By
Eq.~\eqref{eq:cosmology-ESP-canonical-equation}, the momentum $P$
measures the force mismatch accumulated since the first turn, so small
momentum permits brief departures from instantaneous balance.
Combining the scalar and geometric equations eliminates the explicit
trace term: for $J=H+\Pi/(3\sqrt2M_{\rm Pl})$, the $W$ dependence
cancels from $\dot J$
(Appendix~\ref{app:terminal-numerical-check}). Small velocity and
field displacement can therefore keep the geometry close to the
reference even during the response's initial adjustment.
At an exact zero of $H$, the Friedmann and Raychaudhuri equations give
\be
 6M_{\rm Pl}^2\dot H=\rho_\chi+W-\Pi^2.
 \label{eq:cosmology-strict-bounce}
\ee
Since $W\geq0$, the condition $\Pi^2<\rho_\chi$ ensures a strict
bounce from contraction to expansion.

\begin{figure}[tbp]
\centering
\includegraphics[width=0.96\textwidth]{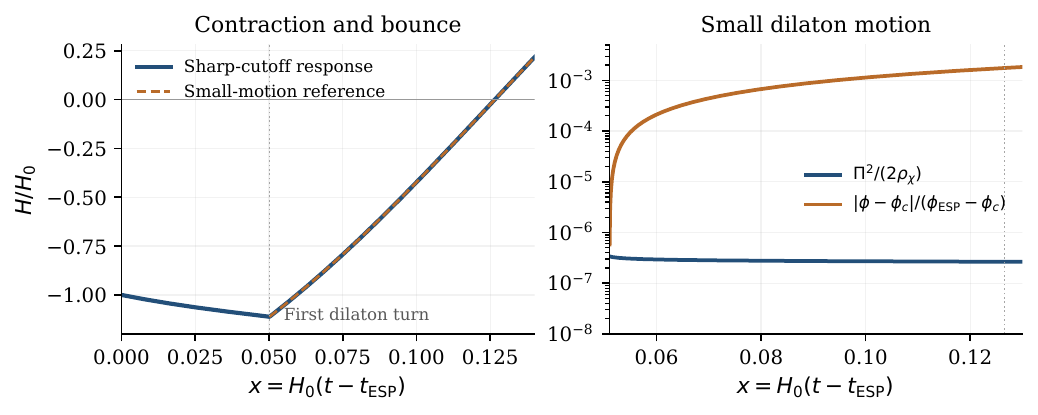}
\caption{ESP-mediated bounce in the sharp-cutoff binary reference,
including the persistent remnant and descendant response. Left: the
retarded Hubble evolution (solid) and nearly fixed-dilaton reference
(dashed), matched at the first dilaton turn without a later fit.
Right: kinetic fraction and field drift on the settled comparison
interval, with the drift measured from $\phi_c=\phi(x=0.051)$. Numerical definitions, convergence tests, and
parameter variations are given in
Appendix~\ref{app:terminal-numerical-check}.}
\label{fig:terminal-bounce-control}
\end{figure}

The full retarded evolution in the sharp-cutoff binary model keeps
the dilaton close enough to rest that the reference fixed at the first
turn predicts the bounce without a subsequent fit
(Fig.~\ref{fig:terminal-bounce-control}). On the settled comparison interval,
the field moves by only about $0.18\%$ of its distance to the ESP
at the interval's start, and the Hubble histories differ by less than
$4\times10^{-4}H_0$, where $H_0=-H(t_{\rm ESP})$.
This calculation includes the
persistent remnant and descendant response through the terminal
crossover, with the full optical history and traceless terminal
products.

Appendix~\ref{app:terminal-numerical-check} gives the matching,
convergence checks, and residual-motion diagnostics. The homogeneous
bounce also survives the tested variations of coupling, terminal
cutoff, and descendant normalization. These tests use the sharp-cutoff
binary response; non-dilute entrance and finite-hierarchy corrections
remain outside this benchmark.

\FloatBarrier

\subsection{Expansion and a slowly changing potential slope}
\label{subsec:cosmology-dark-energy}

During expansion, the same force balance can sustain an IFS contribution
to dark energy \cite{Itzhaki:2024vaw,Itzhaki:2025cyclic}. Consider a
negative potential slope with negligible ESP force and sufficiently
slow positive rolling that Eq.~\eqref{eq:cosmology-potential-balance-trace}
holds and dilaton work is small. A nearly constant slope then maintains
nearly constant $W$. Expansion dilutes the signed radiation component,
leaving $\rho_{\rm IFS}\simeq-p_{\rm IFS}\simeq W_{V,\rm bal}/4$.
The negative component that can drive the contracting geometry toward
a bounce becomes less important on the expanding branch.

When the slope changes, the dual-fluid prescription describes how the
IFS stress follows it. For approximately constant $\epsilon_{V'}<2$,
the radiation term decays relative to the IDE contribution. Once it is
negligible,
\be
 \rho_{\rm IFS}\simeq\rho_{\rm IDE}\simeq
       \frac{W_{V,\rm bal}}{4-2\epsilon_{V'}},\qquad
 \frac{p_{\rm IFS}}{\rho_{\rm IFS}}
       \simeq-1+\frac{2\epsilon_{V'}}3.
 \label{eq:cosmology-varying-slope-stress}
\ee
A decreasing slope magnitude gives positive IFS null energy; an
increasing magnitude gives negative null energy, even after the freely
redshifting term has diluted. Thus the departure from vacuum-like
stress measures the evolution of the force being balanced.
Neglecting dilaton kinetic energy, the combined density is
\be
 \rho_{\rm DE}\simeq V+\frac{W_{V,\rm bal}}{4-2\epsilon_{V'}}
 =V-\frac{M_{\rm Pl}}{\sqrt2(2-\epsilon_{V'})}V_{,\phi}.
 \label{eq:cosmology-slope-corrected-energy}
\ee
For a nearly constant slope this becomes
$V-M_{\rm Pl}V_{,\phi}/(2\sqrt2)$. The extra positive contribution
is controlled by the slope, so the combined density can be positive
even where $V<0$ \cite{Itzhaki:2024vaw}.

These expressions retain the leading slope variation while neglecting
work corrections of relative order $|\Pi|/(M_{\rm Pl}H)$.
When dilaton work is comparable to the slope-variation correction,
the string-frame formulation in
Appendix~\ref{app:cosmology-expansion-estimates} gives the corresponding
stress and energy transfer. A varying exponent also sources
$\rho_R$ through Eq.~\eqref{eq:cosmology-dual-fluid-transfer}; its
accumulated contribution must then be kept in the total stress.

Section~\ref{subsec:cosmological-optical-response} estimates the
rolling rate required to sustain this branch. Whether the branch
attracts solutions of the full retarded system remains to be established.

\subsection{How the cascade supplies the balancing trace}
\label{subsec:cosmological-optical-response}

The bounce and expanding branch require a large trace with little
dilaton motion. The cascade can provide this response because an IFS
continues to contribute long after its first splits. In the stationary
problem of Sec.~\ref{sec:terminal-decay}, the persistent remnant
emits an order-one number of descendant trees per logarithmic age
interval. Each tree has a finite time-integrated trace, giving the
late event response $w_{{\rm ev},s}(t)\simeq\mathcal C_s/(t-s)$
for birth time $s$. Each semiclassical node contributes an integrated response of order
$g_s^{-2}$, and the hierarchy contains $O(g_s^{-\kappa_{\rm IFS}})$
nodes. Their combined response therefore scales as
$\mathcal C_{\rm desc}\propto g_s^{-2-\kappa_{\rm IFS}}
=g_s^{-2\sqrt2}$.
The cosmological response follows by transporting these events in the
evolving background and summing over their production history.

The worldsheet production calculation \cite{Hashimoto:2022iqp}, in
the four-dimensional normalization of Ref.~\cite{Itzhaki:2025cyclic},
gives $\beta\Pi_+^2$ births per unit physical volume and time, with
$\Pi_+=\max(\Pi,0)$ and $\beta=1/(32\pi^6)$. If
$w_{{\rm ev},s}$ is the spatially integrated trace of one event,
including its remnant and descendants, the population trace is
\be
 W(t)=\int_{-\infty}^{t}ds\,
 \left[\frac{a(s)}{a(t)}\right]^3
 \beta\Pi_+^2(s)w_{{\rm ev},s}(t).
 \label{eq:cosmological-retarded-response}
\ee
The integral covers the full production history and is assumed to
converge; for a finite onset it starts at that onset. The volume ratio converts the event density at birth to its value at
observation, while $w_{{\rm ev},s}$ retains each event's decay history. Together with the energy boundary condition and the
other matter sources, this response closes the homogeneous equations
for independently evolving cascades.

The event response can be derived from the common time-dilation
factor in a yo-yo's trace and splitting rate. Let $m$ be its constant
string-frame mass between splits. Oscillation averaging gives an
integrated trace $m/\gamma$ and a splitting rate
$\lambda_s=\Gamma_t m/(2\mu\gamma)$, where
$\Gamma_t=\GammaSplit^{(s)}(t)$. With $a_s=\Omega a$,
$dt_s=\Omega\,dt$, and $\Omega=g_s/g_0$, the Einstein-frame
trace is $w_E=\Omega m/\gamma$. Spatial homogeneity conserves
$k=a_sp_s$, so at high boost $\gamma\simeq k/(a_sm)$ and
\be
 w_E\simeq a\Omega^2\frac{m^2}{k},\qquad
 \lambda_sdt_s\simeq\frac{m^2}{2\mu k}
                         a\Omega^2\Gamma_t\,dt.
 \label{eq:cosmology-trace-splitting-clock}
\ee
The two expressions share the factor $a\Omega^2$. Absorbing this
and the splitting rate into an optical age gives
\be
 Q_s(t)\equiv\bar Q(t)-\bar Q(s)
       =\int_s^t du\,a(u)\Omega^2(u)\Gamma_u.
 \label{eq:cosmology-optical-generator}
\ee
Per unit $\bar Q$, the splitting rate is $m^2/(2\mu k)$, with no
explicit background dependence. The mass and momentum distribution
therefore evolves with the same leading high-boost branching law,
while each live string's trace carries the observation factor
$\dot Q_s/\Gamma_t$. The local small-string conditions and the
finite-boost corrections are given in
Appendix~\ref{app:cosmological-high-boost}.

To follow the response through terminal conversion, we transport the
stationary sharp-cutoff crossover of
Sec.~\ref{subsec:terminal-crossover} in this clock. Matching the
remnant's emissions to their stationary source requires little coupling
variation during the formation intervals that supply the observed
trace; subsequent background evolution is retained in $Q_s$.
Let $\mathscr F_s(Q)$ denote the accumulated event response, including
the remnant and descendants, normalized at birth. The instantaneous
Einstein-frame trace is
\be
 w_{{\rm ev},s}(t)\simeq\frac{\Gamma_s}{\Gamma_t}
       \frac{d}{dt}\mathscr F_s\!\bigl(Q_s(t)\bigr)
 =\frac{\Gamma_s}{\Gamma_t}\dot Q_s(t)\mathscr F_s'\!\bigl(Q_s(t)\bigr).
 \label{eq:app-sharp-benchmark-response}
\ee
Here $\Gamma_s$ is the splitting rate at the IFS birth. The derivative
acts on $\mathscr F_s(Q_s(t))$; the rate ratio converts its birth
normalization to the trace at observation. The full function
$\mathscr F_s$ retains the buildup and crossover of the response.
For the sharp-cutoff reference used in the bounce calculation,
\be
 \mathscr F_s^{\rm sharp}(Q)
 =\mathcal A_s\Psi_{\rm sharp}\!\left(\frac{Q}{Q_{\times,s}}\right)
  +\frac{4\mu}{\Gamma_s}\Xi\!\left(\frac{Q}{Q_{*,s}}\right).
 \label{eq:cosmology-accumulated-event-response}
\ee
The two terms are the accumulated descendant and remnant responses.
The scales $Q_{\times,s}$ and $Q_{*,s}$ mark terminal conversion and
first splitting, respectively. Their normalization and the remnant
function $\Xi$ are specified in
Appendix~\ref{app:terminal-numerical-check}; the youngest-cohort
continuation is tested in Appendix~\ref{app:bounce-numerics}.

At late optical age, the logarithmic accumulation derived in
Sec.~\ref{sec:terminal-decay} gives
\be \mathscr F_s(Q)=\mathcal C_s\ln(Q/Q_{\rm ref,s})+O(1),\ee
with $Q_{\rm ref,s}$ a fixed optical scale. The corresponding tail
$\mathscr F_s'(Q)\sim\mathcal C_s/Q$ yields
\be
 w_{{\rm ev},s}^{\rm late}(t)\simeq
 \mathcal C_s\frac{\Gamma_s}{\Gamma_t}\frac{\dot Q_s(t)}{Q_s(t)}
 =\frac{\mathcal C_s\Gamma_s\,a(t)\Omega^2(t)}
        {\displaystyle\int_s^tdu\,a(u)\Omega^2(u)\Gamma_u}.
 \label{eq:cosmology-terminal-optical-response}
\ee
In the sharp reference, $\mathcal C_s=2\mathcal A_s+4\mu/\Gamma_s$;
more generally the descendant term is twice the source-averaged
integrated tree response. At constant coupling, the expression reduces
to $\mathcal C_s a(t)/\int_s^t a(u)du$, recovering
$\mathcal C_s/(t-s)$ in Minkowski space. The background affects both
the rate at which a cascade ages and its trace at observation.
The late expression applies after the crossover; the full kernel
\eqref{eq:app-sharp-benchmark-response} is needed during the buildup
of production.

For a slowly varying expanding branch, the convolution gives a simple
estimate of the rolling rate. With negligible inherited trace and
nearly constant production over the contributing history,
\be
 W\simeq A_{\rm cas}\Pi_+^2,\qquad
 \Pi_*^2\simeq\frac{W_{V,\rm bal}}{A_{\rm cas}},\qquad
 A_{\rm cas}>0.
\ee
For nearly constant $H>0$ and coupling, the event tail becomes
$w_{{\rm ev},s}^{\rm late}\simeq\mathcal C_s H/[1-e^{-H(t-s)}]$.
For $t_\times\ll t-s\ll H^{-1}$, successive logarithmic age
intervals contribute equally to the population trace.
Expansion suppresses older cohorts through
the volume factor in Eq.~\eqref{eq:cosmological-retarded-response}.
When $Ht_\times\ll1$, this gives
\be
 A_{\rm cas}\simeq\beta\mathcal C(g_s)
       \left[\ln\frac1{Ht_\times}+O(1)\right],\qquad
 \mathcal C(g_s)\propto g_s^{-2\sqrt2}.
 \label{eq:cosmology-expanding-log-response}
\ee
Here $t_\times=2\mu/(\Omega\Gamma_t m_{\rm term})$, with
string-frame splitting rate $\Gamma_t$; the integral is evaluated in
Appendix~\ref{app:cosmology-expansion-estimates}.
At fixed balancing trace,
$\Pi_*\propto g_s^{\sqrt2}/\sqrt{\ln[1/(Ht_\times)]}$.
In this regime, the cascade enhancement suppresses the rolling rate
required to sustain the balancing trace.
The estimate applies while $Ht_\times\ll1$ and production and coupling
vary slowly over a Hubble time. The string-frame crossover time
$t_\times^{(s)}=\Omega t_\times$ scales as $g_s^{-2}$ at fixed
string-frame scales. Thus the weak-coupling limit is not uniform at
fixed $H/\Omega$: the condition $Ht_\times\ll1$ must still hold,
and the full convolution is required outside this regime.

Once production ceases, an inherited population can lose its influence
on the dilaton despite the persistence of individual cascades.
At approximately constant coupling, sufficiently late in a radiation
era, its trace scales as
$W_{\rm old}\propto a^{-3}t^{-1}\propto a^{-5}$
(Appendix~\ref{app:cosmology-expansion-estimates}). Population dilution
and decay of the event response both contribute. The traceful force
then fades faster than radiation, while the signed traceless stress
can retain an $a^{-4}$ mode. Thus the surviving remnants need not maintain an appreciable dilaton
force, even when an $a^{-4}$ energy component remains.

\subsection{The complete cosmological null-energy balance}
\label{subsec:cosmology-energy-conditions}

At the homogeneous level, the affine null-energy balance is fixed by
the geometry. Along a radial null ray with $dt/d\lambda=1/a$,
\be
 I_\gamma\equiv\int_\gamma d\lambda\,
 T_{\mu\nu}k^\mu k^\nu
 =\int_\gamma\frac{\rho+p}{a}\,dt,
 \qquad k^\mu=\frac{dx^\mu}{d\lambda},
 \label{eq:cosmology-anec-definition}
\ee
where $\rho$ and $p$ include the dilaton, ESP particles, and the full
IFS cascade. ANEC requires $I_\gamma\geq0$ on a complete ray.
Using $\rho+p=-2M_{\rm Pl}^2\dot H$ gives
\be
\begin{aligned}
 \int_{t_i}^{t_f}\frac{\rho+p}{a}\,dt
 &=-2M_{\rm Pl}^2\left[\frac Ha\right]_{t_i}^{t_f}
   -2M_{\rm Pl}^2\int_{t_i}^{t_f}\frac{H^2}{a}\,dt,\\
 I_n&=-2M_{\rm Pl}^2\int_{t_n}^{t_{n+1}}\frac{H^2}{a}\,dt<0,
 \qquad H(t_n)=H(t_{n+1})=0,\quad H\not\equiv0.
\end{aligned}
\label{eq:cosmology-flat-frw-anec}
\ee
Between successive bounces, the unweighted integral
$\int(\rho+p)dt$ vanishes, but the affine weight $1/a$ leaves a net
deficit. Thus, if the cascade-supported bounce is embedded in the
cyclic cosmology of Ref.~\cite{Itzhaki:2025cyclic}, every
cycle contributes negative affine null energy, including its
positive-NEC phases. Growth of the scale factor from one cycle to the
next does not change this sign.

For a regular, nonstatic, two-sided null-complete flat-FRW history
satisfying the Einstein equations, with finite limits of $H/a$
at both ends, the complete affine integral is negative, possibly
$-\infty$ \cite{Burwig:2026anec}.
Since complete radial rays in spatially flat FLRW with
$\mathbb{R}^3$ spatial sections are achronal, a self-consistent
semiclassical realization of such a history would violate the
proposed achronal ANEC \cite{Graham:2007va}.
The net-expanding cyclic solution of Ref.~\cite{Itzhaki:2025cyclic},
however, is past null-incomplete when extrapolated through
infinitely many cycles, so its negative cycle integrals do not
constitute a counterexample to this conjecture.
The earlier bounce construction of Ref.~\cite{Itzhaki:2024vaw}
uses an approximately constant effective potential over the
bounce and does not establish the global completeness of the
full evolution.

ANEC violations on prescribed curved backgrounds
\cite{Yurtsever:1995,UrbanOlum:2010} do not by themselves
establish violations by the total stress sourcing a
self-consistent geometry. Positive
curved-space results include the massless-scalar theorem in two
dimensions \cite{WaldYurtsever:1991} and weak-curvature scalar-field
results assuming null convergence
\cite{KontouOlum:2013,Kontou:2015mkt}. The latter assumption fails
at the bounce. 
Wall derives achronal ANEC from the generalized second law
for minimally coupled quantum fields, at leading order
in $\hbar$ about a classical NEC-respecting background
\cite{Wall:2010}. This links the consistency question to horizon
entropy as well as to the stress tensor. For a review of these
distinctions, see Ref.~\cite{KontouSanders:2020}.

\section{Discussion }
\label{sec:discussion-outlook}

The main result of this paper is that IFS decay never terminates.
The original remnant persists indefinitely, continuously seeding a
self-similar fragmentation cascade that produces infinitely many
strings. Remarkably, the complete cascade simultaneously:\\
(i) saturates the ANEC;\\
(ii) has vanishing total energy and momentum;\\
(iii) yet retains negative integrated pressure at every finite time.

The coexistence of these features may have important implications for several proposed applications of IFSs. Having discussed some of the cosmological implications in the previous section, we now turn briefly to two other applications.

\medskip
\noindent
{\it (i) Black holes.}
IFSs have been proposed to play an important role in black-hole interiors \cite{Itzhaki:2018xik,Giveon:2019gfk}. In particular, for near-extremal black fivebranes, Ref.~\cite{Giveon:2020xxh} proposed that their backreaction stabilizes the dilaton, giving rise to a constant-dilaton $\mathrm{AdS}_2$ interior and a horizon shock that obstructs infalling matter.

The decay cascade studied here opens a possible route toward a microscopic understanding of the black-hole information puzzle. The negative-energy folds can strongly affect the propagation of the interior partners of Hawking particles, while the cascade continuously produces correlated descendants and compensating positive-energy flux. Interactions between Hawking partners and this evolving string population, together with the resulting backreaction, could modify the entanglement structure of the black-hole state and potentially provide a mechanism for information to be encoded in the outgoing Hawking radiation.

A natural first step is to determine the backreaction of the complete cascade in the black-hole geometry. In $1+1$ dimensions, the stress-tensor trace and dilaton source,
together with appropriate state and boundary data, determine the
remaining stress-tensor components through the diffeomorphism
Ward identities. Understanding the subsequent interactions of Hawking partners with the cascade could then bring us closer to a microscopic derivation of the Page curve \cite{Page:1993wv}.

\medskip
\noindent
{\it (ii) Wavefunction collapse.}
In Ref.~\cite{Itzhaki2026}, it was argued that, if the present cosmic acceleration is driven by IFSs, their decay could give rise to a Di\'osi--Penrose-like mechanism for objective wavefunction collapse \cite{Diosi:1989dp,Penrose:1996cv}.
The much more detailed understanding of the IFS decay cascade developed here provides an opportunity to revisit this proposal, sharpen its predictions, and explore possible experimental tests.

\section*{Acknowledgments}
This work was supported in part by the ISF through Grant No. 256/22.
U.P. was supported in part by the John Templeton Foundation under
grant number 63749. The opinions expressed in this publication are
those of the author(s) and do not necessarily reflect the views of the
John Templeton Foundation. We thank Paul Steinhardt, Alek Bedroya and Nicolas Patino for helpful discussions.

\appendix

\section{Staircase stress tensors and yo-yo kinematics}
\label{app:staircase-stress-tensor}

This appendix gives the explicit distributional stress tensors associated
with the staircase constructions used in Sec.~\ref{sec:cascade_algorithm}.
The formulas are local to the displayed sampling rectangle.  For the
original IFS remnant, additional records outside the finite rectangle
complete the full infinite remnant.

\subsection{Unbounded staircase of the original IFS remnant}
\label{app:open-staircase-stress}

Let \(p_i=(u_i,v_i)\), \(i=1,\ldots,N\), be the
causal-minimal records, ordered by
\(u_1<\cdots<u_N\) and \(v_1>\cdots>v_N\).
Set \((u_0,v_0)=(u_a,v_b)\) and \(u_{N+1}=u_b\).
The surviving bulk is
\be
 \mathcal S_0=\bigcup_{i=0}^{N}
       [u_i,u_{i+1}]\times[v_a,v_i].
\ee
Writing \(\chi_{[a,b]}(z)=\Theta(z-a)\Theta(b-z)\), its
stress and the positive folds on the stepped boundary are
\begin{align}
 T_{uv}^{\rm stair}(u,v)
 &=\muIFS\sum_{i=0}^{N}
       \chi_{[u_i,u_{i+1}]}(u)\chi_{[v_a,v_i]}(v),
 \label{eq:staircase-common-bulk}\\
 T_{uu}^{\rm stair}(u,v)
 &=\muIFS\sum_{i=1}^{N}(v-v_i)
       \chi_{[v_i,v_{i-1}]}(v)\delta(u-u_i),
 \label{eq:open_staircase_Tuu}\\
 T_{vv}^{\rm stair}(u,v)
 &=\muIFS\sum_{i=1}^{N}(u-u_i)
       \chi_{[u_i,u_{i+1}]}(u)\delta(v-v_i).
 \label{eq:open_staircase_Tvv}
\end{align}
Each fold starts with zero null energy at its splitting point and grows
as it removes bulk from the parent IFS.  The original negative-energy
folds remain attached to the two persistent null arms:
\begin{align}
 T_{uu}^{\rm IFS}
 &=-\muIFS(v-v_a)\Theta(v-v_a)\delta(u-u_a),
 \label{eq:open_staircase_IFS_Tuu}\\
 T_{vv}^{\rm IFS}
 &=-\muIFS(u-u_a)\Theta(u-u_a)\delta(v-v_a).
 \label{eq:open_staircase_IFS_Tvv}
\end{align}
The full zeroth-generation remnant stress tensor is therefore
\be
 \begin{aligned}
 T_{uv}^{(0)}&=T_{uv}^{\rm stair},\qquad
 T_{uu}^{(0)}=T_{uu}^{\rm stair}+T_{uu}^{\rm IFS},\\
 T_{vv}^{(0)}&=T_{vv}^{\rm stair}+T_{vv}^{\rm IFS}.
 \end{aligned}
 \label{eq:zeroth_generation_tensor}
\ee
The remnant is an emitting subsystem.  Inside the sampling patch
its distributional divergence is
\be
 \partial_vT_{uu}^{(0)}+\partial_uT_{uv}^{(0)}=-J_u,
 \qquad
 \partial_uT_{vv}^{(0)}+\partial_vT_{uv}^{(0)}=-J_v,
\ee
with positive null-momentum transfer currents
\begin{align}
 J_u(u,v)&=\muIFS\sum_{i=1}^{N-1}\Delta v_i\,
       \delta(u-u_{i+1})\delta(v-v_i),\\
 J_v(u,v)&=\muIFS\sum_{i=1}^{N-1}\Delta u_i\,
       \delta(u-u_{i+1})\delta(v-v_i).
 \label{eq:remnant-exchange-currents}
\end{align}
Here \(\Delta u_i=u_{i+1}-u_i\) and
\(\Delta v_i=v_i-v_{i+1}\).  These identities are understood against
test functions supported in the interior of the sampling rectangle;
its artificial boundaries are supplied by continuation of the
staircase, not by additional physical sources.

For example, differentiating the upper endpoint of the positive
\(T_{uu}\) fold at \(u=u_{i+1}\) gives
\(-\muIFS\Delta v_i\delta(u-u_{i+1})\delta(v-v_i)\).
The corresponding endpoint of \(T_{vv}\) gives the
\(-\muIFS\Delta u_i\) term.  Along each null arm the bulk and fold
derivatives cancel; at the original split the fold starts with zero
energy.  At \(b_i=(u_{i+1},v_i)\), the daughter inherits the two
incoming fold momenta.  Consequently the stress tensor of all detached
descendants, including their later decay products, obeys
\be
 \partial_vT_{uu}^{\rm desc}+\partial_uT_{uv}^{\rm desc}=J_u,
 \qquad
 \partial_uT_{vv}^{\rm desc}+\partial_vT_{uv}^{\rm desc}=J_v.
 \label{eq:descendant-exchange-currents}
\ee
Transfers between later generations cancel within this sector.
Adding the remnant therefore restores the source-free conservation
equations for the complete cascade, including every birth vertex.

\subsection{Bounded staircase within a yo-yo oscillation}
\label{app:yoyo-staircase-stress}

For a parent yo-yo with oscillation rectangle
\(\mathcal D_Y=[u_a,u_b]\times[v_a,v_b]\), include both turning
points as boundary vertices:
\(p_0=(u_a,v_b)\), \(p_{N+1}=(u_b,v_a)\).
The bulk \(T_{uv}^{Y,\rm stair}\) is given by
Eq.~\eqref{eq:staircase-common-bulk}.  The outgoing components
\(T_{uu}^{Y,\rm stair}\) and \(T_{vv}^{Y,\rm stair}\) have the
forms \eqref{eq:open_staircase_Tuu} and
\eqref{eq:open_staircase_Tvv}, with the summation ranges extended to
\(i=1,\ldots,N+1\) and \(i=0,\ldots,N\), respectively.
The added endpoint terms are the outgoing folds from the two turning
points.

The incoming folds from \((u_a,v_a)\) carry positive null
energy, which decreases as the string bulk expands:
\begin{align}
 T_{uu}^{Y,\rm bottom}(u,v)
 &=\muIFS(v_b-v)\chi_{[v_a,v_b]}(v)\delta(u-u_a),
 \label{eq:yoyo_bottom_Tuu}\\
 T_{vv}^{Y,\rm bottom}(u,v)
 &=\muIFS(u_b-u)\chi_{[u_a,u_b]}(u)\delta(v-v_a).
 \label{eq:yoyo_bottom_Tvv}
\end{align}
The complete stress tensor of the parent during this oscillation is
\be
 \begin{aligned}
 T_{uv}^{Y}&=T_{uv}^{Y,\rm stair},\qquad
 T_{uu}^{Y}=T_{uu}^{Y,\rm stair}+T_{uu}^{Y,\rm bottom},\\
 T_{vv}^{Y}&=T_{vv}^{Y,\rm stair}+T_{vv}^{Y,\rm bottom}.
 \end{aligned}
 \label{eq:general_split_yoyo_tensor}
\ee
For \(N=0\), the formulas reduce to the unsplit yo-yo stress tensor,
\begin{align}
 T_{uv}^{Y}
 &=\muIFS\chi_{[u_a,u_b]}(u)\chi_{[v_a,v_b]}(v),
 \label{eq:unsplit_yoyo_tensor_uv}\\
 T_{uu}^{Y}
 &=\muIFS\big[(v_b-v)\delta(u-u_a)
             +(v-v_a)\delta(u-u_b)\big]\chi_{[v_a,v_b]}(v),\\
 T_{vv}^{Y}
 &=\muIFS\big[(u_b-u)\delta(v-v_a)
             +(u-u_a)\delta(v-v_b)\big]\chi_{[u_a,u_b]}(u).
 \label{eq:unsplit_yoyo_tensor}
\end{align}

\subsection{Component form of the binary transfer map}
\label{app:binary-component-map}
These expressions implement the conservation-based trace transfer
used in Section~\ref{subsec:finite_boost_map}.

Using the \(u\leftrightarrow v\) symmetry, take
\(\Delta u\geq\Delta v\) and define
\be
    r=\frac{\Delta u}{\Delta v}
    =
    \left(
        \gamma+\sqrt{\gamma^2-1}
    \right)^2
    \geq1.
    \label{eq:r-def}
\ee
The daughter boosts are
\be
    \gamma_1
    =
    \frac{r\alpha_U+\alpha_V}
         {2\sqrt{r\alpha_U\alpha_V}},
    \qquad
    \gamma_2
    =
    \frac{
        r(1-\alpha_U)+(1-\alpha_V)
    }{
        2\sqrt{
            r(1-\alpha_U)(1-\alpha_V)
        }
    }.
    \label{eq:daughter-gamma}
\ee
We denote \(\ell_{\gamma,A}=\ln\gamma_A\), \(A=1,2\).

The fractions of the parent trace weight inherited by the daughters are
\begin{align}
    q_1(\gamma;\alpha_U,\alpha_V)
    &\equiv
    \frac{W_1}{W}
    =
    \frac{
        (r+1)\alpha_U\alpha_V
    }{
        r\alpha_U+\alpha_V
    },
    \nonumber\\
    q_2(\gamma;\alpha_U,\alpha_V)
    &\equiv
    \frac{W_2}{W}
    =
    \frac{
        (r+1)(1-\alpha_U)(1-\alpha_V)
    }{
        r(1-\alpha_U)+(1-\alpha_V)
    }.
    \label{eq:q-fractions}
\end{align}
The equivalent fractional trace deficit is
\be
    1-q_1-q_2
    =\frac{r(\alpha_U-\alpha_V)^2}
       {(r\alpha_U+\alpha_V)
        [r(1-\alpha_U)+(1-\alpha_V)]}\geq0.
    \label{eq:W-deficit}
\ee

\section{From the geometric cascade to the dilute entrance population}
\label{appendix:first_generation_source}

The remnant supplies the cascade with strings of fixed characteristic
mass and increasing boost. We derive their emission statistics and
follow them through the initial, non-dilute splittings to a fixed small
mass threshold. The population crossing that threshold supplies the
source for the late-time calculation in
Appendix~\ref{appendix:chronological-renewal}.

\subsection{The first-generation source}
\label{app:direct-source-measure}

The first-generation yo-yos are emitted where positive folds from
consecutive records \(p_i=(u_i,v_i)\) and
\(p_{i+1}=(u_{i+1},v_{i+1})\) meet.  The emission vertex,
time, and null extents are
\begin{align}
 b_i&=(u_{i+1},v_i),\qquad
 t_0=\frac{u_{i+1}+v_i}{2},
 \label{eq:first_generation_emission_time}\\
 \Delta u_i&=u_{i+1}-u_i,\qquad
 \Delta v_i=v_i-v_{i+1}.
 \label{eq:first_generation_extents}
\end{align}
The stationary record variables
\be
 A_i=\GammaSplit u_iv_i,\qquad
 X_i=\GammaSplit v_i(u_{i+1}-u_i),\qquad
 Y_i=\frac{v_{i+1}}{v_i}
\ee
are independent, with \(A_i,X_i\sim\mathrm{Exp}(1)\) and
\(Y_i\sim\mathcal U(0,1)\).  Set
\be
 B=A_i+X_i=\GammaSplit u_{i+1}v_i,\qquad
 Q=\frac{X_i}{B},\qquad S=1-Y_i.
 \label{eq:first_generation_source_variables}
\ee
The variable \(B\) fixes the invariant position of the emission
vertex, while \(Q\) and \(S\) determine the horizontal and vertical
sizes of the emitted yo-yo.  The sum and ratio of the two
independent exponentials give
\be
 \rho_B(B)=Be^{-B},\quad B>0,\qquad
 Q,S\sim\mathcal U(0,1)
 \label{eq:B_distribution}
\ee
independently of \(B\).

At fixed \(B,t_0\), the product
\(u_{i+1}v_i=B/\GammaSplit\) and sum \(u_{i+1}+v_i=2t_0\)
give the two reflection-related branches
\be
 v_\sigma=t_0+\sigma\sqrt{t_0^2-B/\GammaSplit},\qquad
 u_\sigma=\frac{B}{\GammaSplit v_\sigma},\qquad \sigma=\pm.
 \label{eq:source_branches}
\ee
Since \(-\log Y_i\sim\mathrm{Exp}(1)\), the stationary
record process has unit mean density per \(d\log v\).
Changing variables from \((B,\log v)\) to \((B,t_0)\) gives
the source measure on each branch:
\be
 d\dot N_{1,\sigma}
 =\frac{Be^{-B}}{\sqrt{t_0^2-B/\GammaSplit}}\,dB\,dQ\,dS,
 \qquad 0<B<\GammaSplit t_0^2.
 \label{eq:first_generation_source_measure}
\ee
This measure determines the birth statistics before any further
fragmentation.  Its conditional averages are
\be
 \langle F\rangle_{t_0}
 =\frac{1}{\dot N_1(t_0)}\sum_{\sigma=\pm}
       \int d\dot N_{1,\sigma}\,F(B,Q,S;\sigma).
 \label{eq:source_average_definition}
\ee
The extents and kinematic quantities entering these averages are
\begin{align}
 \Delta u_\sigma&=\frac{BQ}{\GammaSplit v_\sigma},\qquad
 \Delta v_\sigma=v_\sigma S,
 \label{eq:source_extents}\\
 A_1&=\frac{BQS}{\GammaSplit},\qquad
 m_1=\frac{2\muIFS}{\sqrt{\GammaSplit}}\sqrt{BQS},
 \label{eq:source_mass}\\
 \gamma_1&=\frac{BQ+h_\sigma S}{2\sqrt{h_\sigma BQS}},\qquad
 h_\sigma=\GammaSplit v_\sigma^2.
 \label{eq:source_boost}
\end{align}
Write \(z=\GammaSplit t_0^2\) and define
\be
 \mathcal M_\nu(z)=\int_0^z dB\,B^\nu e^{-B}
                   \left(1-\frac Bz\right)^{-1/2}.
 \label{eq:crossing_moment_integral}
\ee
Integrating over \(Q,S\) and the two branches gives the rate
and area moments,
\begin{align}
 \dot N_1(t_0)&=\frac{2}{t_0}\mathcal M_1(z)
 =\frac{2}{t_0}\left[1+\frac1z+\mathcal O(z^{-2})\right],
 \label{eq:exact_direct_source_rate}\\
 \langle A_1^p\mid t_0\rangle
 &=\frac{1}{\GammaSplit^p}
   \frac{\mathcal M_{p+1}(z)}{(p+1)^2\mathcal M_1(z)}.
 \label{eq:source_area_moments}
\end{align}
For the mean boost, the intermediate average is
\(\langle\gamma_1\mid B,t_0\rangle
=4\sqrt{\GammaSplit}t_0/(3\sqrt B)\).
The first moments and their late-time limits are therefore
\begin{align}
 \langle A_1\mid t_0\rangle
 &=\frac{\mathcal M_2(z)}{4\GammaSplit\mathcal M_1(z)}
 \longrightarrow\frac{1}{2\GammaSplit},
 \label{eq:mean_source_area}\\
 \langle m_1\mid t_0\rangle
 &=\frac{8\muIFS}{9\sqrt{\GammaSplit}}
   \frac{\mathcal M_{3/2}(z)}{\mathcal M_1(z)}
 \longrightarrow\frac{2\sqrt\pi}{3}
                 \frac{\muIFS}{\sqrt{\GammaSplit}},
 \label{eq:mean_source_mass}\\
 \langle\gamma_1\mid t_0\rangle
 &=\frac43\sqrt z\,\frac{\mathcal M_{1/2}(z)}{\mathcal M_1(z)}
 =\frac{2\sqrt\pi}{3}\sqrt z
       \left[1-\frac{1}{4z}+\mathcal O(z^{-2})\right].
 \label{eq:exact_mean_birth_boost}
\end{align}
Thus the remnant emits two yo-yos per logarithmic interval of
late time, with a fixed invariant-mass scale and boosts growing
linearly with \(t_0\).

The trace average has a logarithmic subleading term that is absent
from the area and mean-boost expansions. On the branch \(u\leq v\),
averaging \(W=4\muIFS uvQS/(uQ+vS)\) first over \(Q,S\) gives
\be
 \frac{\langle W\mid u,v\rangle}{4\muIFS}
 =\frac u2-\frac{u^2}{v}\int_0^1 Q^2
                   \ln\!\left(1+\frac{v}{uQ}\right)dQ.
 \label{eq:first-generation-trace-integral}
\ee
Expanding this integral and averaging with the conditional source measure
\eqref{eq:first_generation_source_measure}, with \(z=\GammaSplit t_0^2\),
yields
\be
 \langle W_1\mid t_0\rangle
 =\frac{2\muIFS}{\GammaSplit t_0}
 \left[1+\frac{2-\gamma_E/2-\ln2-\tfrac12\ln z}{z}
              +O\!\left(\frac{\ln z}{z^2}\right)\right],
 \label{eq:first-generation-trace-expansion}
\ee
where \(\gamma_E\) is Euler's constant. This gives the leading law
\eqref{eq:mean_first_generation_W} and its logarithmic relative
correction. Multiplying by Eq.~\eqref{eq:exact_direct_source_rate}
gives Eq.~\eqref{eq:first_generation_W_source} with the same
\(O(\ln z/z)\) relative accuracy; there is no leading logarithm in
these source laws.

\paragraph{Raw-source moments for chronological matching.}
The same variables verify the source assumptions
in Appendix~\ref{app:pure-power-proof}. With
$c=\GammaSplit/(2\muIFS)$, $t_*=\GammaSplit^{-1/2}$, and the emitted
yo-yo's signed momentum rapidity $y_0$, its marks are
\be
 s_0=cm_0t_*=\sqrt{BQS},\qquad
 U_0=cm_0e^{y_0}u_b=BS,\qquad
 V_0=cm_0e^{-y_0}v_b=BQ.
 \label{eq:raw-source-clock-marks}
\ee
Changing from the birth-event rapidity to $y_0$ only shifts rapidity
by a function of the invariant marks, preserving its unit intensity.
The raw mass moment per unit rapidity is therefore
\be
 \mathscr S_0(p)=\int s_0^p\,d\nu_0
   =\frac{\Gamma(2+p/2)}{(1+p/2)^2},\qquad p>-2.
 \label{eq:raw-source-mass-moment}
\ee
Here $\Gamma$ in the numerator is Euler's gamma function. Positive
polynomial birth-coordinate moments and the mass-weighted negative
logarithmic moments near the selected exponent are finite. The
dimensionless birth radius is
$\sqrt{U_0V_0}/s_0=\sqrt B$, so its mass-weighted large-radius tail
is exponentially suppressed. The following estimates propagate these source moments through the
non-dilute stages to the mass-threshold entrance population.

\subsection{Daughter mass moments and the branching hierarchy}
\label{app:geometric-mass-hierarchy}

Let $s=cmt_*$, with $c=\GammaSplit/(2\muIFS)$ and
$t_*=\GammaSplit^{-1/2}$. The exact staircase law controls the
mass weight passed to later generations, even before the binary
approximation applies.

\paragraph{A bound from the exact staircase daughter law.}
Condition on $n\geq1$ Poisson candidates in the occupied rectangle.
Rescale it to the unit square, retain the causal-minimal records, and
include the boundary vertices $(0,1)$ and $(1,0)$. If consecutive
vertices have gaps $a_i=\Delta u_i$ and $b_i=-\Delta v_i$, the daughter
mass fractions are $z_i=\sqrt{a_i b_i}$. Thus
$\sum_i a_i=\sum_i b_i=1$ and $\sum_i z_i\leq1$ in every realization.
More precisely, with $h=1+p/2$, their conditional moment is
\be
 K_n(p)\equiv\mathbb{E}\!\left[\sum_i z_i^p\mid n\right]
 =\frac{\Gamma(n+1)\Gamma(h)}{h\Gamma(n+h)}
   \left[2+\psi(n+h)-\psi(h+1)\right],\qquad p>0,
 \label{eq:exact-staircase-offspring-moment}
\ee
where $\psi$ is the digamma function. To obtain this expression, the
two boundary gaps contribute $2n\mathrm B(h,n)/h$. For an internal
gap, let $x$ be its right endpoint and $y$ its upper endpoint. The
other $n-2$ candidates must avoid the rectangle of area $xy$.
Integrating the two gap lengths gives
\be
 \frac{n(n-1)}{h^2}\int_0^1(-\ln z)z^h(1-z)^{n-2}dz,
\ee
with this contribution absent for $n=1$; adding the boundary terms
gives Eq.~\eqref{eq:exact-staircase-offspring-moment}.

For $p\geq\kappa_{\rm IFS}$, $K_n(p)$ is nonincreasing in $n$.
Indeed, setting $L_n=2+\psi(n+h)-\psi(h+1)$ gives
$K_{n+1}/K_n=(n+1)[1+1/((n+h)L_n)]/(n+h)$.
The required inequality is
$(h-1)(n+h)L_n\geq n+1$. At $n=1$ it follows from $h^2\geq2$;
its left-minus-right difference increases with $n$, since its
increment is at least $3h-4>0$. Averaging over the conditional
Poisson count therefore gives, at every $\Lambda>0$,
\be
 \mathbb{E}\!\left[\sum_i s_i^{\kappa_{\rm IFS}}\right]
 \leq s^{\kappa_{\rm IFS}},
 \qquad \mathbb{E}\!\left[\sum_i s_i\right]\leq\frac89s.
 \label{eq:geometric-mass-contractions}
\ee
The first two candidate counts have
$K_1(\kappa_{\rm IFS})=K_2(\kappa_{\rm IFS})=1$;
larger counts lower this moment. The exact geometric mass hierarchy
therefore has the same nonincreasing mass weight as the binary
model, although its full offspring distribution and clock differ.

\paragraph{The binary node-count benchmark.}
\label{app:mass-hierarchy-count}

For comparison, the binary mass hierarchy can be solved explicitly.
For a binary hierarchy stopped at mass \(m_{\rm cut}\), define
\(\mathcal C=m/m_{\rm cut}\).  Each daughter reduces this ratio
by its mass fraction:
\be
 \mathcal C_A=z_A\mathcal C,\qquad A=1,2.
 \label{eq:cascade-depth-map}
\ee
Let \(\mathcal N(\mathcal C)\) be the expected number of
nodes above the cutoff.  Counting the parent and its two descendant
hierarchies gives
\be
 \mathcal N(\mathcal C)
 =1+\left\langle\mathcal N(z_1\mathcal C)
                  +\mathcal N(z_2\mathcal C)\right\rangle,
 \qquad \mathcal N(\mathcal C)=0\quad(\mathcal C\leq1).
 \label{eq:cascade-depth-renewal}
\ee
Set $r=\ln\mathcal C$, $n(r)=\mathcal N(e^r)$ and
$\ell_A=-\ln z_A>0$. Then $n(r)=1+\langle n(r-\ell_1)+n(r-\ell_2)\rangle$,
with $n(r)=0$ for $r\leq0$. A shift by $\ell_A$ multiplies the
Laplace transform by $z_A^p$, giving
$\widehat n(p)=1/p+M(p)\widehat n(p)$ and hence
\be
 \widehat n(p)\equiv\int_0^\infty e^{-pr}n(r)\,dr
 =\frac{1}{p[1-M(p)]}.
 \label{eq:malthusian-transform}
\ee
Its unique leading pole is at \(M(\kappa_{\rm IFS})=1\),
so \(n(r)\sim C_Ne^{\kappa_{\rm IFS}r}\), yielding
Eq.~\eqref{eq:malthusian-node-growth}.  Scale-local changes to the
stopping rule affect the coefficient but leave this pole unchanged.
The exponent describes the ensemble-averaged, multiplicity-weighted
cascade of Sec.~\ref{subsec:large-boost-selection}. It also determines
the expected active worldsheet area: every counted node sweeps mean
area $\GammaSplit^{-1}$ before its next split.

\subsection{Mass-threshold entrance into the dilute regime}

\label{app:threshold-entrance-checks}

The first-generation law can be carried through the non-dilute layer
without specifying that layer's full distribution. Fix a dimensionless
mass $s_d>0$, and stop each branch when its mass
$s=cm t_*$ first satisfies $s\leq s_d$. Here
$t_*=\GammaSplit^{-1/2}$ and $c=\GammaSplit/(2\muIFS)$, so the
mean number of candidates in a full oscillation is $\Lambda=s^2$.
The threshold defines a stopping population using only the past
evolution, so the subsequent splitting law is unchanged. We use complete
oscillation rectangles of the unabsorbed geometric model.

\paragraph{Entrance masses and birth times.}
Starting from one raw root of mass $s_0$, the expected sum of masses
over all nodes is at most $9s_0$. Consequently the expected number
of nodes before threshold entrance is at most $9s_0/s_d$.
Moreover, stopping the first inequality in
Eq.~\eqref{eq:geometric-mass-contractions} gives
\be
 \mathbb{E}\!\left[\sum_{e\in\mathcal E_d}s_e^{\kappa_{\rm IFS}}\right]
 \leq s_0^{\kappa_{\rm IFS}}.
 \label{eq:threshold-entrance-mass-bound}
\ee
Both bounds follow first for a finite number of generations and
then by monotone convergence or Fatou's lemma. The first-generation
law has finite polynomial moments, so these bounds are integrable
per unit source rapidity.
Raw roots already below the threshold have positive intensity, so
the entrance moment at $\kappa_{\rm IFS}$ is strictly positive.

Timing requires a separate estimate. Use the invariant birth
coordinates $U=cm e^y u$, $V=cm e^{-y}v$ of
Appendix~\ref{app:pure-power-proof}, and set $H=1+U+V$.
For a parent of mass $s$, couple the number of empty oscillations to
$\epsilon\sim{\rm Exp}(1)$ as $K=\lfloor\epsilon/s^2\rfloor$.
The occupied rectangle lies within a displacement
$s^2(K+1)\leq\epsilon+s^2$ in each parent clock coordinate.
Every daughter therefore obeys
$H_i\leq H+2(\epsilon+s_0^2)$, since masses never increase.
The empty-cycle count is independent of the occupied staircase.
For any fixed $r\geq1$, applying
$(H+D)^r\leq(1+\varepsilon)H^r+C_{r,\varepsilon}D^r$
with $(1+\varepsilon)8/9<1$ yields
\be
 \mathbb{E}\!\left[\sum_{\text{pre-entrance }Y}s_YH_Y^r\right]
 \leq C_r s_0\left[H_0^r+(1+s_0^2)^r\right].
 \label{eq:threshold-birth-coordinate-bound}
\ee
At an occupied split the expected number of daughters is bounded by
$\mathbb{E}[n+1\mid n\geq1]\leq s_0^2+2$.
Using $s_Y>s_d$ before entrance, the same bound controls the
unweighted entrance count and its $H^r$ moments, with constants
depending on $s_d$. Together with the raw-source marks
\eqref{eq:raw-source-clock-marks}, this proves
\be
 \int s^\sigma(1+U+V)^r d\nu_d<\infty
 \qquad(\sigma>0,\ r\geq0)
 \label{eq:threshold-entrance-moments}
\ee
for the threshold entrance measure $d\nu_d$.
Mass-weighted negative logarithms follow by using a slightly smaller
positive $\sigma$. For a daughter entering at a parent's occupied
rectangle, its birth radius obeys
$r_e\leq[H_Y+2(\epsilon+s_0^2)]/(2s_d)$.
The direct raw roots have $r_0=\sqrt B$.
Thus the entrance radial moments are finite as well, implying the
injection-tail condition used in
Eq.~\eqref{eq:entrance-mass-tail-conditions}.

\paragraph{Late-time injection into the dilute cascade.}
For one boost orbit with marks $(s,U,V)$, its laboratory birth time
is $T_b(y)=(Ue^{-y}+Ve^y)/(2s)$ and its injected dimensionless trace
is $s/\cosh y$. If $T_b>T$, at least one of
$Ue^{-y}>sT$ and $Ve^y>sT$ holds. Integrating the corresponding
two rapidity tails gives
\be
 J_{>}(T)\equiv\int dy\,d\nu_d\,
       \frac{s}{\cosh y}\,\mathbf1_{\{T_b(y)>T\}}
 \leq\frac2T\int(U+V)d\nu_d=O(T^{-1}).
 \label{eq:threshold-future-injection-bound}
\ee
Thus the trace still awaiting injection is subleading to the
$T^{\kappa_{\rm IFS}-1}$ cascade tail. The same estimate applied to
the end of each pre-entrance occupied
rectangle, using Eq.~\eqref{eq:threshold-birth-coordinate-bound},
gives $ct_*W_{\rm pre}(t_*T)=O(T^{-1})$. Within an oscillation the
instantaneous trace is at most twice its period average, so this
bound also controls the geometric pre-entrance stress.

Together these estimates supply the entrance hypotheses of
Appendix~\ref{appendix:chronological-renewal}: the stopped mass moment
is finite and positive, Eq.~\eqref{eq:threshold-entrance-moments}
controls the inherited clock, and
Eq.~\eqref{eq:threshold-future-injection-bound} makes delayed injection
and pre-entrance trace subleading. Appendix~\ref{app:geometric-matching}
then follows the geometric evolution with these inherited birth times.
The mass asymptotics use the additional pre-entrance mass condition
stated in App.~\ref{appendix:moment-normalization}.

\section{Chronological renewal equation and the large-age limit}
\label{appendix:chronological-renewal}

Repeated fragmentation both multiplies the strings and lengthens their
lifetimes. We show how their competition fixes the surviving trace,
$W\propto T^{\kappa-1}$, and its normalized boost distribution. The
source sets a common amplitude; the late generations determine the
power and profile. Here $T=t/t_*$ and $q_{\rm IFS}=1-\kappa$.

The calculation starts with the dilute binary process, retaining the
birth time inherited from every ancestor. Its mass-weighted clock loses
memory of the entrance data and approaches a stationary distribution.
Summing generations then produces the pole $M(\kappa)=1$ shared by the
observables below. Monotonicity converts the accumulated response into
the trace on a time slice, while an occupation integral resolves that
trace in boost. Finally, App.~\ref{app:geometric-matching} restores the
geometric splitting clock and multiple-daughter events. Their corrections
are summable; conservation also bounds events close to the observation
slice, establishing the same leading coefficient for the instantaneous
ensemble pressure and the period-averaged live trace.

\subsection{\texorpdfstring{Chronological variables and source assumptions}{Chronological variables and source assumptions}}
\label{app:pure-power-proof}

For the dilute binary reference process, each node propagates along its
center-of-mass worldline until an exponentially distributed split.  Its
proper-time rate is $cm$, with
$c=\GammaSplit/(2\muIFS)$, and the two daughter null fractions are
uniform and independent.  Set
\be
 t_*=\GammaSplit^{-1/2},\qquad \kappa=\kappa_{\rm IFS}.
\ee
For a node of mass $m$, signed rapidity $y$, and birth event $(u,v)$,
define the birth coordinates in its own lifetime units,
\be
 Z=(U,V),\qquad U=cm e^y u,\qquad V=cm e^{-y}v.
 \label{eq:chronological-birth-coordinates}
\ee
After a dimensionless proper waiting time
$\epsilon\sim{\rm Exp}(1)$, its split radius and one daughter's
coordinates are
\begin{align}
 R&\equiv cm\rho=\sqrt{(U+\epsilon)(V+\epsilon)},
 &\rho&=\sqrt{u_{\rm split}v_{\rm split}},
 \label{eq:chronological-split-radius}\\
 z&=\sqrt{ab},
 &Z'&=\bigl(a(U+\epsilon),b(V+\epsilon)\bigr),
 \label{eq:chronological-affine-recursion}
\end{align}
with the second daughter obtained from $(1-a,1-b)$.  In these coordinates a proper wait $\epsilon/(cm)$ adds $\epsilon$
to both entries. The daughter's mass and rapidity then rescale the two
entries by $a$ and $b$. Thus $Z$ expresses the position inherited from
all preceding waits and boost changes in the current node's lifetime units.

\paragraph{Entrance source.}
The strings that first enter a fixed dilute threshold form a stopping
population with boost-invariant expected measure
\be
 dy_0\,d\nu(s,U_0,V_0),\qquad s=cm_0t_*>0,
 \label{eq:chronological-invariant-source}
\ee
and source Mellin factor
\be
 \mathscr S(\zeta)=\int s^\zeta d\nu.
 \label{eq:chronological-source-mellin}
\ee
For some $0<\delta<1-\kappa$ and sufficiently small $\eta>0$, assume
\be
 0<\mathscr S(\kappa)<\infty,\qquad
 \int s^\sigma(1+U_0^\delta+V_0^\delta)d\nu<\infty,
 \qquad |\sigma-\kappa|\leq\eta,
 \label{eq:chronological-matching-moments}
\ee
with $\kappa+\eta+\delta<1$.  The expected future injection satisfies
$J_>(T)=O(T^{-1})$ for $T=t/t_*$.  Appendix~\ref{app:threshold-entrance-checks}
verifies these conditions, together with a subleading pre-entrance
trace, for the geometric mass-threshold population.

\subsection{A common branching resolvent}
\label{app:chronological-contraction}

The mass fraction $z$ is correlated with the inherited clock, so the
generation sum must keep both.  For real $p$ near $\kappa$, selecting
one daughter with the mass weight gives
\be
 d\mathbb P_p(a,b)
 =\frac{2(ab)^{p/2}}{M(p)}\,da\,db,
 \qquad 0<a,b<1,
 \label{eq:main-weighted-path-law}
\ee
and therefore
\be
 \left\langle\sum_{|Y|=n}
       \left(\frac{m_Y}{m_0}\right)^p R_Y^{-p}\right\rangle
 =M(p)^n\,\mathbb{E}_{p,Z_0}[R_n^{-p}].
 \label{eq:main-weighted-generation-clock}
\ee
At $p=\kappa$, $M(p)=1$, so this weighted generation sum is itself a
probability expectation.

The transfer operator advances one generation with this mass weight;
$h_\zeta$ averages the final split radius over the last waiting time:
\begin{align}
 h_\zeta(U,V)
 &=\int_0^\infty d\epsilon\,e^{-\epsilon}
 [(U+\epsilon)(V+\epsilon)]^{-\zeta/2},\\
 (\mathscr T_\zeta f)(U,V)
 &=2\int_0^1da\int_0^1db\,(ab)^{\zeta/2}
   \int_0^\infty d\epsilon\,e^{-\epsilon}
   f\bigl(a(U+\epsilon),b(V+\epsilon)\bigr).
 \label{eq:chronological-transfer-operator}
\end{align}
Then $\mathscr T_\zeta1=M(\zeta)$.  Define
\[
 d_\delta(Z,Z')=|U-U'|^\delta+|V-V'|^\delta,\qquad
 [f]_\delta=\sup_{Z\ne Z'}\frac{|f(Z)-f(Z')|}{d_\delta(Z,Z')},\qquad
 w(Z)=U^\delta+V^\delta.
\]
Short waiting times make
$h_\zeta$ non-Lipschitz at the origin, but for
$\sigma=\operatorname{Re}\zeta$ near $\kappa$ it is H\"older:
\be
 |h_\zeta(Z)|\leq\Gamma(1-\sigma),\qquad
 |h_\zeta(Z)-h_\zeta(Z')|
 \leq C_\zeta\Gamma(1-\sigma-\delta)
 \bigl(|U-U'|^\delta+|V-V'|^\delta\bigr).
 \label{eq:chronological-negative-moment-bound}
\ee
Evolving two copies with the same $a,b,\epsilon$ multiplies their
coordinate differences by $a,b$, giving
\be
 [\mathscr T_\zeta f]_\delta
 \leq\vartheta_\delta(\sigma)[f]_\delta,
 \qquad
 \vartheta_\delta(\sigma)
 =\frac{2}{(1+\sigma/2)(1+\sigma/2+\delta)}<1
 \label{eq:chronological-holder-contraction}
\ee
in a sufficiently narrow strip around $\kappa$.

The same contraction applies to any analytic observable weight
$b_\zeta(Z)$ with these boundedness and H\"older estimates.  With
$f_n=\mathscr T_\zeta^n b_\zeta$, separate
$f_n=a_n+g_n$ with $a_n=f_n(0,0)$.  The coordinate-dependent part
contracts geometrically,
\be
 |g_n(Z)|\leq C_\zeta\theta^n w(Z),\qquad
 |d_n(\zeta)|\leq C'_\zeta\theta^n,
 \quad d_n=(\mathscr T_\zeta g_n)(0,0),\quad \theta<1,
 \label{eq:chronological-remainder-bound}
\ee
and the scalar recurrence
$a_{n+1}=M(\zeta)a_n+d_n$ gives
\be
 \sum_{n\geq0}(\mathscr T_\zeta^n b_\zeta)(Z)
 =\frac{\mathscr A_b(\zeta)}{1-M(\zeta)}
  +\mathscr R_{b,\zeta}(Z),
 \label{eq:chronological-resolvent}
\ee
where $\mathscr A_b$ and $\mathscr R_{b,\zeta}$ are analytic across the
Malthusian line after source integration.  Equivalently, the analytic
pieces may be written as
\be
 \mathscr A_b(\zeta)=b_\zeta(0,0)+\sum_{n\geq0}d_n(\zeta),
 \qquad
 \mathscr R_{b,\zeta}(Z)=\sum_{n\geq0}g_n(Z).
 \label{eq:chronological-analytic-remainders}
\ee

For $b=h$, the $\kappa$-weighted recursion is a Markov chain.  Its birth
coordinates approach a stationary law $\pi$ with finite first moments,
and
\be
 \mathscr A_h(\kappa)=\pi(h_\kappa),
 \qquad 0<\pi(h_\kappa)\leq\Gamma(1-\kappa)<\infty.
\ee
Thus the radial split-event transform
\be
 \mathscr J(\zeta)
 \equiv\int d\nu\,s^\zeta
       \sum_{n\geq0}(\mathscr T_\zeta^n h_\zeta)(Z_0)
 =\frac{\mathscr S(\zeta)\mathscr A_h(\zeta)}{1-M(\zeta)}
  +\mathscr H_{\rm split}(\zeta)
 \label{eq:chronological-meromorphic-transform}
\ee
has one simple pole on $\operatorname{Re}\zeta=\kappa$, with positive
radial residue
\be
 R_*\equiv\mathop{\rm Res}_{\zeta=\kappa}\mathscr J(\zeta)
 =\frac{\mathscr S(\kappa)\pi(h_\kappa)}{-M'(\kappa)}>0.
\ee
The corresponding laboratory-time residue is
\be
 D\equiv
 {\rm B}\!\left(\frac\kappa2,\frac12\right)R_*>0.
 \label{eq:chronological-positive-residue}
\ee
The second zero of $1-M(\zeta)$ lies to the left, so subtracting the
pole leaves a function analytic across the whole boundary line.

\paragraph{From the accumulated response to a time slice.}
The pole fixes the response integrated up to time $T$. To recover the
live trace, we use its decrease under splitting: after including future
injection, the value on a time slice is bounded by averages over adjacent
time intervals. For a nonnegative live observable $Q(T)$ whose transform
has residue $D_Q$
and for which adding the future source tail makes
$Q(T)+J_>(T)$ nonincreasing, Wiener--Ikehara applied to its primitive
gives\footnote{We use the $N=0$ single-pole case of the
Laplace--Stieltjes Wiener--Ikehara theorem, Theorem 5.4 of
Ref.~\cite{Debruyne_2019}, with $x=\ln T$. The finite-radius contribution
is analytic at the boundary.}
\be
 \int_1^TQ(s)ds\sim\frac{D_Q}{\kappa}T^\kappa.
 \label{eq:chronological-integrated-limit}
\ee
The compensated quantity
\be
 Q_{\rm c}(T)=Q(T)+J_>(T)
 \label{eq:chronological-monotone-compensator}
\ee
has the same primitive to leading order because $J_>=O(T^{-1})$.
Writing \(G(T)=\int_1^TQ_{\rm c}(s)ds\), for any fixed
$\lambda>1$ monotonicity gives
\be
 \frac{G(\lambda T)-G(T)}{(\lambda-1)T}
 \leq Q_{\rm c}(T)\leq
 \frac{G(T)-G(T/\lambda)}{(1-1/\lambda)T},
 \label{eq:chronological-density-sandwich}
\ee
so $Q(T)\sim D_QT^{\kappa-1}$. We use this corollary for occupation
observables, including the unmarked binary trace. The exact geometric
instantaneous stress will instead use the positive split measure together
with the endpoint bound in Eq.~\eqref{eq:geometric-Abel-endpoint-bound}.

\subsection{\texorpdfstring{Radial split measure and instantaneous pressure}{Radial split measure and instantaneous pressure}}
\label{app:chronological-pointwise-limit}

Let $\mathscr N(dr)$ be the expected binary split-event measure per unit source
rapidity, expressed in the dimensionless invariant radius
$r=\rho/t_*$.  By construction its Mellin transform is
$\mathscr J(\zeta)$.  The positive-measure Wiener--Ikehara theorem
therefore gives
\be
 \mathscr N((0,R])\sim\frac{R_*}{\kappa}R^\kappa.
 \label{eq:chronological-radial-split-count}
\ee
This is the chronological version of the Malthusian node count: the
clock changes the residue, but not the power.

To convert radial events to a Minkowski-time slice, write
$u=\rho e^\varphi$, $v=\rho e^{-\varphi}$, so $t=\rho\cosh\varphi$.  The boost
orbit contributes
\be
 K_\rho(t)=\int_{\mathbb R}d\varphi\,\delta(t-\rho\cosh\varphi)
 =\frac{2\Theta(t-\rho)}{\sqrt{t^2-\rho^2}},
 \qquad
 \int_r^\infty\frac{2T^{-\zeta}\,dT}{\sqrt{T^2-r^2}}
 ={\rm B}\!\left(\frac\zeta2,\frac12\right)r^{-\zeta}.
 \label{eq:chronological-boost-mellin}
\ee
The binary splitting-rate identity then gives the exact Abel relation
\be
 \mathcal W(T)
 =2\int_{(0,T]}\frac{\mathscr N(dr)}{\sqrt{T^2-r^2}},
 \qquad
 \mathcal W(T)=ct_*W_{\rm live}(t_*T).
 \label{eq:chronological-trace-from-events}
\ee
The boost projection enhances events near $r=T$. In the binary model,
monotonicity controls their contribution: energy and momentum conservation give
\(\sum_i m_i^2/E_i\leq m^2/E\) at a split, so
\(\mathcal W+J_>\) is nonincreasing. Its Mellin transform is
\({\rm B}(\zeta/2,1/2)\mathscr J(\zeta)\), with residue \(D\).
Equations~\eqref{eq:chronological-integrated-limit}--
\eqref{eq:chronological-density-sandwich} therefore give
\be
 \mathcal W(T)
 \sim D T^{\kappa-1},
 \qquad
 W_{\rm live}(t)
 \sim2D\muIFS\GammaSplit^{-1/2}
       (\sqrt{\GammaSplit}t)^{\kappa-1}.
 \label{eq:chronological-pure-power-result}
\ee
Thus the binary pressure has the pure power in
Eq.~\eqref{eq:pressure-law}, with coefficient fixed by the stationary
clock and the entrance mass moment.

\subsection{\texorpdfstring{Boost distribution at fixed time}{Boost distribution at fixed time}}
\label{app:chronological-snapshot-profile}

The boost profile uses the same resolvent with a different reward.  Let
\[
 \mathcal W(T,y)=ct_*\,\frac{dW_{\rm live}(t_*T)}{dy}
\]
for signed velocity rapidity $y$.  For
\be
 L_y(Z,\epsilon)
 =\frac{(U+\epsilon)e^{-y}+(V+\epsilon)e^y}{2},
 \label{eq:snapshot-lab-split-time}
\ee
and a bounded velocity test $\psi$, define the occupation reward
\be
 h_{\zeta,\psi}(Z)
 =\int_{\mathbb R}dy\,\psi(\tanh y)
   \int_0^\infty d\epsilon\,e^{-\epsilon}
        L_y(Z,\epsilon)^{-\zeta}.
 \label{eq:snapshot-occupation-kernel}
\ee
The factor $e^{-\epsilon}$ here is the probability of surviving to
proper age $\epsilon$: averaging an integral over a node's lifetime gives
the displayed integral over its survival probability. The reward obeys
the same H\"older estimates as $h_\zeta$, with the additional finite factor $\int dy(\cosh y)^{-\sigma}$.  The common resolvent
therefore gives
\be
 \int_0^\infty dT\,T^{-\zeta}
     \int dy\,\psi(\tanh y)\mathcal W(T,y)
 =\frac{\mathscr S(\zeta)\mathscr A_\psi(\zeta)}{1-M(\zeta)}
  +\mathscr H_\psi(\zeta),
 \label{eq:snapshot-marked-transform}
\ee
with residue
\be
 D_\psi
 =\frac{\mathscr S(\kappa)}{-M'(\kappa)}
   \int_{\mathbb R}dy\,\psi(\tanh y)\mathscr P_\kappa(y),
 \qquad
 \mathscr P_\kappa(y)
 =\mathbb{E}_{Z\sim\pi,\epsilon}
       [L_y(Z,\epsilon)^{-\kappa}].
 \label{eq:snapshot-residue-profile}
\ee
Integrating over $y$ first gives
\be
 \mathcal Z_\kappa\equiv\int_{\mathbb R}\mathscr P_\kappa(y)dy
 ={\rm B}(\kappa/2,1/2)\pi(h_\kappa),
 \qquad D_1=D.
 \label{eq:snapshot-total-pressure-check}
\ee

To recover a fixed-time distribution, use
\be
 \psi_b(v)=\frac{1}{1-bv},\qquad
 W_b(t)=\sum_Y\frac{m_Y^2}{E_Y-bP_Y},\qquad |b|<1.
 \label{eq:snapshot-monotone-test-family}
\ee
Each split decreases $W_b$ by the same Cauchy--Schwarz argument used
for the unmarked trace.  Since its future-injection tail is still
$O(T^{-1})$, the fixed-time corollary above gives
\be
 T^{1-\kappa}\int dy\,\psi_b(\tanh y)\mathcal W(T,y)
 \longrightarrow D_{\psi_b}.
 \label{eq:snapshot-test-function-limit}
\ee
On the compact velocity interval $[-1,1]$, any subsequential limit
has the displayed integrals against every $\psi_b$. The expansion
$\psi_b(v)=\sum_{n\geq0}b^nv^n$, uniformly convergent for $|b|<1$,
then fixes all its moments and hence the measure. The candidate limit has
no mass at $v=\pm1$, so this also gives weak convergence of the normalized
rapidity measure.  In
terms of $\gamma=\cosh y$,
\be
 \frac{\rho_W(t,\gamma)d\ln\gamma}{W_{\rm live}(t)}
 \Longrightarrow\mathcal R_\kappa(\gamma)d\ln\gamma,
 \qquad
 \mathcal R_\kappa(\gamma)
 =\frac{2\mathscr P_\kappa(y)}{\mathcal Z_\kappa\tanh y},
 \quad y=\operatorname{arcosh}\gamma.
 \label{eq:snapshot-normalized-boost-limit}
\ee

For the finite-boost shape, let $A=U+\epsilon$, $B=V+\epsilon$, and
$w=(1+\tanh y)/2$.  Then
\be
 \mathscr P_\kappa(y)
 = (\cosh y)^{-\kappa}\mathcal F_\kappa^{\rm occ}(w),
 \qquad
 \mathcal F_\kappa^{\rm occ}(w)
 =\mathbb{E}[((1-w)A+wB)^{-\kappa}].
 \label{eq:snapshot-occupation-correction}
\ee
Exchange symmetry gives
$\mathcal F_\kappa^{\rm occ}(w)=\mathcal F_\kappa^{\rm occ}(1-w)$.
Its endpoint value follows from the stationary
${\rm Gamma}(\sqrt2,1)$ marginal,
\be
 \mathcal F_\kappa^{\rm occ}(1)
 =\frac{\Gamma(\sqrt2+1-\kappa)}{\Gamma(\sqrt2+1)}.
 \label{eq:snapshot-tail-constant}
\ee
Consequently
\begin{align}
\mathcal R_\kappa(\gamma)
 &\sim\frac{2\mathcal F_\kappa^{\rm occ}(1)}{\mathcal Z_\kappa}
       \gamma^{-\kappa}, &&\gamma\to\infty,
 \label{eq:snapshot-high-boost-tail}\\
 \mathcal R_\kappa(\gamma)
 &\sim\frac{\sqrt2\,\mathcal F_\kappa^{\rm occ}(1/2)}
             {\mathcal Z_\kappa\sqrt{\gamma-1}},
 &&\gamma\to1.
 \label{eq:snapshot-low-boost-jacobian}
\end{align}
A fixed finite boost bin therefore has a well-defined asymptotic mass;
no uniform approximation up to the moving source edge is assumed.

\subsection{Matching the binary and geometric cascades}
\label{app:geometric-matching}

As a string fragments, its worldsheet area per oscillation decreases.
Both the error in the exponential waiting-time approximation and the
probability of multiple candidates in one oscillation therefore become
small. Their mass-weighted sum over the tree is finite, preserving the
branching exponent and normalized boost profile.

The argument first establishes this summability and then applies it to the live population and to the physical split-event measure. The latter is proportional to
the instantaneous bulk trace and gives the pressure directly. For the
occupation calculation a parent is removed at its first physical split;
daughters enter at their actual birth events. The occupied-period bound
below also covers the snapshot convention used in the simulations.

\paragraph{A common geometric perturbation bound.}
Write $s=cmt_*$, $Z=(U,V)$, $H=1+U+V$, and
$\Lambda=s^2$.  The exact generation operator is
\be
 (\mathscr G_{s,\zeta}f)(Z)
 =\mathbb{E}\!\left[\sum_i z_i^\zeta f(Z_i^{\rm G})\mid s,Z\right].
 \label{eq:exact-geometric-transfer-operator}
\ee
Couple the number of empty oscillations to a unit exponential by
$K=\lfloor\epsilon/\Lambda\rfloor$.  Every event in the occupied
rectangle lies within one period of the exponential clock,
\be
 |\Lambda(K+\theta)-\epsilon|\leq\Lambda,
 \qquad 0\leq\theta\leq1.
 \label{eq:oscillation-clock-coupling}
\ee
Conditional on one candidate, the daughter fractions are exactly the
binary ones and the inherited-coordinate discrepancy is $O(\Lambda)$;
the probability of more than one candidate is $O(\Lambda)$.
Choose $a$ and a narrow strip about $\kappa$ such that
$0<a<1-\kappa-\eta$ and $2a>\eta$.  With the anchored H\"older norm
\be
 \|f\|_{a,*}=|f(0,0)|+[f]_a,
 \qquad |f(Z)|\leq\|f\|_{a,*}(1+U^a+V^a),
 \label{eq:geometric-weighted-function-space}
\ee
these facts give
\be
 |(\mathscr G_{s,\zeta}-\mathscr T_\zeta)f(Z)|
 \leq C s^{2a}H^a\|f\|_{a,*}.
 \label{eq:geometric-kernel-convergence}
\ee
The clock weight can also be controlled generation by generation.
Fix $s_0\leq s_d$. The exact staircase law gives
$\mathbb{E}\bigl[\sum_i z_i^p\bigr]\leq M(p)<1$ for $p>\kappa$.
Every daughter obeys $H_i\leq H+2(\epsilon+s_d^2)$, as in
App.~\ref{app:threshold-entrance-checks}. The empty-cycle clock is
independent of the occupied staircase. For $0<r\leq1$, subadditivity
therefore bounds the conditional sum of $z_i^pH_i^r$ by
$M(p)(H^r+C_r)$. Iterating this inequality, while the unweighted
$p$-mass decreases as $M(p)^g$, gives
\be
 \mathbb{E}\!\left[\sum_{|Y|=g}
 \left(\frac{s_Y}{s_0}\right)^pH_Y^r\right]
 \leq C_{p,r}(g+1)M(p)^gH_0^r,
 \qquad 0\leq r\leq1.
 \label{eq:geometric-weighted-generation-bound}
\ee
Define the weighted exact generation operator by
\be
 (\mathscr P_g^{\rm G}f)(Z_0)
 =\mathbb{E}_{s_0,Z_0}\!\left[\sum_{|Y|=g}
       (s_Y/s_0)^\zeta f(Z_Y^{\rm G})\right],\qquad
 \mathscr P_0^{\rm G}=\mathrm{Id}.
 \label{eq:geometric-weighted-generation-definition}
\ee
Hence the generation error
$\mathscr E_g=\mathscr P_{g+1}^{\rm G}-
\mathscr P_g^{\rm G}\mathscr T_\zeta$ obeys
\be
 |(\mathscr E_gf)(Z_0)|
 \leq Cs_0^{2a}(g+1)M(\sigma+2a)^gH_0^a\|f\|_{a,*},
 \label{eq:geometric-generation-error-bound}
\ee
which is summable, including after entrance-source integration.

Now let $b_\zeta$ be any binary observable weight covered by the common resolvent,
and $b^{\rm G}_{s,\zeta}$ its geometric counterpart.  If
\be
 |b^{\rm G}_{s,\zeta}(Z)-b_\zeta(Z)|
 \leq Cs^{2a}H^a,
 \label{eq:geometric-reward-error}
\ee
then the Duhamel expansion
\be
 \mathscr P_n^{\rm G}
 =\mathscr T_\zeta^n+
   \sum_{j=0}^{n-1}\mathscr E_j\mathscr T_\zeta^{n-1-j}
 \label{eq:geometric-duhamel-expansion}
\ee
and the bounds above give
\be
 \widetilde{\mathcal O}_{\rm G}(\zeta)
 =\frac{\mathscr S_{\rm eff}(\zeta)\mathscr A_b(\zeta)}
        {1-M(\zeta)}+\mathscr H_{{\rm G},b}(\zeta),
 \label{eq:geometric-meromorphic-transform}
\ee
with analytic remainder. To see why the source factor is common,
sum the Duhamel identity over $n$ and insert
Eq.~\eqref{eq:chronological-resolvent}. Its singular part is constant
in $Z$, so each generation error acts on $1$, independently of the
chosen reward:
\be
 \mathscr S_{\rm eff}(\zeta)
 =\int d\nu\,s_0^\zeta
    \left[1+\sum_{j\geq0}(\mathscr E_j1)(Z_0)\right].
 \label{eq:geometric-effective-source}
\ee
The coordinate-dependent resolvent remainder and the direct reward
errors have summable analytic contributions. For the constant reward
the clock drops out: the $O(s^2)$ probability of extra candidates gives
$|\mathscr E_j1|\leq Cs_0^2M(\sigma+2)^j$. At $\zeta=\kappa$,
this implies $\mathscr S_{\rm eff}(\kappa)
=\mathscr S(\kappa)[1+O(s_d^2)]>0$ for sufficiently dilute entrance.
The residue is
\be
 D_{{\rm G},b}
 =\frac{\mathscr S_{\rm eff}(\kappa)}{-M'(\kappa)}
   \pi(b_\kappa).
 \label{eq:geometric-common-residue}
\ee
Thus the early geometric evolution enters these observables through a
common amplitude. Their pole and normalized residue ratios are those
of the binary process.

\paragraph{String occupation and the boost profile.}
For a node of rapidity $y$, put
$B_y=(Ue^{-y}+Ve^y)/2$ and $L_y(Z,r)=B_y+r\cosh y$.  If
$\ell_s(y)$ is its exact lifetime to the first physical split, the
geometric occupation reward is
\be
 b^{\rm G}_{s,\zeta,\psi}(Z)
 =\int_{\mathbb R}dy\,\psi(\tanh y)
   \mathbb{E}\!\left[\int_0^{\ell_s(y)}L_y(Z,r)^{-\zeta}\,dr\right].
 \label{eq:geometric-observation-kernel}
\ee
Since $|\ell_s(y)-\epsilon|\leq\Lambda$, the endpoint discrepancy
satisfies
\be
 |b^{\rm G}_{s,\zeta,\psi}-h_{\zeta,\psi}|
 \leq C\|\psi\|_\infty\Lambda,
 \label{eq:geometric-observation-error}
\ee
so Eq.~\eqref{eq:geometric-reward-error} applies.  The normalized
residues therefore give exactly the binary profile
$\mathcal R_\kappa$.

The finite interval between a split and daughter formation contributes a subleading trace.  Count a daughter
as pending between its parent's first split and its actual birth.  The
probability that a fixed time lies in the occupied period is
\be
 (1-e^{-\Lambda})e^{-\Lambda\lfloor r/\Lambda\rfloor}
 \leq C\Lambda e^{-r},
 \label{eq:geometric-occupied-period-probability}
\ee
and the same boost envelope as in the binary source bound gives
\be
 |\mathcal P_\psi(T)|\leq\frac{C\|\psi\|_\infty}{T}
 \int d\nu_d\,\mathbb{E}\!\left[\sum_Ys_Y^2H_Y\right]=O(T^{-1}).
 \label{eq:geometric-pending-weight-bound}
\ee
Adding this pending term and the $O(T^{-1})$ future source makes the
monotone test rewards $W_b$ nonincreasing, so the fixed-time corollary
of Eq.~\eqref{eq:chronological-density-sandwich} yields
\be
 \mathcal W_{{\rm G},\psi}(T)
 \sim D_{{\rm G},\psi}T^{\kappa-1}
 \qquad(\psi=1\ \hbox{or}\ \psi_b).
 \label{eq:geometric-fixed-time-limit}
\ee
The trace retained by the numerical parent convention differs only in
the same occupied period and is bounded by the unsplit parent's weight.
Equation~\eqref{eq:geometric-pending-weight-bound} therefore also bounds
the difference between the two snapshot conventions by \(O(T^{-1})\).
For physical bulk stress use the factor-two bound
\eqref{eq:instantaneous-yoyo-trace-bound}. This proves the same normalized
fixed-bin boost profile for both geometric occupation conventions.

\paragraph{Physical split events and the instantaneous pressure.}
\label{app:instantaneous-pressure-events}
The same active area carries string tension and admits splitting.
One sheet has $T_{uv}=\muIFS$ and physical split intensity
$\GammaSplit\,du\,dv$.  Thus, for the developed descendant cascade,
\be
 d\langle N_{\rm phys}\rangle
 =\frac{\GammaSplit}{\muIFS}\langle T_{uv}\rangle\,du\,dv.
 \label{eq:physical-split-intensity}
\ee
With $u=t_*re^\varphi$, $v=t_*re^{-\varphi}$ define the radial physical
split measure per unit $d\varphi$,
\be
 \mathscr N_{\rm G}(dr)
 =\frac{2r}{\muIFS}\langle T_{uv}(t_*^2r^2)\rangle dr.
 \label{eq:radial-physical-split-measure}
\ee
The exact instantaneous trace on a Minkowski-time slice is therefore
\be
 \mathcal W_{\rm inst,G}(T)
 =2\int_{(0,T]}\frac{\mathscr N_{\rm G}(dr)}{\sqrt{T^2-r^2}}.
 \label{eq:instantaneous-trace-from-split-measure}
\ee

To match the physical split measure, let
\(R_a^{\rm G}=cm\rho_a\) be the dimensionless invariant radii of
the causal-minimal records in the first occupied rectangle, in the
parent's lifetime units, and
\be
 e^{\rm G}_{s,\zeta}(Z)
 =\mathbb{E}\!\left[\sum_a(R_a^{\rm G})^{-\zeta}\mid s,Z\right].
 \label{eq:geometric-split-event-kernel}
\ee
For a one-candidate rectangle, the coupling above moves each rescaled
null coordinate by at most $\Lambda$; the H\"older estimate
\eqref{eq:chronological-negative-moment-bound} therefore gives an
$O(\Lambda^a)$ difference from $h_\zeta$.  Conditional on occupation,
the expected number of additional candidates is $O(\Lambda)$.  For
occupied-period index $K\geq1$ their inverse-radius moment is bounded by
$C(K\Lambda)^{-\sigma}$, while the exceptional first period contributes
$O(\Lambda^{2-\sigma})$.  Summing over
$\Pr(K=k)=(1-e^{-\Lambda})e^{-k\Lambda}$ gives
\be
 |e^{\rm G}_{s,\zeta}(Z)-h_\zeta(Z)|
 \leq C_\zeta\Lambda^a=C_\zeta s^{2a},
 \label{eq:geometric-split-event-error}
\ee
and logarithmically weighted variants obey the same locally uniform
bound.  Thus the common geometric reward theorem applies to the actual
split measure.  Its Mellin transform has the same pole,
\be
 \mathscr J_{\rm G}(\zeta)
 =\frac{\mathscr S_{\rm eff}(\zeta)\mathscr A_h(\zeta)}
        {1-M(\zeta)}+\mathscr H^{\rm split}_{\rm G}(\zeta),
 \label{eq:geometric-radial-split-transform}
\ee
so the positive-measure theorem gives
\[
 \mathscr N_{\rm G}((0,R])\sim\frac{R_{{\rm G},*}}{\kappa}R^\kappa,
 \qquad
 R_{{\rm G},*}=\frac{\mathscr S_{\rm eff}(\kappa)\pi(h_\kappa)}{-M'(\kappa)}.
\]
The boost kernel enhances events close to the observation slice,
$r=T$. Conservation and positivity bound this local contribution, allowing
the radial asymptotic to pass through the Abel integral.

Include the remnant and pre-entrance stages and denote their full radial
measure by \(\mathscr N_{\rm full}\). Their expected event count per unit
rapidity is finite by the raw-source law and the pre-entrance node and
candidate bounds in App.~\ref{app:threshold-entrance-checks}; hence their
cumulative contribution is \(O(1)\). Thus
\(F(R)=\mathscr N_{\rm full}((0,R])\sim(R_{{\rm G},*}/\kappa)R^\kappa\).
For the full conserved geometric stress, put
\(f(z)=\langle T_{uv}\rangle\) and \(I(z)=\int_0^z f(w)dw\).
Positivity of its regular null stress gives
\(0\leq zf(z)\leq I(z)\), by the conservation identity
\eqref{eq:app-reference-null-components}. At every finite radius,
this implies
\be
 F(R)=\frac{\GammaSplit}{\muIFS}I(t_*^2R^2),\qquad
 0\leq F'(R)\leq\frac{2F(R)}R.
 \label{eq:geometric-radial-density-bound}
\ee
For \(0<\varepsilon<1/2\) and sufficiently large \(T\), this implies
\be
 2\int_{(1-\varepsilon)T}^T
       \frac{\mathscr N_{\rm G}(dr)}{\sqrt{T^2-r^2}}
 \leq2\int_{(1-\varepsilon)T}^T
       \frac{dF(r)}{\sqrt{T^2-r^2}}
 \leq C T^{\kappa-1}\sqrt\varepsilon.
 \label{eq:geometric-Abel-endpoint-bound}
\ee
On \([0,(1-\varepsilon)T]\), the rescaled Abel kernel is bounded and
cumulative regular variation gives the limiting integral. Taking first
\(T\to\infty\), then \(\varepsilon\to0\), gives
\(2R_{{\rm G},*}\int_0^1 x^{\kappa-1}(1-x^2)^{-1/2}dx
={\rm B}(\kappa/2,1/2)R_{{\rm G},*}\). Therefore
\be
 \mathcal W_{\rm inst,G}(T)
 \sim D_{{\rm G},1}T^{\kappa-1}.
 \label{eq:instantaneous-geometric-trace-limit}
\ee
The equality of the coefficient with the live-weight residue follows
from
\[
 \pi(h_{\kappa,1})
 ={\rm B}(\kappa/2,1/2)\pi(h_\kappa).
\]
The exact instantaneous ensemble pressure and the period-averaged live
trace therefore have the same leading asymptotic. The remnant and
pre-entrance contributions are $O(T^{-1})$ and remain subleading.

\section{Energy and summed-rest-mass asymptotics}
\label{appendix:moment-normalization}

The detached strings carry an energy that grows with time, while their
summed invariant mass approaches a plateau. The difference reflects the
increasing boosts of newly emitted strings. We derive the exact energy
sum rule and then calculate the plateau, its leading deficit, and its
relation to the surviving trace.

The mass calculation is a prediction of the binary process with the
source moments specified below. Applying its leading deficit to the
complete geometric population additionally requires the pre-entrance
mass condition \eqref{eq:pre-entrance-mass-remainder} and matching of
the mass-loss observable. The comparison with geometric simulations is
given in App.~\ref{app:time-domain-fits}.

The energy normalization follows directly from conservation.  A
first-generation yo-yo emitted at time \(t_0\) has
\be
    \left\langle E_1\mid t_0\right\rangle
    =
    \muIFS t_0,
    \label{eq:mean-source-energy}
\ee
while the persistent remnant emits first-generation yo-yos at the rate
\(\dot N_1(t_0)=2/t_0+o(t_0^{-1})\).  Subsequent splittings only
redistribute this energy.  Consequently,
\be
    \frac{\dd}{\dd T}
    \left\langle E_{\rm yoyo}(T)\right\rangle
    =
    \muIFS T\,\dot N_1(T),
    \label{eq:yoyo-energy-injection}
\ee
and the exact source measure of
App.~\ref{appendix:first_generation_source} gives
\be
    \left\langle E_{\rm yoyo}(T)\right\rangle
    =
    2\muIFS T
    -\frac{2\muIFS}{\GammaSplit T}
    +\mathcal O(T^{-3}).
    \label{eq:yoyo-energy-asymptotic}
\ee
The energy therefore approaches $2\muIFS T$ with relative correction
of order $(\sqrt{\GammaSplit}T)^{-2}$. Its leading value includes
the moving high-boost source as well as the finite-boost population.

\paragraph{The invariant-mass plateau.}
For the entrance population, assume \(U_0,V_0>0\) almost
everywhere and the mass-weighted bound
\be
 \int d\nu\,s\left[1+U_0+V_0
       +(-\ln U_0)_++(-\ln V_0)_+\right]<\infty,
 \qquad s=cm_0t_*.
 \label{eq:entrance-mass-log-moments}
\ee
Here \((z)_+=\max(z,0)\).  Until the final extension below,
\(M_{\rm inv}\) denotes the mass of this entrance population and its
descendants.
For a fixed node history, vary the common boost of the source
and ask which copies are alive on the observation slice.  Its mass
contribution is \(m\) times the total length of that range of source
rapidities.  With \(c=\GammaSplit/(2\muIFS)\), the rescaled birth
and split radii are \(r_b=\sqrt{UV}\),
\(r_s=\sqrt{(U+\epsilon)(V+\epsilon)}\).
An event of rescaled radius \(r\) lies before time \(t\) on
a rapidity interval of length \(2a_+(cm t/r)\), where
\(a_+(z)=\Theta(z-1)\operatorname{arcosh}z\).
Subtracting copies that have already split from those already
born gives
\be
 M_{\rm node}(t)
 =2m\left[a_+\!\left(\frac{cm t}{r_b}\right)
          -a_+\!\left(\frac{cm t}{r_s}\right)\right].
 \label{eq:node-boost-integrated-mass}
\ee
Although each individual copy eventually splits, increasingly
boosted copies are delayed by time dilation.  Their net rapidity
length approaches a constant.  For each node,
\be
 \begin{split}
 M_{\rm node}(t)&\longrightarrow2m\log(r_s/r_b),\\
 0\leq M_{\rm node}(t)&\leq2m\log(2r_s/r_b).
 \end{split}
 \label{eq:node-mass-plateau-bound}
\ee
It remains to establish that the infinitely many nodes have a finite total contribution.  The expected sum of daughter
invariant masses is only \(M(1)=8/9\) of the parent mass, so successive
generations carry a geometrically decreasing mass weight.  We must
also bound the logarithmic occupation factor.  The normalized operator
\(\mathscr T_1/M(1)\) has independent
\({\rm Beta}(3/2,1)\) multipliers and unit exponential waiting times.
The coordinate means are uniformly bounded along this chain.  Its
negative logarithmic moments are also uniformly bounded after the
first step, since \(U'=a(U+\epsilon)\geq a\epsilon\), and similarly
for \(V'\).  Finally,
\(\log(1+\epsilon/U)\leq\log(1+\epsilon)+(-\log U)_+\).
Thus the expected logarithmic factor in
Eq.~\eqref{eq:node-mass-plateau-bound} is uniformly bounded by an
integrable function of the root marks.  The remaining generation sum
is geometric, \(\sum_n(8/9)^n\); the root integral is finite by
Eq.~\eqref{eq:entrance-mass-log-moments}.  Dominated convergence yields
\be
 M_{\rm inv}(t)\longrightarrow M_\infty
 =\int d\nu\,m_0\sum_{n\geq0}
      (\mathscr T_1^n\ell)(Z_0),
 \qquad
 \ell(U,V)=\mathbb{E}_\epsilon\!\left[
       \log\frac{(U+\epsilon)(V+\epsilon)}{UV}\right],
 \label{eq:chronological-mass-plateau}
\ee
with \(0<M_\infty<\infty\).
This proves existence of the plateau in the unabsorbed binary model.
The following radial mass balance determines its leading approach.

\subsection{\texorpdfstring{The leading mass deficit and its relation to the trace}{The leading mass deficit and its relation to the trace}}
\label{app:mass-deficit-proof}

The plateau measures the mass integrated over complete boost orbits.
Its approach is controlled by the mass still awaiting fragmentation at
large invariant radius.  We express that remaining mass directly
in terms of the chronological branching process.  The resulting boost integral determines both the correction exponent
and the sign of the deficit.  Put
\(T=\widehat t\), \(\kappa=\kappa_{\rm IFS}\), \(q=1-\kappa\),
\(\bar m=ct_*m\), and \(\mathcal M(T)=ct_*M_{\rm inv}(t_*T)\).
Here \(ct_*=\sqrt{\GammaSplit}/(2\muIFS)\), as in the trace normalization.

\paragraph{Mass remaining at invariant radius.}
A cut at fixed invariant radius intersects the live nodes of
the tree.  Denote their expected remaining invariant mass, in the
dimensionless units above and per unit source rapidity, by
\(\mathscr F(r)\).  Its balance has two terms: root births add mass,
and fragmentation reduces the sum of descendant invariant masses.
Let \(\mathscr I(dr)\) be the root-mass injection measure and
\(\mathscr D(dr)\) the expected invariant-mass loss at descendant splits,
per unit source rapidity.  For a general entrance root, its
dimensionless birth radius is
\(r_0=\sqrt{U_0V_0}/s\).  The measures are
\begin{align}
 \mathscr I(dr)&=\int d\nu\,s\,\delta_{r_0}(dr),\\
 \mathscr D(dr)&=d_m\int d\nu\,
       \mathbb{E}\!\left[\sum_Y\bar m_Y\,\delta_{r_Y}(dr)\right],
 &d_m&=1-M(1)=\frac19.
 \label{eq:radial-mass-loss-measures}
\end{align}
In addition to Eq.~\eqref{eq:entrance-mass-log-moments}, assume,
for \(\sigma\) near \(\kappa\) and some \(\varepsilon>0\),
\be
 \int d\nu\,s^\sigma(1+U_0+V_0)^{1+\delta}<\infty,
 \qquad
 \mathscr I((r,\infty))=O(r^{-q-\varepsilon})\quad(r\to\infty).
 \label{eq:entrance-mass-tail-conditions}
\ee
These bounds control mass-weighted occupations and delayed injection;
they are additional to the assumptions for the normalized trace profile.
The sum includes the root's split and all subsequent descendant splits;
\(r_Y=\rho_Y/t_*\) is the event's invariant radius.  The mean loss at a
split is \(d_m\bar m_Y\), since daughter fractions are independent of
the parent's previous history and final waiting time.  The geometric
generation sum gives
\(\mathscr D((0,\infty))=d_m\int d\nu\,\bar m_0
\sum_{n\geq0}M(1)^n=\mathscr I((0,\infty))\).
Thus the mass on the cut is the injected mass minus the loss
already incurred, or equivalently the future loss minus future
injection:
\be
 \mathscr F(r)=\mathscr I((0,r])-\mathscr D((0,r])
             =\mathscr D((r,\infty))-\mathscr I((r,\infty))\geq0.
 \label{eq:radial-surviving-mass}
\ee
Positivity follows because \(\mathscr F(r)\) counts the mass
of live strings on the cut.

An event at radius \(r\) has occurred by time \(T\) on a rapidity
interval of length \(2\operatorname{arcosh}(T/r)\).  Grouping each
parent's removal with its daughters' births in the occupation sum,
and then integrating by parts, yields the exact identity
\be
 \mathcal M(T)
 =2\int_0^T\operatorname{arcosh}(T/r)
                [\mathscr I-\mathscr D](dr)
 =2\int_0^T\frac{T\mathscr F(r)}{r\sqrt{T^2-r^2}}\,dr.
 \label{eq:exact-radial-mass-balance}
\ee
Thus the remaining radial mass determines the time dependence through
the boost projection.

\paragraph{The branching pole fixes the radial tail.}
For birth coordinates \(Z=(U,V)\) and an independent unit exponential
waiting time \(\epsilon\), write
\(R=\sqrt{(U+\epsilon)(V+\epsilon)}\), so that \(r=R/\bar m\).
The transform of the positive loss measure is
\begin{align}
 H_M(\zeta)&\equiv\int_0^\infty r^{1-\zeta}\mathscr D(dr)=d_m\int d\nu\,\bar m_0^\zeta
        \sum_{n\geq0}(\mathscr T_\zeta^n k_\zeta)(Z_0),
 &k_\zeta(Z)&=\mathbb{E}_\epsilon[R^{1-\zeta}].
 \label{eq:mass-loss-transform}
\end{align}
The identity \(\bar m r^{1-\zeta}=\bar m^\zeta R^{1-\zeta}\)
gives this representation in its convergence strip just to the right
of \(\kappa\).  We apply Eq.~\eqref{eq:chronological-resolvent}
with \(b_\zeta=k_\zeta\).  Unlike the pressure kernel, this positive
moment grows with the birth coordinates, so its contraction requires
the following separate check.

To extend the contraction argument to this observable, let
\(\sigma=\operatorname{Re}\zeta\) and choose
\(0<\delta<(1-\sigma)/2\) in a narrow strip about \(\kappa\).
The power in \(k_\zeta\) obeys the weighted H\"older bound
\be
 |k_\zeta(Z)-k_\zeta(Z')|
 \leq C_\zeta\,d_\delta(Z,Z')
              (1+U+V+U'+V'),
 \label{eq:mass-positive-moment-holder}
\ee
where \(d_\delta\) is defined in
App.~\ref{app:chronological-contraction}.
Under \(\mathscr T_\sigma/M(\sigma)\), the coordinate multipliers are
independent \({\rm Beta}(1+\sigma/2,1)\) variables.  Their contracting
products and the exponential waiting times give uniformly bounded
coordinate second moments, with polynomial dependence on the initial
coordinates. Coupling to the chain started at zero and applying
Cauchy--Schwarz therefore gives
\begin{align}
 |\mathscr T_\zeta^n k_\zeta(Z)
       -\mathscr T_\zeta^n k_\zeta(0)|
 &\leq C_\zeta\theta_\delta(\sigma)^n(1+U+V)^{1+\delta},\\
 \theta_\delta(\sigma)
 &=M(\sigma)\sqrt{\frac{1+\sigma/2}{1+\sigma/2+2\delta}}<1.
 \label{eq:mass-positive-moment-contraction}
\end{align}
The strict inequality holds throughout a sufficiently narrow strip.
Equation~\eqref{eq:entrance-mass-tail-conditions} supplies the
required weighted polynomial moments of the entrance source.
The first-step moments also bound \(d_n\), so the hypotheses
of Eq.~\eqref{eq:chronological-resolvent} hold.  Its pole has residue
\be
 \mathcal A_M\equiv\mathop{\rm Res}_{\zeta=\kappa}H_M(\zeta)
 =\frac{d_m\mathscr S(\kappa)\,
              \mathbb{E}_{\pi,\epsilon}[R^q]}{-M'(\kappa)}>0.
 \label{eq:mass-loss-positive-residue}
\ee
Here \(\pi\) is the same stationary birth-coordinate law as in the
pressure calculation.  The expectation includes the final waiting time.

Apply the positive-measure Wiener--Ikehara theorem in the form used in
App.~\ref{app:chronological-pointwise-limit} to
\(S_D(R)=\int_{(1,R]}r\mathscr D(dr)\).  Removing \(r<1\) changes
the transform by an analytic function near the Malthusian line, since
\(\mathscr D\) is finite.  The theorem gives
\(S_D(R)\sim(\mathcal A_M/\kappa)R^\kappa\).
Stieltjes integration by parts then gives
\(\mathscr D((r,\infty))\sim(\mathcal A_M/q)r^{-q}\);
the injection tail is \(o(r^{-q})\) by
Eq.~\eqref{eq:entrance-mass-tail-conditions}, so
\be
 \mathscr F(r)\sim K_Mr^{-q},\qquad K_M=\mathcal A_M/q>0.
 \label{eq:radial-mass-tail}
\ee

\paragraph{The boost integral fixes the sign of the deficit.}
Using the plateau and Eq.~\eqref{eq:exact-radial-mass-balance},
\begin{align}
 \mathcal M_\infty-\mathcal M(T)
 &=2\int_T^\infty\frac{\mathscr F(r)}r\,dr\\
 &\quad+2\int_0^T\frac{\mathscr F(r)}r
          \left(1-\frac{T}{\sqrt{T^2-r^2}}\right)dr.
 \label{eq:mass-deficit-subtracted-kernel}
\end{align}
The first term is positive: radii beyond \(T\) have not yet
appeared anywhere on the time slice, so their plateau contribution is
missing.  The second is negative: radii just inside the slice receive
an enhanced rapidity weight from the boost Jacobian.  The deficit is
the balance of these two geometric effects; discarding either would
give the wrong coefficient.  Substituting the regularly varying
tail \eqref{eq:radial-mass-tail} into the combined expression yields
\begin{align}
 \mathcal M_\infty-\mathcal M(T)
 &\sim K_MT^{-q}
 \left[\frac2q+2\int_0^1u^{-q-1}
                 \left(1-\frac1{\sqrt{1-u^2}}\right)du\right]\\
 &=-K_M\mathrm B(-q/2,1/2)T^{-q}.
 \label{eq:mass-deficit-beta-coefficient}
\end{align}
The subtracted kernel is \(O(u^2)\) at zero and has an integrable
singularity at one; the bounded-radius contribution is \(O(T^{-2})\).
These facts justify the substitution.  Since \(0<q<1\),
\(-\mathrm B(-q/2,1/2)>0\).  In physical mass units,
\be
 M_{\rm inv}(t)=M_\infty-B\widehat t^{-q}
                         +o(\widehat t^{-q}),\qquad
 B=\frac{-K_M\mathrm B(-q/2,1/2)}{ct_*}>0.
 \label{eq:three-moment-mass-summary-main}
\ee
The positive coefficient fixes the leading approach from below.

\paragraph{A common coefficient relation.}
The mass left to lose and the surviving tension probe the same
late generations, with different occupation weights.  Their residues
contain different moments of the stationary clock, but the same
source strength and branching factor.  The pressure residue in
Eq.~\eqref{eq:chronological-positive-residue} contains the same
\(\mathscr S(\kappa)\) and branching derivative; dividing the two
amplitudes therefore cancels both:
\be
 \mathfrak a_\kappa
 \equiv\lim_{t\to\infty}
       \frac{M_\infty-M_{\rm inv}(t)}{W_{\rm live}(t)}
 =\frac{-\mathrm B(-q/2,1/2)}{9q\,\mathrm B(\kappa/2,1/2)}
       \frac{\mathbb{E}_{\pi,\epsilon}[R^q]}
            {\mathbb{E}_{\pi,\epsilon}[R^{-\kappa}]}
 \simeq3.08.
 \label{eq:mass-trace-amplitude-ratio}
\ee
The value $\mathfrak a_\kappa\simeq3.08$ follows from the stationary
binary recursion, independently of the time-series fit. Both clock moments
are finite; the source factor cancels for entrance laws obeying the stated
bounds. Extending this coefficient to the geometric cascade requires
matching the growing mass-loss weight in the comparison of
App.~\ref{app:geometric-matching}.

For the complete detached population, suppose also that the mass
still in pre-entrance stages satisfies
\be
 M_{\rm pre}(t_*T)=M_{{\rm pre},\infty}+o(T^{-q}).
 \label{eq:pre-entrance-mass-remainder}
\ee
Its plateau then adds to $M_\infty$, leaving the leading deficit and
amplitude ratio unchanged once the geometric mass-loss coefficient is
matched. This is a separate condition on mass: a highly boosted string
carries mass $\gamma W$ even when its trace $W$ is small.

\section{Stress-tensor moment diagnostics}
\label{app:stress-moments-terminal-anec}

Equal-time moments diagnose the spatial distribution of the surviving tension and the location of the negative energy relative to the compensating positive energy.  Boost invariance
relates the trace moments to the total pressure, and conservation then
fixes the energy and pressure moments.  Their powers and coefficient
ratios test the same physical picture as the null integral, using
different measurements of the cascade.

\subsection{The semiclassical spatial distribution}
\label{subsec:moment_diagnostics}

Define the ensemble-averaged complete stress moments by
\begin{align}
 M_n^{(p)}(t)&=\int dx\,x^n\langle T_{11}(t,x)\rangle,
 \label{eq:pressure-moments}\\
 M_n^{(E)}(t)&=\int dx\,x^n\langle T_{00}(t,x)\rangle,
 \label{eq:energy-moments}\\
 M_n^{(W)}(t)&=M_n^{(E)}(t)-M_n^{(p)}(t).
 \label{eq:trace-weight-moments}
\end{align}
Here $x$ is the position on a constant-time slice, and $M_0^{(W)}$
includes the remnant and all live descendants.

The trace defines a positive spatial measure. Its normalized
moments describe localization. The complete energy density is
signed, so its moments diagnose cancellation and cannot be read as
moments of a probability distribution for size.

For the complete IFS decay state, the total energy and the first energy
moment vanish,
\be
    M_0^{(E)}(t)=0,
    \qquad
    M_1^{(E)}(t)=0.
    \label{eq:E0-E1-vanish}
\ee
The first identity is total energy conservation.  The second is
conservation of the center-of-energy moment, with the origin chosen at
the center of the initial IFS and total momentum zero.  More generally,
stress-tensor conservation implies
\be
    \frac{d^2}{dt^2}M^{(E)}_n(t)
    =
    n(n-1)M^{(p)}_{n-2}(t),
    \label{eq:moment-conservation-identity}
\ee
assuming that the boundary terms vanish.
In particular, \(\ddot M_2^{(E)}=2p_{\rm tot}\).
The pressure theorem fixes \(M_0^{(W)}=-p_{\rm tot}\); boost invariance
then determines the higher trace moments, and conservation determines
their energy and pressure counterparts.

\paragraph{Boost invariance and trace-moment coefficients.}
Boost invariance fixes the relative coefficients as well as the powers.
For example, \(M_2^{(W)}/(t^2M_0^{(W)})\) measures the mean of
\((x/t)^2\) in the positive trace distribution.  Its semiclassical
limit \(1/(2-q_{\rm IFS})\simeq0.547\) resolves a profile spread
across the wedge.  The terminal limit of one derived below means that
the surviving weight is instead concentrated near \(x=\pm t\).

All moments in the following identities are ensemble averages, with
brackets suppressed.  In the boost-invariant flat-space model, write
\(\langle T_{uv}(u,v)\rangle=f(uv)\), so the complete traceful density
inside the future wedge is
\(\langle T_{00}-T_{11}\rangle=4f(t^2-x^2)\).
The original null folds and traceless terminal products do not
contribute to this density.  Changing variables to \(z=t^2-x^2\) gives
\be
 M_{2k}^{(W)}(t)
 =4\int_0^{t^2}(t^2-z)^{k-1/2}f(z)\,dz,
 \qquad k\geq0.
 \label{eq:trace-moments-invariant-profile}
\ee
For \(k\geq1\), the upper-endpoint term vanishes, and differentiation
gives the exact relation
\be
 \frac{d}{dt}M_{2k}^{(W)}(t)
 =(2k-1)tM_{2k-2}^{(W)}(t).
 \label{eq:trace-moment-recursion}
\ee
Local boundedness of the trace profile makes the derivative integrable.

Define the dimensionless semiclassical coefficients by
\be
 M_{2k}^{(W)}(t)
 \sim C_{2k}^{(W)}\,\muIFS\GammaSplit^{-k-1/2}
       (\sqrt{\GammaSplit}t)^{2k-q_{\rm IFS}}.
 \label{eq:trace-moment-coefficients}
\ee
Starting from the established zeroth-moment tail and integrating
Eq.~\eqref{eq:trace-moment-recursion} successively yields
\be
 C_{2k}^{(W)}
 =\frac{2k-1}{2k-q_{\rm IFS}}\,C_{2k-2}^{(W)},
 \qquad k\geq1.
 \label{eq:trace-coefficient-recursion}
\ee
Finite-time integration constants are subleading.  The zeroth
coefficient is \(C_0^{(W)}=C_P\); the zeroth- and second-moment
predictions are collected in
Eq.~\eqref{eq:joint-moment-targets}.

\paragraph{Energy and pressure coefficients.}
The same inputs fix the leading coefficients of all even energy and
pressure moments.  Define their signed coefficients using the
normalization of Eq.~\eqref{eq:trace-moment-coefficients},
\be
 M_{2k}^{(A)}(t)
 \sim C_{2k}^{(A)}\,\muIFS\GammaSplit^{-k-1/2}
       (\sqrt{\GammaSplit}t)^{2k-q_{\rm IFS}},
 \qquad A=E,p,\quad k\geq1.
 \label{eq:energy-pressure-moment-coefficients}
\ee
Combining Eqs.~\eqref{eq:moment-conservation-identity} and
\eqref{eq:trace-weight-moments} gives the exact hierarchy
\be
 \frac{d^2}{dt^2}M_{2k}^{(E)}
 =2k(2k-1)\bigl(M_{2k-2}^{(E)}-M_{2k-2}^{(W)}\bigr).
 \label{eq:energy-trace-moment-hierarchy}
\ee
Starting from \(M_0^{(E)}=0\), integrate this hierarchy twice
at each step and use Eq.~\eqref{eq:trace-coefficient-recursion}.
The leading power \(t^{2k-q_{\rm IFS}}\) dominates the constant and
linear integration terms because \(0<q_{\rm IFS}<1\).
Induction gives
\be
 \begin{split}
 C_{2k}^{(E)}
 &=-\frac{2k}{1-q_{\rm IFS}}\,C_{2k}^{(W)}
   =-k(1+\sqrt2)\,C_{2k}^{(W)},\\
 C_{2k}^{(p)}
 &=-\frac{2k+1-q_{\rm IFS}}{1-q_{\rm IFS}}\,C_{2k}^{(W)}.
 \end{split}
 \label{eq:energy-pressure-coefficient-ratios}
\ee
Thus \(C_0^{(W)}=C_P\) fixes every leading even-moment coefficient.
The signed moments include the original negative-energy folds, and
conservation closes their hierarchy. Odd moments vanish by ensemble
symmetry. The numerical comparison uses orders zero and two, with
$\muIFS=\GammaSplit=1$.

For the numerical test, set \(q=q_{\rm IFS}\) and remove the
predicted time powers:
\be
 X_0=\frac{(\sqrt{\GammaSplit}t)^q M_0^{(W)}}
                    {\muIFS\GammaSplit^{-1/2}},\qquad
 X_2^{(A)}=\frac{(\sqrt{\GammaSplit}t)^{q-2}M_2^{(A)}}
                    {\muIFS\GammaSplit^{-3/2}},\quad A=W,E.
 \label{eq:joint-moment-normalization}
\ee
The three fitted observables have the limits
\be
 (X_0,X_2^{(W)},X_2^{(E)})
 \longrightarrow C_P\left(1,\frac{1}{2-q},
                         -\frac{2}{(1-q)(2-q)}\right).
 \label{eq:joint-moment-targets}
\ee
Pressure supplies a conservation check but no independent
coordinate, since \(M^{(p)}=M^{(E)}-M^{(W)}\).
Coefficient ratios are formed from ensemble means of signed moments.
The higher even moments follow from the same hierarchy.

\begin{figure}[t!]
 \centering
 \includegraphics[width=0.98\textwidth]{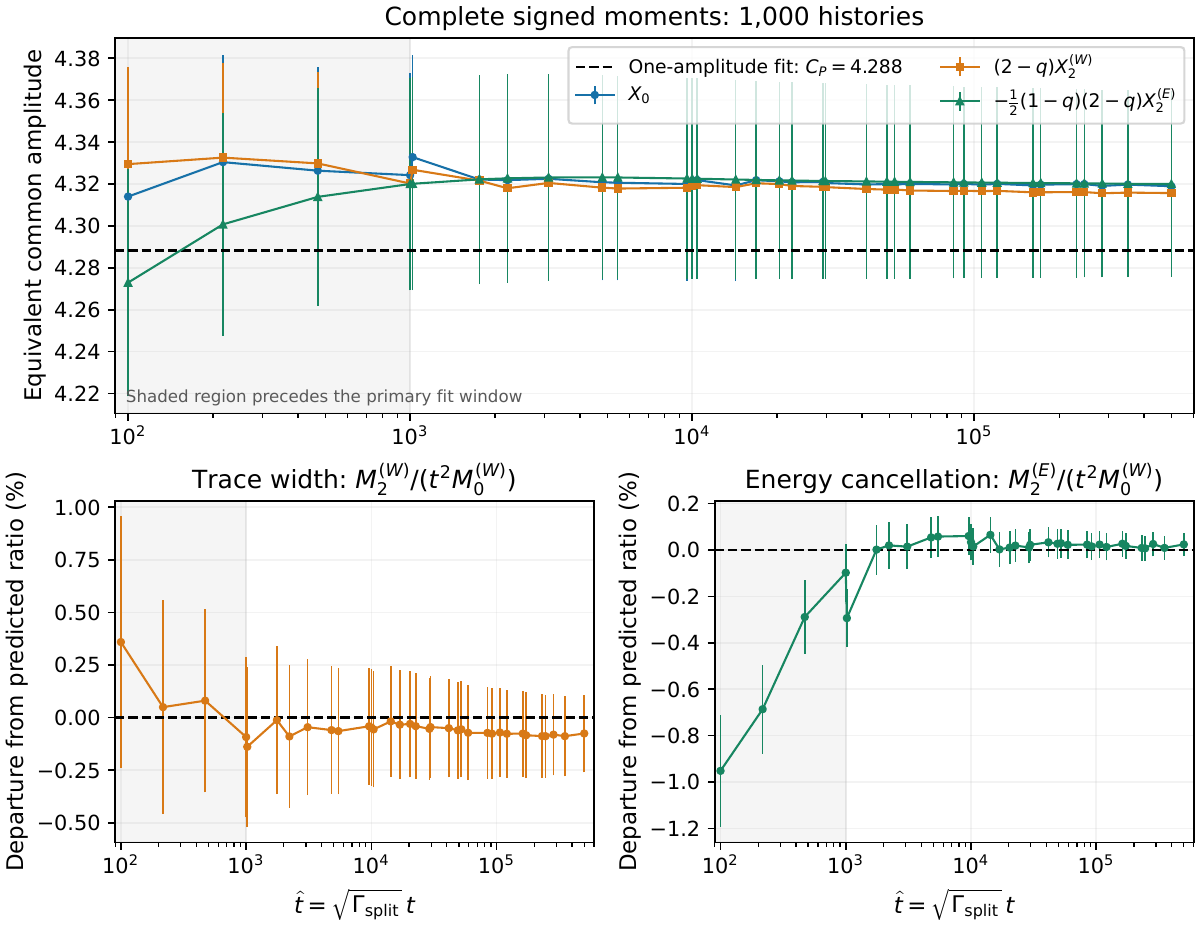}
 \caption{Complete signed spatial moments from 1,000 unabsorbed
 histories. Top: rescaling each observable to the same predicted
 limiting amplitude; the dashed line is the fiducial fit. Bottom:
 departures of ratios of signed ensemble means from the predicted
 ratios. Error bars are marginal standard errors; ratio errors use
 linear error propagation with the within-history pairing retained.
 Shading marks times before the fiducial fitting window. These errors do not
 display the full covariance used in the joint tests.}
 \label{fig:spatial-moments-1000}
\end{figure}

\paragraph{Comparison with the geometric cascade.}
Figure~\ref{fig:spatial-moments-1000} compares the zeroth and second
moments in 1,000 geometric histories with their predicted limits.
The ratios lie close to the asymptotic values, but their correlated
departures are statistically resolved. A correction proportional to
$1/t$ in the total trace improves the joint fit while leaving further
finite-time structure. Appendix~\ref{app:spatial-moment-fits} gives the
fits, their uncertainties, and the conservation checks.

\subsection{Localization after terminal decay}
The same exact moment identity also fixes the terminal limit.  Under
the finite-response assumptions of Sec.~\ref{sec:terminal-decay}, let
\be
 M_0^{(W)}(t)\sim\frac{\mathcal C_{\rm term}}{t},
 \qquad \mathcal C_{\rm term}=4I_\infty>0.
\ee
Successive integration of Eq.~\eqref{eq:trace-moment-recursion} gives,
for every fixed \(k\geq0\),
\be
 M_{2k}^{(W)}(t)\sim\mathcal C_{\rm term}t^{2k-1},
 \qquad
 \frac{M_{2k}^{(W)}(t)}{t^{2k}M_0^{(W)}(t)}\longrightarrow1.
 \label{eq:terminal-trace-moment-coalescence}
\ee
In the dimensionless convention of
Eq.~\eqref{eq:trace-moment-coefficients}, with terminal power
\(2k-1\), this means
\(C_{2k}^{(W),\rm term}=\GammaSplit\mathcal C_{\rm term}/\muIFS\)
for every fixed \(k\).  Positivity of the trace turns the
second-moment limit into a physical localization statement: its
normalized weight outside any fixed neighborhood of $x/t=\pm1$
vanishes. This describes localization relative to the light-cone size.

The corresponding energy and pressure moments have a logarithmic
factor.  Starting from
\(\ddot M_2^{(E)}=-2M_0^{(W)}\sim-2\mathcal C_{\rm term}/t\)
and integrating Eq.~\eqref{eq:energy-trace-moment-hierarchy}
successively yields, for each fixed \(k\geq1\),
\be
 \begin{split}
 M_{2k}^{(E)}(t)
 &\sim-2k\,\mathcal C_{\rm term}\,
         t^{2k-1}\log(t/t_{\rm ref}),\\
 M_{2k}^{(p)}(t)
 &\sim-2k\,\mathcal C_{\rm term}\,
         t^{2k-1}\log(t/t_{\rm ref}).
 \end{split}
 \label{eq:terminal-energy-pressure-moment-hierarchy}
\ee
The logarithm comes from twice integrating the terminal pressure
$-\mathcal C_{\rm term}/t$; the hierarchy fixes the higher coefficients.
The reference time $t_{\rm ref}>0$ affects only subleading terms.

\begingroup

\let\IFSoriginalfootnote\footnote
\renewcommand{\footnote}[1]{\IFSoriginalfootnote{#1}}
\section{Finite smearing for a general null generator}
\label{app:finite-transverse-smearing}

We bound the residual after transverse smearing by separating the
distance of the longitudinal cascade from the original folds from
the additional spreading caused by terminal decay. At each terminal
conversion, replace the outgoing products by two collinear null
packets with the same total four-momentum. This reference remains in
the original $(t,x)$ plane and gives the decomposition
\be
    \mathcal A_{k,L}
    =\mathcal A^{\rm col}_{k,L}
     +\delta\mathcal A^{\rm term}_{k,L}.
    \label{eq:app-reference-decomposition}
\ee
The first-split bound controls the first spatial moment of the
reference; mass conservation controls the terminal contribution.
Together they give a sufficient smearing scale
$L_{\rm smear}=\GammaSplit^{-1/2}\sim\ell_s/g_s$, without requiring
finite integrated trace.

\subsection{Setup and result}
\label{subsec:app-smearing-setup}

We work at fixed weak coupling in flat spacetime. The ensemble is
invariant under boosts along the original string and under
$x\mapsto-x$. The preterminal evolution is the collinear splitting
cascade, with no joining, and only the two original folds carry
negative energy. Each terminal conversion removes a positive-energy
yo-yo from its tree and replaces it locally by free causal products
with the same total four-momentum. Their momenta obey
$p_j^0>0$ and $p_j^2=-m_j^2\leq0$; their masses, multiplicity, and
angular distribution are otherwise unrestricted. The conversion law
is boost covariant. Massive free products are allowed here because
the trace bound is applied to the collinear reference. Their actual
trace need not vanish or have finite integrated response. The
dilaton-response model of Appendix~\ref{app:terminal-response}
instead assumes traceless products.
Free propagation is used for the quantitative smearing estimate;
the direct completion proof in
Appendix~\ref{subsec:app-complete-null-flux} needs only the conservation
and positivity assumptions stated there.

Let $W_*\equiv\sup_{t>0}\langle W^{\rm col}_{\rm tot}(t)\rangle$
be the uniform bound on the reference trace weight, and let
$\mathcal M_{\rm term}$ be the terminal-parent rest-mass intensity
per unit momentum rapidity, defined in
Eq.~\eqref{eq:app-terminal-event-measure}. They determine two lengths:
\be
    \ell_{\rm col}\equiv\frac{W_*}{4\muIFS},
    \qquad
    \ell_{\rm ang}\equiv\frac{\mathcal M_{\rm term}}{\muIFS}.
    \label{eq:app-two-spreading-lengths}
\ee
We will prove
\be
    \sup_{\mathbf b_0}|\mathcal A_{k,L}(s,\mathbf b_0)|
    \leq C_K\frac{\mathscr F_k(s)}{L^d}
               \frac{\ell_{\rm col}+\ell_{\rm ang}}{L},
    \qquad L>0,
    \label{eq:app-general-moment-bound}
\ee
and bound both lengths by $O(\GammaSplit^{-1/2})$.
Here $\mathcal A_{k,L}$ uses the ensemble-averaged stress tensor,
$d=D-2$, and $C_K$ depends only on bounds on the kernel and its first
two derivatives. It is independent of $s,L,k$ and of terminal decay
times and angular distributions. The residual can have either sign.

As in Sec.~\ref{subsec:finite-transverse-smearing},
$k=(1,\widehat{\mathbf n})$ is normalized and
$s=-k\cdot x>0$ labels the null planes. Write
$c=\widehat{\mathbf n}\cdot\widehat{\mathbf x}$.
A fold on $x=\sigma t$ contributes $-2\muIFS s$ whenever
$1-\sigma c>0$. Thus the magnitude of the negative fold flux is
\be
    \mathscr F_k(s)=
    \begin{cases}
        4\muIFS s, & |c|<1,\\
        2\muIFS s, & |c|=1.
    \end{cases}
    \label{eq:app-negative-fold-flux}
\ee
At tangency the parallel fold carries no test flux; only the other
fold contributes.

\subsection{The collinear reference stays close to the folds}
\label{subsec:app-collinear-reference}

For a terminal parent with momentum $P=(E,P_x,\mathbf0_\perp)$,
the reference packets move along $\pm x$ with nonnegative energies
$(E+P_x)/2$ and $(E-P_x)/2$. Their four-momenta sum to $P$.
Keeping the original folds and all preterminal evolution defines a
conserved reference stress with vanishing total energy and momentum.
The positive reference packets preserve the branch-charge bound of
Sec.~\ref{sec:introduction-summary}: in each history,
$W^{\rm col}_{\rm tot}(t)\leq8\muIFS t_s$, where $t_s$ is the first
split time. Since the unbroken worldsheet up to $t$ has null area
$2t^2$, $\Pr(t_s>t)=e^{-2\GammaSplit t^2}$, and hence
\be
    W_*\leq8\muIFS\langle t_s\rangle
    =\frac{2\sqrt{2\pi}\,\muIFS}{\sqrt{\GammaSplit}}.
    \label{eq:app-first-split-trace-bound}
\ee
No period averaging or assumption about subsequent decay is used.

Suppressing the transverse delta function, boost invariance gives
the ensemble-averaged trace
\be
    T^{\rm col}_{uv}=f_{\rm col}(z),\qquad
    z=uv,\qquad u=t-x,\quad v=t+x,
    \label{eq:app-reference-trace}
\ee
where $f_{\rm col}\geq0$ and $f_{\rm col}(0)=\muIFS$. Define
\be
    I_{\rm col}(z)=\int_0^z f_{\rm col}(w)\,dw,
    \qquad I_\infty^{\rm col}=\int_0^\infty f_{\rm col}(w)\,dw.
    \label{eq:app-integrated-reference-trace}
\ee
The latter may be infinite. Equation~\eqref{eq:intro-Tuv-boundary}
applied to the reference gives
\be
    \frac{I_{\rm col}(z)}z
    \leq\frac{\langle W^{\rm col}_{\rm tot}(\sqrt z)\rangle}
                  {4\sqrt z}
    \leq\frac{W_*}{4\sqrt z}.
    \label{eq:app-integrated-trace-dilution}
\ee
Conservation fixes the regular null components:
\be
    T^{\rm col,reg}_{uu}
    =\frac{I_{\rm col}(z)-zf_{\rm col}(z)}{u^2},\qquad
    T^{\rm col,reg}_{vv}
    =\frac{I_{\rm col}(z)-zf_{\rm col}(z)}{v^2}.
    \label{eq:app-reference-null-components}
\ee
Their positivity implies $zf_{\rm col}\leq I_{\rm col}$ and
$(I_{\rm col}/z)'\leq0$. Together with
$I_{\rm col}(z)/z\to\muIFS$ as $z\to0$, this yields
\be
    0\leq f_{\rm col}(z)\leq\frac{I_{\rm col}(z)}z
    \leq\muIFS\qquad(z>0).
    \label{eq:app-uniform-trace-bound}
\ee
Away from the two negative folds, the positive energy density is
\be
\begin{split}
    \rho_+(t,x)
    &=[I_{\rm col}(z)-zf_{\rm col}(z)](u^{-2}+v^{-2})
      +2f_{\rm col}(z)\\
    &=\partial_x\!\left[\frac{2xI_{\rm col}(t^2-x^2)}{t^2-x^2}\right].
\end{split}
    \label{eq:app-reference-density}
\ee

Let $r_{\rm f}=t-|x|$ be the distance to the nearer fold.
Integration by parts on each half of the string gives
\be
\begin{split}
    \int_{-t}^{t}dx\,r_{\rm f}\rho_+
    &=2\int_0^{t^2}\frac{I_{\rm col}(z)}z\,dz
      \leq W_*t,\\
    \int_{-t}^{t}dx\,\rho_+&=4\muIFS t.
\end{split}
    \label{eq:app-reference-fold-moment}
\ee
The boundary terms in the first identity vanish because
$I_{\rm col}(z)/z$ is bounded at $z=0$. The inequality follows from
Eq.~\eqref{eq:app-integrated-trace-dilution}. The second identity
also follows from zero total energy and the two fold energies
$-2\muIFS t$. By reflection symmetry, each half carries exactly the
positive energy needed to cancel its nearest fold. In particular,
the mean distance in this positive cancelling measure satisfies
\be
    \langle r_{\rm f}\rangle_+
    \equiv\frac{\int_{-t}^{t}r_{\rm f}\rho_+\,dx}{4\muIFS t}
    \leq\ell_{\rm col}.
    \label{eq:app-reference-mean-width}
\ee

For a test direction perpendicular to the string, the null plane
meets the reference worldsheet at $t=s$. Pair each positive energy
element with its nearest fold. Their kernel values differ by at most
$\|\nabla K\|_\infty r_{\rm f}/L^{d+1}$, so
\be
\begin{split}
    \sup_{\mathbf b_0}|\mathcal A^{\rm col}_{k_\perp,L}|
    &\leq\frac{\|\nabla K\|_\infty}{L^{d+1}}
          \int_{-s}^{s}dx\,r_{\rm f}\rho_+\\
    &\leq C_K\frac{\mathscr F_{k_\perp}(s)}{L^{d+1}}
                  \ell_{\rm col}.
    \label{eq:app-reference-smearing}
\end{split}
\ee
For $|c|<1$, a boost of velocity $c$ along the string takes $k$ to
$\sqrt{1-c^2}\,k_\perp$. It induces an isometry on the screen, and
both the light-ray integral and the fold flux scale linearly under
rescaling of $k$. Boost invariance and the time-uniform bound above
therefore give the same estimate for every $|c|<1$.
At $c=\pm1$ the collinear reference cancels directly on the complete
null ray. Including the fold,
\be
    \int_{-\infty}^{U}du\,T^{\rm col}_{uu}(u,v)
    =-\frac{I_{\rm col}(Uv)}{U}\longrightarrow0,
    \qquad U\longrightarrow\infty,
    \label{eq:app-tangent-reference}
\ee
since its magnitude is at most $(W_*/4)\sqrt{v/U}$ by
Eq.~\eqref{eq:app-integrated-trace-dilution}. The same holds after
$u\leftrightarrow v$. No singular boost limit or finite
$I_\infty^{\rm col}$ is required at exact tangency.

\subsection{Complete null-plane flux and the continuing source}
\label{subsec:app-complete-null-flux}

We now establish completion directly for the actual stress tensor.
The argument uses conservation, causal support inside the future light
cone of nucleation, and the positivity properties of the sector other
than the two original folds. Writing $\rho=T^{00}$ and $j^i=T^{0i}$,
this sector obeys $\rho\geq|\mathbf j|$, the null energy condition,
and $T^{00}-T^{xx}\geq0$. Positive string stress, positive folds, and
causal particles have these properties. The original folds retain
their prescribed null trajectories and energies. For longitudinal
completion we also use boost invariance of the ensemble and two
separately conserved descendant lineages whose positive sectors obey
the same inequalities. Free propagation after terminal
conversion is not needed for this argument; any further interactions
must preserve these conservation and positivity properties.

\paragraph{Nonparallel null directions.}
Fix $k=(1,\widehat{\mathbf n})$ and
$H_s=\{t-\widehat{\mathbf n}\cdot\mathbf x=s\}$, with $s>0$.
For $|c|<1$, where $c=\widehat{\mathbf n}\cdot\widehat{\mathbf x}$,
both negative folds cross $H_s$ at finite times $s/(1\mp c)$.
Choose $T_0>s/(1-|c|)$. The region ahead of the plane and its
remaining null charge are
\be
    R_t=\{\mathbf x:\widehat{\mathbf n}\cdot\mathbf x>t-s\},
    \qquad
    Q_{\rm rem}(t)=\int_{R_t}d^{D-1}x\,
                    (\rho-\widehat{\mathbf n}\cdot\mathbf j).
    \label{eq:app-remaining-null-charge}
\ee
Let $\Phi_k(s;T)$ denote the flux through $H_s$ with Minkowski-time
cutoff $t\leq T$, integrated over the complete screen. Conservation
and the absence of stress before nucleation give
\be
    \frac{dQ_{\rm rem}}{dt}
    =-\int_{H_s\cap\{t\}}d^{D-2}b\,T_{\mu\nu}k^\mu k^\nu,
    \qquad \Phi_k(s;T)=-Q_{\rm rem}(T).
    \label{eq:app-remaining-charge-flux}
\ee
For $t\geq T_0$ the remaining region contains only positive stress,
so $Q_{\rm rem}\geq0$ and $dQ_{\rm rem}/dt\leq0$.
Introduce its positive energy moment
\be
    M_{\rm rem}(t)=\int_{R_t}d^{D-1}x\,
       (s-t+\widehat{\mathbf n}\cdot\mathbf x)\rho.
    \label{eq:app-remaining-energy-moment}
\ee
Energy conservation gives $dM_{\rm rem}/dt=-Q_{\rm rem}$, since the
weight vanishes at the moving boundary. Integration and monotonicity
of $Q_{\rm rem}$ yield
\be
    0\leq Q_{\rm rem}(T)
    \leq\frac{M_{\rm rem}(T_0)}{T-T_0}
    \leq\frac{4\muIFS sT_0}{T-T_0}\longrightarrow0.
    \label{eq:app-remaining-charge-tail}
\ee
Indeed, causal support bounds the weight by $s$, and zero total
energy fixes the positive energy to $E_+(T_0)=4\muIFS T_0$.
The estimate holds history by history and also controls the ensemble
limit. It proves completion for each fixed nonparallel direction,
but is not uniform as $|c|\to1$.

\paragraph{Longitudinal null directions.}
For the parallel directions, integrate the actual stress over the
coordinates $\mathbf y$ transverse to the original string:
$\overline T_{\alpha\beta}(u,v)
\equiv\int d^{D-2}y\,T_{\alpha\beta}(u,v,\mathbf y)$.
The transverse divergence in
$\partial_vT_{uu}+\partial_uT_{uv}=\tfrac12\partial_aT_{au}$
integrates to zero, leaving the same conservation equation as in
$1+1$ dimensions. The relevant positive weight is the longitudinal
projection
\be
    W_\parallel(t)\equiv\int d^{D-1}x\,(T^{00}-T^{xx})
    =4\int dx\,\overline T_{uv}.
    \label{eq:app-longitudinal-projected-weight}
\ee
It differs from the full scalar trace weight by
$\int d^{D-1}x\sum_aT^{aa}$. In particular, an oblique massless
particle contributes $p_\perp^2/E$ to $W_\parallel$ despite having
zero scalar trace.

After the first split at $(u_s,v_s)$, label the two complete lineages
by their original outer folds, rather than their spatial positions.
Their conserved charges are $Q_R=(E-P_x)_R=2\muIFS u_s$ and
$Q_L=(E+P_x)_L=2\muIFS v_s$. Each negative outer fold contributes
zero to its lineage charge and to $W_\parallel$. The positive sector
obeys, for example,
$2Q_R-W_{\parallel,R}
=\int_R d^{D-1}x\,(T^{00}+T^{xx}-2T^{0x})\geq0$;
the analogous left-lineage inequality has the opposite sign of
$T^{0x}$. Thus
\be
    0\leq W_\parallel(t)\leq2(Q_R+Q_L)=8\muIFS t_s.
    \label{eq:app-projected-first-split-bound}
\ee
Before the first split $W_\parallel(t)=8\muIFS t\leq8\muIFS t_s$,
so the same bound holds at every time.

Boost invariance gives
$\langle\overline T_{uv}\rangle=f_\parallel(uv)\geq0$.
With $I_\parallel(z)=\int_0^z f_\parallel(w)\,dw$, the time-slice
identity of Sec.~\ref{sec:introduction-summary} yields
\be
    \frac{I_\parallel(z)}z
    \leq\frac{\langle W_\parallel(\sqrt z)\rangle}{4\sqrt z}
    \leq\frac{2\muIFS\langle t_s\rangle}{\sqrt z}.
    \label{eq:app-projected-trace-dilution}
\ee
Consequently, at fixed $v>0$,
\be
    \int d^{D-2}y\int_{-\infty}^{U}du\,
       \langle T_{uu}(u,v,\mathbf y)\rangle
    =-\frac{I_\parallel(Uv)}{U}\longrightarrow0,
    \label{eq:app-actual-longitudinal-completion}
\ee
with the analogous result after $u\leftrightarrow v$.
The normalized $k$-flux is twice this expression and also vanishes.

Together the two arguments establish, for every fixed normalized
null direction and $s>0$,
\be
    \int d^{D-2}b\,\mathcal A_k(s,\mathbf b)=0.
    \label{eq:app-complete-null-zero-mode}
\ee
After the visible negative folds have crossed, all additional flux
is nonnegative. Its complete integral is therefore finite and equals
$\mathscr F_k(s)$. No terminal lifetime, finite integrated trace, or
collinear replacement is needed for this completion statement.

\paragraph{Longitudinal rays and positional averaging.}
\label{app:longitudinal-completion}
For the collinear cascade, conservation gives, history by history,
$\mathcal A(v;U)=-\int_0^vT_{uv}(U,w)dw\leq0$.
After the original fold is crossed, the residual increases with $U$
and is bounded below by $-\muIFS v$.
Equation~\eqref{eq:app-integrated-trace-dilution} gives
$I(z)=o(z)$, so its ensemble limit vanishes. Bounded convergence
then implies saturation with
probability one for each fixed $v$. Intersect these probability-one
events for positive rational $r$. For any $r>v$,
$\mathcal A(r;U)\leq\mathcal A(v;U)\leq0$, which extends the
result to all rays using the cumulative-integral representative.
The argument applies both to the unabsorbed semiclassical cascade
and to its traceless terminal completion; finite $I_\infty$ is needed
only for the stronger positional estimate below.

\label{app:positional-average}
For \(N\) IFSs with origins
\((u_0,v_0)=(a,-a)\) averaged uniformly over \(-R\leq a\leq R\),
the residual on \(v=0\), cut off at \(u=U>0\), is
\[
 \overline{\mathcal A}_{N,R}(U)
 =-\frac{N}{2R}\int_0^{\min(R,U)}
       \frac{I(a(U-a))}{U-a}\,da.
\]
Now assume $I_\infty<\infty$. The trace bound
\eqref{eq:app-uniform-trace-bound} gives
\(I(z)\leq\min(\muIFS z,I_\infty)\).  Splitting the integral at
\(U-a=I_\infty/(\muIFS U)\) when \(U\leq2R\), and using
\(U-a\geq U-R\) when \(U>2R\), yields
\[
 \sup_{U>0}|\overline{\mathcal A}_{N,R}(U)|
 \leq\frac{N I_\infty}{2R}
       \left[1+\ln\!\left(1+\frac{4\muIFS R^2}{I_\infty}\right)\right]
 \longrightarrow0
 \qquad(R\to\infty).
\]
This is a bound for the normalized positional average of the
\(1+1\)-dimensional cascade at fixed finite \(N\), uniform
in the affine cutoff. This stronger uniformity does not follow
from bounded $W_{\rm tot}$ alone and is not used in the
transverse-smearing bound.

\subsection{Terminal products of arbitrary mass}
\label{subsec:app-terminal-angular-spreading}

Consider one actual terminal packet and its collinear replacement.  Let the parent have momentum $P=mU$ and define its
positive $k$-flux and Doppler factor by
\be
    F=-k\cdot P=m\omega,
    \qquad
    \omega=-k\cdot U>0.
    \label{eq:app-parent-flux}
\ee
For an outgoing product of momentum $p_j$ with $f_j>0$, let
\be
    f_j=-k\cdot p_j,
    \qquad
    \mathbf a_j
    =
    \Pi_k\!\left(\frac{p_j}{f_j}-\frac{P}{F}\right).
    \label{eq:app-packet-definitions}
\ee
Here $\Pi_k$ projects the spatial part onto the screen perpendicular
to $\widehat{\mathbf n}$. Along a trajectory, $p_j/f_j$ is its
velocity with respect to the null clock $s$, so $\mathbf a_j$ is
the relative screen velocity. If conversion occurs at $X_e$, write
$s_e=-k\cdot X_e$. For $s>s_e$ a free product with $f_j>0$ crosses
the plane once, at $X_e+(s-s_e)p_j/f_j$, and contributes
flux $f_j$. A parallel product has $f_j=0$ and carries no test flux.

Four-momentum conservation gives the two identities which drive the
argument:
\be
    \sum_j f_j=F,
    \qquad
    \sum_j f_j\mathbf a_j=0,
    \qquad
    \sum_j f_j|\mathbf a_j|^2
    \leq \frac{m^2}{F}
    =\frac{F}{\omega^2}.
    \label{eq:app-centered-packet}
\ee
The middle identity says that the packet is centered in positive
$k$-flux. To derive the last bound for massive products, write
$h_j=E_j+\widehat{\mathbf n}\cdot\mathbf p_j$.
The mass shell gives $|\mathbf p_{j\perp}|^2=f_jh_j-m_j^2$.
Expanding the centered square and using total four-momentum
conservation therefore gives
\be
 \sum_{f_j>0}f_j|\mathbf a_j|^2
 =\frac{m^2}{F}
  -\sum_{f_j>0}\frac{m_j^2}{f_j}
  -2E_\parallel
 \leq\frac{m^2}{F}.
 \label{eq:app-massive-packet-identity}
\ee
Here sums involving $\mathbf a_j$ omit zero-flux products,
and $E_\parallel$ is the total energy of massless products exactly
parallel to $k$; these have zero screen momentum and zero test flux.
Massive products have $f_j>0$. At fixed parent momentum, the massless
case therefore gives the largest allowed centered moment.

Suppose conversion occurs on slice $s_e$.  On a later slice $s$, the
relative displacement is
$(s-s_e)\mathbf a_j$.  Taylor expanding the smeared kernel around the
parent continuation, the centered identity cancels the linear term and
the last identity bounds the quadratic remainder:
\be
    |\delta\mathcal A^{\rm packet}_{k,L}|
    \leq
    C_K\frac{F}{L^d}
    \frac{(s-s_e)^2}{L^2\omega^2}.
    \label{eq:app-packet-taylor-bound}
\ee
There is also the trivial bound
$|\delta\mathcal A^{\rm packet}_{k,L}|
\leq C_KF/L^d$, because both packets are positive and have the same
total flux.  We apply the better of the two bounds only after summing
the complete packet:
\be
    |\delta\mathcal A^{\rm packet}_{k,L}|
    \leq
    C_K\frac{m\omega}{L^d}
    \min\!\left\{1,
       \frac{(s-s_e)^2}{L^2\omega^2}
    \right\}.
    \label{eq:app-single-packet-bound}
\ee
Both the actual and collinear packets obey this estimate relative to the
same fictitious parent continuation.  Their mutual difference therefore
obeys Eq.~\eqref{eq:app-single-packet-bound}, with a larger $C_K$.
This capped form is important: a very old or highly transverse packet
can contribute at most its total positive flux.

It remains to sum over all terminal events.  Let $m_j$ and $\chi_j$ be
the mass and longitudinal momentum rapidity of terminal parent $j$.
Boost stationarity defines its mass intensity by
\be
    \left\langle
       \sum_{j\in{\rm term}}m_j h(\chi_j)
    \right\rangle
    =
    \mathcal M_{\rm term}
    \int_{-\infty}^{\infty}d\chi\,h(\chi)
    \label{eq:app-terminal-event-measure}
\ee
for every nonnegative integrable test function $h$.  The sum includes
the complete conversion history.  This measure retains correlations between mass, conversion time, and
ancestry. The uniform source intensity in rapidity is preserved by the
random shift from each root to its descendants.

For a parent of rapidity $\chi$,
\be
    \omega
    =
    \cosh\chi-c\sinh\chi.
    \label{eq:app-terminal-doppler-factor}
\ee
All terminal events lie in the future cone, so a packet observed on
slice $s$ has $0\leq s_e\leq s$.  In the positive upper bound
\eqref{eq:app-single-packet-bound} we may replace $s-s_e$ by $s$ and
then include the full event measure, even events with $s_e>s$.  Hence,
with $a=s/L$,
\be
\begin{split}
    \sup_{\mathbf b_0}
    |\delta\mathcal A^{\rm term}_{k,L}|
    &\leq
    C_K\frac{\mathcal M_{\rm term}}{L^d}J_c(a),
\\
    J_c(a)
    &\equiv
    \int_{-\infty}^{\infty}d\chi\,
    \omega\min\!\left\{1,\frac{a^2}{\omega^2}\right\}.
    \label{eq:app-angular-rapidity-integral}
\end{split}
\ee

This integral is finite uniformly in $c$.  For $|c|<1$, set
$\epsilon=\sqrt{1-c^2}$ and
$y=\chi-\operatorname{arctanh}c$, so that
$\omega=\epsilon\cosh y$.  On either half-line put
$v=\epsilon e^{|y|}$.  Since
$v/2\leq\omega\leq v$ and $|d\chi|=dv/v$,
\be
    J_c(a)
    \leq
    2\int_0^\infty dv\,
    \min\!\left\{1,\frac{2a^2}{v^2}\right\}
    =4\sqrt2\,a.
    \label{eq:app-uniform-rapidity-bound}
\ee
At exact tangency, $\omega=e^{\mp\chi}$ and direct use of $\omega$ as
the integration variable gives $J_{\pm1}(a)=2a$.  The grazing limit is
therefore harmless: the large propagation distance of nearly parallel
products is compensated by their small $k$-flux.  Combining
Eqs.~\eqref{eq:app-angular-rapidity-integral} and
\eqref{eq:app-uniform-rapidity-bound}, and using
$\mathscr F_k(s)\geq2\muIFS s$, gives
\be
    \sup_{\mathbf b_0}
    |\delta\mathcal A^{\rm term}_{k,L}|
    \leq
    C_K\frac{\mathscr F_k(s)}{L^d}
    \frac{\ell_{\rm ang}}{L}.
    \label{eq:app-angular-smearing}
\ee

\subsection{Mass conservation and the smearing scale}
\label{subsec:app-mass-budget-smearing}

It remains to bound the terminal mass intensity. At a splitting of
a positive-energy yo-yo, the sum of daughter rest masses is at most
the parent mass, by four-momentum conservation. Terminal parents are
disjoint leaves of the descendant tree. Therefore a first-generation
yo-yo of mass $m_1$ obeys
\be
    \sum_{j\in{\rm term}(1)}m_j\leq m_1.
    \label{eq:app-terminal-tree-mass-budget}
\ee
The inequality holds for any finite set of terminal leaves, and thus
for the full set by monotone convergence, even if some branches never
terminate. It also follows directly from the exact staircase bound
$\sum_i z_i\leq1$ in Appendix~\ref{app:geometric-mass-hierarchy}.

To pass to intensities, write each terminal rapidity as
$\chi_j=\chi_1+\delta\chi_j$. Boost covariance makes the law of the
relative rapidities and masses independent of the root rapidity
$\chi_1$. Integrating a test function $h(\chi_1+\delta\chi_j)$ over
the stationary root rapidity therefore gives $\int h(\chi)\,d\chi$
for every descendant. Equation~\eqref{eq:app-terminal-tree-mass-budget}
then bounds $\mathcal M_{\rm term}$ by the first-generation mass
intensity $\mathcal M_1$. Using the raw source moment
\eqref{eq:raw-source-mass-moment} at $p=1$ gives
\be
    \mathcal M_{\rm term}\leq\mathcal M_1
    =\frac{2\muIFS}{\sqrt{\GammaSplit}}\mathscr S_0(1)
    =\frac{2\sqrt\pi}{3}\frac{\muIFS}{\sqrt{\GammaSplit}}.
    \label{eq:app-source-mass-budget}
\ee
This uses the first-generation source law, not a count or lifetime
estimate for later generations.

Equations~\eqref{eq:app-first-split-trace-bound} and
\eqref{eq:app-source-mass-budget} imply
\be
    \ell_{\rm col}\leq\sqrt{\frac\pi2}\,\GammaSplit^{-1/2},
    \qquad
    \ell_{\rm ang}\leq\frac{2\sqrt\pi}{3}\,\GammaSplit^{-1/2}.
    \label{eq:app-source-scale-lengths}
\ee
Inserting these bounds in Eq.~\eqref{eq:app-general-moment-bound}
and absorbing numerical constants into $C_K$ yields
\be
    \sup_{\mathbf b_0}|\mathcal A_{k,L}(s,\mathbf b_0)|
    \leq C_K\frac{\mathscr F_k(s)}{L^d}
               \frac{1}{L\sqrt{\GammaSplit}},
    \qquad s>0,\quad\text{every normalized }k.
    \label{eq:app-final-smearing-bound}
\ee
Thus a sufficient scale is
\be
    L_{\rm smear}\equiv\GammaSplit^{-1/2}
    \sim\ell_s g_s^{-1}.
    \label{eq:app-sufficient-smearing-scale}
\ee
Accuracy $\varepsilon$ requires
$L\gtrsim C_KL_{\rm smear}/\varepsilon$. This is a sufficient
resolution scale, not an optimal-width claim. It applies to sharp or
smooth terminal conversion within the stated assumptions, and also
to the unabsorbed collinear cascade, where $\mathcal M_{\rm term}=0$.

\subsection{Stronger terminal response and second-moment estimates}
\label{subsec:app-malthusian-moments}

Finite integrated trace gives additional information, including the
$O(U^{-1})$ longitudinal cutoff rate in
Sec.~\ref{sec:terminal-decay}. We record the sharp absorbing-model
estimate separately. For $m_{\rm term}=O(m_s)$, define
\be
    \xi_*\equiv\frac{2\muIFS}{m_{\rm term}\sqrt{\GammaSplit}}
    \sim g_s^{-1}.
    \label{eq:app-terminal-source-hierarchy}
\ee
Under the entrance-weight assumptions of
Sec.~\ref{sec:terminal-decay}, the source-averaged internal-node and
leaf counts satisfy
$\langle N_{\rm int}\rangle,\langle N_{\rm leaf}\rangle
=O(\xi_*^{\kappa_{\rm IFS}})$.
Each internal node contributes expected integrated trace
$2\muIFS/\GammaSplit$. Since $dt\,dx=\tfrac12 dz\,dy$ and
$T^{\rm col}_{00}-T^{\rm col}_{xx}=4f_{\rm col}$, the integrated
trace per unit spacetime rapidity is $2I_\infty^{\rm col}$.
The source intensity therefore gives
\be
    \frac{I_\infty^{\rm col}}{\muIFS}
    \lesssim\frac{1+\xi_*^{\kappa_{\rm IFS}}}{\GammaSplit}.
    \label{eq:app-collinear-malthusian-moment}
\ee
The constant term includes the remnant and source-scale stages.

This controls a second spatial moment. On the right half of the
string, $r_{\rm f}=u$ and $r=u/v\in[0,1]$, so
$r_{\rm f}^2\rho_+=I_{\rm col}(1+r^2)-zf_{\rm col}(1-r)^2
\leq2I_\infty^{\rm col}$. Reflection and integration give
\be
    \langle r_{\rm f}^2\rangle_+
    \leq\frac{I_\infty^{\rm col}}{\muIFS}
    \equiv\ell_2^2,
    \qquad
    \ell_2\lesssim\ell_s g_s^{-1-\kappa_{\rm IFS}/2}
    =\ell_s g_s^{-\sqrt2}.
    \label{eq:app-reference-rms-width}
\ee
Under the same finite-response assumptions, the late pressure law
proved in Appendix~\ref{app:terminal-pressure-endpoint} gives
\be
    \ell_2^2
    =\frac{1}{4\muIFS}\lim_{t\to\infty}
       t\langle W^{\rm col}_{\rm tot}(t)\rangle
    =-\frac{1}{4\muIFS}\lim_{t\to\infty}
       t\langle p^{\rm col}_{\rm tot}(t)\rangle.
    \label{eq:app-rms-pressure-tail}
\ee
Thus the coefficient of the late negative-pressure tail also sets
this upper bound on the RMS separation of positive energy from the
folds.

The leaf count also refines the angular estimate in this model:
\be
    \mathcal M_{\rm term}\lesssim
       m_{\rm term}\xi_*^{\kappa_{\rm IFS}},\qquad
    \ell_{\rm ang}\lesssim\ell_s g_s^{-\kappa_{\rm IFS}}.
    \label{eq:app-angular-malthusian-moment}
\ee
For a smooth crossover, these additional estimates require the
complete-tree response and terminal-mass matching of
Sec.~\ref{sec:terminal-decay}. They need not hold for a prolonged
preterminal phase. The first-moment smearing bound
\eqref{eq:app-final-smearing-bound} requires neither of them.

\endgroup

\section{Gravitational emission from semiclassical yo-yos}
\label{app:gravitational-emission}

A yo-yo loses energy gradually through gravitational radiation and
redistributes it abruptly when it splits. We estimate the fraction
radiated before the next split in weak, locally flat gravity. The
long-wavelength calculation tests the neglect of radiation during the
semiclassical cascade.

\paragraph{Radiation before the next split.}
In the rest frame set \(\Delta u=\Delta v=L\), and write
\(\sigma=2\muIFS\) for the tension of the doubled string.  Its mass,
period, and period-averaged splitting rate in our model are
\be
 m=\sigma L,\qquad \mathcal T=L,\qquad
 \lambda_{\rm split}=\GammaSplit L.
 \label{eq:grav-yoyo-scales}
\ee
For \(\ls\ll L\ll\GammaSplit^{-1/2}\), many oscillations precede a
split.  If \(P_{<\Omega}\) is the period-averaged power below frequency
\(\Omega\), the leading perturbative estimate of the fractional loss is
\be
 \varepsilon_{<\Omega}
 \equiv\frac{P_{<\Omega}}{m\lambda_{\rm split}}
 =\frac{P_{<\Omega}}{\sigma\GammaSplit L^2}.
 \label{eq:grav-loss-before-split}
\ee
This uses the unperturbed motion and splitting clock; small
\(\varepsilon_{<\Omega}\) makes that treatment self-consistent for
the frequency band considered.

\paragraph{The classical spectrum.}
First consider four-dimensional gravity, with Newton constant \(G_4\).
Let \(\chi\) be the cosine of the emission angle relative to the string,
\(\omega_n=2\pi n/L\), and define the period-normalized Fourier tensor
\(T_n^{\mu\nu}=L^{-1}\int_0^Ldt\int d^3x\,
e^{i\omega_n(t-\hat{\mathbf k}\cdot\mathbf x)}T^{\mu\nu}\).
Only its \(00,01,11\) components are nonzero.  Conservation gives
\(T_n^{00}=\chi^2T_n^{11}\) and \(T_n^{01}=\chi T_n^{11}\), so the
graviton polarization sum reduces to
\(T_n^{\mu\nu *}T_{n\mu\nu}-|T_n^\mu{}_{\mu}|^2/2
=|T_n^\mu{}_{\mu}|^2/2\).
The null folds are essential for conservation, although their trace
vanishes.  The remaining bulk trace is \(-2\sigma\) inside the diamond
\(0<u,v<L\); its Fourier transform has magnitude
\be
 |T_n^\mu{}_{\mu}|=
 \frac{4\sigma L}{\pi^2n^2(1-\chi^2)}
 \sin^2\!\left[\frac{\pi n}{2}(1-\chi)\right].
 \label{eq:grav-trace-fourier}
\ee
Using the weak-field radiation formula
\(dP_n/d\Omega_{\hat{\mathbf k}}=G_4\omega_n^2
|T_n^\mu{}_{\mu}|^2/(2\pi)\) (see, e.g.,
Ref.~\cite{MartinVilenkin:1997radiation}), the angular integral becomes
\be
 P_n=\frac{16G_4\sigma^2}{\pi n}
 \int_0^{\pi n/2}
 \frac{\sin^4 y\,dy}{y^2[1-y/(\pi n)]^2}
 \sim\frac{4G_4\sigma^2}{n}.
 \label{eq:grav-harmonic-power}
\ee
The asymptotic expression refers to the ideal classical source.
Using $[1-y/(\pi n)]^{-2}\leq4$ and
$\int_0^\infty\sin^4y\,dy/y^2=\pi/4$ gives the bound
\be
 P_{<\Omega}\leq16G_4\sigma^2H_{N_\Omega},\qquad
 \varepsilon_{<\Omega}\leq
 \frac{16G_4\sigma}{\GammaSplit L^2}H_{N_\Omega},
 \quad N_\Omega=\left\lfloor\frac{\Omega L}{2\pi}\right\rfloor,
 \label{eq:grav-soft-bound}
\ee
where \(H_N=\sum_{n=1}^Nn^{-1}\leq1+\ln N\) for \(N\geq1\).
Choose \(L^{-1}\ll\Omega\ll\ls^{-1}\), within the four-dimensional
effective regime: $\Omega$ must also lie below the first relevant
compactification excitation scale.
At fixed compactification data,
\(G_4\sigma=O(g_s^2)\) and \(\GammaSplit=O(g_s^2/\ls^2)\), so
\(\varepsilon_{<\Omega}
=O((\ls/L)^2[1+\ln(\Omega L)])\).
Thus a macroscopic yo-yo splits before losing an appreciable fraction
of its mass to these modes.  The logarithm comes from radiation
concentrated near the null-fold directions.
For a fixed small \(\Omega\ls\), its growth is \(\ln(L/\ls)+O(1)\).

\paragraph{The ultraviolet contribution.}
The ideal four-dimensional spectrum has a logarithmically divergent
harmonic sum. The total loss therefore depends on the quantum spectrum
and any modes resolving the compact dimensions. Iengo's maximal-angular-momentum folded-string example agrees with the
classical spectrum at low emitted energy and falls off more rapidly at
higher energy \cite{Iengo:2006radiation}; extending this
behavior to longitudinal yo-yos requires a calculation for that state.

The kinematic emission limit is
$\omega=(m^2-m'^2)/(2m)\leq m/2$, potentially well above
$\ell_s^{-1}$. If the quantum power is bounded by an envelope of order
$G_4\sigma^2/n$, even extending it to
$n_{\max}=O(mL)=O((L/\ell_s)^2)$ gives only
$\ln n_{\max}=2\ln(L/\ell_s)+O(1)$. The suppression of radiative
losses then survives. Establishing that envelope, including the narrow
fold-direction cones, is the remaining ultraviolet input. The controlled
long-wavelength bound supports neglecting gravitational losses in its
frequency range.

The splitting rate used in this comparison is the local IFS law
\eqref{eq:local_split_rate}, motivated by Ref.~\cite{Itzhaki:2024cxh}.
The relative importance of radiative and splitting channels depends on the
string state; see Ref.~\cite{Iengo:2006nn}.

\section{Terminal response and the crossover}
\label{app:terminal-response}

Terminal decay cuts off the hierarchy of massive strings, producing
the crossover from the semiclassical power to the late $1/t$ trace.
The crossover shape is first calculated in the large mass-hierarchy limit. We then hold the microscopic hierarchy fixed and study how the continuing source builds its logarithmic accumulated response. The first limit
determines the crossover kernel; the second determines the delay caused
by a finite source onset.

\subsection{Terminal kernels and the crossover}
\label{app:terminal-kernel}

A local terminal law first determines the total response, which is then resolved in invariant radius. Throughout this subsection
the background parameters are fixed, final conversion products are
traceless, and \(\chi=\GammaSplit/(2\muIFS)\),
\(\kappa=\kappa_{\rm IFS}\).

\paragraph{Rest-frame conversion and the integrated response.}
Suppose conversion at proper-time rate \(\nu_{\rm dec}(m)\) competes with
the binary rate \(\chi m\). Put \(\eta(m)=\nu_{\rm dec}(m)/(\chi m)\).
Both rates acquire the same factor \(\gamma^{-1}\) in the laboratory
frame, while \(\int Wdt=m\Delta\tau\). A node therefore contributes
\(1/[\chi(1+\eta)]\) in expectation and splits with probability
\(1/(1+\eta)\). First-event conditioning gives the mass renewal equation
\be
 [1+\eta(m)]H(m)=1+2\int_0^1da\int_0^1db\,
 H(m\sqrt{ab}),\qquad
 H(m)=\chi\mathcal R_{\rm tree}(m).
 \label{eq:terminal-rate-renewal}
\ee
The physical solution is the minimal nonnegative one. For example,
a constant competing ratio $\eta>1$ below the terminal scale gives
mean offspring number $2/(1+\eta)<1$ and response $H=1/(\eta-1)$ there.

A second example isolates the response accumulated after fragmentation
stops. Stop binary
splitting below \(m_{\rm term}\), but let each terminal yo-yo convert
to traceless products at rest-frame rate
\(\nu_{\rm term}(m)=\eta\GammaSplit m/(2\muIFS)\), with fixed
\(\eta>0\). Its expected integrated trace is
\(m/\nu_{\rm term}=2\muIFS/(\eta\GammaSplit)\), independent of boost:
the time-dilation factor cancels between its trace and its lifetime.
A stopped binary tree has one more terminal leaf than internal nodes,
so this completion gives
\be
 \mathcal R_{\rm tree}^{(\eta)}(\xi)
 =\frac{2\muIFS}{\GammaSplit}
 \left[(1+\eta^{-1})\mathcal N_{\rm int}(\xi)+\eta^{-1}\right].
 \label{eq:terminal-finite-rate-example}
\ee
It preserves the hierarchy exponent while quantifying the extra trace
accumulated before conversion.

\paragraph{Sharp-cutoff radial kernel.}
Use the chronological coordinates and weighted path law of
Eqs.~\eqref{eq:chronological-birth-coordinates}--
\eqref{eq:main-weighted-generation-clock}. Under \(\mathbb P_\kappa\),
the fractions \(a,b\) are independent \({\rm Beta}(h,1)\) variables with
\(h=\sqrt2\). The affine clock approaches its stationary law \(\pi\),
whose marginal coordinates \(U,V\) have \({\rm Gamma}(h,1)\) laws.
They remain correlated through their common exponential waits.
At large boost the occupation mark is \(\gamma(V+\epsilon)\).
Trace weighting multiplies the \({\rm Gamma}(h+1,1)\) density of
\(V+\epsilon\) by \((V+\epsilon)^{-\kappa}\); the resulting shape is
\(h+1-\kappa=3-\sqrt2\). Keeping masses above \(m_{\rm term}\)
requires \(V+\epsilon\ge T/\gamma\), giving
Eq.~\eqref{eq:terminal-moving-cutoff}.
Let \(S_n=\ln(m_0/m_n)\), \(L=\ln(m_0/m_{\rm term})\), and
\(\bar X=\mathbb{E}_\kappa[S_{n+1}-S_n]=1/h\).
For the expected number of above-cutoff splits inside radius \(\rho\),
the many-to-one identity gives
\be
 e^{-\kappa L}\mathbb{E}[N_{\rm rad}(\rho)]
 =\sum_{n\ge0}\mathbb{E}_\kappa\!\left[
 e^{-\kappa(L-S_n)}\mathbf1_{S_n<L}
 \mathbf1_{R_ne^{S_n-L}\le z}\right],\qquad
 z=\chi m_{\rm term}\rho.
 \label{eq:terminal-stopped-radial-renewal}
\ee
Here \(R_n\) is the split mark in
Eq.~\eqref{eq:chronological-split-radius}.
The marked renewal limit replaces the sum by
\(\bar X^{-1}\int_0^\infty e^{-\kappa u}
\Pr_\pi(R\le ze^u)du\). One direct justification is to truncate the
clock to its last \(K\) generations. The remaining ancestral mass sum
is independent of that finite block; its renewal density is
\(h(1-e^{-4hs})\), since each mass step is gamma distributed with
shape two and rate \(2h\). First let \(L\to\infty\), then \(K\to\infty\).
The affine contraction restores the stationary clock, and the integrable
weight \(e^{-\kappa u}\) controls the large-\(u\) tail. Thus
\be
 \begin{aligned}
 F_{\rm sharp}(z)&=\mathbb{E}_\pi\bigl[\min\{1,(z/R)^\kappa\}\bigr],\qquad
 R=\sqrt{(U+\epsilon)(V+\epsilon)},\\
 e^{-\kappa L}\mathbb{E}[N_{\rm rad}(\rho)]
 &\longrightarrow\frac{F_{\rm sharp}(z)}{\kappa\bar X}.
 \end{aligned}
 \label{eq:terminal-radial-cdf}
\ee
Equivalently, \(F_{\rm sharp}\) is the distribution function of
\(Z_{\rm r}=Re^{-E_\kappa}\), where
\(E_\kappa\sim{\rm Exp}(\kappa)\) is independent of the stationary
clock and \(\epsilon\sim{\rm Exp}(1)\).

Integrating the boost orbit of a radial event gives
\(2\operatorname{arcosh}(t/\rho)\mathbf1_{\rho<t}\). Hence
\be
 \Psi_{\rm sharp}(T)
 =2\mathbb{E}\!\left[\operatorname{arcosh}(T/Z_{\rm r})
                     \mathbf1_{Z_{\rm r}<T}\right]
 =2\int_0^1\frac{F_{\rm sharp}(Tu)}{u\sqrt{1-u^2}}\,du.
 \label{eq:terminal-psi-kernel}
\ee
The unbounded logarithmic reward is controlled by a negative moment of
any order \(0<r<\kappa\): \(R\ge\epsilon\) gives
\(\mathbb{E}[R^{-r}]\le\Gamma(1-r)\), and the renewal weight becomes
\(e^{-(\kappa-r)u}\). The same limit therefore applies to the accumulated
response. For the fixed scaled entrance law of
Eq.~\eqref{eq:chronological-invariant-source}, source integration gives
\be
 \begin{aligned}
 \mathcal A_{\rm term}
 &=\frac{2\muIFS}{\GammaSplit}
   \frac{\xi_*^\kappa\mathscr S(\kappa)}{\kappa\bar X},\\
 C_{\rm early}&={\rm B}(\kappa/2,1/2)\mathbb{E}[R^{-\kappa}],\qquad
 C_{\rm sharp}=2\ln2-2\mathbb{E}[\ln R]+\frac2\kappa.
 \end{aligned}
 \label{eq:terminal-kernel-normalization}
\ee
The stopped node-count bound provides domination by the entrance mass
moment $\mathscr S(\kappa)$. The limit first sends
$\xi_*=2\muIFS/(m_{\rm term}\sqrt{\GammaSplit})$ to infinity
at fixed $T$, with the scaled entrance law held fixed. The small- and
large-$T$ limits then describe this limiting kernel. At fixed string scale,
$2\muIFS/\GammaSplit\sim g_s^{-2}$ and $\xi_*\sim g_s^{-1}$,
so $\mathcal A_{\rm term}\sim g_s^{-2-\kappa}$: the response per
node and the number of nodes supply separate enhancements. Numerical evaluation of this limiting kernel is described in
App.~\ref{app:bounce-numerics}.

\paragraph{The late response for a general terminal law.}
For a positive finite radial trace measure, normalize its weight
\(\sum_j m_jd\tau_j\) to one and again denote the sampled dimensionless
radius by \(Z_{\rm r}\). Equation~\eqref{eq:terminal-psi-kernel} then holds
with its corresponding distribution function, irrespective of the local
conversion mechanism. If \(F(z)\sim az^\kappa\) at small \(z\), the
early power is \(a{\rm B}(\kappa/2,1/2)T^\kappa\). For the late limit,
assume \(\mathbb{E}\bigl[|\ln Z_{\rm r}|\bigr]<\infty\), set
\(C_{\rm term}=2\ln2-2\mathbb{E}[\ln Z_{\rm r}]\), and subtract the
logarithm inside the expectation. The exact remainder is
\be
 \begin{gathered}
 \Psi(T)-2\ln T-C_{\rm term}=\mathbb{E}[D(Z_{\rm r}/T)],\\
 D(x)=\begin{cases}
 2\ln[(1+\sqrt{1-x^2})/2],&0<x<1,\\
 2\ln(x/2),&x\ge1.
 \end{cases}
 \end{gathered}
 \label{eq:terminal-remainder-function}
\ee
Since $D(x)=O(x^2)$ at zero and grows logarithmically at infinity,
any finite radial moment of order $0<p\leq2$ gives
\be
 |\Psi(T)-2\ln T-C_{\rm term}|\leq
 \frac{K_p\mathbb{E}[Z_{\rm r}^p]}{T^p}.
 \label{eq:terminal-radial-moment-bound}
\ee
For the sharp law, the gamma marginals of $U+\epsilon,V+\epsilon$
give all positive moments of $R$, hence of $Z_{\rm r}$.
Using $\mathbb{E}[UV]=[h/(h+1)]^2(\mathbb{E}[UV]+2h+2)$ and
$D(x)=-x^2/2+O(x^4)$ yields
\be
 \begin{aligned}
 \mathbb{E}[Z_{\rm r}^2]
 &=\frac{\kappa}{\kappa+2}\mathbb{E}[R^2]
 =\frac{8+5\sqrt2}{7},\\
 \Psi_{\rm sharp}(T)&=2\ln T+C_{\rm sharp}
 -\frac{8+5\sqrt2}{14T^2}+O(T^{-4}).
 \end{aligned}
 \label{eq:terminal-sharp-remainder}
\ee
The terminal law enters the offset and convergence through these radial
moments, while the finite total response fixes the logarithmic slope.

\paragraph{Pressure on late time slices.}
\label{app:terminal-pressure-endpoint}
The accumulated response also determines the instantaneous terminal
pressure. For the collinear massless completion,
Eq.~\eqref{eq:app-uniform-trace-bound} gives
$0\leq zf(z)\leq I(z)\leq I_\infty$.
Writing $S=t^2$ in Eq.~\eqref{eq:pressure-abel-trace},
\[
 \frac{tW(t)}4=\int_0^S\frac{f(z)}{\sqrt{1-z/S}}\,dz.
\]
For fixed $0<\varepsilon<1$, dominated convergence on
$0\leq z\leq(1-\varepsilon)S$ gives $I_\infty$ as $S\to\infty$.
The endpoint contribution is at most
$2I_\infty\sqrt\varepsilon/(1-\varepsilon)$, by $f(z)\leq I_\infty/z$.
Taking $S\to\infty$ and then $\varepsilon\to0$ proves
$W(t)\sim4I_\infty/t$, including the remnant.

\subsection{Logarithmic emission and the accumulated response}
\label{app:logarithmic-response}

The persistent remnant emits a constant mean number of trees per
logarithmic interval of age. Their finite responses therefore add to
a constant response per logarithmic interval at late times, even when
individual trees have long and unequal lifetimes. We derive this
statement at fixed microscopic parameters, including the effect of a
finite source onset.

Let $Q>0$ be the optical age measured from the IFS birth, with
$\dot Q>0$, and start the stationary source at $Q_0>0$:
$d\langle N_1\rangle=\nu\,dQ_e/Q_e$, with $\nu=2$ for the two arms.
Scale covariance gives the source-averaged tree response
\be
 d\mathscr R_{\rm tree}=\mathcal F(y)dy,\qquad
 y=Q/Q_e-1\geq0,\qquad
 R_{\rm tree}=\int_0^\infty\mathcal F(y)dy<\infty,
 \quad\mathcal F\geq0.
 \label{eq:app-log-tree-profile}
\ee
In logarithmic age and delay, $x=\ln(Q/Q_0)$ and
$z=\ln(Q/Q_e)$, define
\be
 g(z)=e^z\mathcal F(e^z-1),\qquad
 \int_0^\infty g(z)dz=R_{\rm tree}.
 \label{eq:app-log-delay-profile}
\ee
Convolving the constant emission rate with this delay profile yields
\begin{align}
 \mathcal J_{\rm desc}(x)
 &=\nu\int_0^x g(z)dz\longrightarrow\nu R_{\rm tree},\nonumber\\
 w_{\rm desc}(t)
 &=\frac{\dot Q}{Q}\mathcal J_{\rm desc}(x)
 \sim\nu R_{\rm tree}\frac{\dot Q}{Q},
 \label{eq:app-log-response-rate}\\
 \mathscr R_{\rm desc}(Q)
 &=\nu\int_0^x(x-z)g(z)dz
 \sim\nu R_{\rm tree}\ln(Q/Q_0).
 \label{eq:app-log-cumulative-response}
\end{align}
The last limit follows by dominated convergence applied to
$(1-z/x)_+g(z)$. These are large-optical-age limits,
$Q/Q_0\to\infty$. The direct remnant adds its coefficient $C_0$
to the logarithmic slope $\nu R_{\rm tree}$.

The delay controls the offset from this limiting logarithm:
\be
 \nu R_{\rm tree}x-\mathscr R_{\rm desc}
 =\nu\left[\int_0^x zg(z)dz+x\int_x^\infty g(z)dz\right].
 \label{eq:app-log-response-deficit}
\ee
If $M_1=\int_0^\infty zg(z)dz<\infty$, the deficit tends to
$\nu M_1$. A finite first moment in physical delay gives a rate of
convergence, as the following estimate shows for sharp absorption.

\paragraph{Delay moments in the stopped binary cascade.}
In a stationary flat background, let a root of mass $m_0$ and energy
$E_0$ be emitted at $t_e>0$. Stop each branch at
$m\leq m_{\rm term}$, and put $\xi=m_0/m_{\rm term}$ and
$c=\GammaSplit/(2\muIFS)$. For $p>\kappa_{\rm IFS}$,
$M(p)=2/(1+p/2)^2<1$. If $N_n$ counts above-cutoff nodes at
generation $n$, comparison with the unrestricted mass tree gives
\be
 \mathbb{E}[N_n]\leq\xi^pM(p)^n,\qquad
 \mathbb{E}\!\left[\sum_{n\geq0}(n+1)N_n\right]
 \leq\frac{\xi^p}{[1-M(p)]^2}.
 \label{eq:app-stopped-depth-bound}
\ee
Indeed, each retained node obeys $1\leq(m_v/m_{\rm term})^p$.
Energy conservation within the positive-energy tree also gives
$E_v\leq E_0$, so every retained node has
\be
 \lambda_v=c\frac{m_v^2}{E_v}\geq\frac{cm_{\rm term}^2}{E_0},
 \qquad \lambda_v^{-1}\leq\tau_0
 \equiv\frac{E_0}{cm_{\rm term}^2}.
 \label{eq:app-stopped-lifetime-bound}
\ee
Conditional on the mass and momentum genealogy, the clocks are
independent exponentials. For a node born at delay $b_v$ and living
for $\tau_v$, $W_v/\lambda_v=1/c$ gives
\[
 \mathbb{E}\!\left[W_v\int_0^{\tau_v}(b_v+s)ds
                 \,\middle|\,\text{genealogy}\right]
 =\frac1c\left(\sum_{a\prec v}\lambda_a^{-1}
                         +\lambda_v^{-1}\right).
\]
Summing over nodes and using the depth bound yields
\be
 \int_0^\infty T\,\langle W_{\rm tree}(T)\rangle dT
 \leq\frac{E_0\xi^p}{c^2m_{\rm term}^2[1-M(p)]^2}.
 \label{eq:app-tree-first-delay-bound}
\ee
The source satisfies $E_0\leq2\muIFS t_e$ and has finite mass
moments \eqref{eq:raw-source-mass-moment}. Hence its averaged profile
has a finite first delay moment,
\be
 K_1\equiv\int_0^\infty y\mathcal F(y)dy
 =\left\langle\frac1{t_e}\int_0^\infty
          T W_{\rm tree}(T)dT\right\rangle_{\rm src}
 \leq\frac{2\muIFS\langle\xi^p\rangle_{\rm src}}
 {c^2m_{\rm term}^2[1-M(p)]^2}<\infty.
 \label{eq:app-source-first-delay-bound}
\ee
This bound uses binary evolution from the root. An initial geometric
stage adds its own delay moment and the birth delays of the entrance
roots: $D_{\rm root}=D_{\rm pre}+\mathbb{E}\bigl[\sum_j(b_jR_j+D_j)\bigr]$,
where $R_j,D_j$ are their integrated response and first delay moment
measured from birth. Finiteness of this sum, and of any added terminal
delay moment, supplies the corresponding extension.

\paragraph{The delayed logarithmic onset.}
Since $\ln(1+y)\leq y$, $M_1\leq K_1$. Moreover,
$K=\int_0^\infty e^zg(z)dz=R_{\rm tree}+K_1<\infty$, so the
exact convolution gives
\begin{align}
 0\leq\nu R_{\rm tree}-\mathcal J_{\rm desc}(x)
 &\leq\nu K e^{-x},\nonumber\\
 0\leq\mathscr R_{\rm desc}-\nu R_{\rm tree}x+\nu M_1
 &=\nu\int_x^\infty(z-x)g(z)dz
 \leq\frac{\nu K}{e}e^{-x}.
 \label{eq:app-log-strong-remainder}
\end{align}
The second inequality uses $u e^{-u}\leq e^{-1}$. Equivalently,
\be
 \begin{aligned}
 \mathscr R_{\rm desc}(Q)
 &=\nu R_{\rm tree}\ln(Q/Q_0)-\nu M_1+O(Q_0/Q),\\
 w_{\rm desc}
 &=\frac{\dot Q}{Q}[\nu R_{\rm tree}+O(Q_0/Q)].
 \end{aligned}
 \label{eq:app-log-sharp-asymptotics}
\ee
Thus the finite mean delay shifts the onset scale to
$Q_{\rm eff}=Q_0\exp(M_1/R_{\rm tree})$, leaving the logarithmic
slope unchanged. The constants depend on the microscopic hierarchy;
these estimates are at fixed coupling. With only finite total response,
Eq.~\eqref{eq:app-log-cumulative-response} still gives the leading
logarithm, while the offset and convergence rate require delay moments.

A bounded emission rate $\nu(x)\to\nu$ gives the same limiting
slope by dominated convergence in
$\mathcal J_{\rm desc}(x)=\int_0^x\nu(x-z)g(z)dz$. An integrable
source transient changes the cumulative response by a constant when
$M_1<\infty$; a finite-response initial population contributes a bounded
term. The instantaneous tail additionally uses the source-tail control
of Sec.~\ref{subsec:terminal-persistent-source}. Throughout this argument
the tree profile is stationary in logarithmic delay. An evolving remnant
source is retained explicitly in the cosmological transport of
App.~\ref{app:cosmological-transport}.

\section{Cosmological transport and a sharp-cutoff bounce}
\label{app:cosmological-transport}
\label{app:terminal-bounce-control}

The homogeneous response combines transport of the emitted strings with the
source supplied by the remnant. The transport and the criterion for birth-time
source matching are formulated first, followed by an application of the
sharp-cutoff response to the ESP background. The local string and binary
approximations are the same as in the stationary calculation; the long
propagation interval is retained in the optical clock.

\subsection{High-boost transport in a homogeneous background}
\label{app:cosmological-high-boost}

Section~\ref{subsec:cosmological-optical-response} introduces the
optical clock from the common background dependence of the trace and
splitting rate. Here we derive its branching generator, retain the
remnant's emission history, and specify the conditions for source
matching. The integrated trace measure $w_Edt=w_sdt_s=m\,d\tau_s$
is preserved by the local frame conversion
$g_{\mu\nu}^{(s)}=\Omega^2g_{\mu\nu}^{(E)}$.

The small-string approximation gives the oscillation-averaged worldline
action \(-m\int d\tau_s\), with constant string-frame mass between
splits. Spatial homogeneity conserves \(k=a_sp_s\), even when the
Einstein-frame mass \(\Omega m\) varies. Combining this motion with
\(\lambda_s=\Gamma_t m/(2\mu\gamma)\) and the clock
\eqref{eq:cosmology-optical-generator} gives
\be
 \lambda_sdt_s=
 \frac{m^2}{2\mu\sqrt{k^2+a_s^2m^2}}\,d\bar Q
 =\frac{m^2}{2\mu k}[1+O(\gamma^{-2})]d\bar Q.
 \label{eq:app-cosmology-optical-splitting-law}
\ee
The optical clock retains the history of $a_s\Gamma_t$; finite boost
changes the rate by the factor $[1+(a_sm/k)^2]^{-1/2}$.
For the live descendants of an IFS born at $s$, the total splitting
activity and integrated trace are
\be
 \mathcal J_s(Q)=\left\langle\sum_{\rm live}
 \frac{m^2}{2\mu k}\right\rangle,
 \qquad
 w_{{\rm desc},s}(t)\simeq
 \frac{2\mu}{\Gamma_t}\dot Q_s(t)\mathcal J_s(Q_s(t)).
 \label{eq:cosmology-optical-response-measure}
\ee

On an outward high-boost branch the two daughters have, to leading order,
\[
 (m_1,k_1)=(m\sqrt{ab},ak),\qquad
 (m_2,k_2)=(m\sqrt{(1-a)(1-b)},(1-a)k),
\]
with independent uniform \(a,b\). For an additive test observable, the
branching generator per unit \(Q\) is consequently
\be
 \begin{aligned}
 (\mathcal L F)(m,k)
 &=\frac{m^2}{2\mu k}\Bigl\{
 \int_0^1\!da\int_0^1\!db\,
 \bigl[F(m\sqrt{ab},ak)\\
 &\hspace{18mm}+F(m\sqrt{(1-a)(1-b)},(1-a)k)\bigr]-F(m,k)\Bigr\}.
 \end{aligned}
 \label{eq:app-high-boost-generator}
\ee
Away from absorption, \(F=m^2/k\) satisfies \(\mathcal LF=0\):
high-boost splitting conserves this trace moment. Its leading loss is
therefore set by terminal conversion, while cosmological redshift supplies
the common observation weight.
Set \(F=0\) below a fixed absorbing mass to obtain
\(\mathcal L_{\rm term}\). More generally, a competing proper-time
conversion rate
\(\nu_{\rm dec}^{(s)}=\Gamma_t m\eta(m/m_s)/(2\mu)\)
adds \(-m^2\eta F/(2\mu k)\). The generator remains autonomous if
\(\eta\) is independent of the background. A terminal rate with a
different coupling dependence instead leaves a time-dependent \(\eta\)
in the operator. Traceless conversion products contribute no further trace.

Let \(n_s(dm\,dk;Q_s)\) be the expected live descendant measure from
an IFS born at \(s\), and \(\mathcal S_s(Q_s)\) its remnant's emission
measure per unit optical time. The leading homogeneous transport closes as
\be
 \partial_{Q_s}n_s=\mathcal L_{\rm term}^*n_s+\mathcal S_s(Q_s),\qquad
 n_s(Q)=e^{Q\mathcal L_{\rm term}^*}n_s(0)
 +\int_0^Q e^{(Q-q)\mathcal L_{\rm term}^*}\mathcal S_s(q)\,dq.
 \label{eq:app-high-boost-transport}
\ee
The trace is obtained by integrating \(m^2/(2\mu k)\) against this
measure and applying Eq.~\eqref{eq:cosmology-optical-response-measure}.
The leading transport retains the time dependence of redshift and
coupling through the optical clock and the remnant's emission history. The remnant source is also explicit in the
same near-null approximation.

\paragraph{Remnant records and emitted marks.}
Use string-frame conformal null coordinates,
\(ds_s^2=-a_s^2du\,dv\), on the IFS worldsheet. Near an outgoing
arm set \(z=2u\). To leading order across the thin strip,
\(\Gamma_t a_s^2du\,dv=dz\,dQ\). Thus its surviving width has record
rate \(z\) per unit \(Q\), and a record replaces \(z\) by \(Yz\),
with \(Y\) uniform on \((0,1)\). The density satisfies
\be
 \partial_Qp(z,Q)=-zp(z,Q)+\int_z^\infty p(z',Q)\,dz',
 \qquad p(z,Q)\sim Qe^{-Qz}.
 \label{eq:app-remnant-record-law}
\ee
Consequently \(\langle z\rangle\sim Q^{-1}\), giving
\(d\langle N_1\rangle/dQ\sim2/Q\) for two arms and
\(w_{{\rm rem},s}\sim4\mu\Omega a_s/Q_s\). The source onset and
strip-width corrections are separate from this late record law.

The second state variable is the comoving fold impulse \(K\) accumulated
since the previous record. The tension supplies
\(dK=2\mu a_sdt_s=2\mu\,dQ/\Gamma(Q)\). Together, the deterministic
evolution and record jump are
\be
 \begin{aligned}
 &\frac{dz}{dQ}=0,\qquad \frac{dK}{dQ}=\frac{2\mu}{\Gamma(Q)},
       \qquad \text{jump rate}=z,\\
 &(z,K)\longmapsto(Yz,0),\qquad
   k_{\rm em}=K,\quad m_{\rm em}^2=2\mu K(1-Y)z .
 \end{aligned}
 \label{eq:app-remnant-marked-process}
\ee
The mass formula combines the long fold impulse with the locally short
opposite null momentum. In particular,
\(m_{\rm em}^2/(2\mu k_{\rm em})=(1-Y)z\): the injected trace moment
equals the lost width, independently of the long impulse history.
The expected jump measure of the two arms is precisely
\(\mathcal S_s\) in Eq.~\eqref{eq:app-high-boost-transport}.
It is the first-generation source; any non-dilute entrance stage must
still be propagated or matched as in
Appendix~\ref{appendix:first_generation_source}.

Over a record interval \([Q_i,Q_{i+1}]\), put
\be
 \frac1{\Gamma_{\rm eff}}
 =\frac1{\Delta Q}\int_{Q_i}^{Q_{i+1}}\frac{dq}{\Gamma(q)},\qquad
 X=z\Delta Q,\qquad
 m_{\rm em}^2=\frac{4\mu^2X(1-Y)}{\Gamma_{\rm eff}}.
 \label{eq:app-remnant-effective-rate}
\ee
At constant coupling the marks in \(Q\) are the flat-space marks even
if the scale factor changes substantially. When the coupling varies,
the only additional history in these leading marks is the interval
average \(\Gamma_{\rm eff}^{-1}\). A fixed birth-time source coefficient is therefore appropriate when
$\Gamma_{\rm eff}$ stays close to the birth value during formation.
The later transport uses the full optical interval $Q(t)-Q(t_e)$.

Terminal absorption selects the high-boost population, with
$\gamma\sim\Delta t_s/t_\times^{(s)}$ in the stationary reference.
Here $\Delta t_s$ is the string-frame age and
$t_\times^{(s)}\equiv2\mu/(\Gamma_t m_{\rm term})=\Omega t_\times$
is the corresponding string-frame crossover time. Local matching
requires small $|H_s|t_\times^{(s)}$,
$|\partial_{t_s}\ln g_s|t_\times^{(s)}$, and
$\sqrt{|\partial_{t_s}H_s|}t_\times^{(s)}$. The worldline action also
requires negligible background variation across an oscillation in the
string's rest frame, including the boost of the local gradients. Within
this regime, $Q$ retains the subsequent background history. Corrections
come from the trace outside the high-boost population and from accumulated
finite-boost effects; a quantitative error bound for them remains to be
established.

For the stationary optical source, the normalized trace moment has the
late form $\mathcal J_s\sim A_s/Q_s$, with constant-background
descendant coefficient $\mathcal C_{{\rm desc},s}=2\mu A_s/\Gamma_s$.
The remnant strip has width proportional to $Q_s^{-1}$ and the same
observation weight $\Omega a_s$. Together they give the late response
\eqref{eq:cosmology-terminal-optical-response}, with
$\mathcal C_s=4\mu/\Gamma_s+2R_{{\rm tree},s}$, where
$R_{{\rm tree},s}$ is the source-averaged integrated response of a
finite tree. Each semiclassical node contributes a response of order
$g_s^{-2}$, and the hierarchy contains $O(g_s^{-\kappa_{\rm IFS}})$
nodes, giving $\mathcal C_{\rm desc}\propto g_s^{-2\sqrt2}$.

Formation fixes this source normalization, while subsequent propagation
is retained by $Q_s(t)-Q_s(t_e)$ for emission time $t_e$.
The birth-time coefficient applies when the splitting rate varies
little over the formation intervals supplying most of the observed
response. Later propagation may include substantial background
evolution while the local transport conditions hold. Under this
matching, the accumulated response in
Sec.~\ref{subsec:cosmological-optical-response} satisfies
$d\mathscr F_s(Q_s)=(\Gamma_t/\Gamma_s)w_{{\rm ev},s}dt$.
A varying source uses Eqs.~\eqref{eq:app-remnant-marked-process}
and \eqref{eq:app-high-boost-transport}. The following estimate
identifies the formation intervals relevant to old IFS cohorts.

\subsection{Emission times and local source matching}
\label{app:source-age-localization}

The terminal tail is carried mainly by descendants of first-generation
yo-yos emitted early in the IFS's history. Source matching concerns
their formation interval \(t_e-s\), whereas the optical clock transports
their response over the subsequent interval $t-t_e$. We identify the
emission window that supplies the terminal trace, formulate its matching
condition, and derive the window's scaling in the binary reference.

\paragraph{The response-weighted emission window.}
At fixed parameters, changing variables from emission time to
\(y=Q/Q_e-1\) in Eq.~\eqref{eq:app-log-tree-profile} gives
\be
 \frac{d\mathscr R_{\rm desc}}{dQ}
 =\nu\int_{Q_0}^Q\frac{dQ_e}{Q_e^2}\,
          \mathcal F(Q/Q_e-1)
 =\frac{\nu}{Q}\int_0^{Q/Q_0-1}\mathcal F(y)\,dy.
 \label{eq:app-source-age-measure}
\ee
Once the stationary tail is reached, the fraction supplied by any range
of emission times is therefore measured by
$\mathcal F(y)dy/R_{\rm tree}$. This weights emissions by their
contribution to the observed trace.

Write $D=Q/Q_e$ for the ratio of observation to emission age.
For a fixed small $\varepsilon>0$, choose its lower quantile in the
response-weighted measure. In the sharp-cutoff binary reference,
\be
 \Pr_{\rm resp}(D<d_\varepsilon)=\varepsilon,\qquad
 d_\varepsilon(\xi_*)=\xi_*^{1+o(1)},\qquad \xi_*\to\infty.
 \label{eq:app-source-age-quantile}
\ee
Thus a fraction $1-\varepsilon$ of the terminal trace comes from trees
emitted at optical ages $Q_e\leq Q/d_\varepsilon$. The finite-coupling
quantile specifies the accuracy; its leading power identifies the
parametric separation. We formulate source matching in terms of this
quantile and then derive its scaling.

\paragraph{Matching the source in an evolving background.}
For an IFS born at \(s\), define \(t_{e,\varepsilon}\) by
\be
 Q_s(t_{e,\varepsilon})=\frac{Q_s(t)}{d_\varepsilon},\qquad
 \epsilon_{\rm src}(s,t;\varepsilon)
 =\sup_{s\leq u\leq t_{e,\varepsilon}}
       \left|\ln\frac{\Gamma_u}{\Gamma_s}\right|.
 \label{eq:app-source-locality-condition}
\ee
The stationary reference places a fraction \(1-\varepsilon\) of
the terminal trace in this early emission window. For those emissions,
\(\epsilon_{\rm src}\ll1\) bounds every record's
\(\Gamma_{\rm eff}/\Gamma_s\) between
\(e^{-\epsilon_{\rm src}}\) and \(e^{\epsilon_{\rm src}}\).
At fixed record marks, $m^2$ and $k$ change together, leaving the
optical splitting rate $r_0=m^2/(2\mu k)$ unchanged. The coupled binary histories thus have the same
clock and differ only by a mass hierarchy factor within
\(e^{\pm\epsilon_{\rm src}/2}\). Their terminal trace is bracketed
by these two cutoff histories, whose large-hierarchy integrated
normalizations differ by \(O(\kappa_{\rm IFS}\epsilon_{\rm src})\).
The excluded source tail can be evaluated with
Eq.~\eqref{eq:app-remnant-marked-process}; a uniform comparison bound
\(M\) on its actual versus reference contribution gives total relative
error \(O(\epsilon_{\rm src}+M\varepsilon)\).
The source-onset truncation in Eq.~\eqref{eq:app-source-age-measure}
must also be small before using the stationary quantile.

Only the formation interval \(t_e-s\) enters this matching condition.
The subsequent interval $t-t_e$ can contain substantial evolution of
$g_s$ and $H$, retained by the high-boost generator through
$Q_s(t)-Q_s(t_e)$ under the local conditions of
App.~\ref{app:cosmological-high-boost}.

If \(|\partial_t\ln\Gamma|\lesssim t_{\rm bg}^{-1}\) near birth,
a sufficient source condition is
\(t_{e,\varepsilon}-s\ll t_{\rm bg}\). When \(\dot{\bar Q}\)
is approximately constant over that short formation window this reads
\be
 \frac{Q_s(t)}{\dot{\bar Q}(s)}\ll d_\varepsilon t_{\rm bg}.
 \label{eq:app-source-optical-age-bound}
\ee
If the full optical age is also comparable to
\(\dot{\bar Q}(s)(t-s)\), the familiar estimate is
\(t-s\ll g_s(s)^{-1+o(1)}t_{\rm bg}\). The optical condition is the appropriate criterion when the full background
history is substantial.
Expansion can suppress older cohorts through population dilution, whereas
contraction supplies no such bound. The source condition should therefore
be checked in optical age with the actual production history.
For the ESP example below, the measured coupling variation is small across
the whole production history, so source matching holds directly over that interval.

\paragraph{Emission times selected by the terminal tail.}
For a root of mass ratio \(\xi=m_0/m_{\rm term}>1\) and rate
\(r_0=m_0^2/(2\mu k_0)\), follow a trace-weighted daughter at each
high-boost split. Its null fractions have density \(2b\,da\,db\), so
\(A=-\ln a\) and \(C=-\ln b\) are independent exponentials of
rates 1 and 2. Write
\be
 \begin{gathered}
 L=\ln\xi,\qquad S_n=\frac12\sum_{j=1}^n(A_j+C_j),\qquad
 V_n=\sum_{j=1}^n C_j,\qquad N_L=\inf\{n:S_n\geq L\},\\
 \Theta_L=\sum_{n<N_L}\epsilon_n e^{V_n},\qquad
 \epsilon_n\sim\mathrm{Exp}(1).
 \end{gathered}
 \label{eq:app-tag-terminal-clock}
\ee
The root's expected normalized trace at delay \(\Delta Q\) is
\(r_0\Pr(\Theta_L>r_0\Delta Q)\). Thus a delay sampled from its
integrated response has \(Z=r_0\Delta Q\) distributed as
\be
 p_L(z)=\frac{\Pr(\Theta_L>z)}{\mathbb{E}[\Theta_L]},\qquad
 \mathbb{E}_{\rm resp}[Z^r]
 =\frac{\mathbb{E}[\Theta_L^{1+r}]}{(1+r)\mathbb{E}[\Theta_L]},
 \quad -1<r<1.
 \label{eq:app-response-delay-moments}
\ee
Sampling a time from the integrated trace weights each tagged lifetime
by its duration.

A short moment argument suffices to determine its leading scale.
The joint increment transform determines the renewal exponent:
\be
 \mathbb{E}\bigl[e^{qC-p(A+C)/2}\bigr]
 =\frac{2}{(1+p/2)(2-q+p/2)},\qquad
 \vartheta(q)=q-3+\sqrt{(q-1)^2+8}.
 \label{eq:app-source-delay-exponent}
\ee
For \(q\) near 1 and any \(\eta>0\),
\be
 \mathbb{E}[\Theta_L^q]\leq C_{q,\eta}
        e^{[\vartheta(q)+\eta]L},\qquad
 \mathbb{E}[\Theta_L]\asymp e^{\kappa_{\rm IFS}L}.
 \label{eq:app-source-delay-moment-bound}
\ee
Indeed, for \(p>\vartheta(q)\),
\(\mathbb{E}\bigl[e^{qV_n};\,S_n<L\bigr]\leq
 e^{pL}(\mathbb{E}\bigl[e^{qC-p(A+C)/2}\bigr])^n\), a summable geometric
bound. Subadditivity proves the first estimate for \(q\leq1\).
For \(q>1\), H\"older's inequality with normalized weights
proportional to \(e^{-\delta n}\) inserts only
\(e^{\delta(q-1)n}\); choosing \(\delta\) sufficiently small
preserves summability. The mean is the expected above-cutoff node count,
since every node contributes unit expected integrated normalized trace;
its exponent is \(\kappa_{\rm IFS}=\vartheta(1)\).
Finally \(\vartheta'(1)=1\). Applying Markov's inequality to
Eq.~\eqref{eq:app-response-delay-moments} with small positive and
negative \(r\) proves, for each fixed \(\delta>0\),
\be
 \Pr_{\rm resp}\{Z\notin[\xi^{1-\delta},\xi^{1+\delta}]\}
 \leq C_\delta\xi^{-c_\delta},\qquad c_\delta>0.
 \label{eq:app-response-delay-concentration}
\ee
For example, the tail exponent is bounded by
\(\sqrt{r^2+8}-2\sqrt2-|r|\delta+\eta<0\) after choosing
\(|r|\) and then \(\eta\) small enough. 

This leading power survives the first-generation source average.
Relabel its two uniform fractions as \(U,V\), with \(V\) the short
null fraction. The late marks obey
\(\xi=\xi_*\sqrt{BUV}\) and \(r_0Q_e=BV\), where
\(B\) has density \(Be^{-B}\).
Equation~\eqref{eq:raw-source-mass-moment} and the corresponding mixed
moments with \((BV)^{-r}\) are finite near \(r=0\).
Averaging the numerator and denominator of
Eq.~\eqref{eq:app-response-delay-moments} with these marks therefore
gives the same concentration for \(D=Q/Q_e=1+Z/(BV)\), with
\(\xi\) replaced by \(\xi_*\). This proves Eq.~\eqref{eq:app-source-age-quantile}. A different entrance
or terminal law uses its own response quantile; the same leading power
follows when the additional delay and response weights obey these
neighboring moment bounds.

\subsection{A sharp-cutoff ESP bounce}
\label{app:terminal-numerical-check}
\label{app:terminal-background}

We apply the crossover calculated in Sec.~\ref{subsec:terminal-crossover}
in the homogeneous ESP equations. The calculation uses the large-hierarchy binary reference from its
first-generation source, with traceless terminal products and the local
small-string transport of App.~\ref{app:cosmological-high-boost}.
The source fixes the amplitude. The reference model neglects non-dilute entrance and finite-hierarchy corrections.

\paragraph{The response entering the background equations.}
Let \(m_s=1/\sqrt{\alpha'}\), \(m_{\rm term}=c\,m_s\), and use
\be
 \mu=\frac1{2\pi\alpha'},\qquad
 \Gamma_s=\frac{g_s(s)^2}{\alpha'},\qquad
 \xi_s=\frac{2\mu}{\sqrt{\Gamma_s}m_{\rm term}}
      =\frac1{\pi c\,g_s(s)}.
 \label{eq:app-sharp-benchmark-normalization}
\ee
Both the tension and splitting rate carry the same inverse string area.
For binary evolution from the first-generation source,
Eq.~\eqref{eq:raw-source-mass-moment} fixes
\(\mathscr S(\kappa)=\Gamma(2+\kappa/2)/(1+\kappa/2)^2\simeq0.62691\),
with \(\kappa=\kappa_{\rm IFS}\) and \(\bar X=1/\sqrt2\). Define
\be
 \mathcal A_s=\frac{2\mu}{\Gamma_s}
      \frac{\xi_s^\kappa\mathscr S(\kappa)}{\kappa\bar X},\qquad
 Q_{\times,s}=\frac{2\mu a(s)\Omega(s)}{m_{\rm term}},\qquad
 Q_{*,s}=a(s)\Omega(s)\sqrt{\Gamma_s}.
 \label{eq:app-sharp-benchmark-amplitude}
\ee
The remnant function obeys $\Xi(0)=0$ and $\Xi'(z)=2D(z)$,
where $D(z)=e^{-z^2}\int_0^z e^{y^2}dy$ is Dawson's integral.

With these normalizations, the accumulated response
\eqref{eq:cosmology-accumulated-event-response} gives the event trace
through Eq.~\eqref{eq:app-sharp-benchmark-response}. Its remnant term recovers
the stationary result $8\mu D(\sqrt\Gamma t)/\sqrt\Gamma$,
including both the early growth and the late $4\mu/(\Gamma t)$ tail.
The convolution is evaluated using the cumulative kernel and its
first moment. The continuation to the youngest cohorts and its
numerical sensitivity are described in
Appendix~\ref{app:bounce-numerics}.

The full-history integral \eqref{eq:cosmological-retarded-response}
can be initialized at a finite time by evolving any earlier event
population separately and adding its trace. The energy boundary
condition also retains any earlier traceless products. In the present
benchmark both populations are absent at the initial ESP crossing.

\paragraph{Background and initial data.}
We evolve the scalar equation with the full optical history and use
the shifted Hubble variable \(J=H+\Pi/(3\sqrt2M_{\rm Pl})\).
Combining the Friedmann constraint, Raychaudhuri equation, and scalar
equation eliminates the explicit trace force from its evolution:
\be
 \dot J=-2H^2+\frac{\rho_\chi+W_{\chi,\rm bal}}{6M_{\rm Pl}^2}
 -\frac{H\Pi}{\sqrt2M_{\rm Pl}}-\frac{\Pi^2}{6M_{\rm Pl}^2}.
 \label{eq:cosmology-shifted-Hubble-evolution}
\ee
This identity is exact within the negligible-potential approximation;
the scalar equation retains the computed trace rather than imposing
\(W=W_{\chi,\rm bal}\).
The initial ESP crossing has \(a=1\), \(H=-H_0\),
\(\Pi=-\sqrt6M_{\rm Pl}H_0\), and no earlier IFS or additional
radiation. The ordinary potential is neglected and the ESP particles are
stable on the interval. The fiducial parameters are
\[
 g_s(\mathrm{ESP})=g_0=10^{-4},\quad
 H_0=10^{-22}M_{\rm Pl},\quad H_0\ell_s=10^{-17},\quad
 \lambda_\chi=10^{-3},\quad c=1.
\]
Thus \(\alpha'M_{\rm Pl}^2=10^{10}\) and
\(\mu/M_{\rm Pl}^2=1.59155\times10^{-11}\).
The ESP production prescription fixes
\(n_{\chi,*}=(\lambda_\chi|\Pi_{\rm ESP}|)^{3/2}/(2\pi)^3
=4.88735\times10^{-40}M_{\rm Pl}^3\).
The IFS production coefficient is \(\beta=1/(32\pi^6)\). For the numerical
results below, define \(x\equiv H_0(t-t_{\rm ESP})\).

Appendix~\ref{app:bounce-numerics} describes the response integration
and convergence checks.

\paragraph{Small motion and the bounce.}
The first dilaton turn occurs at \(x_m\simeq0.050119\).
The reference solution \eqref{eq:cosmology-ESP-two-fluid} is matched there
using \((a_m,J_m,\phi_m)\), with no subsequent fit. To measure the settled
motion, set \(\phi_c=\phi(x=0.051)\) and use \(I:0.051\leq x\leq0.13\):
\be
 \epsilon_K=\sup_I\frac{\Pi^2}{2\rho_\chi},\qquad
 \epsilon_\phi=\sup_I
 \frac{|\phi-\phi_c|}{\phi_{\rm ESP}-\phi_c}.
 \label{eq:cosmology-small-motion}
\ee
The fiducial calculation gives \(\epsilon_K\simeq3.4\times10^{-7}\) and
\(\epsilon_\phi\simeq1.8\times10^{-3}\), with a bounce at
\(x_b\simeq0.12653\). The frozen-dilaton reference bounces at
\(x_b^{(0)}\simeq0.12651\); on \(I\),
\(\max|H-H^{(0)}|/H_0<4\times10^{-4}\).
The sampled range in \(\ln\Gamma\) from the turn through \(x=0.13\)
is \(1.7\times10^{-4}\), so source matching is locally appropriate
over the whole production history. The maximum \(|H|t_\times\) is
below \(5\times10^{-10}\). The direct remnant is retained, although
descendants supply more than 99\% of the trace at the bounce.

\begin{table}[ht!]
\centering
\renewcommand{\arraystretch}{1.14}\small
\begin{tabular}{ccrrr}
\toprule
\(g_0\) & \(c=m_{\rm term}/m_s\) & \(x_b\) & \(\epsilon_K\) & \(\epsilon_\phi\)\\
\midrule
\(10^{-4}\) & 1 & 0.12653 & \(3.4\times10^{-7}\) & \(1.8\times10^{-3}\)\\
\addlinespace[0.22em]
\(10^{-4}\) & 0.5 & 0.12653 & \(2.0\times10^{-7}\) & \(1.4\times10^{-3}\)\\
\addlinespace[0.22em]
\(10^{-4}\) & 2 & 0.12654 & \(5.8\times10^{-7}\) & \(2.4\times10^{-3}\)\\
\addlinespace[0.22em]
\(5\times10^{-5}\) & 1 & 0.12652 & \(5.2\times10^{-8}\) & \(7.1\times10^{-4}\)\\
\addlinespace[0.22em]
\(2\times10^{-4}\) & 1 & 0.12657 & \(2.2\times10^{-6}\) & \(4.7\times10^{-3}\)\\
\bottomrule
\end{tabular}
\caption{Sharp-cutoff response: fiducial case and parameter variations.
All other physical inputs are held fixed; the frame normalization is
\(g_0=g_s(\mathrm{ESP})\) in each run. Motion diagnostics use the interval
specified in the text.}
\label{tab:sharp-bounce-sensitivity}
\end{table}

The variations in Table~\ref{tab:sharp-bounce-sensitivity} preserve the
bounce while changing the motion needed to supply its trace. Increasing
the coupling weakens the cascade response and increases the motion;
lowering the terminal threshold allows more descendants and decreases it.

To test the uncertainty in the first-generation entrance normalization,
we independently multiply only the descendant term of
Eq.~\eqref{eq:cosmology-accumulated-event-response} by a constant $f_{\rm desc}$,
leaving the direct remnant and all background and production parameters
fixed. For $f_{\rm desc}=0.3,1,3$, the bounce times are respectively
$x_b=0.126551,0.126532,0.126523$; the settled kinetic fractions are
$\epsilon_K=(1.13\times10^{-6},3.39\times10^{-7},1.13\times10^{-7})$,
and the settled relative field drifts are
$\epsilon_\phi=(3.34,1.83,1.06)\times10^{-3}$. In all three cases
$\max_I |H-H^{(0)}|/H_0<7\times10^{-4}$, with the reference still
fixed at the first dilaton turn. Including the initial adjustment,
the displacement from that turn through the bounce stays below
$3.3\times10^{-3}$ of the initial distance to the ESP. Halving the
integration step from $5\times10^{-5}$ to $2.5\times10^{-5}$ changes
the two extreme bounce times by less than $5\times10^{-9}$ in $x$.
The kinetic fraction is approximately inversely proportional to
$f_{\rm desc}$, as expected from near force balance. The sweep isolates
the overall response normalization; it does not calculate changes in
the microscopic entrance dynamics or crossover shape.

\subsection{Dual-fluid evolution and expanding limits}
\label{app:cosmology-expansion-estimates}

\paragraph{String-frame scaling and the dual-fluid transfer.}
The dilaton work in Eq.~\eqref{eq:ifs-ward-cosmology} is absorbed
into ordinary continuity in the string frame. With $\Omega=g_s/g_0$,
the scale factor and time are $a_s=\Omega a$ and $dt_s=\Omega\,dt$.
The energy conversion $E_E=\Omega E_s$ and the volume conversion
$V_s=\Omega^3V_E$ give
$\rho^{(s)}=\Omega^{-4}\rho_{\rm IFS}$ and $W^{(s)}=\Omega^{-4}W$.
Write $q_\phi=\dot\Omega/\Omega=\Pi/(\sqrt2M_{\rm Pl})$ and
$\Theta_s=d\ln a_s/dt=H+q_\phi=\Omega H_s$. Then
\[
 \frac{d\rho^{(s)}}{dt}+4\Theta_s\rho^{(s)}
 =\Omega^{-4}(\dot\rho_{\rm IFS}+4H\rho_{\rm IFS})
 =\Theta_s W^{(s)},
\]
or, on an interval with $\Theta_s\ne0$,
\be
 \frac{d\rho^{(s)}}{d\ln a_s}+4\rho^{(s)}=W^{(s)},\qquad
 \bar\epsilon_W\equiv-\frac12\frac{d\ln W^{(s)}}{d\ln a_s}.
 \label{eq:cosmology-string-frame-scaling}
\ee
In Einstein-frame variables,
\be
 \bar\epsilon_W=-\frac12\frac{\dot W/W-4q_\phi}{H+q_\phi},\qquad
 \bar\epsilon_W\simeq
 \frac{H\epsilon_{V'}+2q_\phi}{H+q_\phi}
 \quad\hbox{in potential force balance}.
 \label{eq:app-cosmology-barred-exponent}
\ee
This reduces to $\epsilon_{V'}$ for slow coupling evolution. Constant
Einstein-frame $W$ does not by itself give $\bar\epsilon_W=0$.

For constant $\bar\epsilon_W\ne2$, integration gives
$\rho^{(s)}=W^{(s)}/(4-2\bar\epsilon_W)+C_s/a_s^4$.
Transforming back yields the dual-fluid prescription including dilaton
work,
\be
\begin{aligned}
 \rho_{\rm IFS}&=\rho_{\rm IDE}+\rho_R,
 &p_{\rm IFS}&=p_{\rm IDE}+\frac{\rho_R}{3},\\
 \rho_{\rm IDE}&=\frac{W}{4-2\bar\epsilon_W},
 &p_{\rm IDE}&=\left(-1+\frac{2\bar\epsilon_W}{3}\right)\rho_{\rm IDE}.
\end{aligned}
 \label{eq:app-cosmology-local-dual-fluid}
\ee
The homogeneous radiation mode scales as $a^{-4}$ because
$\Omega^4a_s^{-4}=a^{-4}$.
The trace/null split and this prescription coincide for
$\bar\epsilon_W=0$; in general,
$\rho_N=\rho_R+\bar\epsilon_W\rho_{\rm IDE}/2$.

When $\bar\epsilon_W$ varies, differentiating the same local
definition of $\rho_{\rm IDE}$ gives
\be
 \dot\rho_{\rm IDE}+4H\rho_{\rm IDE}
 =(H+q_\phi)W+
       \frac{2W\dot{\bar\epsilon}_W}{(4-2\bar\epsilon_W)^2}.
 \label{eq:app-cosmology-transfer-derivation}
\ee
Using $3p_{\rm IDE}=\rho_{\rm IDE}-W$, the Bianchi equations for
the two components are
\be
\begin{aligned}
 \dot\rho_{\rm IDE}+3H(\rho_{\rm IDE}+p_{\rm IDE})
 &=\frac{\Pi W}{\sqrt2M_{\rm Pl}}-\mathcal Q_R,\\
 \dot\rho_R+4H\rho_R&=\mathcal Q_R,\qquad
 \mathcal Q_R=-\frac{\dot{\bar\epsilon}_W}{2-\bar\epsilon_W}\rho_{\rm IDE}
            =-\frac{2W\dot{\bar\epsilon}_W}{(4-2\bar\epsilon_W)^2}.
\end{aligned}
 \label{eq:cosmology-dual-fluid-transfer}
\ee
Their sum is the IFS Ward identity. For $W>0$, a decreasing exponent
sources positive $\rho_R$, and an increasing exponent sources a
negative contribution. The accumulated radiation-like stress is
\be
 \rho_R(t)=\frac{1}{a^4(t)}\left[
 a_*^4\rho_R(t_*)+\int_{t_*}^{t}du\,a^4(u)\mathcal Q_R(u)\right].
 \label{eq:app-cosmology-sourced-radiation}
\ee
Here $\mathcal Q_R$ measures transfer between effective components
of the same IFS stress. The microscopic radiation from terminal
conversion is already included in that stress and is not counted
separately by $\mathcal Q_R$.

\paragraph{Regular evolution through the onset of production.}
For arbitrary $W(t)$, the total IFS density instead has the regular
integral representation
\be
 \rho_{\rm IFS}(t)=\frac{1}{a^4(t)}\left[
 a_*^4\rho_{{\rm IFS},*}+\int_{t_*}^{t}du\,a^4(u)
 \left(H(u)+\frac{\Pi(u)}{\sqrt2M_{\rm Pl}}\right)W(u)\right].
 \label{eq:app-cosmology-general-trace-integral}
\ee
This expression and the trace/null split require neither division by
$W$ nor a monotone string-frame scale factor. By contrast, the local
IDE parametrization uses
\[
 4-2\bar\epsilon_W=
 \frac{\dot W+4HW}{\Theta_s W},
 \qquad
 \rho_{\rm IDE}=\frac{\Theta_s W^2}{\dot W+4HW}.
\]
The logarithmic exponent is undefined at $W=0$ or $\Theta_s=0$,
and $\bar\epsilon_W=2$ makes the local scaling particular solution
singular. These are limitations of the parametrization, not necessarily
of the total stress. For a constant exponent equal to $2$, the
string-frame solution contains
$a_s^{-4}[C_s+W_*^{(s)}a_{s,*}^4\ln(a_s/a_{s,*})]$ instead.

A local onset model illustrates this transfer without specifying
the full production history. As an example, let $W=A\tau^\nu[1+O(\tau)]$, with
$\tau=t-t_0>0$, $A>0$, $\nu>0$, smooth background evolution,
$\Theta_s(t_0)=\Theta_0\ne0$, and no inherited IFS stress.
Then the exact energy integral and the local IDE definition give
\be
 \begin{aligned}
 \rho_{\rm IFS}&=\frac{\Theta_0 A}{\nu+1}\tau^{\nu+1}
                  +O(\tau^{\nu+2}),\\
 \rho_{\rm IDE}&=\frac{\Theta_0 A}{\nu}\tau^{\nu+1}
                  +O(\tau^{\nu+2}),
 &\rho_R&=-\frac{\Theta_0 A}{\nu(\nu+1)}\tau^{\nu+1}
                  +O(\tau^{\nu+2}).
 \end{aligned}
 \label{eq:app-cosmology-production-onset}
\ee
Thus $\mathcal Q_R=-\Theta_0 A\tau^\nu/\nu+O(\tau^{\nu+1})$
builds a radiation-like correction even with zero initial density.
Its sign depends on $\Theta_0$, whereas
$\rho_N=-W/4+O(\tau^{\nu+1})$ is negative in either case.
These are local asymptotics for the assumed trace growth; the ESP
calculation instead evolves its full retarded response. Through a later
zero of $4-2\bar\epsilon_W$, the total energy and trace/null equations
provide the regular evolution variables.

Finally, in the absence of ESP particles, set
$\rho_{\phi+\rm IFS}=\Pi^2/2+V+\rho_{\rm IFS}$. The scalar and
IFS equations give the exact conserved-sector identity
\be
 \dot\rho_{\phi+\rm IFS}+4H\rho_{\phi+\rm IFS}
 =H(4V+W-\Pi^2).
 \label{eq:app-cosmology-combined-trace}
\ee
In potential force balance, $q_\phi W\simeq-\dot V$:
the work gained by the strings is lost by the dilaton. The dual-fluid
transfer cancels internally and does not alter this accounting.

\paragraph{Dilution of an inherited cohort.}
Consider an older, finite comoving population after IFS production has
ceased. At approximately constant coupling, the late one-event trace
is proportional to $\dot Q_s/Q_s$ and the density dilutes as $a^{-3}$.
If radiation domination begins at $t_r$, the optical age is
$Q_s(t)=Q_s(t_r)+C_Q\int_{t_r}^t a(u)\,du$, with
$C_Q=\Omega^2\Gamma$ constant in this approximation.
With $a(t)\propto t^{1/2}$ (using the radiation-era time origin),
its late behavior, once this increment dominates the inherited age, is
\be
 Q_s(t)\propto t^{3/2},
 \qquad \frac{\dot Q_s}{Q_s}\simeq\frac{3}{2t}.
 \label{eq:app-old-cohort-radiation-clock}
\ee
The terminal trace consequently scales as
$W_{\rm old}\propto a^{-3}t^{-1}\propto a^{-5}$, faster than
$\rho_{\rm rad}\propto a^{-4}$. This suppresses the inherited
traceful dilaton force in a sufficiently developed radiation era, but
it does not fix the amplitude of the signed $a^{-4}$ mode in
Eq.~\eqref{eq:ifs-ward-cosmology}. If the
inherited optical age remains important or the coupling evolves,
the general $a^{-3}(\Gamma_s/\Gamma_t)\dot Q_s/Q_s$ response
must instead be retained.

\paragraph{Logarithmic response on the expanding force-balance branch.}
Suppose $H>0$, the coupling and production rate vary slowly over a
Hubble time, and $Ht_\times\ll1$. Writing $u=t-s$ and taking
$a(t-u)/a(t)\simeq e^{-Hu}$, the optical clock for nearly constant
coupling gives $\dot Q_s/Q_s\simeq H/(1-e^{-Hu})$. Thus the terminal
response and the population dilution give
\be
 \begin{aligned}
 A_{\rm cas}
 &\simeq\beta\mathcal C(g_s)\left[
      \int_{t_\times}^\infty du\,
        \frac{H e^{-3Hu}}{1-e^{-Hu}}+O(1)\right]\\
 &=\beta\mathcal C(g_s)
      \left[\ln\frac1{Ht_\times}+O(1)\right].
 \end{aligned}
 \label{eq:app-expanding-response-integral}
\ee
The finite crossover contribution and the part at $Hu=O(1)$
affect only the $O(1)$ term; the logarithm comes from
$t_\times\ll u\ll H^{-1}$.
At fixed $W_{V,\rm bal}$, the weak-coupling coefficient
$\mathcal C(g_s)\propto g_s^{-2\sqrt2}$ gives
$\Pi_*\propto g_s^{\sqrt2}/\sqrt{\ln[1/(Ht_\times)]}$.
This scaling requires $Ht_\times\ll1$ throughout. At fixed
string-frame scales, $t_\times^{(s)}=\Omega t_\times\propto g_s^{-2}$,
so it is not a uniform $g_s\to0$ limit at fixed $H/\Omega$.

\FloatBarrier
\section{Numerical methods and validation}
\label{appendix:statistical-methods}

This appendix collects the numerical tests of the cascade and the homogeneous bounce. It begins with the common treatment of ensemble averages and correlated errors, followed by the implementation and results for each observable. The analytical predictions are derived in the
preceding sections and appendices.

\subsection{Ensemble averages and statistical tests}
\label{app:ensemble-covariance}

Each of \(N\) independent histories supplies a complete
observable vector \(X_s\). We use \(N=1{,}000\) for the histogram,
live-trace, invariant-mass, ANEC, and spatial-moment fits.
Coordinates may be boost bins, time-interval averages, sampled snapshots, or
paired observables; correlations within a history are retained. Time-domain
fits use exact live-history integrals as specified in
App.~\ref{app:time-domain-fits}. For the ANEC grid, integrate
piecewise-linear interpolation in each logarithmic coordinate and
divide by the interval width. This gives nonnegative weights
\(H_{ji}\), with \(\sum_iH_{ji}=1\); use their tensor product
in \(\ln U\) and \(\ln v\) for both data and prediction.
The histogram uses direct bin integrals. Estimate the mean and its
covariance by
\be
 y=\frac1N\sum_sX_s,\qquad
 \widehat\Sigma=\frac1{N(N-1)}
     \sum_s(X_s-y)(X_s-y)^T.
 \label{eq:time-fit-covariance}
\ee
For the time-domain, ANEC, and spatial-moment fits, use this full
covariance without shrinkage or eigenmode removal. For the histogram,
write \(\widehat\Sigma=DRD\), where \(R\) is the empirical
correlation matrix and \(D_{ii}=\sqrt{\widehat\Sigma_{ii}}\), and use
\(C=D[(1-\lambda)R+\lambda I]D\).  This preserves the marginal
variances of the mean.  The histogram estimates \(\lambda\) by
Ledoit--Wolf.

With the applicable covariance \(C\) and model vector
\(m(\theta)\), fit the free parameters by minimizing
\be
 \chi^2(\theta)=[y-m(\theta)]^{\mathsf T}C^{-1}[y-m(\theta)].
 \label{eq:ensemble-quadratic-fit}
\ee
Quoted degrees of freedom count coordinates minus fitted parameters
and are nominal: the probabilities below are calibrated by resampling.

\paragraph{Bootstrap calibration.}
\label{app:shared-fit-calibration}

For goodness of fit, draw \(N\) centered vectors \(X_s-y\) with
replacement and add the fitted null mean \(m(\widehat\theta)\).
Each bootstrap sample re-estimates the covariance with the same
prescription, including any estimated shrinkage, and refits the model.
We use \(B_{\rm boot}=5{,}000\) samples for the time-domain and
spatial-moment tests, and \(9{,}999\) for the histogram, ANEC, and
paired pressure--ANEC comparisons. We report
\(p_{\rm boot}=(1+n_{\geq})/(B_{\rm boot}+1)\), where \(n_{\geq}\)
counts bootstrap statistics at least as large as the observed one.
Nested tests generate samples under the constrained model, refit both
models, and calibrate the improvement in the quadratic statistic.

Confidence intervals instead resample uncentered histories.
Paired observables use the same seed indices in every resample,
preserving their correlations in amplitude ratios and joint fits.
Rebinning and window checks reuse the same histories and are correlated
diagnostics.

Error bars show marginal standard errors. The interpretation
of the coherent trace fluctuations and residuals is given in
App.~\ref{app:source-fluctuations}.

\subsection{Independent chronological binary benchmark}
\label{app:binary-chronological-benchmark}

To test the binary prediction separately from geometric matching, we
simulate both daughters at every split, with independent uniform null
fractions and exponential proper waits of rate \(cm\). Lab-frame birth times are inherited along each explicitly simulated
tree, independently of the stationary-clock evaluation. Roots have mass
\(m_0\), are born at the origin, and have unit entrance intensity per signed
rapidity. We integrate over \(|y_0|\leq16.1181\), using 40,000 roots in
100 independent, rapidity-stratified batches. There is no generation cap;
the 142.4 million processed nodes reach generation 33. These runs form a separate binary ensemble.

The omitted rapidity tail has a uniform analytic bound. Conservation of
four-momentum at a binary split, together with convexity of $p^2/E$, implies
$\sum_i m_i^2/E_i\leq m_{\rm parent}^2/E_{\rm parent}$, so the total
trace of a tree never exceeds its root trace $m_0/\cosh y_0$.
Here $p$ denotes longitudinal momentum.
For unit entrance intensity per signed rapidity, the ensemble trace
$W_{>Y}(t)$ omitted by the cutoff therefore obeys
\be
 0\leq W_{>Y}(t)\leq
 m_0\int_{|y_0|>Y}\frac{dy_0}{\cosh y_0}
 =4m_0\arctan(e^{-Y})\leq4m_0e^{-Y}.
 \label{eq:binary-benchmark-rapidity-tail}
\ee
At $Y=16.1181$, this is less than $4\times10^{-7}m_0$ at every
snapshot. It is an absolute truncation bound, separate from the sampling
uncertainties quoted below.

\begin{figure}[ht!]
\centering
\includegraphics[width=0.98\textwidth]{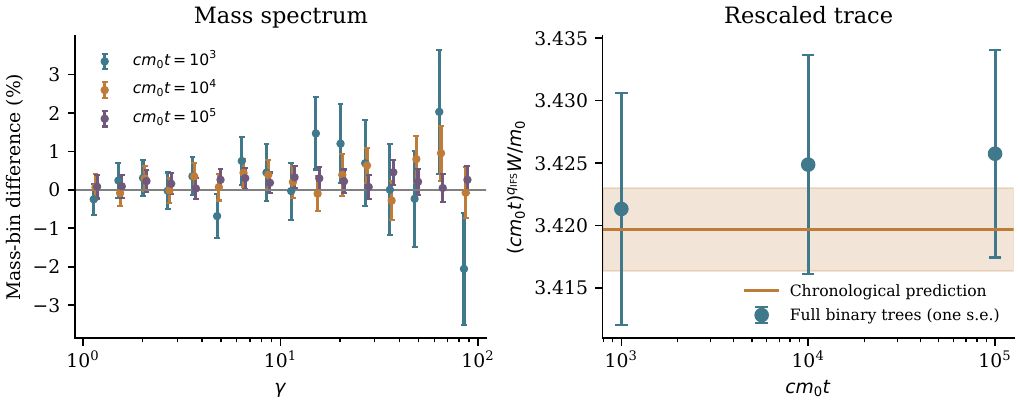}
\caption{Independent full-tree binary benchmark. Left: differences from
absolute predicted mass-bin integrals in 16 logarithmic bins over
\(1\leq\gamma\leq100\); bars combine simulation and reference-integration
standard errors. Right: the rescaled total trace and the predicted
constant, with its reference-integration uncertainty shaded.
No amplitude is fitted.}
\label{fig:binary-chronological-benchmark}
\end{figure}

Figure~\ref{fig:binary-chronological-benchmark} compares snapshots at
\(cm_0t=10^3,10^4,10^5\) with the absolute prediction of
App.~\ref{app:chronological-snapshot-profile}, with normalization fixed
by the specified root population.
The reference reward is evaluated with eight independent samples of
65,536 stationary clocks. All 48 mass-bin residuals are within 1.6
combined standard errors. At the latest time, the maximum bin discrepancy
is 0.46\%, and the total trace differs by 0.18\%.
Uncertainties use independent batches, retaining sibling and ancestral
correlations; no normalization or exponent is fitted to the profile.

The effective trace exponent increases from 0.17112 over the first decade
to 0.17146 over the second. The latter has a paired-batch bootstrap
95\% interval \([0.17121,0.17170]\), containing
\(q_{\rm IFS}=0.171573\). These checks test the binary chronological prediction at finite time.

\subsection{Mass-spectrum implementation}
\label{app:mass-histogram-methods}
In the plotted units $\widehat M=\sqrt{\GammaSplit}M/(2\muIFS)$,
the density template is
\be
 \frac{d\widehat M_{\rm fit}}{d\ln\gamma}
 =A_{\rm fit}\,
 \frac{\gamma^{q_{\rm IFS}}}{\sqrt{1-\gamma^{-2}}}
 \frac{\mathcal F_\kappa^{\rm occ}(w_\gamma)}
      {\mathcal F_\kappa^{\rm occ}(1)}.
 \label{eq:mass-profile-fit-template}
\ee
The low-boost singularity is integrable; the comparison uses its
integral over each measured bin.

For Fig.~\ref{fig:renewal-mass-profile}, \(h_{ri}\) denotes the
live invariant mass in bin \(i\) of history \(r\), in the plotted
\(\widehat M\) units.  Fit \(\bar h_i=N^{-1}\sum_r h_{ri}\) to
\(af_i\), where
\be
 f_i=\int_{\gamma_i}^{\gamma_{i+1}}
       \gamma\mathcal R_\kappa(\gamma)\,d\ln\gamma.
 \label{eq:chronological-histogram-gls}
\ee
The amplitude in Eq.~\eqref{eq:mass-profile-fit-template} is
\(A_{\rm fit}=2a\mathcal F_\kappa^{\mathrm{occ}}(1)/\mathcal Z_\kappa\).
Only \(a\) is fitted, using Eq.~\eqref{eq:ensemble-quadratic-fit}
and the histogram covariance in App.~\ref{app:ensemble-covariance}.

We evaluate the stationary recursion with eight independently
scrambled Sobol samples of \(2^{18}\) points, truncated at 64 generations.
Rapidity integration removes the endpoint singularity; changing the
scramble shifts the fiducial 50-bin \(\chi^2\) by less than 0.085.
The geometric simulation uses \(\muIFS=\GammaSplit=1\) with terminal
absorption disabled, a generation cap of 50, and at most \(5\times10^6\)
explicit layers per history.  A parent leaves the snapshot at its earliest geometric
daughter-birth time; each daughter enters at its own birth time.
These settings specify the finite-time comparison with the dilute binary
limit of App.~\ref{app:chronological-snapshot-profile}.
Across all 1,000 histories, the maximum realized generation was 38 and the
maximum number of explicit layers was 534,349; neither computational cap was
reached. Bootstrap intervals quantify sampling
uncertainty in this finite-time geometric comparison.

Over $1\leq\gamma\leq10^5$, the fiducial 50-bin fit gives
$\chi^2=39.51$ for 49 nominal degrees of freedom and
$p_{\rm boot}=0.486$. A finer 63-bin comparison, with a half-width final bin, gives $53.81/62$
and $p_{\rm boot}=0.404$. Using the full sample covariance for the
50-bin fit gives $51.63/49$ and $p_{\rm boot}=0.473$. Each comparison
fits one normalization to the same 1,000 histories.

\subsection{Trace and invariant-mass fits}
\label{app:time-domain-fits}

The ensemble contains paired live-trace and invariant-mass data
for 1,000 seeds evolved without terminal absorption to
\(\widehat t=10^6\). The fiducial fit uses 20 logarithmic intervals
from \(10^3\) to \(10^6\).
The same histories enter all time-domain fits. Exact interval
integration makes the convention at the final time immaterial.

\paragraph{Time averages and covariance.}
The time-domain observables are exact integrals of each string's
live contribution against \(d\ln\widehat t\), obtained from its
birth and death times. We pool complete fine intervals with their
logarithmic widths as weights. This uses the whole live history,
including changes between sampled snapshots. The model columns are
therefore exact interval averages:
\be
 f_j(q)=\frac{a_j^{-q}-b_j^{-q}}{q\ln(b_j/a_j)},\qquad
 g_j=\frac{a_j^{-1}-b_j^{-1}}{\ln(b_j/a_j)},
 \label{eq:averaged-time-fit-columns}
\ee
where \([a_j,b_j]\) is the \(j\)th time interval.
The leading and corrected trace models are \(Af_j\) and
\(Af_j+Cg_j\); the mass model is \(M_\infty-Bf_j\).
Concatenating the paired trace and mass intervals gives a 40-coordinate
joint vector, with the full covariance of
App.~\ref{app:ensemble-covariance}.
The mass is fitted directly as $M_\infty-B\widehat t^{-q}$; the figure
displays its deficit from the fitted plateau. The displayed deficit
errors include the plateau uncertainty and its correlation with the
measured mass. All quoted fit uncertainties are statistical.

\paragraph{Amplitudes and the mass--trace relation.}
The separate corrected-trace and mass fits give
$A=4.30\,[4.23,4.37]$, $C=-3.48\,[-5.14,-1.88]$,
$B=12.84\,[11.94,13.67]$, and $M_\infty=13.83\,[13.61,14.02]$
in units $\muIFS=\GammaSplit=1$, with 95\% bootstrap intervals
in brackets. Thus $\widehat M_\infty=6.91$ in dimensionless mass units.
The free plateau and deficit amplitude limit the mass fit's sensitivity
to changes in the exponent.

The joint fit with independent \(A,B,C,M_\infty\) gives
\(\chi^2=41.01\) for 36 nominal degrees of freedom,
\(p_{\rm boot}=0.329\), and
\be
 B/A=3.004,\qquad B/A\in[2.774,3.236]
 \quad(95\%\text{ interval}).
 \label{eq:time-fit-amplitude-ratio}
\ee
Imposing the stationary prediction
\(\mathfrak a_\kappa=3.0774\) leaves \(A,C,M_\infty\) free and
gives the joint row of Table~\ref{tab:direct-time-predictions}.
The increase \(\Delta\chi^2=0.440\) has nested-bootstrap
\(p_\Delta=0.528\).

Adding $C/\widehat t$ lowers $\chi^2$ by 16.46, with
nested-bootstrap $p_\Delta=0.0002$. The fitted correction is negative
and about $0.26\%$ of the leading trace at $\widehat t=10^3$,
decreasing thereafter. It describes this finite window; the
asymptotic subleading power remains undetermined.

\paragraph{Source fluctuations and correlated residuals.}
\label{app:source-fluctuations}
Early remnant fluctuations leave a persistent imprint on the late trace.
For each history, averaging the remnant trace divided by its exact mean
$8D(\widehat t)$ against $d\ln\widehat t$ over $0.1\leq\widehat t\leq10$ gives an onset
coefficient correlated with the last live-trace interval at
$r=0.88$ (95\% paired-bootstrap interval $[0.86,0.90]$).
A nearly constant amplitude mode accounts for $99.7\%$ of the
fractional trace covariance. Consequently the overall level is less
precise than differences between times: marginal error bars can obscure
well-resolved changes of shape, and a covariance-weighted amplitude fit
can have residuals of one sign. All fiducial fits use the ensemble averages with these correlations retained.

\subsection{Cutoff ANEC fits}
\label{app:anec-fit-calibration}

The ANEC data consist of $\mathcal A_r(v_b;U_a)$ on logarithmic grids of
1,000 cutoffs $1\leq U_a\leq10^6$ and 301 ray coordinates
$0.5\leq v_b\leq100$, for the same 1,000 histories as the pressure fit.
All integrals in the analysis range are negative, and the full fitting
window lies within the simulated cutoff range. Each cell uses the tensor
product of the
logarithmic averaging weights in App.~\ref{app:ensemble-covariance},
one for $U$ and one for $v$.
For $\alpha=-(1+q)/2$ and $\beta=(1-q)/2$, the averaged data and
prediction are
\begin{align}
 y_{r,ji}&=-\sum_{a,b}H^{(U)}_{ia}H^{(v)}_{jb}
                         \mathcal A_r(v_b;U_a),\\
 F_{ji}(q)&=
       \left(\sum_aH^{(U)}_{ia}U_a^\alpha\right)
       \left(\sum_bH^{(v)}_{jb}v_b^\beta\right).
 \label{eq:anec-averaged-fit-template}
\end{align}
We fit $C_AF(q)$ by the common full-covariance procedure. We use a
$5\times6$ grid for the fiducial fit; a finer $6\times8$ grid displays the
same histories and checks sensitivity to the binning.
At fixed $q=3-2\sqrt2$, the fiducial fit gives $C_A=0.731$,
$\chi^2=22.99$ for 29 nominal degrees of freedom and
$p_{\rm boot}=0.797$. The finer grid gives $p_{\rm boot}=0.870$.
Changing the starting cutoff between $10^2$ and $10^4$ and varying the
grid leaves the normalization within $0.729$--$0.733$, with compatible
goodness of fit. These checks reuse the same histories.

Tests and confidence intervals use
App.~\ref{app:shared-fit-calibration}, with the fixed-$q$ null for the
nested comparison. The free-$q$ fiducial fit has
$\chi^2=22.18$ for 28 nominal degrees of freedom and
$p_{\rm boot}=0.795$, with the boost relation between $\alpha$ and
$\beta$ imposed. Starting at $U=10^2$ or $10^4$
gives $q=0.17148$ or $0.17163$, with respective 95\% intervals
$[0.17097,0.17196]$ and $[0.17106,0.17218]$.

The pressure--ANEC ratio pairs the ANEC and 20-interval pressure
fits over \(10^3\leq\widehat t\leq10^6\) under this resampling,
using the corrected trace model
$A_W\widehat t^{-q}+C/\widehat t$ to extract its leading amplitude.
Its interval tests the predicted relation between the leading
asymptotic amplitudes.

\subsection{Spatial-moment fits}
\label{app:spatial-moment-fits}
The hierarchy predicts leading powers and limiting ratios; a finite-window
fit also tests whether subleading terms are negligible at the measured
precision. We use complete signed moments from 1,000 unabsorbed histories
with $\muIFS=\GammaSplit=1$, selecting twelve times approximately uniform
in $\ln t$ over $10^3\leq t\leq5\times10^5$.
The 36-coordinate vector $(X_0,X_2^{(W)},X_2^{(E)})$ uses the common
full-covariance and whole-history bootstrap procedure of App.~\ref{app:ensemble-covariance}.
With independent leading amplitudes, the measured ratios are
\be
 \begin{split}
 C_2^{(W)}/C_0^{(W)}&=0.54680\quad[0.54656,0.54703],\\
 C_2^{(E)}/C_0^{(W)}&=-1.32046\quad[-1.32053,-1.32039],
 \end{split}
\ee
with 95\% bootstrap intervals, compared with predictions $0.546918$
and $-1.320377$. The joint fit resolves small departures from these limits because
common amplitude fluctuations cancel in differences between moments
and times.

\paragraph{Finite-window corrections and sensitivity.}
A sensitivity test adds \(D/t\) to the zeroth trace moment.
The exact moment identities then require, in the same units,
\be
 \begin{split}
 M_0^{(W)}&=C_Pt^{-q}+D/t,\\
 M_2^{(W)}&=\frac{C_P}{2-q}t^{2-q}+Dt+B_W,\\
 M_2^{(E)}&=-\frac{2C_P}{(1-q)(2-q)}t^{2-q}
       -2Dt\ln(t/t_{\rm ref})+B_Et+C_E .
 \end{split}
 \label{eq:spatial-moment-correction-test}
\ee
The integration constants allow for the history before the fitting
window; changing \(t_{\rm ref}=10^4\) is absorbed in \(B_E\).
This five-parameter finite-time model respects the exact moment
identities. Its comparison with the leading model gives
\begin{center}
\begin{tabular}{@{}lrrr@{}}
\toprule
Model & $C_P$ (95\% interval) & $\chi^2$/nominal d.o.f. & $p_{\rm boot}$\\
\midrule
Leading & $4.288\ [4.213,4.346]$ & $65.24/35$ & 0.0028\\
With $D/t$ & $4.301\ [4.226,4.360]$ & $43.65/31$ & 0.0862\\
\bottomrule
\end{tabular}
\end{center}
The correction improves the fiducial fit ($p_\Delta=0.00040$), but
freeing the leading ratios improves even this model ($p_\Delta=0.0072$).
Window and time-grid checks likewise leave resolved departures. Thus the
ratios are close to their predicted limits, while the tested correction
does not fully describe their finite-time approach.

The covariance matrix of the fiducial fit has condition number $2.4\times10^9$.
Independent singular-value whitening reproduces its quadratic statistics within $2\times10^{-7}$
without removing modes. We exclude $t=10^6$ because the half-open interval convention does not
include folds terminating exactly at that endpoint, and use
$t\leq t_{\max}/2$ without imposing conservation numerically. Zeroth- and first-energy residuals
are below $5\times10^{-10}$ of $4\muIFS t$ and $4\muIFS t^2$,
respectively. The quoted intervals measure sampling uncertainty;
systematic errors in the signed second moments are not bounded by these
zeroth- and first-moment checks.

\subsection{Crossover sampling and cosmological integration}
\label{app:bounce-numerics}

The sharp-cutoff kernel uses 600,000 independent stationary-clock
samples after 70 contraction steps; the exponential variable
$E_\kappa$ is integrated out analytically. The same samples produce
Fig.~\ref{fig:terminal-crossover} and the response used in the
homogeneous evolution of App.~\ref{app:terminal-numerical-check}.

\paragraph{Youngest cohorts.}
For numerical convolution we extend \(\Psi_{\rm sharp}\) to the origin.
Its derivative behaves as \(T^{-q_{\rm IFS}}\), with \(q_{\rm IFS}<1\),
so the youngest-event integral is finite. This extension is used only for
the convolution; the microscopic early-time behavior is fixed instead by
the exact onset. The remnant starts as \(8\mu t\), while detached
descendants require two splits. Equation~\eqref{eq:exact_direct_source_rate}
gives
\(N_1(t)=\tfrac23\Gamma^2t^4+O(\Gamma^3t^6)\), so the unabsorbed
geometric descendant trace starts at order \(\mu\Gamma^2t^5\),
and its cumulative response at order \(t^6\).
Terminal absorption can suppress this onset further.

We test entrance sensitivity by replacing the cumulative descendant
response below \(T_{*,s}=Q_{*,s}/Q_{\times,s}=\xi_s^{-1}\) with
\(\Psi_{\rm sharp}(T_{*,s})(T/T_{*,s})^p\), for \(p=4,6\),
while preserving it above that scale. The sixth power has the unabsorbed
descendant onset; both choices serve only as matching prescriptions.
The primitive and its first moment are changed consistently in the
convolution. The bounce and motion diagnostics agree to numerical precision for both
replacements and after doubling the sixth-power matching scale. The fiducial
calculation therefore uses the unmodified cumulative kernel together with
the exact remnant response.

\paragraph{Numerical convergence.}
The source is integrated against the
calculated cumulative kernel and its first moment, including the youngest
birth interval. Variable-step BDF2 evolves the background.

With \(x=H_0(t-t_{\rm ESP})\), maximum steps
\(10^{-4},5\times10^{-5},2.5\times10^{-5}\) give convergent results.
The last two runs differ in bounce time by less than \(4\times10^{-8}\)
relatively and in momentum by less than \(5\times10^{-6}\) relatively
on the common comparison interval. Reducing the stationary-clock sample changes the directly evaluated response
quantities by less than \(2\times10^{-4}\) on the response grid. These
comparisons establish numerical convergence for the large-hierarchy binary
response used here. The physical approximations are stated
in App.~\ref{app:terminal-numerical-check}.
\clearpage
\bibliographystyle{ieeetr}
\bibliography{IFS_references}

\begin{thebibliography}{10}

\bibitem{Itzhaki:2018xik}
N.~Itzhaki, ``{Stringy Instability Inside the Black Hole},'' {\em JHEP},
  vol.~10, p.~145, 2018.
\newblock \href{https://arxiv.org/abs/1808.02259}{arXiv:1808.02259}.

\bibitem{Giveon:2019gfk}
A.~Giveon and N.~Itzhaki, ``{Stringy Black Hole Interiors},'' {\em JHEP},
  vol.~11, p.~014, 2019.

\bibitem{Giveon:2019twx}
A.~Giveon and N.~Itzhaki, ``{Stringy Information and Black Holes},'' {\em
  JHEP}, vol.~06, p.~117, 2020.

\bibitem{StringyEREPR}
D.~L. Jafferis and E.~Schneider, ``{Stringy ER=EPR},'' {\em JHEP}, vol.~10,
  p.~195, 2022.
\newblock \href{https://arxiv.org/abs/2104.07233}{arXiv:2104.07233}.

\bibitem{EREPR}
J.~Maldacena and L.~Susskind, ``{Cool horizons for entangled black holes},''
  {\em Fortsch. Phys.}, vol.~61, no.~9, pp.~781--811, 2013.
\newblock \href{https://arxiv.org/abs/1306.0533}{arXiv:1306.0533}.

\bibitem{Hashimoto:2022iqp}
A.~Hashimoto, N.~Itzhaki, and U.~Peleg, ``{A Worldsheet Description of Instant
  Folded Strings},'' {\em JHEP}, vol.~02, p.~088, 2023.
\newblock \href{https://arxiv.org/abs/2209.04988}{arXiv:2209.04988}.

\bibitem{Itzhaki:2024cxh}
N.~Itzhaki and U.~Peleg, ``{When Strings Surprise},'' {\em JHEP}, vol.~08,
  p.~172, 2024.
\newblock \href{https://arxiv.org/abs/2404.03215}{arXiv:2404.03215}.

\bibitem{Giribet:2011tn}
G.~Giribet, ``{On the Timelike Liouville Three-Point Function},'' {\em Phys.
  Rev. D}, vol.~85, p.~086009, 2012.
\newblock \href{https://arxiv.org/abs/1110.6118}{arXiv:1110.6118}.

\bibitem{Witten:1991yr}
E.~Witten, ``{On String Theory and Black Holes},'' {\em Phys. Rev. D}, vol.~44,
  p.~314, 1991.

\bibitem{Elitzur:1990ubs}
S.~Elitzur, A.~Forge, and E.~Rabinovici, ``{Some Global Aspects of String
  Compactifications},'' {\em Nucl. Phys. B}, vol.~359, p.~581, 1991.

\bibitem{Mandal:1991tz}
G.~Mandal, A.~M. Sengupta, and S.~R. Wadia, ``{Classical Solutions of
  Two-Dimensional String Theory},'' {\em Mod. Phys. Lett. A}, vol.~6, p.~1685,
  1991.

\bibitem{Dijkgraaf:1991ba}
R.~Dijkgraaf, H.~Verlinde, and E.~Verlinde, ``{String Propagation in a Black
  Hole Geometry},'' {\em Nucl. Phys. B}, vol.~371, p.~269, 1992.

\bibitem{Itzhaki:2024vaw}
N.~Itzhaki and U.~Peleg, ``{Instant Cosmology},'' {\em JHEP}, vol.~05, p.~026,
  2025.

\bibitem{Itzhaki:2025cyclic}
N.~Itzhaki, U.~Peleg, and P.~J. Steinhardt, ``{Instant Folded Strings, Dark
  Energy and a Cyclic Bouncing Universe},'' {\em JCAP}, vol.~05, p.~084, 2026.

\bibitem{Attali:2018goq}
K.~Attali and N.~Itzhaki, ``{The Averaged Null Energy Condition and the Black
  Hole Interior in String Theory},'' {\em Nucl. Phys. B}, vol.~943, p.~114631,
  2019.
\newblock \href{https://arxiv.org/abs/1811.12117}{arXiv:1811.12117}.

\bibitem{Verch:1999nt}
R.~Verch, ``{The Averaged Null energy condition for general quantum field
  theories in two-dimensions},'' {\em J. Math. Phys.}, vol.~41, pp.~206--217,
  2000.

\bibitem{Hartman:2016lgu}
T.~Hartman, S.~Kundu, and A.~Tajdini, ``{Averaged Null Energy Condition from
  Causality},'' {\em JHEP}, vol.~07, p.~066, 2017.
\newblock \href{https://arxiv.org/abs/1610.05308}{arXiv:1610.05308}.

\bibitem{Faulkner:2016mzt}
T.~Faulkner, R.~G. Leigh, O.~Parrikar, and H.~Wang, ``{Modular Hamiltonians for
  Deformed Half-Spaces and the Averaged Null Energy Condition},'' {\em JHEP},
  vol.~09, p.~038, 2016.
\newblock \href{https://arxiv.org/abs/1605.08072}{arXiv:1605.08072}.

\bibitem{Casini:2008cr}
H.~Casini, ``{Relative Entropy and the Bekenstein Bound},'' {\em Class. Quant.
  Grav.}, vol.~25, p.~205021, 2008.
\newblock \href{https://arxiv.org/abs/0804.2182}{arXiv:0804.2182}.

\bibitem{Bardeen:1975gx}
W.~A. Bardeen, I.~Bars, A.~J. Hanson, and R.~D. Peccei, ``{A Study of the
  Longitudinal Kink Modes of the String},'' {\em Phys. Rev. D}, vol.~13,
  pp.~2364--2382, 1976.

\bibitem{Bardeen:1976yt}
W.~A. Bardeen, I.~Bars, A.~J. Hanson, and R.~D. Peccei, ``{Quantum Poincare
  Covariance of the D = 2 String},'' {\em Phys. Rev. D}, vol.~14, p.~2193,
  1976.

\bibitem{Kofman:2004}
L.~Kofman, A.~Linde, X.~Liu, A.~Maloney, L.~McAllister, and E.~Silverstein,
  ``{Beauty is Attractive: Moduli Trapping at Enhanced Symmetry Points},'' {\em
  JHEP}, vol.~05, p.~030, 2004.

\bibitem{KellyWall:2014}
W.~R. Kelly and A.~C. Wall, ``{Holographic proof of the averaged null energy
  condition},'' {\em Phys. Rev. D}, vol.~90, no.~10, p.~106003, 2014.
\newblock Erratum: Phys. Rev. D 91, 069902 (2015).

\bibitem{Rubakov:2014NEC}
V.~A. Rubakov, ``{The Null Energy Condition and its violation},'' {\em Phys.
  Usp.}, vol.~57, no.~2, pp.~128--142, 2014.

\bibitem{BattefeldPeter:2015}
D.~Battefeld and P.~Peter, ``{A Critical Review of Classical Bouncing
  Cosmologies},'' {\em Phys. Rept.}, vol.~571, pp.~1--66, 2015.

\bibitem{BrandenbergerPeter:2017}
R.~Brandenberger and P.~Peter, ``{Bouncing Cosmologies: Progress and
  Problems},'' {\em Found. Phys.}, vol.~47, pp.~797--850, 2017.

\bibitem{Creminelli:2006NEC}
P.~Creminelli, M.~A. Luty, A.~Nicolis, and L.~Senatore, ``{Starting the
  Universe: Stable Violation of the Null Energy Condition and Non-standard
  Cosmologies},'' {\em JHEP}, vol.~12, p.~080, 2006.

\bibitem{IjjasSteinhardt:2016}
A.~Ijjas and P.~J. Steinhardt, ``{Classically Stable Nonsingular Cosmological
  Bounces},'' {\em Phys. Rev. Lett.}, vol.~117, no.~12, p.~121304, 2016.

\bibitem{IjjasSteinhardt:2017}
A.~Ijjas and P.~J. Steinhardt, ``{Fully stable cosmological solutions with a
  non-singular classical bounce},'' {\em Phys. Lett. B}, vol.~764,
  pp.~289--294, 2017.

\bibitem{Kobayashi:2016Horndeski}
T.~Kobayashi, ``{Generic instabilities of nonsingular cosmologies in Horndeski
  theory: A no-go theorem},'' {\em Phys. Rev. D}, vol.~94, no.~4, p.~043511,
  2016.

\bibitem{Burwig:2026anec}
N.~L. Burwig and D.~A. Easson, ``{Affine ANEC Selects the Closed FRW Branch for
  Geodesically Complete Cosmology},'' 2026.
\newblock \href{https://arxiv.org/abs/2605.18965}{arXiv:2605.18965}.

\bibitem{Graham:2007va}
N.~Graham and K.~D. Olum, ``{Achronal Averaged Null Energy Condition},'' {\em
  Phys. Rev. D}, vol.~76, p.~064001, 2007.
\newblock \href{https://arxiv.org/abs/0705.3193}{arXiv:0705.3193}.

\bibitem{Yurtsever:1995}
U.~Yurtsever, ``{Remarks on the averaged null energy condition in quantum field
  theory},'' {\em Phys. Rev. D}, vol.~52, no.~2, pp.~R564--R568, 1995.

\bibitem{UrbanOlum:2010}
D.~Urban and K.~D. Olum, ``{Averaged Null Energy Condition Violation in a
  Conformally Flat Spacetime},'' {\em Phys. Rev. D}, vol.~81, p.~024039, 2010.

\bibitem{WaldYurtsever:1991}
R.~M. Wald and U.~Yurtsever, ``{General Proof of the Averaged Null Energy
  Condition for a Massless Scalar Field in Two-Dimensional Curved Spacetime},''
  {\em Phys. Rev. D}, vol.~44, pp.~403--416, 1991.

\bibitem{KontouOlum:2013}
E.-A. Kontou and K.~D. Olum, ``{Averaged Null Energy Condition in a Classical
  Curved Background},'' {\em Phys. Rev. D}, vol.~87, p.~064009, 2013.

\bibitem{Kontou:2015mkt}
E.-A. Kontou and K.~D. Olum, ``{Proof of the Averaged Null Energy Condition in
  a Classical Curved Spacetime Using a Null-Projected Quantum Inequality},''
  {\em Phys. Rev. D}, vol.~92, p.~124009, 2015.
\newblock \href{https://arxiv.org/abs/1507.00297}{arXiv:1507.00297}.

\bibitem{Wall:2010}
A.~C. Wall, ``{Proving the Achronal Averaged Null Energy Condition from the
  Generalized Second Law},'' {\em Phys. Rev. D}, vol.~81, p.~024038, 2010.

\bibitem{KontouSanders:2020}
E.-A. Kontou and K.~Sanders, ``{Energy Conditions in General Relativity and
  Quantum Field Theory},'' {\em Class. Quant. Grav.}, vol.~37, no.~19,
  p.~193001, 2020.

\bibitem{Giveon:2020xxh}
A.~Giveon, N.~Itzhaki, and U.~Peleg, ``{Instant Folded Strings and Black
  Fivebranes},'' {\em JHEP}, vol.~08, p.~020, 2020.

\bibitem{Page:1993wv}
D.~N. Page, ``{Information in black hole radiation},'' {\em Phys. Rev. Lett.},
  vol.~71, pp.~3743--3746, 1993.

\bibitem{Itzhaki2026}
N.~Itzhaki, ``Wave-function collapse in string theory,'' {\em Physical Review
  D}, vol.~114, p.~046013, Aug. 2026.

\bibitem{Diosi:1989dp}
L.~Di{\'o}si, ``Models for universal reduction of macroscopic quantum
  fluctuations,'' {\em Phys. Rev. A}, vol.~40, pp.~1165--1174, 1989.

\bibitem{Penrose:1996cv}
R.~Penrose, ``{On Gravity's role in Quantum State Reduction},'' {\em Gen. Rel.
  Grav.}, vol.~28, pp.~581--600, 1996.

\bibitem{Debruyne_2019}
G.~Debruyne and J.~Vindas, ``{Complex Tauberian theorems for Laplace transforms
  with local pseudofunction boundary behavior},'' {\em J. Anal. Math.},
  vol.~138, pp.~799--833, 2019.
\newblock \href{https://arxiv.org/abs/1604.05069}{arXiv:1604.05069}.

\bibitem{MartinVilenkin:1997radiation}
X.~Martin and A.~Vilenkin, ``{Gravitational Radiation from Monopoles Connected
  by Strings},'' {\em Phys. Rev. D}, vol.~55, pp.~6054--6060, 1997.
\newblock \href{https://arxiv.org/abs/gr-qc/9612008}{arXiv:gr-qc/9612008}.

\bibitem{Iengo:2006radiation}
R.~Iengo, ``{Massless Radiation from Strings: Quantum Spectrum Average
  Statistics and Cusp-Kink Configurations},'' {\em JHEP}, vol.~05, p.~054,
  2006.
\newblock \href{https://arxiv.org/abs/hep-th/0602125}{arXiv:hep-th/0602125}.

\bibitem{Iengo:2006nn}
R.~Iengo and J.~G. Russo, ``{Handbook on String Decay},'' {\em JHEP}, vol.~02,
  p.~041, 2006.
\newblock \href{https://arxiv.org/abs/hep-th/0601072}{arXiv:hep-th/0601072}.

\end{thebibliography}


\begin{thebibliography}{10}

\bibitem{Itzhaki:2018xik}
N.~Itzhaki, ``{Stringy Instability Inside the Black Hole},'' {\em JHEP},
  vol.~10, p.~145, 2018.
\newblock \href{https://arxiv.org/abs/1808.02259}{arXiv:1808.02259}.

\bibitem{Hashimoto:2022iqp}
A.~Hashimoto, N.~Itzhaki, and U.~Peleg, ``{A Worldsheet Description of Instant
  Folded Strings},'' {\em JHEP}, vol.~02, p.~088, 2023.
\newblock \href{https://arxiv.org/abs/2209.04988}{arXiv:2209.04988};
\href{https://doi.org/10.1007/JHEP02(2023)088}{doi:10.1007/JHEP02(2023)088}.

\bibitem{Itzhaki:2024cxh}
N.~Itzhaki and U.~Peleg, ``{When Strings Surprise},'' {\em JHEP}, vol.~08,
  p.~172, 2024.
\newblock \href{https://arxiv.org/abs/2404.03215}{arXiv:2404.03215}.

\bibitem{Giribet:2011tn}
G.~Giribet, ``{On the Timelike Liouville Three-Point Function},'' {\em Phys.
  Rev. D}, vol.~85, p.~086009, 2012.
\newblock \href{https://arxiv.org/abs/1110.6118}{arXiv:1110.6118};
\href{https://doi.org/10.1103/PhysRevD.85.086009}{doi:10.1103/PhysRevD.85.086009}.

\bibitem{Witten:1991yr}
E.~Witten, ``{On String Theory and Black Holes},'' {\em Phys. Rev. D}, vol.~44,
  pp.~314--324, 1991.
\newblock \href{https://doi.org/10.1103/PhysRevD.44.314}{doi:10.1103/PhysRevD.44.314}.

\bibitem{Elitzur:1990ubs}
S.~Elitzur, A.~Forge, and E.~Rabinovici, ``{Some Global Aspects of String
  Compactifications},'' {\em Nucl. Phys. B}, vol.~359, pp.~581--610, 1991.
\newblock \href{https://doi.org/10.1016/0550-3213(91)90073-7}{doi:10.1016/0550-3213(91)90073-7}.

\bibitem{Mandal:1991tz}
G.~Mandal, A.~M. Sengupta, and S.~R. Wadia, ``{Classical Solutions of
  Two-Dimensional String Theory},'' {\em Mod. Phys. Lett. A}, vol.~6,
  pp.~1685--1692, 1991.
\newblock \href{https://doi.org/10.1142/S0217732391001822}{doi:10.1142/S0217732391001822}.

\bibitem{Dijkgraaf:1991ba}
R.~Dijkgraaf, H.~Verlinde, and E.~Verlinde, ``{String Propagation in a Black
  Hole Geometry},'' {\em Nucl. Phys. B}, vol.~371, pp.~269--314, 1992.
\newblock \href{https://doi.org/10.1016/0550-3213(92)90237-6}{doi:10.1016/0550-3213(92)90237-6}.

\bibitem{Itzhaki:2024vaw}
N.~Itzhaki and U.~Peleg, ``{Instant Cosmology},'' {\em JHEP}, vol.~05, p.~026,
  2025.
\newblock \href{https://arxiv.org/abs/2412.02630}{arXiv:2412.02630}.

\bibitem{Itzhaki:2025cyclic}
N.~Itzhaki, U.~Peleg, and P.~J. Steinhardt, ``{Instant Folded Strings, Dark
  Energy and a Cyclic Bouncing Universe},'' {\em JCAP}, vol.~05, p.~084, 2026.
\newblock \href{https://arxiv.org/abs/2508.09745}{arXiv:2508.09745};
\href{https://doi.org/10.1088/1475-7516/2026/05/084}{doi:10.1088/1475-7516/2026/05/084}.

\bibitem{Attali:2018goq}
K.~Attali and N.~Itzhaki, ``{The Averaged Null Energy Condition and the Black
  Hole Interior in String Theory},'' {\em Nucl. Phys. B}, vol.~943, p.~114631,
  2019.
\newblock \href{https://arxiv.org/abs/1811.12117}{arXiv:1811.12117};
\href{https://doi.org/10.1016/j.nuclphysb.2019.114631}{doi:10.1016/j.nuclphysb.2019.114631}.

\bibitem{Verch:1999nt}
R.~Verch, ``{The Averaged Null Energy Condition for General Quantum Field
  Theories in Two Dimensions},'' {\em J. Math. Phys.}, vol.~41, pp.~206--217,
  2000.
\newblock \href{https://arxiv.org/abs/math-ph/9904036}{arXiv:math-ph/9904036};
\href{https://doi.org/10.1063/1.533130}{doi:10.1063/1.533130}.

\bibitem{Hartman:2016lgu}
T.~Hartman, S.~Kundu, and A.~Tajdini, ``{Averaged Null Energy Condition from
  Causality},'' {\em JHEP}, vol.~07, p.~066, 2017.
\newblock \href{https://arxiv.org/abs/1610.05308}{arXiv:1610.05308};
\href{https://doi.org/10.1007/JHEP07(2017)066}{doi:10.1007/JHEP07(2017)066}.

\bibitem{Casini:2008cr}
H.~Casini, ``{Relative Entropy and the Bekenstein Bound},'' {\em Class. Quant.
  Grav.}, vol.~25, p.~205021, 2008.
\newblock \href{https://arxiv.org/abs/0804.2182}{arXiv:0804.2182};
\href{https://doi.org/10.1088/0264-9381/25/20/205021}{doi:10.1088/0264-9381/25/20/205021}.

\bibitem{Faulkner:2016mzt}
T.~Faulkner, R.~G. Leigh, O.~Parrikar, and H.~Wang, ``{Modular Hamiltonians for
  Deformed Half-Spaces and the Averaged Null Energy Condition},'' {\em JHEP},
  vol.~09, p.~038, 2016.
\newblock \href{https://arxiv.org/abs/1605.08072}{arXiv:1605.08072};
\href{https://doi.org/10.1007/JHEP09(2016)038}{doi:10.1007/JHEP09(2016)038}.

\bibitem{Bardeen:1975gx}
W.~A. Bardeen, I.~Bars, A.~J. Hanson, and R.~D. Peccei, ``{A Study of the
  Longitudinal Kink Modes of the String},'' {\em Phys. Rev. D}, vol.~13,
  pp.~2364--2382, 1976.
\newblock \href{https://doi.org/10.1103/PhysRevD.13.2364}{doi:10.1103/PhysRevD.13.2364}.



\bibitem{Bardeen:1976yt}
W.~A.~Bardeen, I.~Bars, A.~J.~Hanson and R.~D.~Peccei,
``Quantum Poincare Covariance of the D = 2 String,''
Phys. Rev. D \textbf{14} (1976), 2193
doi:10.1103/PhysRevD.14.2193


\bibitem{Burwig:2026anec}
N.~L. Burwig and D.~A. Easson, ``{Affine ANEC Selects the Closed FRW Branch for
  Geodesically Complete Cosmology},''
\href{https://arxiv.org/abs/2605.18965}{arXiv:2605.18965 [gr-qc]} (2026).

\bibitem{Graham:2007va}
N.~Graham and K.~D. Olum, ``{Achronal Averaged Null Energy Condition},''
  {\em Phys. Rev. D}, vol.~76, p.~064001, 2007.
\newblock \href{https://arxiv.org/abs/0705.3193}{arXiv:0705.3193};
\href{https://doi.org/10.1103/PhysRevD.76.064001}{doi:10.1103/PhysRevD.76.064001}.

\bibitem{Giveon:2019gfk}
A.~Giveon and N.~Itzhaki, ``{Stringy Black Hole Interiors},'' {\em JHEP},
  vol.~11, p.~014, 2019.
\newblock \href{https://arxiv.org/abs/1908.05000}{arXiv:1908.05000};
\href{https://doi.org/10.1007/JHEP11(2019)014}{doi:10.1007/JHEP11(2019)014}.

\bibitem{Giveon:2019twx}
A.~Giveon and N.~Itzhaki, ``{Stringy Information and Black Holes},'' {\em
  JHEP}, vol.~06, p.~117, 2020.
\newblock \href{https://arxiv.org/abs/1912.06538}{arXiv:1912.06538};
\href{https://doi.org/10.1007/JHEP06(2020)117}{doi:10.1007/JHEP06(2020)117}.

\bibitem{Giveon:2020xxh}
A.~Giveon, N.~Itzhaki, and U.~Peleg, ``{Instant Folded Strings and Black
  Fivebranes},'' {\em JHEP}, vol.~08, p.~020, 2020.
\newblock \href{https://arxiv.org/abs/2004.06143}{arXiv:2004.06143};
\href{https://doi.org/10.1007/JHEP08(2020)020}{doi:10.1007/JHEP08(2020)020}.

\bibitem{Itzhaki2026}
N.~Itzhaki, ``Wave-function collapse in string theory,'' {\em Phys. Rev. D},
  vol.~114, no.~4, p.~046013, 2026.
\newblock \href{https://doi.org/10.1103/lyyp-3svl}{doi:10.1103/lyyp-3svl}.

\bibitem{Debruyne_2019}
G.~Debruyne and J.~Vindas, ``{Complex Tauberian Theorems for Laplace Transforms
  with Local Pseudofunction Boundary Behavior},'' {\em J. Anal. Math.},
  vol.~138, no.~2, pp.~799--833, 2019.
\newblock \href{https://arxiv.org/abs/1604.05069}{arXiv:1604.05069};
\href{https://doi.org/10.1007/s11854-019-0045-3}{doi:10.1007/s11854-019-0045-3}.

\bibitem{MartinVilenkin:1997radiation}
X.~Martin and A.~Vilenkin, ``{Gravitational Radiation from Monopoles Connected
  by Strings},'' {\em Phys. Rev. D}, vol.~55, pp.~6054--6060, 1997.
\newblock \href{https://arxiv.org/abs/gr-qc/9612008}{arXiv:gr-qc/9612008};
\href{https://doi.org/10.1103/PhysRevD.55.6054}{doi:10.1103/PhysRevD.55.6054}.

\bibitem{Iengo:2006radiation}
R.~Iengo, ``{Massless Radiation from Strings: Quantum Spectrum Average
  Statistics and Cusp-Kink Configurations},'' {\em JHEP}, vol.~05, p.~054,
  2006.
\newblock \href{https://arxiv.org/abs/hep-th/0602125}{arXiv:hep-th/0602125};
\href{https://doi.org/10.1088/1126-6708/2006/05/054}{doi:10.1088/1126-6708/2006/05/054}.

\bibitem{Iengo:2006nn}
R.~Iengo and J.~G. Russo, ``{Handbook on String Decay},'' {\em JHEP}, vol.~02,
  p.~041, 2006.
\newblock \href{https://arxiv.org/abs/hep-th/0601072}{arXiv:hep-th/0601072};
\href{https://doi.org/10.1088/1126-6708/2006/02/041}{doi:10.1088/1126-6708/2006/02/041}.





\end{thebibliography}
\end{document}